\documentclass[11pt]{article}

\usepackage{natbib}

\usepackage{amsmath}
\usepackage{amssymb}
\usepackage{amsthm}
\usepackage{bbm}
\usepackage{bm}

\usepackage{setspace}

\usepackage{graphicx}

\usepackage{apalike}
\usepackage{caption}
\usepackage{subcaption}
\usepackage[left=25.4mm,right=25.4mm,top=38.1mm,bottom=38.1mm,a4paper]{geometry}
\usepackage{booktabs}

\usepackage{color,soul}

\usepackage{hyperref}
\newcommand{\footremember}[2]{
	\footnote{#2}
	\newcounter{#1}
	\setcounter{#1}{\value{footnote}}
}

\theoremstyle{definition}

\newtheorem{definition}{Definition}

\newcommand{\E}{\mathbb E}
\newcommand{\e}{\mathrm e}

\newcommand{\D}{\mathrm{d}}
\newcommand{\F}{\mathcal F}

\newcommand{\PP}{\mathbb P}

\begin{document}
	
\title{Modeling bid and ask price dynamics with an extended Hawkes process and its empirical applications for high-frequency stock market data}


\author{
	Kyungsub Lee\footremember{KLee}{ksublee@yu.ac.kr, Department of Statistics, Yeungnam University, Gyeongsan, Gyeongbuk 38541, Korea.}
	\and Byoung Ki Seo\footremember{BKSeo}{(corresponding author) bkseo@unist.ac.kr, School of Business Administration, Ulsan National Institute of Science and Technology, Ulsan 44919, Korea.}
}
\date{}

\maketitle

\begin{abstract}
This study proposes a versatile model for the dynamics of the best bid and ask prices using an extended Hawkes process. The model incorporates the zero intensities of the spread-narrowing processes at the minimum bid-ask spread, spread-dependent intensities, possible negative excitement, and nonnegative intensities. We apply the model to high-frequency best bid and ask price data from US stock markets. The empirical findings demonstrate a spread-narrowing tendency, excitations of the intensities caused by previous events, the impact of flash crashes, characteristic trends in fast trading over time, and the different features of market participants in the various exchanges.
\end{abstract}

\section{Introduction}

A considerable volume of research on high-frequency trading, quotes, and financial data is available.
High-frequency quotes and trading, which are the main sources of high-frequency financial data, are considered to be (automated) trading generally based on a (mathematical) algorithm that can generate a large number of quotes and trades over a short time horizon. 
However, the precise definition varies across studies and regulatory entities.
Numerous studies consider the impact of high-frequency trading on market quality in terms of liquidity, transaction costs, volatility, price discovery, social benefits, and its role in market crashes. 

Studies of market microstructures and high-frequency data rely on statistical models that can precisely analyze market data such as order duration, transaction time, and price changes.
One effective method is based on the Hawkes process \citep{Hawkes1,Hawkes2,Hawkes&Oakes}, 
a type of point process for estimating the parameters related to the impacts of events.
Researchers first applied Hawkes models to seismology, but presently apply them extensively throughout the natural and social sciences, including finance.
For example, the earliest studies applying the Hawkes process to financial markets include \cite{Hewlett2006}, who introduces a symmetric bivariate Hawkes process to examine the arrival times of trades and price impacts; \cite{Large2007}, who examines market resilience; and
\cite{Bowsher2007}, who analyzes the relationship between trading times and mid-price changes.
Since then, the model has often been applied to financial data. 
\cite{Barcry2015}, \cite{Law2015}, and \cite{Hawkes2018} summarize recent developments in the application of the Hawkes process to finance.

The basic Hawkes model captures several interesting properties of price dynamics at the micro level \citep{Chavez2012, Bacry2014,Fonseca2014} 
such as order clustering, microstructure noise, and the impacts of past events.
The Hawkes process is suitable for modeling the intensities of a series of events
such as price changes in tick structure and order arrivals.
In addition, other studies aim to clarify the complex microstructure of the limit order book and price dynamics using extended models \citep{Zheng2014,morariu2018state,Hainaut2019,Jang2020}.
Our approach is also based on modifying the existing Hawkes model to allow applications to bid and ask price dynamics.
Owing to its versatility, this extended Hawkes process allows us to infer useful information from high-frequency financial data
and understand the market microstructure and high-frequency activities.

Specifically, we develop a rigorous and novel probability model that describes the nature of bid-ask price dynamics.
The proposed model incorporates characteristics that cannot be described by the basic Hawkes model.
Bid-ask spreads are positive and our model guarantees this property.
If the spread widens, competition among market makers to replenish liquidity increases under typical market conditions
and thus the proposed model has spread-level-dependent intensities.

In stock market microstructures, 
a bid or ask price change is expected to increase the likelihood of events such as new order arrivals and successive price changes occurring.
The basic Hawkes model can describe these excitation properties.
However, if price changes occur such that spreads decline, intensity may not increase.
For example, if the spread reduces to a tick size (i.e., it reaches the minimum level), then no further events that reduce the spread occur; 
that is, the associated intensities that narrow the spread become zeros.
Since the basic Hawkes model always has positive intensities, it needs to be modified using, for example, negative excitements.
To add a negative excitement term into the model, the model design must ensure that intensity is nonnegative to satisfy mathematical rigor.

Related ideas can be found in the past literature.
An extended Hawkes process that partially describes non-exciting relationships was proposed in \cite{Bremaud1996}.
Along that line, \cite{Hansen2015} considered multivariate counting processes with inhibitory effects in a non-linear setting whose importance is also mentioned in \cite{Bacry2016}.

We apply this model to high-frequency financial data and make the following interesting findings.
First, we find evidence of a growing number of high-frequency traders even though not all stocks show the same monotone increasing growth pattern over time.
Second, liquidity provision and removal behaviors have different responsiveness.
In all the stocks in our sample, we find higher instantaneous responsiveness in liquidity provision than in removal.
Third, we also observe different features from different exchanges under market-making activity.
Being a highly fragmented market, stock market exchanges have different fee structures, policies, and technology, which leads their participants to have different characteristics.

We also report the dramatic changes in market conditions under special circumstances such as the Flash Crash of 2010. 
Although we perform most of our analyses on a daily basis, because of the nature of high-frequency financial data, sufficient data are available for a few minutes, thereby allowing us to conduct the analysis on a near real-time basis.
\cite{lee2017marked} also show this property.
Using sophisticatedly designed parameters, we can thus observe how the behaviors of high-frequency traders change during the crash, recovery period, and next day.

The remainder of the paper proceeds as follows.
Section~\ref{Sec:lit} reviews the related literature.
Section~\ref{Sec:model} presents the extended Hawkes model of the dynamics of the best bid and ask prices.
The proposed model incorporates the zero intensities of the spread-narrowing processes at the minimum bid-ask spread, spread-dependent intensities, possible negative excitement, and nonnegative intensities.
Section~\ref{Sec:empirical} discusses the findings from our empirical study performed by applying the presented model to high-frequency financial data.
The results demonstrate increasing trends in low-latency quotes and trading, the characteristics of stocks and exchanges, a comparison between liquidity provision and removal, and the interesting behaviors of high-frequency traders during and after the Flash Crash.
Section~\ref{Sec:concl} concludes.

\section{Literature review}~\label{Sec:lit}

This section reviews studies analyzing high-frequency trading and high-frequency financial data.
The main advantage of high-frequency trading, especially quotes, is the supply of liquidity.
Many studies demonstrate that the liquidity provided by active high-frequency quotes and large numbers of outstanding orders reduces the bid-ask spread and that transaction costs decrease accordingly \citep{Bershova2013,CHABOUD2014,stoll2014}.
In addition, \citet{hagstromer2013} and \citet{Menkveld2013} report that many high-frequency traders are liquidity providers.
However, a proportion of this liquidity is an illusion in the abundance of fleeting orders \citep{hasbrouck2009technology}, and some researchers express skepticism about whether it is sustainable during a market crisis \citep{Cvitanic2010},
even though high-frequency trading is not generally considered to be the primary originator of the Flash Crash~\citep{Kirilenko2017}.

The rapidity of high-frequency trading is based on high-end, state-of-the-art computing, network equipment, and costly financial services such as colocation, which involves installing traders' servers as close to trade matching engines as possible.
However, while this rapidity can help provide liquidity \citep{brogaard2015trading}, high-frequency traders armed with new equipment may take advantage of low-frequency traders \citep{Jarrow2012, Kervel2015, Biais2015}.
Theoretical and empirical research thus suggests that the rapidity of high-frequency trading can be harmful as well as beneficial \citep{Foucault2018}.
The general assumption is that the speed of high-frequency trading is a positive aspect of price discovery as it allows the market to rapidly process new information and reflect it in new prices~\citep{Manahov2014}.
However, whether information processing and quote updates that occur just a few seconds or milliseconds earlier are of practical utility remains an open question \citep{Brogaard2014}.

Because volatility is an important indicator of the market conditions, measuring the volatility of prices or returns more accurately using high-frequency financial data has long been a topic of study \citep{Zhou1996,Barndorff2002a,ABDL}.
In addition, the effect of high-frequency trading on volatility is under debate \citep{Bollen2015, Kelejian2016, hasbrouck2018}.
In some ways, high-frequency activities such as the exploitation of arbitrage opportunities in high-frequency trading, crowding effect, flickering quotes, and high order cancellation rates are sources of higher volatility, 
but some studies find no significant statistical relations between high-frequency activities and volatility.
\citet{Chung2016}, \citet{Menkveld2016}, and \citet{Virgilio2019} provide excellent summaries of the arguments in the literature above.

The research interest in high-frequency trading is not limited to the above discussion; it extends to the optimal strategies of market makers and investors based on high-frequency trading \citep{Avellaneda2008, Guilbaud2013, Cartea2013, Cartea2018, choi2021}
and limit order book modeling \citep{Cont2013, Toke2015}.
\citet{bouchaud2002statistical}, \citet{hollifield2004empirical}, and \citet{huang2015simulating} describe the statistical properties of limit order books.

\section{The model}~\label{Sec:model}
\subsection{Definition}\label{subsec:def}
In this section, we model the best bid and ask price dynamics of financial assets using point processes.
The best bid and ask price processes consist of
\begin{align*}
&A^u \textrm{: the number of upward movements in the best ask process,} \\
&A^d \textrm{: the number of downward movements in the best ask process,}\\
&B^u \textrm{: the number of upward movements in the best bid process, and}\\
&B^d \textrm{: the number of downward movements in the best bid process,}
\end{align*}
defined on the probability space $(\Omega, \mathcal{F},\{\mathcal{F}_{t}\}_{t\geq0},\mathbb P)$ satisfying the usual conditions.
Each process is a random counting measure that counts the number of events in any Borel set in the timeline, $\mathbb{R}$.
We can represent each process using a stochastic process (i.e., nondecreasing integer-valued random step functions) by denoting $A^u (t) = A^u (\omega, (0, t])$ for $\omega \in \Omega$ and 
similarly denoting the other counting measures.
We assume the point processes are simple, i.e.,
$$\PP \{N(\{t\}) = 0 \textrm{ or } 1 \textrm{ for all } t  \} = 1$$
for $N = A^u, A^d, B^u$ or $B^d$.
We can categorize the counting processes further as
\begin{align*}
&A^u, B^d \textrm{ : the bid-ask spread-widening processes}\\
&A^d, B^u \textrm{ : the bid-ask spread-narrowing processes}.
\end{align*}

We establish a system of counting processes with the corresponding conditional intensities:
$$ 
\bm{N}_t =
\begin{bmatrix} 
N_1(t) \\
N_2(t) \\
N_3(t) \\
N_4(t) 
\end{bmatrix}
=
\begin{bmatrix} 
A^{u}(t) \\
A^{d}(t) \\
B^{u}(t) \\
B^{d}(t) 
\end{bmatrix}, \quad
\bm{\lambda}_t = 
\begin{bmatrix} 
\lambda_1(t) \\
\lambda_2(t) \\
\lambda_3(t) \\
\lambda_4(t)
\end{bmatrix}
= 
\begin{bmatrix} 
\lambda_A^{u}(t) \\
\lambda_A^{d}(t) \\
\lambda_B^{u}(t) \\
\lambda_B^{d}(t)
\end{bmatrix}.
$$ 
We can express the conditional intensities as the expected number of events over an infinitesimal interval $\D t$:
$$ \lambda_i(t) \D t \approx \E [ N_i(t + \D t) - N_i(t) | \F_{t-}], $$
where $\F_{t-}$ is the $\sigma$-field generated by $\bm{N}$ at times up to but not including $t$. 
Thus, the intensities are left continuous and the counting processes are right continuous.
We have two types of notations, number-based, such as $N_i$ and $\lambda_i$, and Roman-based, $A^u$ and $\lambda_A^u$.
We use them interchangeably depending on the context.

The following model is used in our main empirical study in Section~\ref{Sec:empirical}.
Slightly different versions of the model are also examined in Section~\ref{Sec:math}.
The motivation and property of the model is further explained in Subsection~\ref{subsec:discuss}.

\begin{definition}\label{Def:model}
	The intensities of the up and down movements of the best bid and ask prices are assumed to be
	\begin{equation}
	\bm{\lambda}_t = \bm{\mu}_t + \int_{-\infty}^{t} \bm{h}(t, u) \D \bm{N}_u, \label{eq:intensity}
	\end{equation}
	where
	$$
	\bm{\mu}_t = 
	\begin{bmatrix} 
	\mu \\
	f(\ell(t-)) \\
	f(\ell(t-)) \\
	\mu
	\end{bmatrix},
	$$
	where $\ell$ denotes the level of the bid-ask spread relative to the mid-price and
	\begin{equation}
	f(\ell(t-)) = \eta \ell(t-) \label{eq:linear_f}
	\end{equation}
	with a constant parameter, $\eta > 0$. 
	The relative level is defined by
	$$ \ell(t) = \frac{L(t)}{p(t)}, $$
	where $p(t)$ is the mid-price
	and $L(t) \in \{ 0, 1, 2, \cdots \}$ is the absolute level of the bid-ask spread with $L(t)=0$ implying the minimum level.
	The matrix of excitement terms is based on an exponential kernel:
	$$
	\bm{h}(t, u) =
	\e^{-\beta(t-u)}
	\begin{bmatrix}
	\alpha_{s1} & \alpha_{m} & \alpha_{s2} & 0 \\
	\alpha_{w1} & \alpha_{n1}(u) & \alpha_{n1}(u) & \alpha_{w2} \\
	\alpha_{w2} & \alpha_{n2}(u) & \alpha_{n2}(u) & \alpha_{w1} \\
	0 & \alpha_{s2} & \alpha_{m} & \alpha_{s1} \\
	\end{bmatrix},
	$$
	where $\alpha_{s1}, \alpha_{s2}, \alpha_m, \alpha_{w1},$ and $\alpha_{w2}$ are parameters under the condition that
	\begin{equation}
	\alpha_{s1} + \alpha_{s2} + \alpha_m < \beta \label{eq:condition}
	\end{equation}
	and $\alpha_{n1}$ and $\alpha_{n2}$ are stochastic terms with
	\begin{align}
	\alpha_{n1}(u) = - \sum_{j=1}^4 \lambda_{2j}(u) + \xi \ell(u), \quad \alpha_{n2}(u) = - \sum_{j=1}^4 \lambda_{3j}(u) + \xi \ell(u), \label{eq:linear_an}
	\end{align}
	for constant $\xi \geq 0$ and $\lambda_{ij}$ is a component of $\lambda_i$ such that
 	$$ \lambda_{ij}(t) = \int_{-\infty}^t h_{ij}(t, u) \D N_j(u).$$
	
\end{definition}

We use relative level rather than absolute level in the modeling of $\bm{\mu}$ and $\bm{h}$ for empirical research.
In the US stock market, the tick size is fixed at \$0.01, while the price of a stock varies from stock to stock.
However, for mathematical analysis, the absolute level model is more tractable, which will be discussed in Subsection~\ref{subsec:spread}.

In the definition, we use the left continuous version of $\ell(t-)$ to define $\bm{\mu}$ to ensure that the intensity process is left continuous with a right limit.
The base intensities of the spread-narrowing processes, $A^d$ and $B^u$, increase as the spread increases.
When $\ell(t-) = 0$ (i.e., the spread is at its minimum), the base intensities of $A^d$ and $B^u$ are equal to zero.

The stochastic terms $\alpha_{n1}$ and $\alpha_{n2}$ control the amounts of changes in the spread-narrowing intensities right after the spread narrows.
We introduce these terms as stochastic to ensure mathematical rigor which will be discussed in the next subsection.
We cannot rule out the possibility that $\xi$ has different values for $\alpha_{n1}$ and $\alpha_ {n2}$ in practice, 
but we assume that $\xi$ takes the same value in both for model parsimony.

The coefficients in the top right and left bottom corners of the matrix $\bm{h}$ are intentionally set equal to zero for model parsimony, since the interaction between upward movement in the best ask and downward movement in the best bid can negligible especially in ultra-high-frequency.
The empirical evidence will be discussed in Subsection~\ref{subsec:selection}.

\begin{figure}
	\centering
	\includegraphics[width=\textwidth]{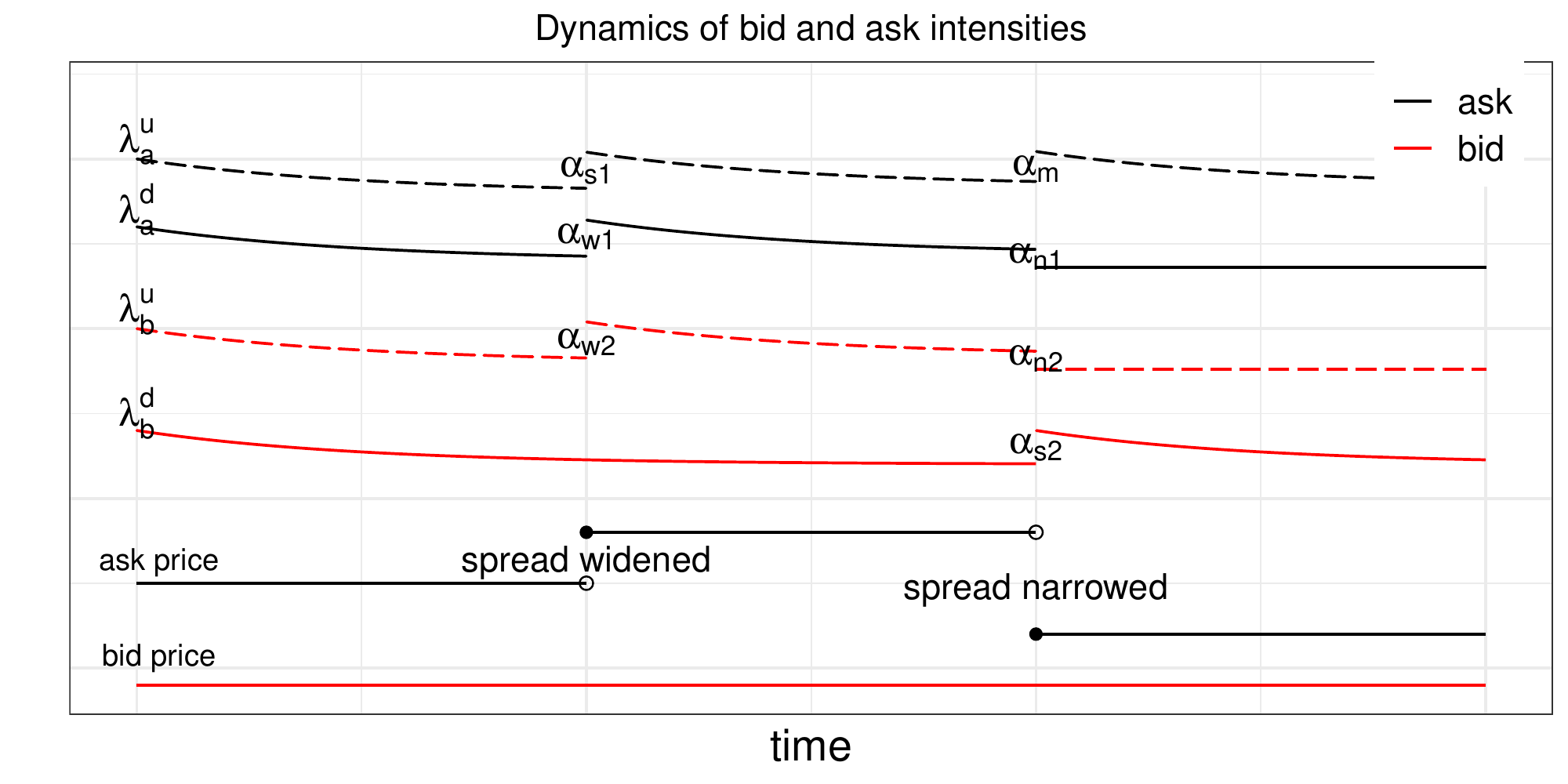}
	\caption{Illustration of the intensity processes in the proposed model}\label{Fig:intensities}
\end{figure}

An example of intensity changes by the events under the model in Definition~{\ref{Def:model}} is shown in Figure~{\ref{Fig:intensities}}.
An hypothetical situation where the bid-ask spread widens first and then decreases to the minimum level is illustrated.
Note that the spread-narrowing intensities $\lambda_b^u$ and $\lambda_a^d$ become zeros right after the spread hits the minimum.

\subsection{Discussion of the model}~\label{subsec:discuss}

Previous studies often assume that the mid-price process follows a self and mutually excited Hawkes process, which has empirical support \citep{Bacryetal2013,Fonseca2014, lee2017modeling}.
Self-excitement means that upward or downward movements of mid-prices are excited by past events in their own processes and are often considered to be due to the order splitting. 
Mutual excitement means that downward movement excites upward movement and vice versa.
Because the mid-price is the mean of the best bid and ask prices,
it is natural to assume that we can use a Hawkes-type model to represent the bid and ask price processes in this context.
The intensities in the Hawkes model consist of the base intensities and excitations by past events.
The typical intensity process in the Hawkes model takes the form of Eq.~\eqref{eq:intensity}
with constant base intensity $\bm{\mu}$ and constant excitation kernel $\bm{h}$.

Our goal of modeling is to find the latent information in the level-one limit order book by modeling the base intensities and excitation matrix appropriately.
To define a model as in Definition~\ref{Def:model}, the followings are considered.
The model is proposed to incorporate the following properties in addition to Hawkes-type features in the mid-price process.

\bigskip\noindent
{\it Zero intensity with minimum spread}

When the spread between the best bid and ask prices reaches the minimum tick size,
the ask price does not fall and the bid price does not rise before the spread widens. 
Mathematically, the intensities of the spread-narrowing processes, $A^{d}(t)$ and $B^{u}(t)$, should be zero at the minimum level of the bid-ask spread until a new event occurs.
Thus, $\lambda_A^{d}(t)$ and $\lambda_B^{u}(t)$ do not follow the intensities of the classical Hawkes model
in which the intensities can be close to zero, but are always greater than zero.
Hence, it is necessary to tweak the model to set both the background intensities and the excitations equal to zero when the spread reaches the minimum.

\bigskip\noindent
{\it Spread-dependent intensity}

When the bid-ask spread is wide, we assume that it tends to narrow depending on its level.
This property may be due to market makers who quote limit orders with aggressive prices between the widened bid and ask orders, providing market liquidity.
As the spread increases, the potential profit for market makers increases, and market making hence becomes more active.
We can observe this tendency in two ways. 
One is the increase in the base intensities, regardless of previous events, and the other comes from the excitations due to past events.
Our model includes both aspects.
The base intensity rates of the spread-narrowing processes, $A^{d}$ and $B^{u}$, increase as the spread widens.
In addition, the overall intensities of the processes are excited immediately after the spread-widening events by the Hawkes-type kernel.

\bigskip\noindent
{\it Possible negative excitement when the spread narrows}

If spread-narrowing events occur, we can assume that these activities can slow the rate of the additional arrivals of aggressive quotes. 
In other words, the intensities of the spread-narrowing processes, $A^{d}(t)$ and $B^{u}(t)$, may diminish, or even vanish, when the spread becomes minimal.
We can express decreases in intensities due to a series of activities as a negative term in the Hawkes-type kernel.

\bigskip\noindent
{\it Avoiding negative intensity}

As mentioned above, negative excitements (i.e., diminishing intensity) can occur, but we should control this amount to ensure that intensity is nonnegative.
Thus, the amount of excitement in the spread-narrowing processes after a spread-narrowing event depends on the intensities of the narrowing processes.

\bigskip\noindent
{\it Simultaneous movement of bid and ask prices}

Theoretically, bid and ask prices can move simultaneously in rare circumstances.
For example, assume 100 volumes of the best ask price in a limit order book.
When a marketable limit order to buy 120 shares at the best ask price arrives, 
it consumes the existing best ask orders and 
the previous second-best ask order becomes the best.
In addition, the remaining market order of 20 volumes converts to the bid order, which becomes the best bid.
This feature is difficult to interpret as a mathematical model, since we assume simple processes in general, 
where only one jump occurs in an infinitesimal time interval.
In practice, we assume a small time interval between two events.
We can model this using the network, computational, and transaction delays in a trading system, and hence from raw data,
or by data preprocessing, as we explain in Section~\ref{subsect:data}.

The parameters in the excitation kernel, $\alpha_{s1}, \alpha_{s2}, \alpha_{m}, \alpha_{w1},$ and $\alpha_{w2}$, increase the intensities after certain events.
Increased intensities induce successive arrivals of corresponding events.
Our empirical study shows that the parameter estimates are large owing to the abundant high-frequency trading in the highly liquid US equity markets.
Next, we discuss the role of each constant parameter.

\begin{figure}[]
	\begin{subfigure}{\textwidth}
		\centering
		\includegraphics[width=0.4\textwidth]{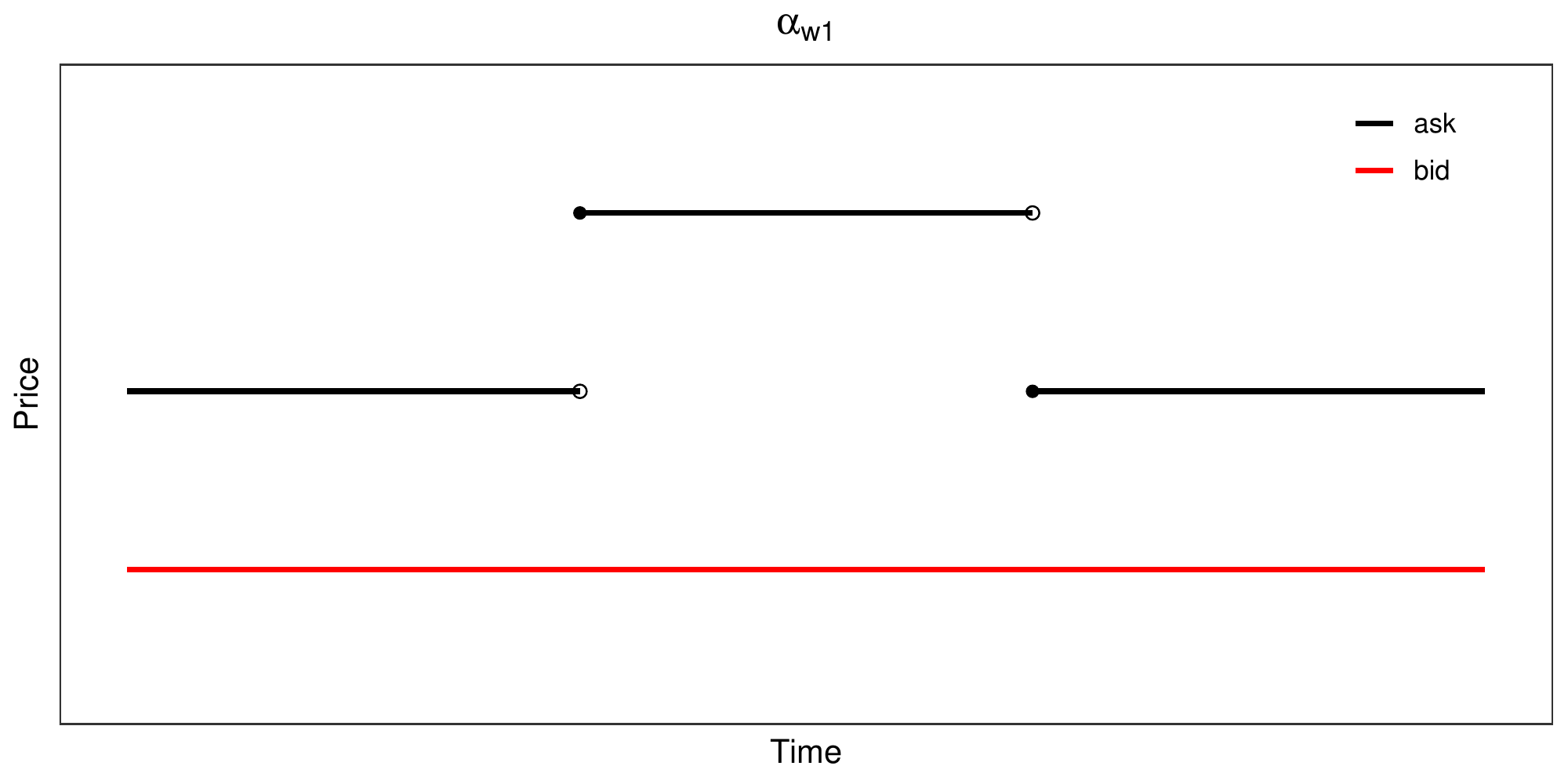}\quad\quad
		\includegraphics[width=0.4\textwidth]{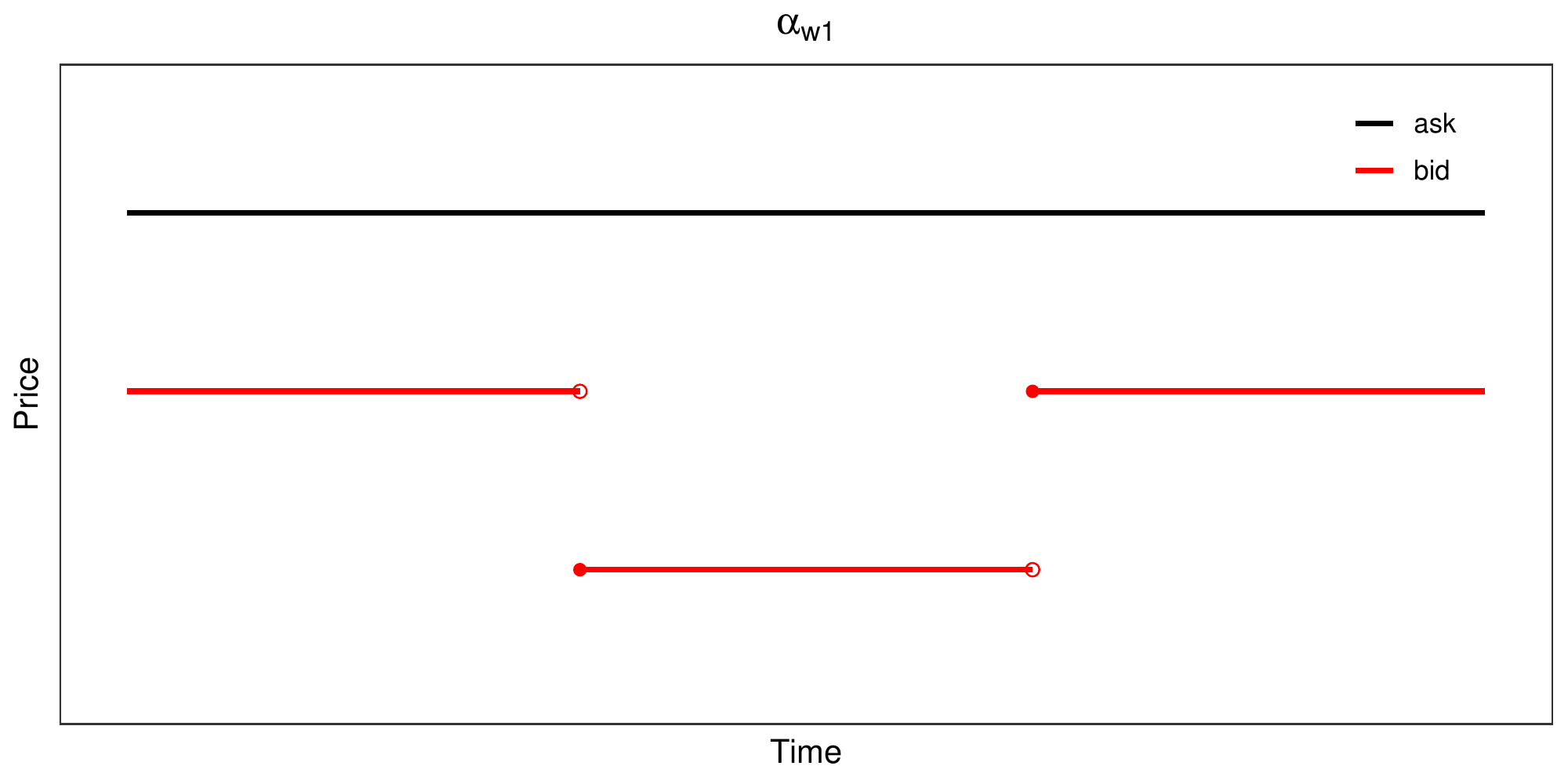}
		\caption{$\alpha_{w1}$: after a widening event, the spread decreases by the process that caused the widening}
		\label{fig:alpha_w1}
	\end{subfigure}
	
	\vspace*{2mm}
	
	\begin{subfigure}{\textwidth}
		\centering
		\includegraphics[width=0.4\textwidth]{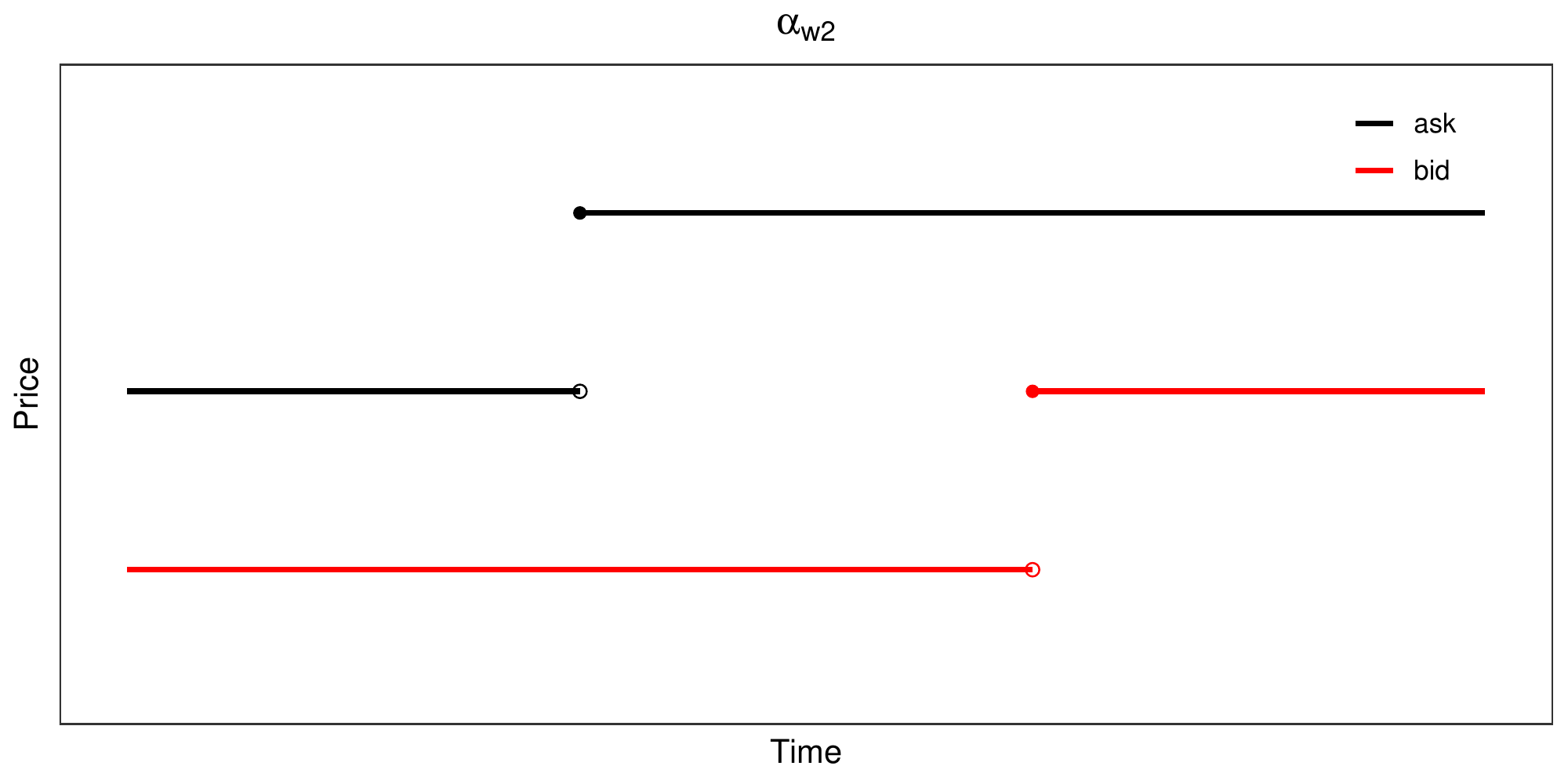}\quad\quad
		\includegraphics[width=0.4\textwidth]{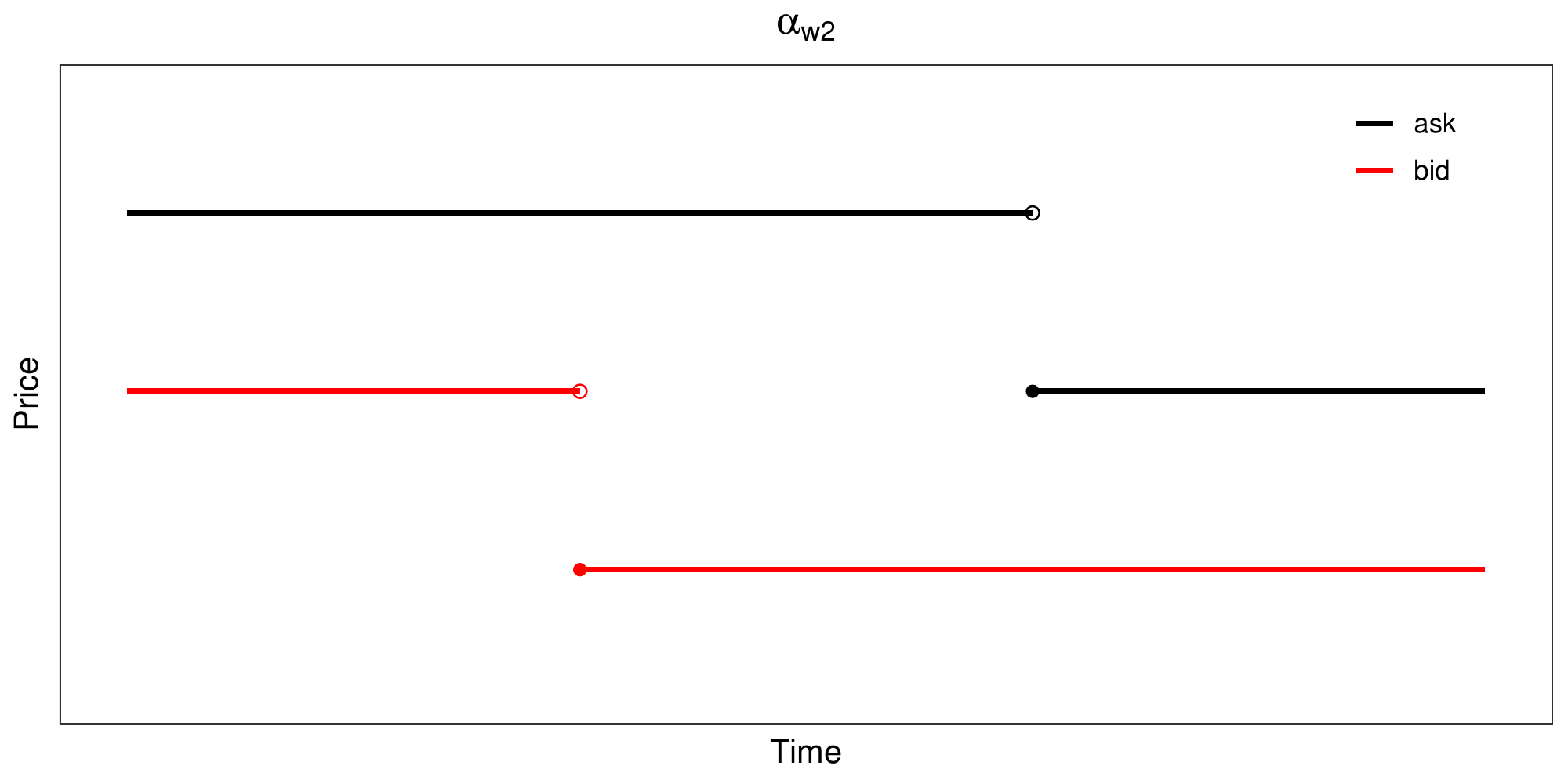}
		\caption{$\alpha_{w2}$: after a widening event, the spread decreases by the process that did not cause the widening}
		\label{fig:alpha_w2}
	\end{subfigure}
	
	\vspace*{2mm}
	
	\begin{subfigure}{\textwidth}
		\centering
		\includegraphics[width=0.4\textwidth]{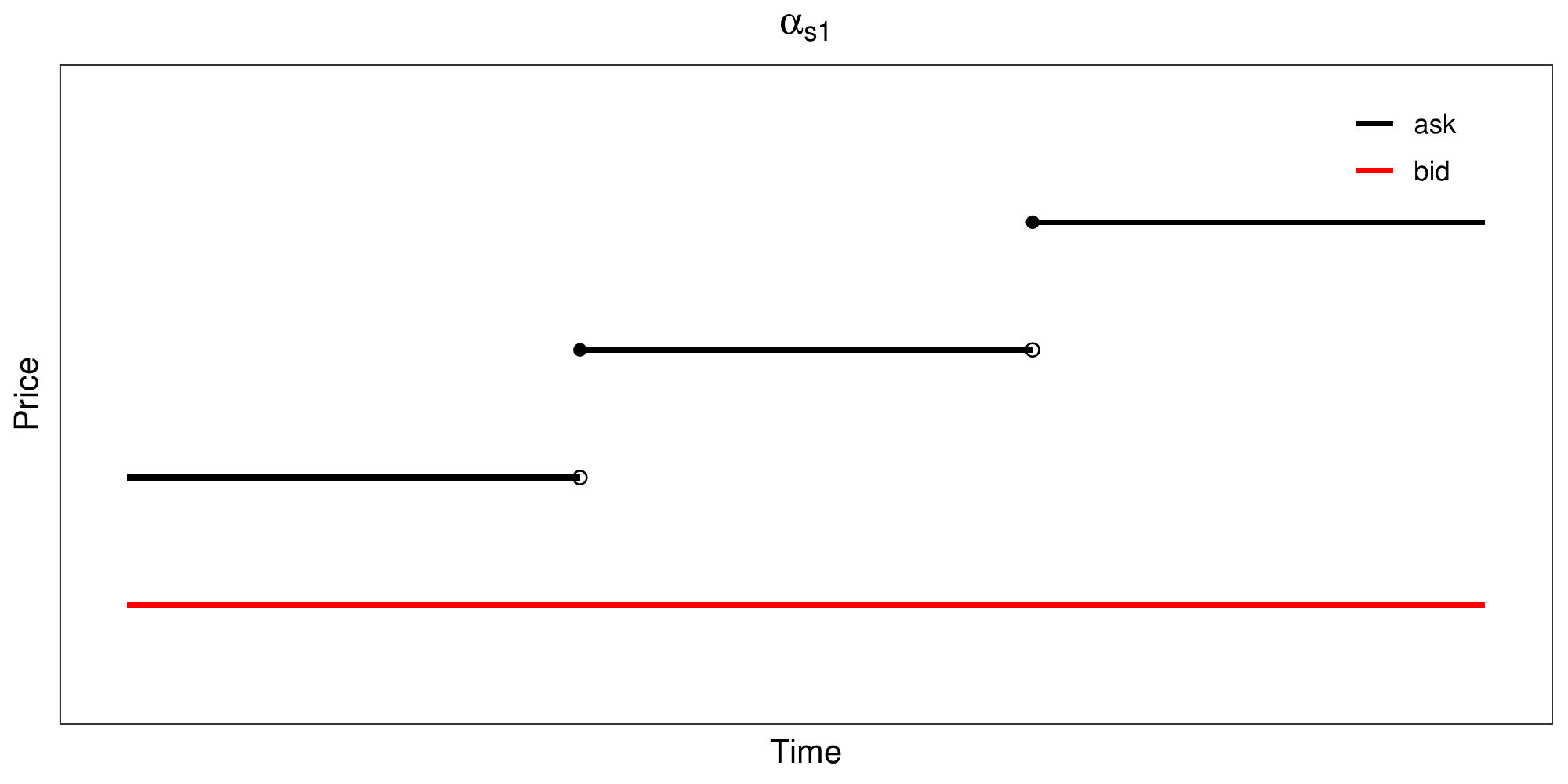}\quad\quad
		\includegraphics[width=0.4\textwidth]{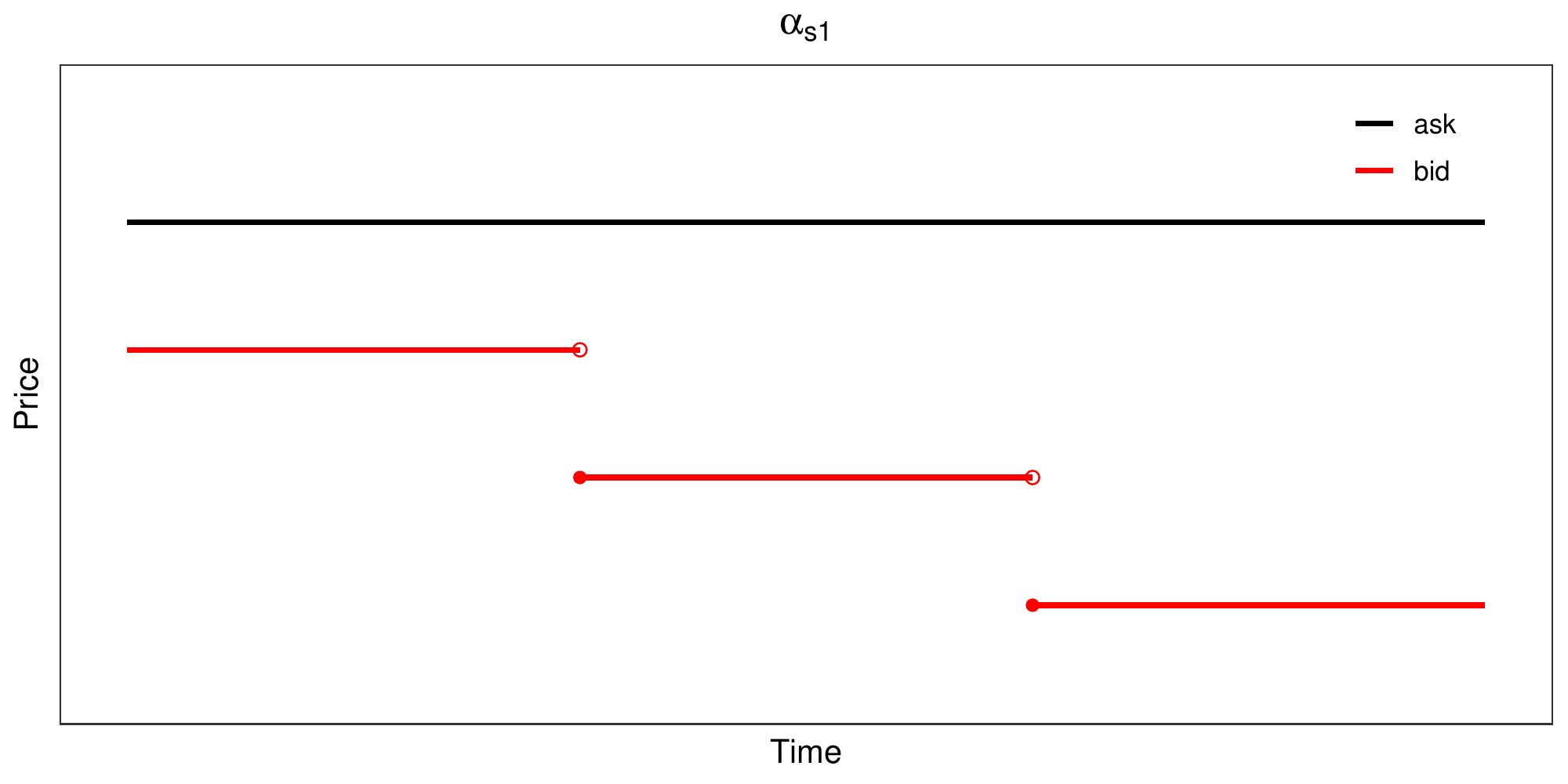}
		\caption{$\alpha_{s1}$: self-excitement in the mid-price by one type of process}
		\label{fig:alpha_s1}
	\end{subfigure}
	
	\vspace*{2mm}
	
	\begin{subfigure}{\textwidth}
		\centering
		\includegraphics[width=0.4\textwidth]{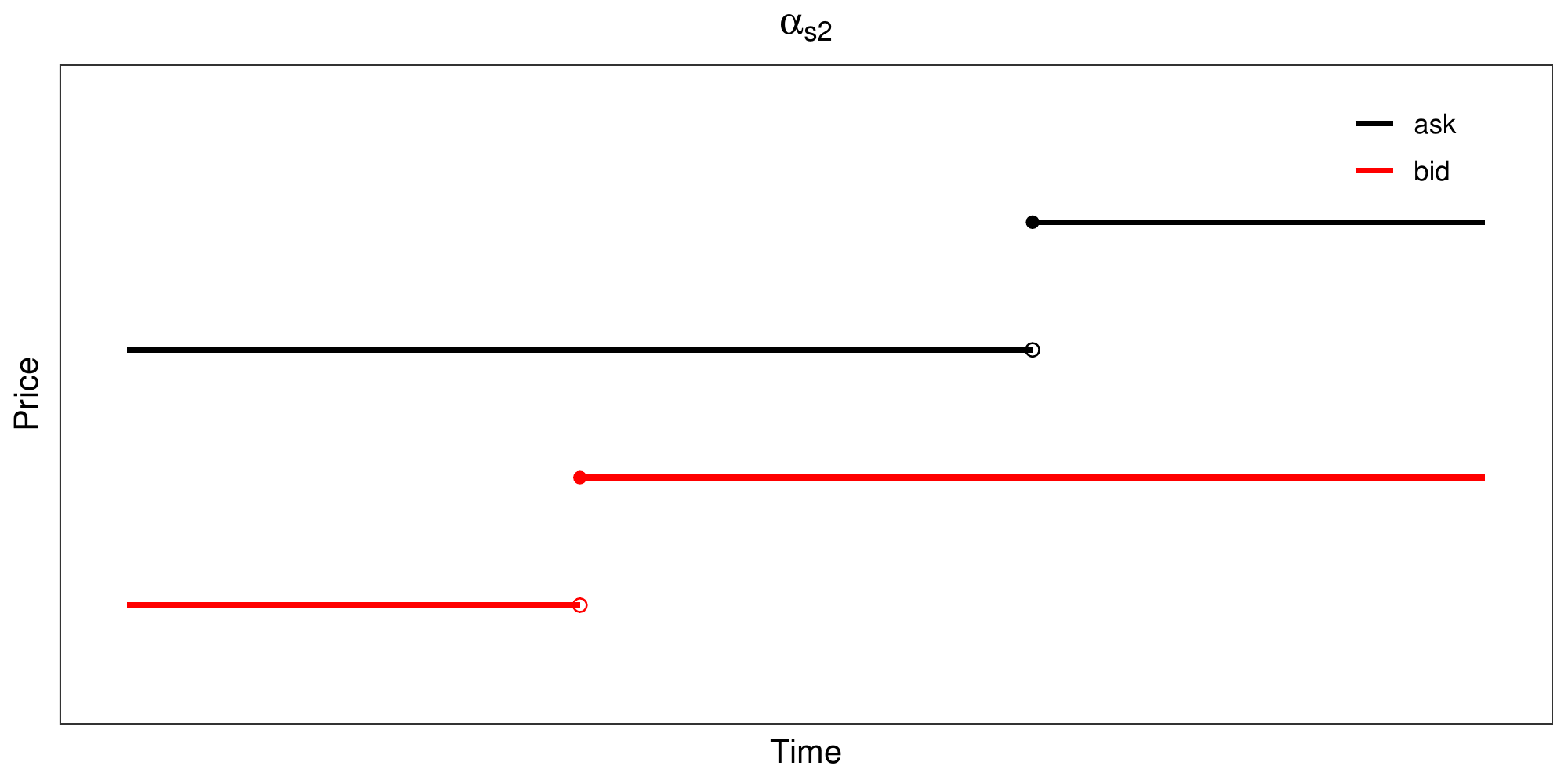}\quad\quad
		\includegraphics[width=0.4\textwidth]{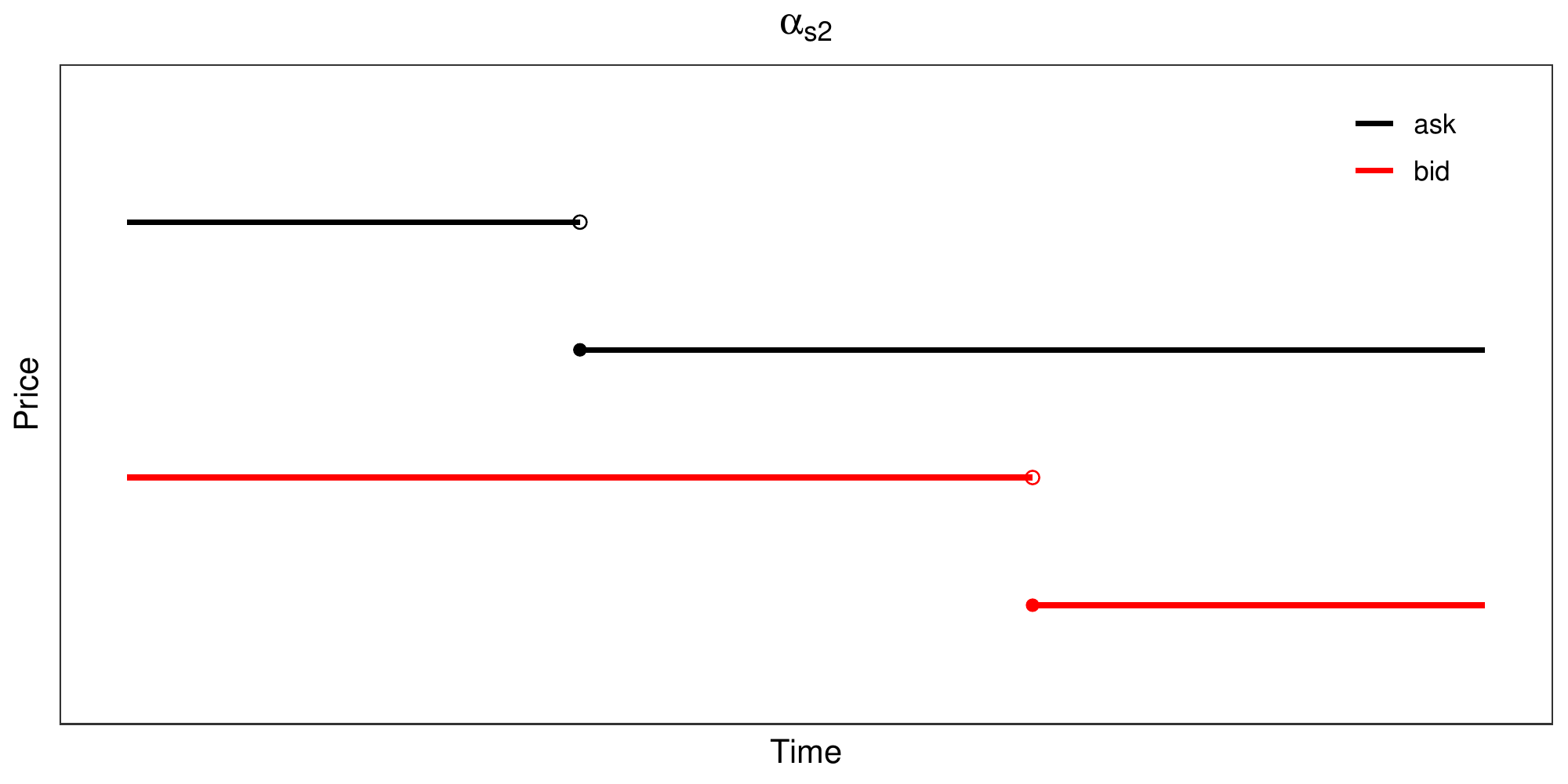}
		\caption{$\alpha_{s2}$: self-excitement in the mid-price by different types of processes}
		\label{fig:alpha_s2}
	\end{subfigure}
	
	\vspace*{2mm}
	
	\begin{subfigure}{\textwidth}
		\centering
		\includegraphics[width=0.4\textwidth]{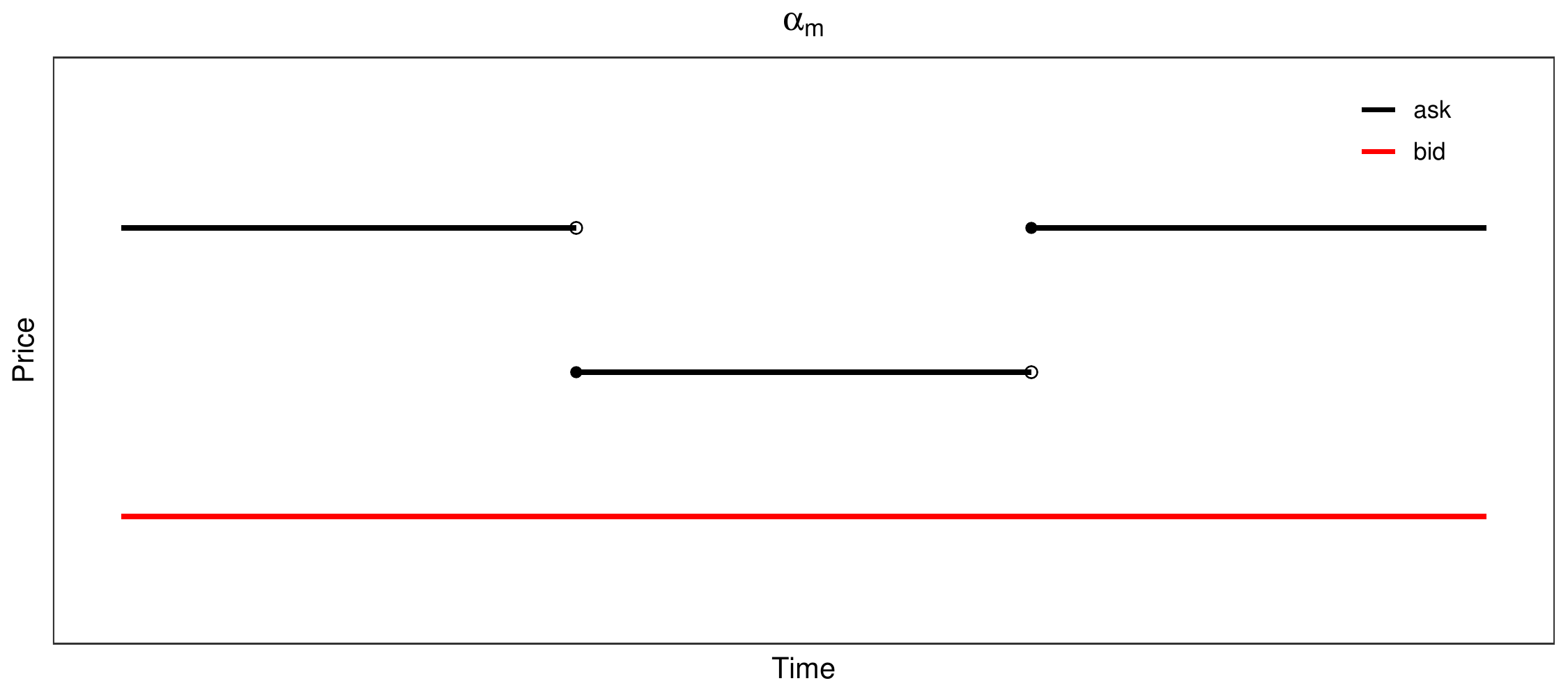}\quad\quad
		\includegraphics[width=0.4\textwidth]{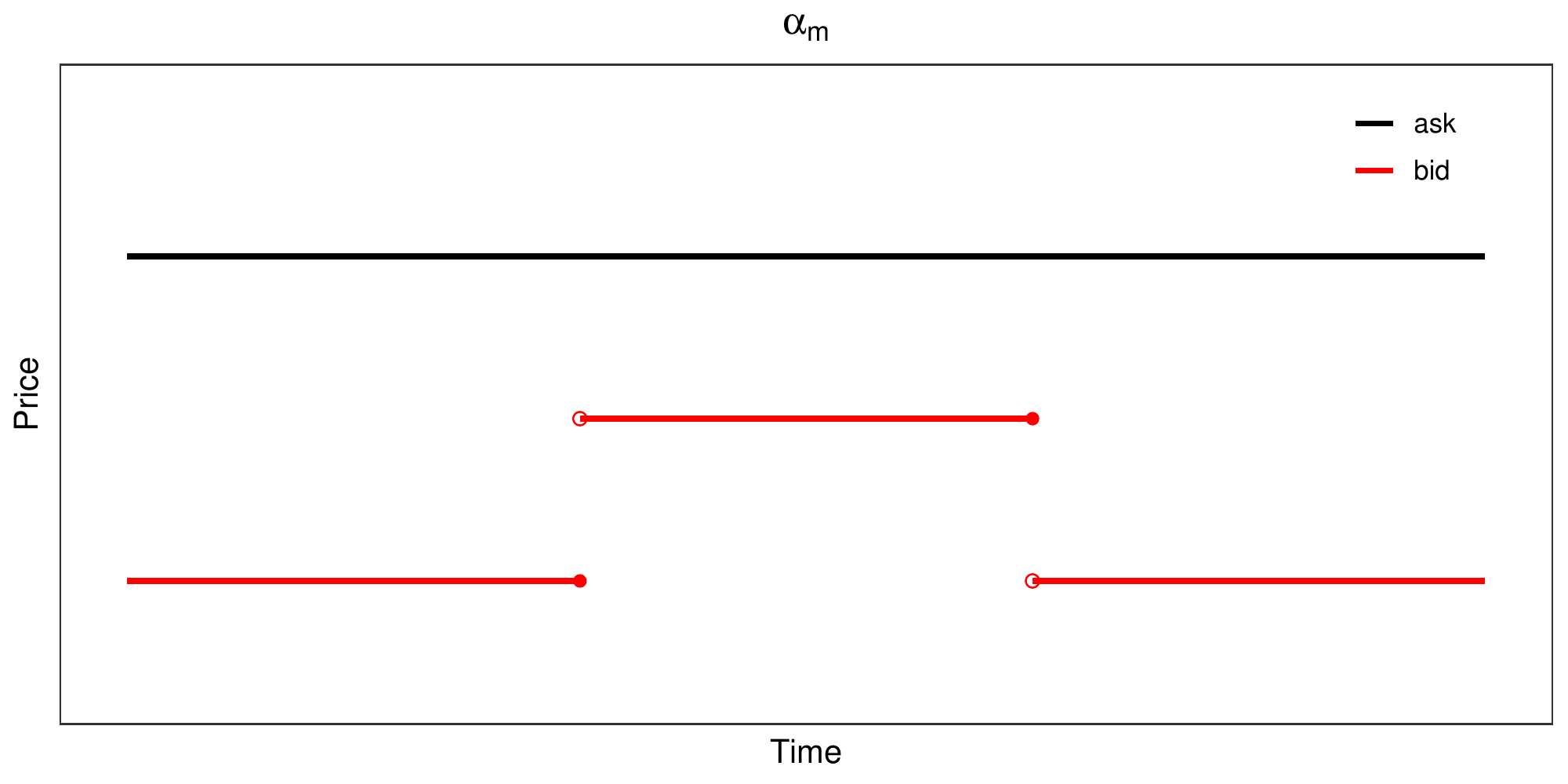}
		\caption{$\alpha_{m}$: mutual excitement in the mid-price by one type of process}
		\label{fig:alpha_m}
	\end{subfigure}
	
	\caption{Possible bid and ask price movements when the corresponding $\alpha$ is large and the other $\alpha$s are close to zero}\label{Fig:alpha}
\end{figure}

\begin{itemize}
	
	\item
	$\alpha_{w1}$: This parameter is related to liquidity provision. 
	If $\alpha_{w1}$ is large and the other parameters are small, then the bid and ask prices are likely to follow the series of movements in Figure~\ref{fig:alpha_w1} when a spread-widening event occurs. 
	From the market-making perspective, this parameter is associated with the replenishment speed for limit orders.
	When a large market order sweeps the existing best limit orders or all the best orders are canceled, the bid-ask spread widens.
	If $\alpha_{w1}$ is sufficiently large, then liquidity providers tend to replenish the depleted limit orders immediately.
	These activities may occur through (automated) market makers or rebate-seekers in maker-taker exchanges to supply liquidity.
	
	\item 
	$\alpha_{w2}$: This parameter is also related to liquidity provision and is similar to $\alpha_{w1}$.
	The relevant bid and ask prices behaviors with large $\alpha_{w2}$ is presented in Figure~\ref{fig:alpha_w2} when a spread-widening event occurs. 
	With large $\alpha_{w2}$, the widened bid-ask gap tends to narrow immediately, but by a different type of limit order from a depleted limit order.
	If these events occur consecutively in the same direction, then significant mid-price changes may eventually result.
	
	\item 
	$\alpha_{s1}$: 	This parameter is related to removing liquidity.
	If $\alpha_{s1}$ is large, then we can observe increments in the ask price or decrements in the bid price by successive limit order depletions, as in Figure~\ref{fig:alpha_s1}. 
	If the order depletion in the same direction happens in a very short time, it may be related to anticipatory order cancellation, which means that when a large market order arrives, the market maker considers it to be an informed trader's action and immediately cancels its limit orders to prevent adverse selection \citep{fox2019new}.
	If the market is sufficiently liquid, as the US equity market then successive depletions and the corresponding widened spread are likely to be replenished immediately with large $\alpha_w$s.
	
	\item 
	$\alpha_{s2}$: This parameter is also related to liquidity removal and related movements are described in Figure~\ref{fig:alpha_s2} when a spread-narrowing event occurs. 
	As in the case of $\alpha_{w2}$, if these events occur consecutively, then price changes may eventually result.
	
	\item 
	$\alpha_{m}$: This parameter is also related to removing liquidity.
	When $\alpha_{m}$ is large, if a better limit order than the current best price arrives, 
	then the new limit order is executed or canceled immediately, and the price rebounds to its original status, as in Figure~\ref{fig:alpha_m}.
	This result may occur because market participants may consider the new order to be good 
	and high-frequency traders execute the market order immediately.
	Alternatively, a trader may issue a limit order aggressively and cancel the order instantly to avoid adverse selection, especially in an uncertain market.
	Order cancellations are abundant in stock markets. 
	\citet{scholtus2014} report that 60\% of all orders are canceled within one second in the S\&P 500 exchange-traded funds traded on the Nasdaq from 2009 to 2011.
	\citet{hasbrouck2009technology} refer to limit orders canceled immediately as fleeting orders.
	Repeated order submissions and cancellations are also known as flickering quotes, a concept introduced more than two decades ago \citep{Harris}.
	
\end{itemize}

The terms $\alpha_n$ can be negative in contrast to the classical Hawkes model in which t he $\alpha$ parameters are nonnegative.
Although $\alpha_{n}(t)$s may be negative, however, they are bounded by 
$- \sum_{j=1}^4 \lambda_{ij}(t)$ below, and hence $\lambda$s are always nonnegative.
More precisely, for $i=3$ or 4, and assume that there is a spread narrowing event at time $u$, then
$$ \lambda_i(u+) = \lambda_i(u) + \alpha_{nm}(u) \geq \lambda_i(u) - \sum_{j=1}^4 \lambda_{ij}(u)  = \eta \ell(u) \geq 0 $$
where $m=1$ for $i=2$ and $m=2$ for $i=3$.
In addition, under the symmetric setting, i.e., $\xi$ is the same in $\alpha_{n1}$ and $\alpha_{n2}$,
we have $\lambda_2 = \lambda_3$ and $\alpha_{n1} = \alpha_{n2}$.

\begin{figure}[t]
	\centering
	\includegraphics[width=0.45\textwidth]{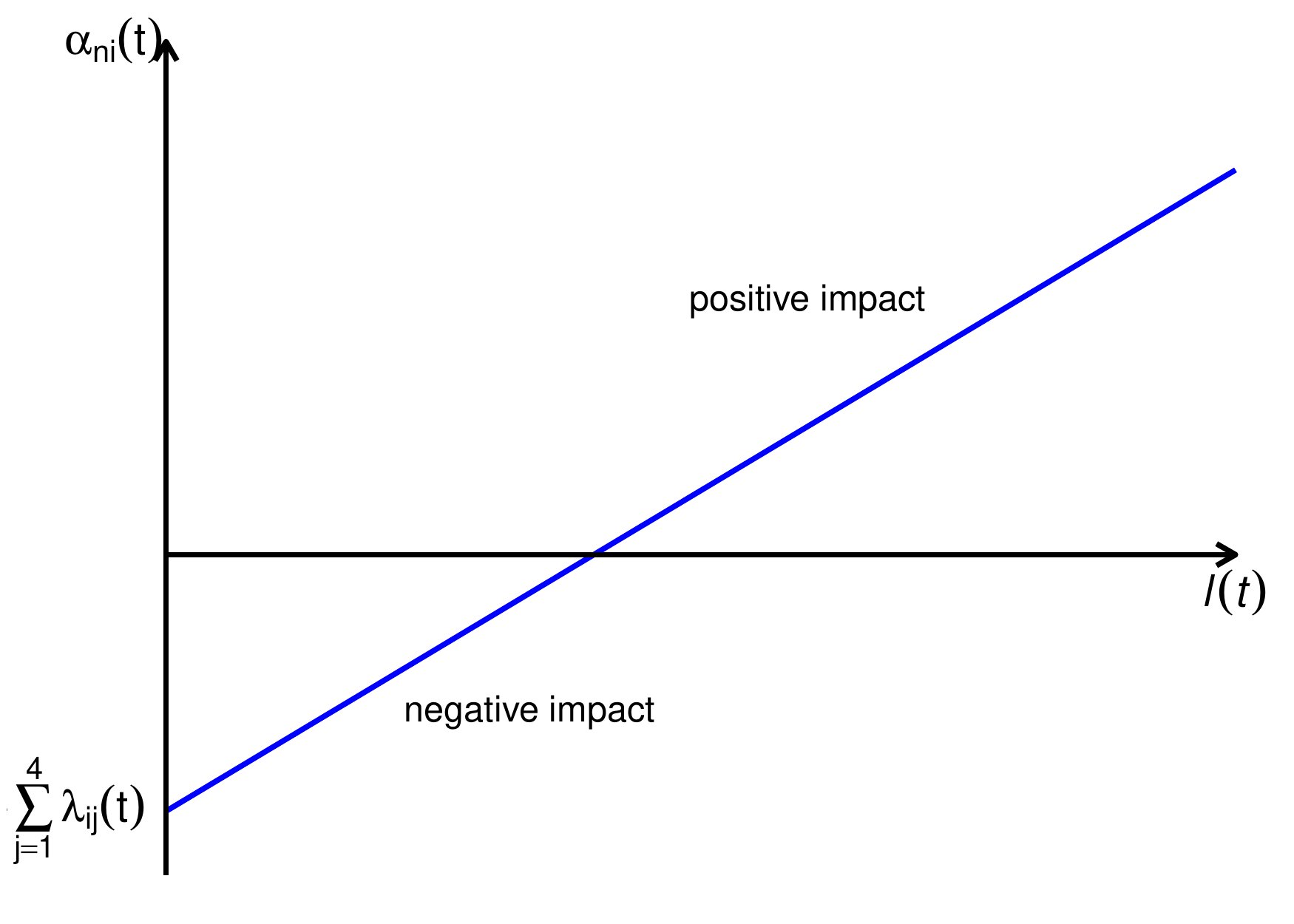}
	\caption{$\alpha_{ni}(t)$ versus the relative spread in a linear model}\label{Fig:impact}
\end{figure}

Figure~\ref{Fig:impact} shows the impact of the spread-narrowing intensities when the spread narrows.
The figure shows that the impact can be positive or negative depending on the level.
If the level is high, even if the spread narrows, the speed at which it narrows can still increase.
If the level is sufficiently low, then the spread-narrowing event reduces the rate at which the spread decreases.
The $y$-intercept of the line is $- \sum_{j=1}^4 \lambda_{ij}(t)$, a $\F_{t-}$-measurable random variable, and the slope $\xi$ is a constant parameter.
This stochastic $y$-intercept ensures that the negative excitement does not make the intensities less than zero.

We can rewrite the intensity processes as
\begin{align*}
\lambda_A^u(t) ={}& \mu + \int_{-\infty}^{t} \alpha_{s1} \e^{-\beta(t-u)} \D N_A^u(u) + \int_{-\infty}^{t} \alpha_{m} \e^{-\beta(t-u)} \D N_A^d(u) + \int_{-\infty}^{t} \alpha_{s2} \e^{-\beta(t-u)} \D N_B^u(u) \\
\lambda_A^d(t) ={}& f(\ell(t-)) + \int_{-\infty}^{t} \alpha_{w1} \e^{-\beta(t-u)} \D N_A^u(u) + \int_{-\infty}^{t} \alpha_{w2} \e^{-\beta(t-u)} \D N_B^d(u) \\
& \phantom{f(\ell(t-))} + \int_{-\infty}^{t} \alpha_{n1}(u) \e^{-\beta(t-u)} (\D N_A^d(u) + \D N_B^u(u)) \\
\lambda_B^u(t) ={}& f(\ell(t-)) + \int_{-\infty}^{t} \alpha_{w2} \e^{-\beta(t-u)} \D N_A^u(u) + \int_{-\infty}^{t} \alpha_{w1} \e^{-\beta(t-u)} \D N_B^d(u) \\
& \phantom{f(\ell(t-))} + \int_{-\infty}^{t} \alpha_{n2}(u) \e^{-\beta(t-u)} (\D N_A^d(u) + \D N_B^u(u)) \\
\lambda_B^d(t) ={}& \mu + \int_{-\infty}^{t} \alpha_{s2} \e^{-\beta(t-u)} \D N_A^d(u) + \int_{-\infty}^{t} \alpha_{m} \e^{-\beta(t-u)} \D N_B^u (u) + \int_{-\infty}^{t} \alpha_{s1} \e^{-\beta(t-u)} \D N_B^d(u).
\end{align*}
In the classical Hawkes model, the parameter $\alpha$s are restricted by $\beta$ for nonexplosive solutions.
Intuitively, a large $\alpha$ causes intensity to increase rapidly.
For example, if $\alpha$ is greater than $\beta$, which describes the decrease in intensity over time, 
then the increment in intensity caused by $\alpha$ is much larger than the decrease caused by $\beta$, and intensity explodes.
By contrast, in our model, $\alpha_w$, which relates to the spread-narrowing processes, can have a larger number than $\beta$.
If $\lambda_A^d$ or $\lambda_B^u$ rapidly increases owing to a large $\alpha_w$, 
then the spread goes to zero and $\lambda_A^d$ or $\lambda_B^u$ vanishes according to the model definition.
This feature makes it possible to estimate $\alpha_w$ without restriction which represents market resiliency.

Note that to fully characterize the mid price and spread processes, we need additional information about the size of movement in the price processes.
The mid-price process is
\[
p(t) = p(0) +  \int_0^t \delta_A^u(u) \D N_A^u(u) + \int_0^t \delta_B^u(u) \D N_B^u(u)  - \int_0^t \delta_A^d(u) \D N_A^d(u) - \int_0^t \delta_B^d(u) \D N_B^d(u)
\]
where $\delta_A^u, \delta_B^u, \delta_A^d$ and $\delta_B^d$ denote the corresponding jump sizes.
Similarly, the spread process is
$$ L(t) = L(0) + \int_0^t \delta_A^u(u) \D N_A^u(u) + \int_0^t \delta_B^d(u) \D N_B^d(u)  - \int_0^t \delta_A^d(u) \D N_A^d(u) - \int_0^t \delta_B^u(u) \D N_B^u(u).$$
However, the inclusion of the jump size to the model make introduce marked point processes, which is beyond the scope of our paper.
We assume that the jump sizes at time $t$ are exogenously given, i.e., $\delta(t)$s are $\mathcal F_t$-measurable but not measurable to the internal history of $\bm{N}$.
In our model, only if the jump sizes are all constant unity, then the mid-price and spread processes are endogenously determined.

\section{Mathematical analysis}\label{Sec:math}

In this section, we investigate the mathematical properties of the model.
However, since the proposed model in Definition~{\ref{Def:model}} is too complicated, slightly different versions are investigated instead in Subsections~{\ref{subsec:mid}}~and~{\ref{subsec:spread}}.
This does not provide a complete proof of the stability of the proposed model, but will indirectly provide insight.

\subsection{Reset property}\label{subsec:reset}

The stochastic terms $\alpha_n$s reset the excitement terms in the spread-narrowing intensities, $\lambda_2$ and $\lambda_3$, to $\xi \ell$ right after a spread-narrowing event.
Let $\{ \tau_{k} \}$ be a sequence of the spread-narrowing event times.
For $i = 2$ and $3$,
$$ 
\lambda_i(\tau_{k}+) = f(\tau_{k}) + \xi \ell(\tau_{k})
$$
where the excitement term is reset to $\xi \ell(\tau_{k})$ which only depends on the current level of spread.
In addition, for $ \tau_{k} < t  < \tau_{k+1}$,
\begin{equation}
\lambda_i(t) = f(\ell(t-)) + \xi \ell(\tau_{k})\e^{-\beta(t-\tau_{k})} +  \int_{\tau_{k}}^{t} \alpha_{w1} \e^{-\beta(t-u)} \D N_1(u) + \int_{\tau_{k}}^{t} \alpha_{w2} \e^{-\beta(t-u)}\D N_4(u).\label{eq:reset}
\end{equation}
The above equation implies that whenever a spread-narrowing event happens, 
the excitement term reset to $\xi \ell$ with decreasing rate $\beta$ 
and up to the next spread-narrowing event, typical constant Hawkes mutually excitement terms, $\alpha_w$s, are added with decreasing rate $\beta$ on the occurrences of $N_1$ and $N_4$.
This is because the excited parts of the spread-narrowing intensities before time $\tau_{k}$ are canceled out by
$$ - \sum_{j=1}^4 \lambda_{ij}(\tau_{k}), \textrm{ for } i = 2 \textrm{ and } 3.$$
Especially, when the spread hits the minimum by a spread-narrowing event, the reset term $\xi \ell$ becomes zero 
and hence the excitements of intensities of the spread-narrowing processes also become zero.

\subsection{Mid-price process under constant model}\label{subsec:mid}

In this subsection, we examine the properties of the mid-price process from a slightly different version of our model.
In the linear model of $f$ and $\alpha_n$ in Definition~{\ref{Def:model}}, there is a rather complicated aspect of the analysis, so let's simplify the model a little bit.
Because the mid-price is the average of the bid and ask prices, we can derive the mid-price process from the bid and ask price processes.
Let $N^u$ and $N^d$ be the counting processes for the upward and downward movements of the mid-price process; then, 
$$ N^u = N^u_A + N^u_B, \quad N^d = N^d_A + N^d_B$$
and
$$ \lambda^u = \lambda^u_A + \lambda^u_B, \quad \lambda^d = \lambda^d_A + \lambda^d_B.$$
Thus, in the symmetric setting (i.e., $\alpha_n := \alpha_{n1} = \alpha_{n2}$),
\begin{align*}
\lambda^u(t) =  \mu + f(\ell(t-))
&+ 
\int_{-\infty}^{t} (\alpha_{s1} + \alpha_{w2}) \e^{-\beta(t-u)} \D N^u_A(u) + 
\int_{-\infty}^{t} (\alpha_{s2} + \alpha_{n}(u)) \e^{-\beta(t-u)} \D N^u_B(u) \\
&+ 
\int_{-\infty}^{t} (\alpha_{m} + \alpha_{n}(u)) \e^{-\beta(t-u)} \D N^d_A(u) +
\int_{-\infty}^{t} \alpha_{w1} \e^{-\beta(t-u)} \D N^d_B(u) \\
\lambda^d(t) =  \mu + f(\ell(t-))
&+
\int_{-\infty}^{t} \alpha_{w1}  \e^{-\beta(t-u)} \D N^u_A(u) + 
\int_{-\infty}^{t} (\alpha_{m} + \alpha_{n}(u))\e^{-\beta(t-u)} \D N^u_B(u) \\
&+
\int_{-\infty}^{t} (\alpha_{s2} + \alpha_{n}(u)) \e^{-\beta(t-u)} \D N^d_A(u) +
\int_{-\infty}^{t} (\alpha_{s1} + \alpha_{w2}) \e^{-\beta(t-u)} \D N^d_B(u).
\end{align*}

As mentioned before, we adopt a simpler model in this subsection, a constant model for $f$ and $\alpha_n$; that is,
$$ f(\ell(t-)) = \bf{1}_{\{ \ell(t-) > 0 \}} \eta $$
and
\begin{align*}
\alpha_{n1}(t) = - \bf{1}_{\{ \ell(t) = 0 \}} \sum_{j=1}^4 \lambda_{2j}(t) +  \bf{1}_{\{ \ell(t) > 0 \}} \xi, 
\quad
\alpha_{n2}(t) = - \bf{1}_{\{ \ell(t) = 0 \}} \sum_{j=1}^4 \lambda_{3j}(t) +  \bf{1}_{\{ \ell(t) > 0 \}} \xi, 
\end{align*}
for some constants $\eta > 0$ and $\xi > 0$.
This model is simpler than the linear model, 
but it consistent with the general concept in Subsection~{\ref{subsec:def}}.
This model guarantees zero intensities of the spread-narrowing processes at the minimum spread as in the linear model.
The base intensities for the spread-narrowing processes become zero right after the spread hit the minimum due to the fact that
$ f(0) = 0 $,
and the excitation parts are canceled out by
$$ - \sum_{j=1}^4 \lambda_{2j}(t).$$
By canceling the exact amount of the existing excitation parts, it guarantees the nonnegative intensities.
Once the spread becomes larger than zero, the model acts like a basic Hawkes model with constant parameters $\eta$ and $\xi$.

Under the special circumstances, the mid-price process of the constant model is similar to that of the classical Hawkes model.
We assume that the spread is sufficiently large; that is, $\mathbb{P}(\ell(t) = 0) $ is close to zero under the constant model.
Then,
\begin{align*}
\lambda^u(t) \approx  \mu + \eta  
&+ 
\int_{-\infty}^{t} (\alpha_{s1} + \alpha_{w2}) \e^{-\beta(t-u)} \D N^u_A(u) + 
\int_{-\infty}^{t} (\alpha_{s2} + \xi) \e^{-\beta(t-u)} \D N^u_B(u) \\
&+ 
\int_{-\infty}^{t} (\alpha_{m} + \xi) \e^{-\beta(t-u)} \D N^d_A(u) +
\int_{-\infty}^{t} \alpha_{w1} \e^{-\beta(t-u)} \D N^d_B(u) \\
\lambda^d(t) \approx  \mu + \eta
&+
\int_{-\infty}^{t} \alpha_{w1}  \e^{-\beta(t-u)} \D N^u_A(u) + 
\int_{-\infty}^{t} (\alpha_{m} + \xi)\e^{-\beta(t-u)} \D N^u_B(u) \\
&+
\int_{-\infty}^{t} (\alpha_{s2} + \xi) \e^{-\beta(t-u)} \D N^d_A(u) +
\int_{-\infty}^{t} (\alpha_{s1} + \alpha_{w2}) \e^{-\beta(t-u)} \D N^d_B(u).
\end{align*}
For further simplification, suppose that
$\alpha_{s1} + \alpha_{w2} = \alpha_{s2} + \xi$ and $\alpha_{w1}  = \alpha_{m} + \xi $,
which is used for demonstration purposes.
Then,
\begin{align*}
\lambda^u(t) &\approx  \mu + \eta  +  
\int_{-\infty}^{t} (\alpha_{s1} + \alpha_{w2}) \e^{-\beta(t-u)} \D N^u(u) + 
\int_{-\infty}^{t} \alpha_{w1} \e^{-\beta(t-u)} \D N^d(u) \\
\lambda^d(t) &\approx  \mu + \eta +
\int_{-\infty}^{t} \alpha_{w1}  \e^{-\beta(t-u)} \D N^u(u) + 
\int_{-\infty}^{t} (\alpha_{s1} + \alpha_{w2}) \e^{-\beta(t-u)} \D N^d(u),
\end{align*}
which is a typical symmetric self- and mutually excited Hawkes process for the mid-price dynamics.
To describe the full dynamics of the mid-price process, we require a jump size distribution.

\subsection{Spread process under absolute level model}\label{subsec:spread}

This subsection investigate the stability in mean of the spread process under a slightly different version.
The spread is the difference between the bid and ask prices; thus, the bid and ask price processes can induce counting processes for the number of upward or downward movements of the spread process.	
Let $N_s^u$ and $N_s^d$ be the counting processes for the upward and downward movements of the spread process, respectively.
Then,
$$ N_s^u = N^u_A + N^d_B, \quad N_s^d = N^d_A + N^u_B$$
and
$$ \lambda_s^u = \lambda^u_A + \lambda^d_B, \quad \lambda_s^d = \lambda^d_A + \lambda^u_B.$$
Thus, in the symmetric setting,
\begin{align}
&\lambda_s^u(t) =  2\mu 
+ 
\int_{-\infty}^{t} \alpha_{s1} \e^{-\beta(t-u)} \D N^u_s(u) + 
\int_{-\infty}^{t} (\alpha_{s2} + \alpha_{m})\e^{-\beta(t-u)} \D N^d_s(u) \label{Eq:spread1}\\
&\lambda_s^d(t) =  2 f(L(t-)) 
+
\int_{-\infty}^{t} (\alpha_{w1} + \alpha_{w2}) \e^{-\beta(t-u)} \D N^u_s(u) + 
\int_{-\infty}^{t} \alpha_{n}(u)\e^{-\beta(t-u)} \D N^d_s(u). \label{Eq:spread2}
\end{align}
The intensity of the upward movements of the spread process is the classical Hawkes model and that of the downward movements is the Hawkes model with stochastic excitement.
As mentioned previously, to describe the full dynamics of the spread process, we require a jump size distribution of up and down price movements,  which is beyond the scope of this study.
In addition, for mathematical analysis, it is more tractable when modeling with absolute level of spread (especially in the base intensity) in Eq.~{\eqref{Eq:spread2}}.

Assume that the jump size is one so that the spread process is represented by the difference between $N_s^u$ and $N_s^d$.
In addition, let us start with a simple case such that
\begin{align}
&\lambda_s^u(t) =  2\mu 
+ 
\int_{-\infty}^{t} \alpha_{s1} \e^{-\beta(t-u)} \D N^u_s(u) + 
\int_{-\infty}^{t} (\alpha_{s2} + \alpha_{m})\e^{-\beta(t-u)} \D N^d_s(u) \label{eq:sp_int1}\\
&\lambda_s^d(t) =  2 \eta L(t-)\label{eq:sp_int2}
\end{align}
i.e., without the excitement terms in $\lambda_s^d(t)$.
Since the differential form of $\lambda_s^u(t)$ can be written as
$$ \D \lambda_s^u(t) = \beta (2\mu - \lambda_s^u(t)) \D t +  \alpha_{s1} \D N_s^u (t)  + (\alpha_{s2} + \alpha _m) \D N_s^d(u), $$
we have
$$ \E[\lambda_s^u(t) | \mathcal F_0] = \lambda_s^u(0) + \int_0^t 2 \{ \beta \mu + (\alpha_{s1} - \beta)\E[\lambda_s^u(u)] + (\alpha_{s2} + \alpha_m)\E[\lambda_s^d(u)] \} \D u.$$
In addition,
$$ \lambda_s^d(t) =  2 \eta L(t-) = 2 \eta (L(0) + N_s^u(t-) - N_s^d(t-)) $$
and
$$ \E[\lambda_s^d(t)| \mathcal F_0] =  2 \eta L(0) +  \int_0^t 2 \eta (\E[\lambda_s^u(u)] - \E[\lambda_s^d(u)]) \D u.$$
Thus, we construct a system of differential equation,
\begin{align*}
&\frac{\D \E[\lambda_s^u (t)| \mathcal F_0]}{\D t} = 2 \beta \mu + (\alpha_{s1} - \beta)\E[\lambda_s^u(t)] + (\alpha_{s2} + \alpha_m)\E[\lambda_s^d(t)] \\
&\frac{\D \E[\lambda_s^d (t)| \mathcal F_0]}{\D t} = 2 \eta (\E[\lambda_s^u(t)] - \E[\lambda_s^d(t)]).
\end{align*}
The matrix of the system of differential equation is
$$ A = \begin{bmatrix} \alpha_{s1} - \beta & \alpha_{s2} + \alpha_m \\ 2\eta & - 2\eta 
\end{bmatrix}.$$
It is known that the stability conditions of the system is that the trace of $A$ is negative and the determinant is positive, i.e.,
$$ \mathrm{tr}(A) = \alpha_{s1} - \beta - 2\eta  <0, \quad  \mathrm{det}(A) =  2 \eta (\beta - \alpha_{s1} - \alpha_{s2} - \alpha_m) > 0.$$
Since $\eta >0$, if Eq.~{\eqref{eq:condition}} holds, then this stability condition satisfies.
In addition, under the steady-state of the differential equation system,
$$
\begin{bmatrix}
\E[\lambda_s^u(t)] \\
\E[\lambda_s^d(t)]
\end{bmatrix}
= - A^{-1} 
\begin{bmatrix}
2 \beta \mu \\
0
\end{bmatrix} 
$$
and this yields that the steady-state average intensity is
$$ \E[\lambda_s^u(t)] = \E[\lambda_s^d(t)] = \frac{2 \beta \mu}{\beta - \alpha_{s1} - \alpha_{s_2} - \alpha_m} $$
and
$$ \E[L(t)] = \frac{\E[\lambda_s^d(t)]}{2\eta} = \frac{\beta \mu}{ \eta (\beta - \alpha_{s1} - \alpha_{s_2} - \alpha_m)}.$$
Therefore, under the Eqs.~{\eqref{eq:sp_int1}}~and~{\eqref{eq:sp_int2}}, the expected intensities and the absolute level of the spread are finite.

Adding exciting terms to $\lambda_s^d$ in Eq.~{\eqref{eq:sp_int2}} does not change much of the discussion above.
As in the previous, let $\{ \tau_{k} \}$ be a sequence of the spread-narrowing event times.
For $\tau_{k} < t < \tau_{k+1}$,
$$ \lambda_s^u(t) = 2 \mu + (\lambda_s^u(\tau_{k}) - 2\mu)\e^{-\beta(t - \tau_{k})} + \int_{\tau_{k}}^{t} \alpha_{s1} \e^{-\beta(t-u)} \D N_s^u(u)$$
and $\lambda_s^u(t)$ only has a self-exciting feature, 
and hence if Eq.~{\eqref{eq:condition}} holds, there are only finite number of events by $N_s^{u}$ during the period $[\tau_{k}, \tau_{k+1}]$ almost surely and $\lambda_s^u(t)$ will not explode to infinity.
In addition, for $\tau_{k} < t < \tau_{k+1}$, by recalling the reset property for $\lambda_s^d = \lambda_2 + \lambda_3$ in Eq~{\eqref{eq:reset}},
$$ \lambda_s^d(t) = 2 \eta L(t-) + \xi L(\tau_{k}) \e^{-\beta(t-\tau_{k})} + \int_{\tau_{k}}^{t} (\alpha_{w1} + \alpha_{w2}) \e^{-\beta(t-u)} \D N^u_s(u)$$
is also finite almost surely, because there are only finite events by $N_s^{u}$ (even when $\alpha_{w1} + \alpha_{w2} > \beta$).

Since the processes are stable for $\tau_{k} < t < \tau_{k+1}$, it is sufficient to check the stability at each time of $\{ \tau_k \}$. 
By recalling the reset property, the intensities right after a spread-narrowing event can be represented by
\begin{align}
\lambda_s^u(\tau_{k}+) &=  2\mu 
+ 
\int_{-\infty}^{\tau_{k}} \alpha_{s1} \e^{-\beta(t-u)} \D N^u_s(u) + 
\int_{-\infty}^{\tau_{k}} (\alpha_{s2} + \alpha_{m})\e^{-\beta(t-u)} \D N^d_s(u) \\
\lambda_s^d(\tau_{k}+) 
&= 2 \eta L(\tau_{k}). 
\end{align}
These are discrete-time versions of Eqs.~{\eqref{eq:sp_int1}}~and~{\eqref{eq:sp_int2}}
and as long as Eq.~{\eqref{eq:condition}} holds, we can expect the stability at times of $\{ \tau_k \}$.

\section{Empirical study}~\label{Sec:empirical}

Our model allows us to examine various aspects of the first level order book in high-frequency.
This section includes the following topics.

\begin{itemize}
	\item 
	Since the intensities are functions of the bid-ask spread level, the model captures the narrowing tendency of the spread when the spread is large, see Subsection~{\ref{subsec:basic}}.
	\item 
	The parameter $\alpha$s measure the responsiveness rates of the market to the previous events, see also Subsection~{\ref{subsec:basic}}.
	\item 
	For each stock, by comparing liquidity providing parameters, $\alpha_w$s, and removing parameters, $\alpha_m$ and $\alpha_{s}$s, 
	we can examine the market-making speed in various stocks, see Subsection~{\ref{subsec:compare}}.
	\item 
	We also found some stocks received particular attentions from ultra-high-frequency liquidity providers, see Subsection~{\ref{subsec:outlier}}.
	\item 
	The estimation results also show that the propensities of participants differ for each stock exchange, see Subsection~{\ref{subsec:exchange}}.
	\item 
	The estimation result of $\eta$ in an almost continuous-time provides another perspective during the Flash Crash, see Subsection~{\ref{subsec:Crash}}.
	
\end{itemize}
Before the main empirical analysis results, we explain the method, data and model selection.

\subsection{Likelihood estimation and simulation study}

We estimate the parameters of the model using the maximum log-likelihood method.
The method to calculate the log-likelihood of our model is similar to that for calculating the log-likelihood of the classical Hawkes process.
For discussions on maximum likelihood estimations for the stationary point process, see \citet{ogata} and \citet{ozaki1979maximum}.
The log-likelihood of the model up to time $T$ is
$$ \log \mathcal{L}(\theta, T) = \sum_{i=1}^{4} \left( \int_0^T \log \lambda_{i}(t)  \D N_{i}(t) - \int_0^T  \lambda_{i}(t) \D t \right),$$
where the intensity functions should be considered to be left continuous and $\theta$ denotes the set of model parameters.
We calculate the location of the maximum log-likelihood using a numerical optimizer in R; for more details, see \citet{nash2014nonlinear}.

We conduct simulations to check whether the numerical optimizer works well in our model.
We generate 500 simulated sample paths using presumed parameter values 
and then estimate the parameters for each sample path.
For the simulation of the Hawkes process, we can apply the thinning algorithm of \citet{ogata1981lewis} or the exact method of \citet{dassios2013exact}.
Thus, we obtain 500 estimation results for each parameter.
Each sample path has 10,000 observations of bid or ask price changes.
The presumed parameter values in the simulation are the estimates from the real stock data in the later empirical study.
Likewise, the distributions of the jump sizes in the price processes are based on empirical distributions.
Table~\ref{Table:simulation} summarizes the results.
The row labeled ``mean'' is the average of the 500 estimates calculated from the 500 sample paths and 
the row labeled ``std.'' is the standard deviation of the estimates.
The results show that the numerical estimator is unbiased with sufficiently small standard deviations.

\begin{table}
	\caption{Simulation study with 500 sample paths}\label{Table:simulation}
	\centering
	\begin{tabular}{cccccccccc}
		\hline
		& $\mu$ & $\eta$ & $\alpha_{s1}$ & $\alpha_{s2}$ & $\alpha_{m}$ & $\alpha_{w1}$ & $\alpha_{w2}$ & $\beta$ & $\xi$\\
		\hline
		True & 0.080 & 0.100 & 4.000 & 26.00 & 5.000 & 11.00 & 7.000 & 50.00 & 2.700 \\
		Mean & 0.080 & 0.100 & 3.993 & 26.02 & 5.003 & 11.02 & 6.973 & 50.03 & 2.705 \\
		Std. & 0.003 & 0.002 & 0.360 & 0.658 & 0.277 & 0.477 & 0.404 & 0.796 & 0.085 \\
		\hline
		True & 0.170 & 0.140 & 200.0 & 250.0 & 150.0 & 300.0 & 330.0 & 1200 & 50.00 \\
		Mean & 0.170 & 0.140 & 199.6 & 252.3 & 151.3 & 301.5 & 332.0 & 1207 & 50.04 \\
		Std. & 0.003 & 0.003 & 8.369 & 8.362 & 7.327 & 9.371 & 9.210 & 7.570 & 2.441 \\
		\hline
	\end{tabular}
\end{table}

Since the log-likelihood function is not convex, it may be questionable whether the numerical optimizer can find the maximum well.
The following simulation study demonstrates that if true $\beta$ is large and the initial value of $\beta$ is also large, 
then sometimes the appropriate convergence for $\alpha$s, $\xi$, and $\beta$ might not occur.
The difficulty can be mitigated with large sample sizes and small initial points of optimization.

\begin{figure}[hbt!]
	\centering
	\includegraphics[width=0.45\textwidth]{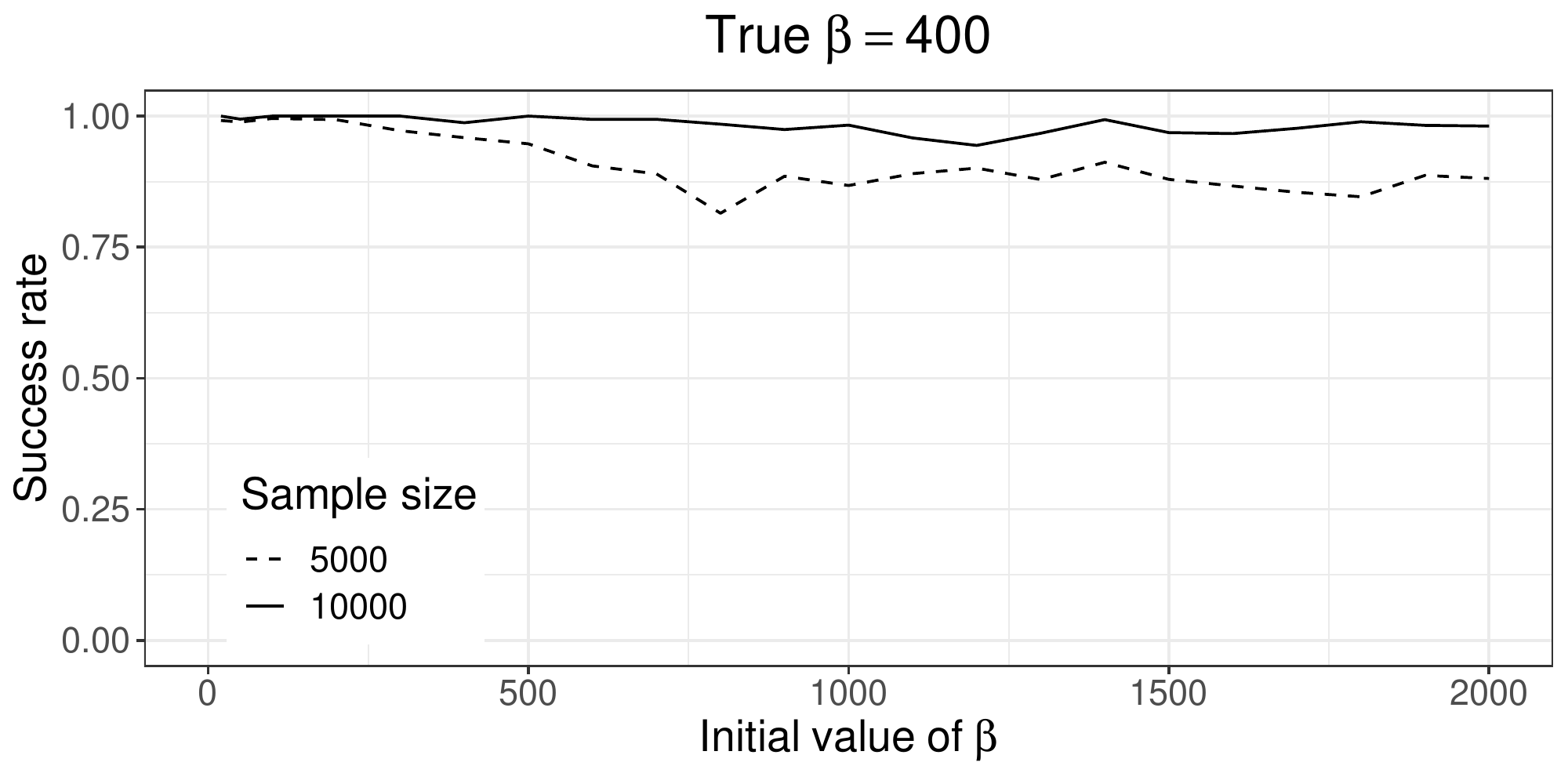} \quad
	\includegraphics[width=0.45\textwidth]{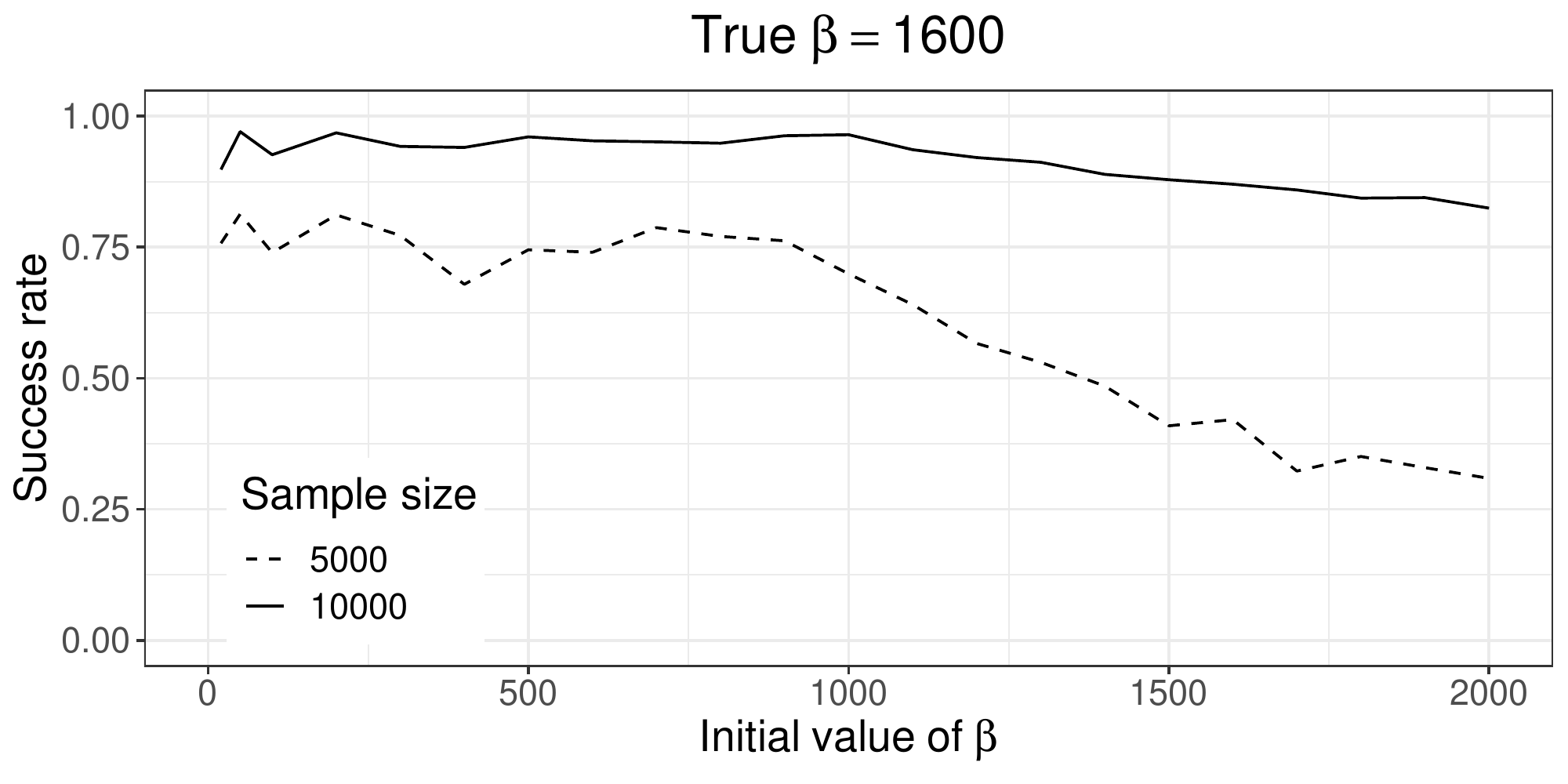}
	\caption{The success rate of estimation to find the global maximum with various initial points (x-axis) with true $\beta = 400$ (left) and $\beta = 1,600$ (right)}
	\label{Fig:success}
\end{figure}

Figure~{\ref{Fig:success}} is for simulation result testing convergence rate with true $\beta = 400$ in the left and true $\beta = 1,600$ in the right.
For other parameters, we use $\mu = 0.1, \eta = 0, \xi = \beta / 10 $ and all $\alpha$s are a quarter of $\beta$, for simplicity.
With various fixed initial $\beta_0$, represented in the x-axis, 
the rate of successful convergences is examined with 500 randomly generated sample paths.
The convergence is defined as success when the root mean squared error between true parameter values and estimates is less than 0.2
The initial values of $\mu_0$ and $\eta_0$ are randomly chosen between 0 and 10, and $\alpha$s and $\xi$ are also randomly chosen between 0 and $\beta_0$.

The left panel of the figure shows that when $\beta$ is relatively small such as 400, in both sample sizes 5,000 and 10,000,
the optimization converges well even with relatively large $\beta_0$.
For relatively large $\beta$ as in the right panel, we recommend relatively small initial points for $\beta$ and $\alpha$s.
Numerical optimization with a rather small initial value of $\beta$ leads to a successful result in most cases.
With sample size 10,000 and true $\beta = 1,600$, 
with reasonable choices of the initial values of $\beta$ such as around from 10 to 1,000,
the optimizations converge well.

\subsection{Data}~\label{subsect:data}

The high-frequency data for the analysis consist of the arrival time stamps and prices of the best bid and ask quotes from various stock exchanges in the US managed by the Consolidated Tape Association or Unlisted Trading Privileges (UTP).
We obtain the limit order information for NYSE-listed stocks from the Consolidated Tape System (CTS)
with time resolutions in milliseconds up to July 2015, microseconds up to around September 2018, and nanoseconds for the rest.
The series of time stamps of the Nasdaq-listed securities' limit order events managed by the UTP Plan are recorded in milliseconds up to July 2015, microseconds up to around October 2016, and nanoseconds for the rest.
The sample for the empirical study ranges from March 2009 to December 2019.
We use data from 10:00 to 15:30 to reduce the intraday seasonal effects of market opening and closing.

We focus on two types of best bid and ask price processes:
the national best bid and offer (NBBO) price processes 
and the best bid and offer (BBO) price processes per stock exchange,
as the US stock market is highly fragmented. 
See Table~\ref{Table:ex} for more details.
More precisely, the NBBO price data consist of the official NBBO data from the NYSE and the quotes completed by Wharton Research Data Services (WRDS).
The quotes supplemented by the WRDS are those that contain both the national best bid and ask prices that the NYSE does not add\footnote{https://wrds-www.wharton.upenn.edu/pages/support/data-overview/wrds-overview-taq/}.
Because only the displayed limit orders generate the quote data, we base our analysis solely on these values.
In addition, since our model depends on the spread, 
we use samples with a sufficient spread (i.e., with prices sufficiently large, around \$100 or higher, relative to the tick size \$0.01).

\begin{table}
	\centering
	\caption{US stock market exchanges}\label{Table:ex}
	\begin{tabular}{ll}
		\hline
		Abbreviation & Exchange \\
		\hline
		NYSE & New York Stock Exchange \\
		NSDQ & Nasdaq Stock Market \\
		Arca & NYSE Arca \\
		BZX & Cboe BZX exchange, formerly Bats Global Markets  \\
		BYX & Cboe BYX exchange, formerly Bats-Y Global Markets \\
		EDGA & Cboe EDGA exchange, formerly Direct Edge Exchange A \\
		EDGX & Cboe EDGX exchange, formerly Direct Edge Exchange X \\
		BSTN & Nasdaq OMX BX, formerly Boston Stock Exchange \\
		PSX & Nasdaq PSX \\
		NSX & National Cincinnati Stock Exchange  \\
		IEX & Investors Exchange \\
		\hline
	\end{tabular}	
\end{table}

Table~\ref{Table:percentage} shows the percentage of NBBO contributions by exchange for the IBM and AMZN stocks.
The NYSE, Arca, NSDQ, and BZX mainly determine the NBBO for the NYSE-listed stock IBM.
The Arca, NSDQ, BZX, and, recently, PSX determine the NBBO for the Nasdaq-listed stock AMZN.
A higher percentage indicates more substantial liquidity provided by the corresponding exchange.

\begin{table}[t]
	\caption{Percentage of BBOs from each exchange in the NBBO}\label{Table:percentage}
	\small
	\centering
	\begin{tabular}{ccccccccccc}
		\hline
		& \multicolumn{10}{c}{IBM}\\
		Year & NYSE & Arca & NSDQ & BZX & BYX & EDGA & EDGX & BSTN & PSX & NSX \\
		\hline
		2009 & 20.1 & 20.3 & 27.6 & 22.9 & 0.0 & 0.0 & 0.0 & 1.8 & 0.0 & 2.8 \\
		2010 & 45.8 & 14.6 & 26.6 & 6.6 & 0.0 & 0.8 & 1.0 & 1.3 & 0.0 & 1.0 \\
		2011 & 20.2 & 11.5 & 20.7 & 8.0 & 10.4 & 7.1 & 3.3 & 11.4 & 2.0 & 5.3 \\
		2012 & 16.2 & 12.1 & 23.2 & 7.8 & 8.8 & 2.1 & 5.3 & 10.7 & 0.4 & 13.0\\
		2013 & 22.5 & 13.0 & 18.8 & 11.1 & 5.8 & 3.8 & 7.2 & 7.6 & 2.3 & 6.8 \\
		2014 & 16.0 & 10.5 & 25.8 & 8.1 & 11.7 & 7.0 & 6.6 & 6.4 & 4.3 & 2.9 \\
		2015 & 25.3 & 10.3 & 25.6 & 7.7 & 9.7 & 5.8 & 5.9 & 6.0 & 3.6 & 0.0 \\
		2016 & 50.1 & 8.8 & 23.4 & 4.8 & 1.9 & 0.6 & 4.8 & 2.1 & 1.1 & 0.0 \\
		2017 & 35.5 & 12.0 & 23.7 & 8.2 & 2.8 & 1.0 & 8.8 & 3.0 & 1.7 & 0.0 \\
		2018 & 27.5 & 6.5 & 18.9 & 16.3 & 2.8 & 1.1 & 6.5 & 3.5 & 2.8 & 2.1 \\
		\hline
		& \multicolumn{10}{c}{AMZN}\\
		2009 & 0.0 & 17.5 & 43.7 & 27.0 & 0.0 & 0.0 & 0.0 & 3.6 & 0.0 & 1.9 \\
		2010 & 0.0 & 18.3 & 52.3 & 11.1 & 0.0 & 0.8 & 1.7 & 1.2 & 0.0 & 10.8 \\
		2011 & 0.0 & 17.7 & 23.3 & 6.6 & 8.2 & 7.3 & 3.2 & 11.9 & 2.0 & 18.1 \\
		2012 & 0.0 & 15.4 & 21.1 & 5.7 & 8.9 & 2.9 & 4.2 & 31.0 & 1.5 & 9.0 \\
		2013 & 0.0 & 9.6 & 21.2 & 5.4 & 7.0 & 6.2 & 6.8 & 21.6 & 8.8 & 12.7 \\
		2014 & 0.0 & 10.4 & 31.1 & 5.8 & 11.6 & 10.0 & 9.8 & 10.6 & 7.8 & 0.7 \\
		2015 & 0.0 & 14.2 & 33.2 & 8.0 & 9.8 & 6.3 & 8.2 & 13.9 & 6.5 & 0.0 \\
		2016 & 0.0 & 15.3 & 31.3 & 8.7 & 3.6 & 1.1 & 7.1 & 1.4 & 8.7 & 0.0 \\
		2017 & 0.0 & 7.3 & 24.9 & 6.1 & 3.2 & 0.4 & 4.0 & 2.4 & 33.8 & 0.0\\
		2018 & 0.0 & 3.2 & 21.0 & 7.5 & 0.8 & 0.2 & 2.6 & 0.4 & 22.4 & 19.1 \\
		\hline
	\end{tabular}
\end{table}

Because it is not possible to apply some of the original data directly to the Hawkes estimation, we apply data preprocessing.
For example, some records of different quotes have the same time stamps.
Because the probability of two or more events occurring at the same time is almost surely zero in the Hawkes-type model, 
these time records require modification.
We assume that the same time quotes are nearly simultaneously released limit orders and 
we relocate the times of these records to equidistant arrivals within the time resolution.

In some cases, the best bid and ask prices change at the same time.
Similar to the previous case, we separate the orders into two successive events 
that we assume occur sequentially within the time resolution.
Whether we assume that the bid or ask price changes first depends on the spread.
For example, if we assume that the bid quote occurs first and the bid price becomes greater than or equal to the ask price as a result, 
then the assumption is false.
In this case, it is reasonable to assume that the ask quote occurs first.
If it does not matter which quote occurs first, then we randomly select the order of the quotes.

Before March 2009, the percentage of data with these issues is as high as 20\%.
From March 2009 (more precisely, February 25, 2009), the quality of the data improves dramatically.
For example, from March 2009, around 2\% of the data have the above issues, 
and in 2018, less than 0.1\% of the data do.
Because the proportion of abnormal data is small,
we assume that the data preprocessing described above does not have a significant impact on our statistical analysis.

In addition, we need to modify the NBBO data before the estimation.
The national best bid and ask prices can be equal or sometimes reversed for reasons such as network delays.
The following is an example of IBM's NBBO on January 3, 2018.
At time 10:03:09.057956, the national best bid price was \$159.33, disclosed by exchange X (Nasdaq PSX).
After about a second, the new national best ask price was \$159.33, disclosed by exchange P (NYSE Arca).

\vspace*{5mm}
\begin{tabular}{lccc}
	\hline
	Time &  Bid & Ask & Exchange \\
	\hline
	10:03:09.057956 & 159.33 & & X \\
	\multicolumn{4}{c}{$\vdots$} \\
	10:03:10.107464 &  & 159.33 & P \\
	10:03:10.107586 & 159.32 & & X \\
	\hline
\end{tabular}
\vspace*{5mm}

To conserve space, we do not show the best ask prices for all the exchanges, 
but \$159.33 was the prevailing ask price across all the exchanges at this moment.
However, because the best bid price was still \$159.33, this best ask price was not logically correct. 
This result may have been due to network or computational delays,
and we consider the following scenario in this case.

The following table lists the transaction details near the time of the event above.
Using these details, we can presume the following scenario, although it is not possible to know the actual process exactly.

\vspace*{5mm}
\begin{tabular}{lcc}
	\hline
	Time & Price & Exchange \\
	\hline
	10:03:09.060686 & 159.33 & P \\
	10:03:09.150237 & 159.33 & K \\
	10:03:10.107383 & 159.33 & K \\ 
	10:03:10.107446 & 159.33 & K \\
	10:03:10.107650 & 159.33 & X \\
	\hline
\end{tabular}
\vspace*{5mm}

A large sell order arrived at exchange P at time 10:03:09.060686.
It consumed all existing bid limit orders at exchange P, but was still not filled completely.
This order was routed to other exchanges, 
and the trading system noticed that the volume was still insufficient across all the exchanges.
For the existing matched quantity, 
the transactions proceeded in parallel on exchanges K and X, 
and the insufficient quantity was posted as an ask limit order on exchange P at time 10:03:10.107464, as in the previous table.
The transaction processes took more time and 
were eventually finalized in exchange X at time 10:03:10.107586. The best bid price was updated
when all the limit bid orders were consumed at \$159.33.
Transaction reporting ended slightly later and finally finished at 10:03.10.107650 (combined with the previous transactions in exchange K).
For physical reasons, this type of delay seems to be inevitable.
For example, \citet{ding2014slow} report that the discrepancy between the NBBOs and synthetic NBBOs calculated using data from the exchanges is around one to two milliseconds. 

Events that violate the assumptions of our model constitute about 1--2\% of all events.
Because the data set is so large, it is impossible to identify the cause of these anomalies in each case.
We assume that these data do not significantly change the statistical properties 
and proceed with the estimation after removing these records.

\subsection{Estimation results with the NBBO}

\subsubsection{Model selection}\label{subsec:selection}

This subsection covers the results of various model selection issues.
First, we discuss the estimation results from the basic Hawkes model for the bid and ask price processes, even though the model is unsuitable because it does not guarantee the positivity and reverting property of the spread process.
Consider the basic Hawkes process with constant parameters:
$$
\textrm{Basic Hawkes model: }
\bm{\mu}_t = 
\begin{bmatrix} 
\mu \\
\mu \\
\mu \\
\mu
\end{bmatrix},
\quad
\bm{h}(t, u) =
\e^{-\beta(t-u)}
\begin{bmatrix} 
\alpha_{11} & \alpha_{12} & \alpha_{13} & \alpha_{14} \\
\alpha_{21} & \alpha_{22} & \alpha_{23} & \alpha_{24} \\
\alpha_{31} & \alpha_{32} & \alpha_{33} & \alpha_{34} \\
\alpha_{41} & \alpha_{42} & \alpha_{43} & \alpha_{44} \\
\end{bmatrix}.
$$
We can obtain some intuitions by fitting the data to the basic Hawkes model.

\begin{figure}[!hbt]
	\begin{subfigure}{.5\textwidth}
		\centering
		\includegraphics[width=0.94\textwidth]{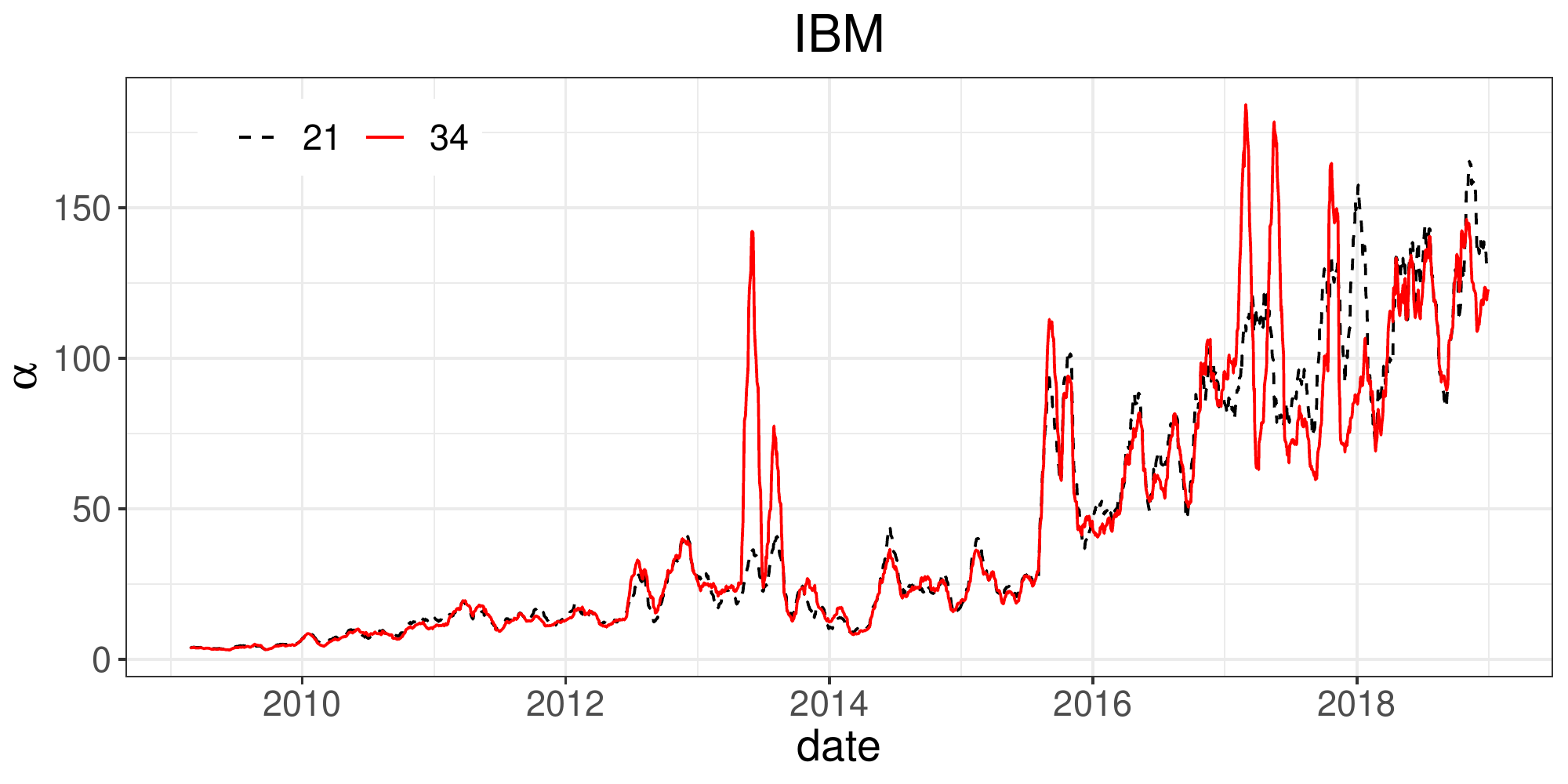}
		\caption{$\alpha_{21}$ and $\alpha_{34}$}
		\label{fig:IBM_NYSE_Hawkes_alpha21_34}
	\end{subfigure}
	\begin{subfigure}{.5\textwidth}
		\centering
		\includegraphics[width=0.94\textwidth]{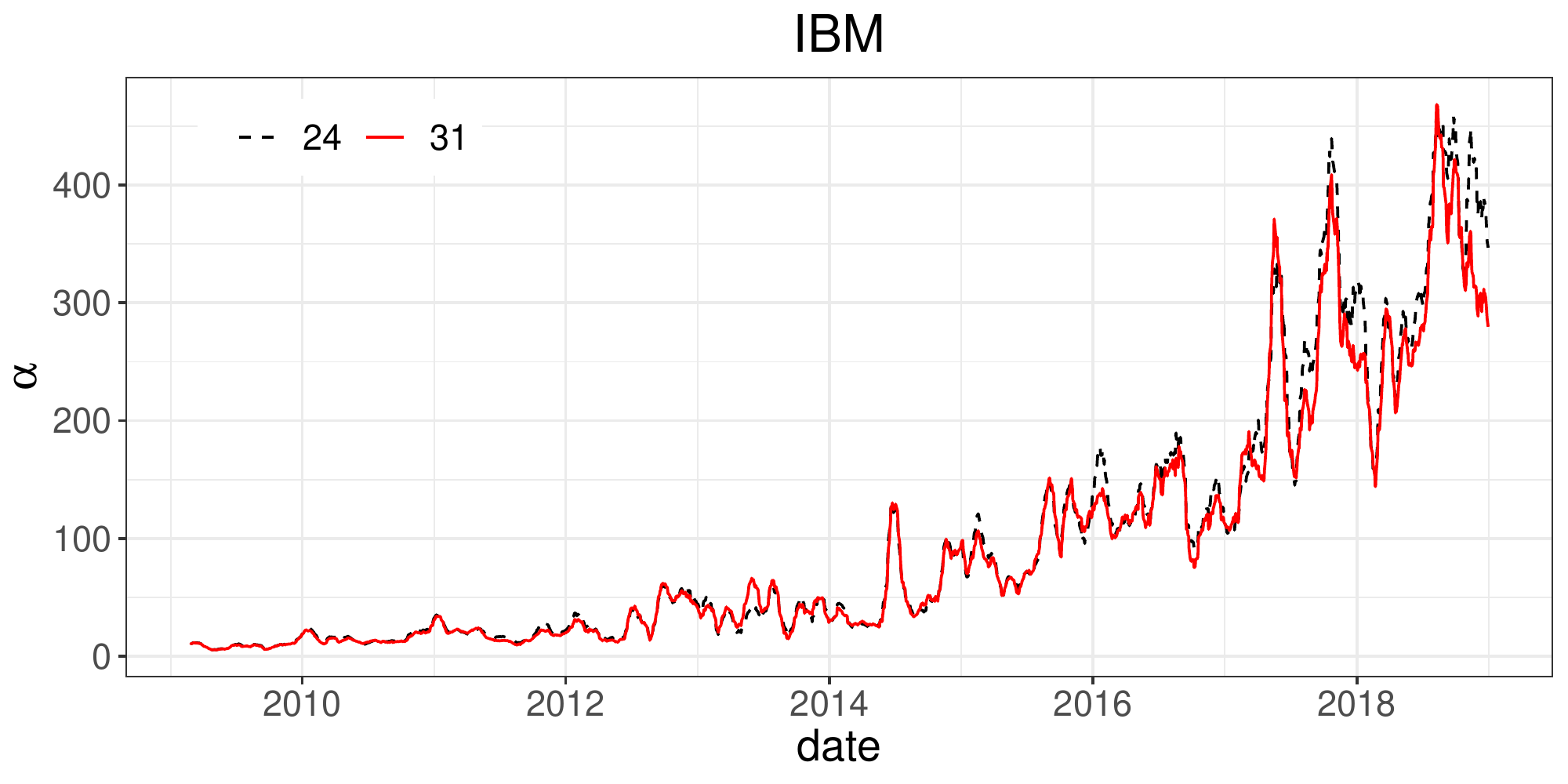}
		\caption{$\alpha_{24}$ and $\alpha_{31}$}
		\label{fig:IBM_NYSE_Hawkes_alpha24_31}
	\end{subfigure}
	
	\begin{subfigure}{.5\textwidth}
		\centering
		\includegraphics[width=0.94\textwidth]{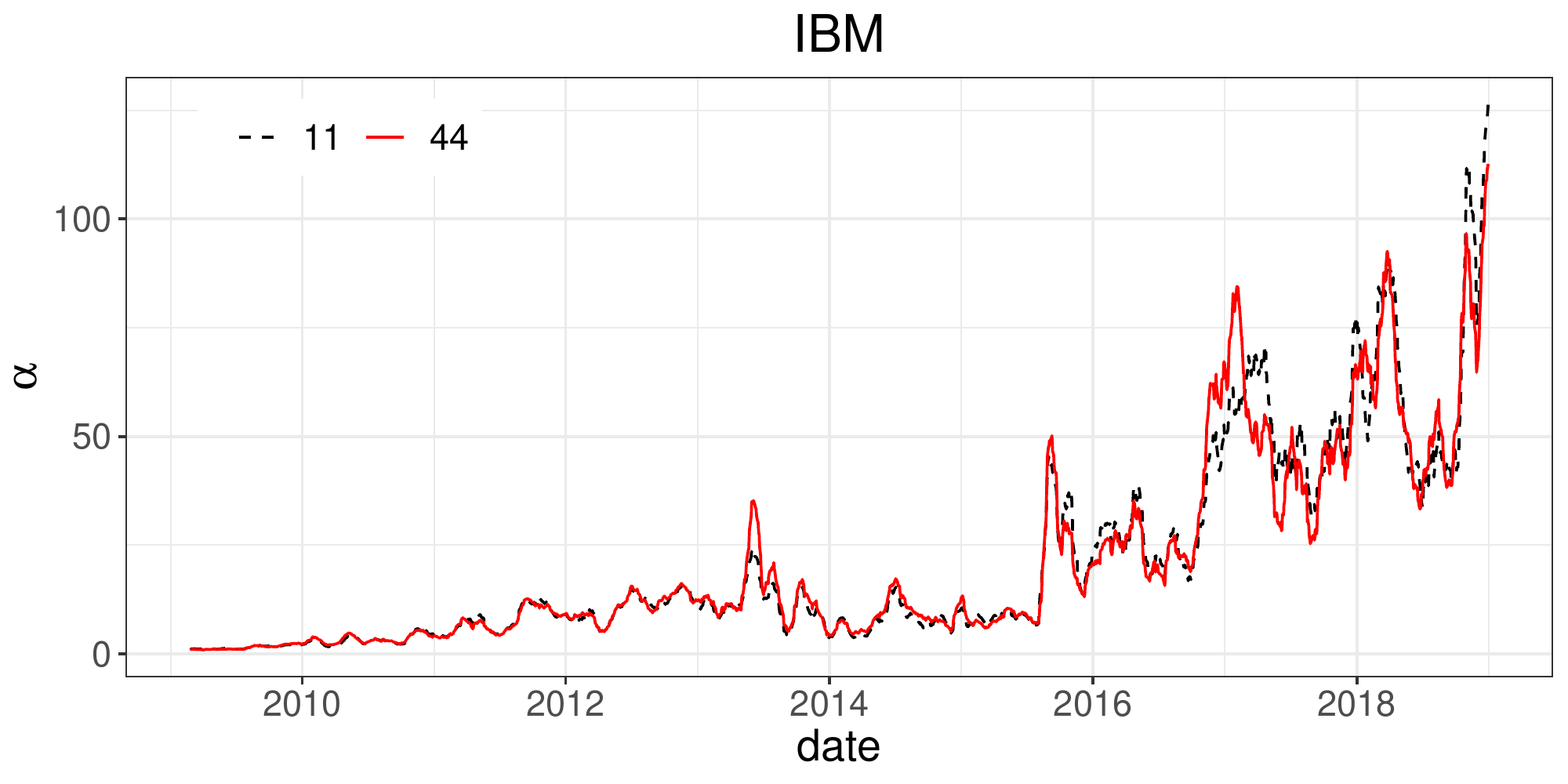}
		\caption{$\alpha_{11}$ and $\alpha_{44}$}
		\label{fig:IBM_NYSE_Hawkes_alpha11_44}
	\end{subfigure}
	\begin{subfigure}{.5\textwidth}
		\centering
		\includegraphics[width=0.94\textwidth]{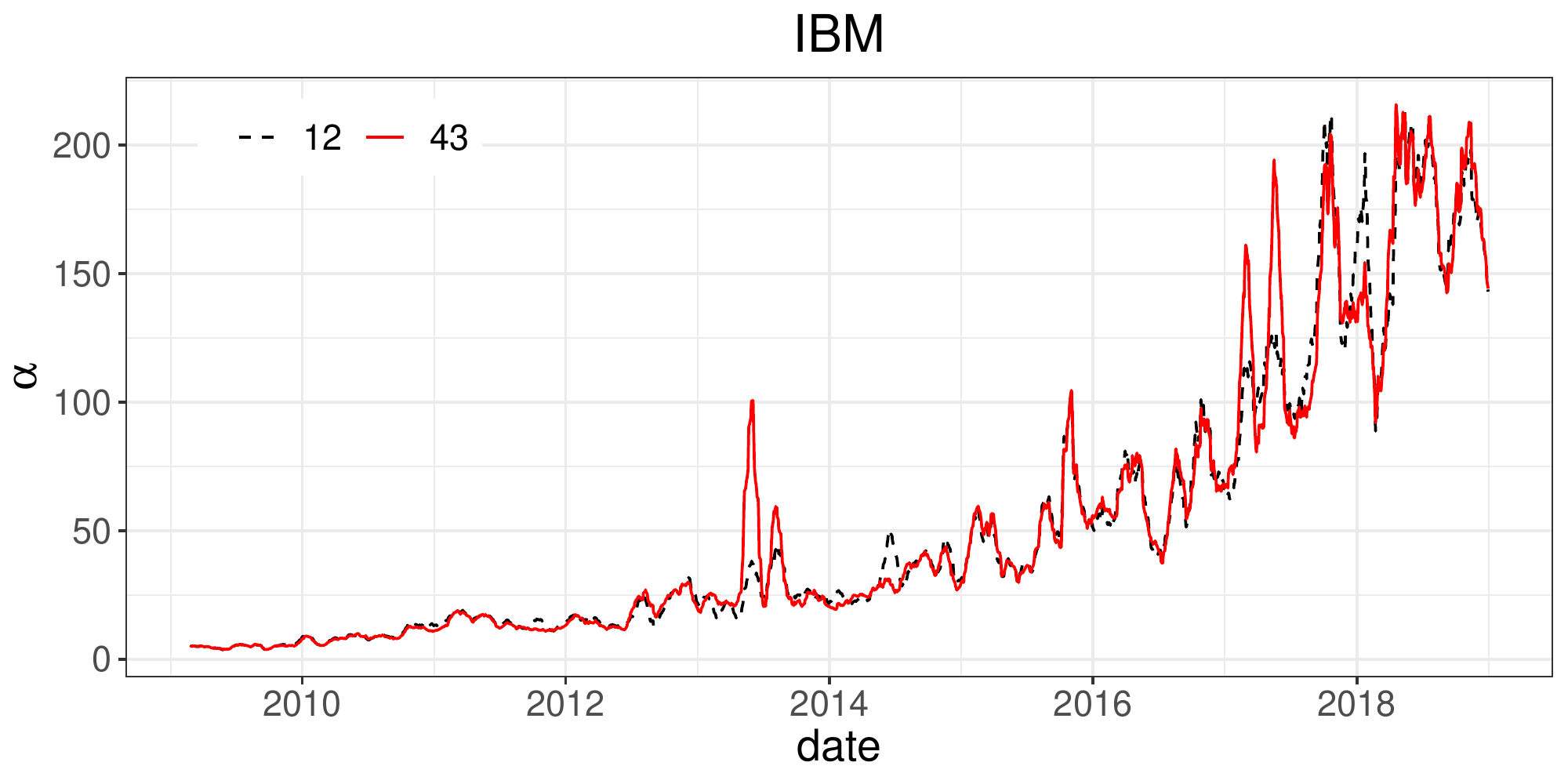}
		\caption{$\alpha_{12}$ and $\alpha_{43}$}
		\label{fig:IBM_NYSE_Hawkes_alpha12_43}
	\end{subfigure}
	
	\begin{subfigure}{.5\textwidth}
		\centering
		\includegraphics[width=0.94\textwidth]{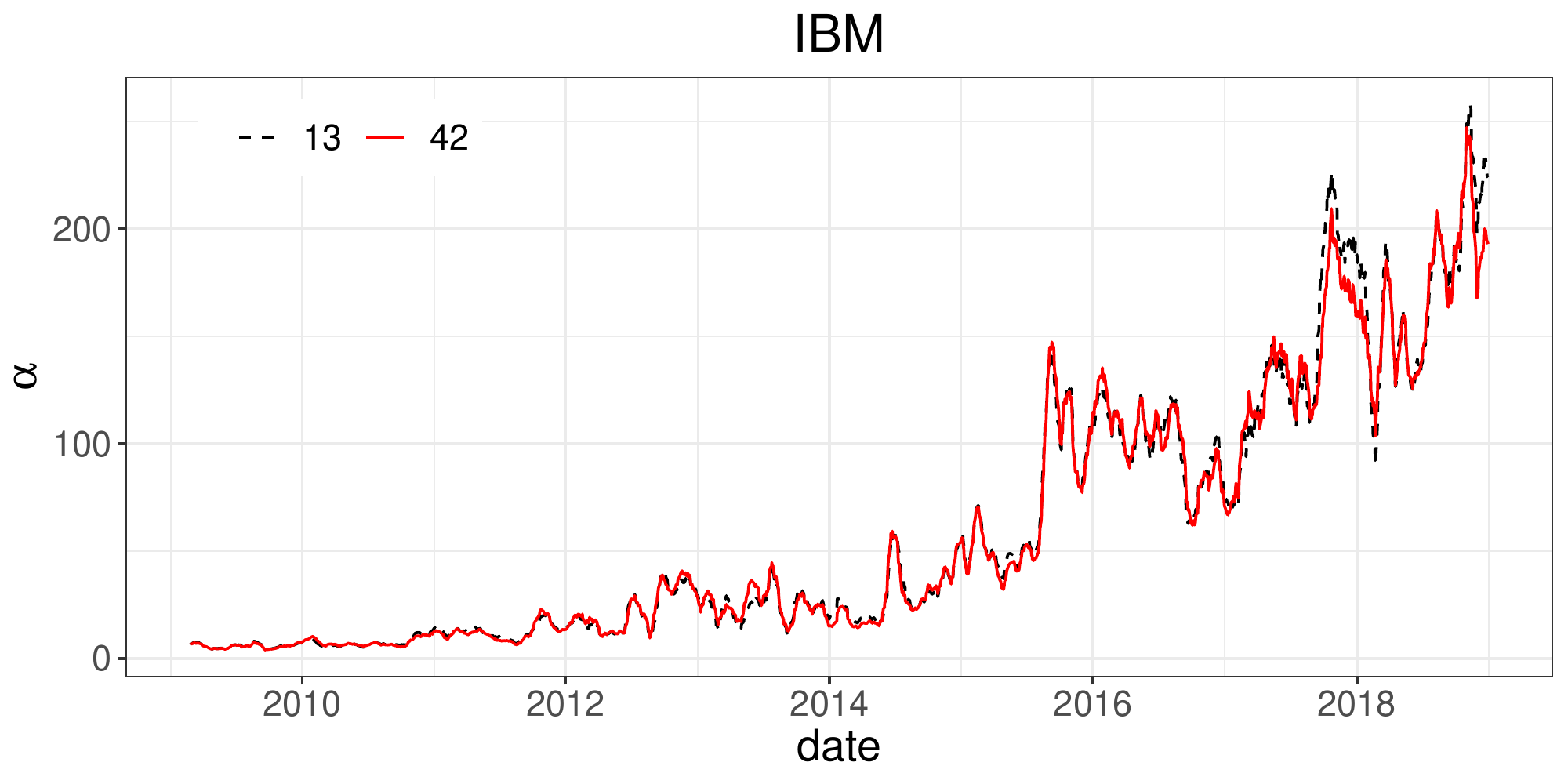}
		\caption{$\alpha_{13}$ and $\alpha_{42}$}
		\label{fig:IBM_NYSE_Hawkes_alpha13_42}
	\end{subfigure}
	\begin{subfigure}{.5\textwidth}
		\centering
		\includegraphics[width=0.94\textwidth]{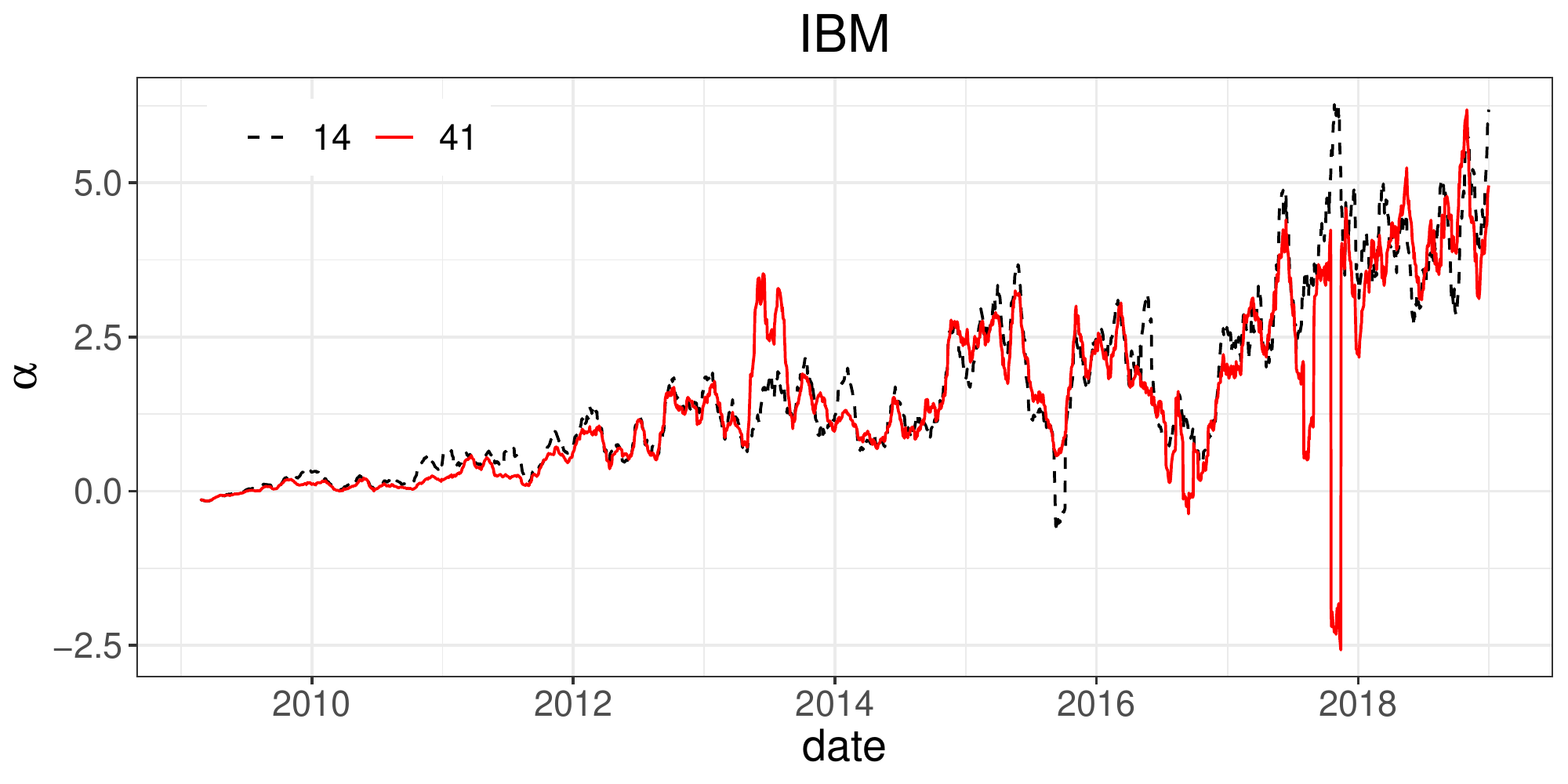}
		\caption{$\alpha_{14}$ and $\alpha_{41}$}
		\label{fig:IBM_NYSE_Hawkes_alpha14_41}
	\end{subfigure}
	
	\caption{Daily estimates for the NBBO of IBM under the basic Hawkes model from 2009 to 2018}
	\label{Fig:IBM_NYSE_Hawkes}
\end{figure}

Figure~\ref{fig:IBM_NYSE_Hawkes_alpha21_34} plots the daily estimated $\alpha_{21}$ and $\alpha_{34}$, while Figure~\ref{fig:IBM_NYSE_Hawkes_alpha24_31} plots $\alpha_{24}$ and $\alpha_{31}$.
These terms represent the excitation terms of the upward movements of the ask price and downward movements of the bid price caused by events that widen the spread (i.e., $A^u$ and $B^d$).
To improve the clarity of the graph, we omit the standard errors, but note that these values are significantly greater than zero.
The estimates of $\alpha_{24}$ and $\alpha_{31}$ are similar,
as are those of $\alpha_{21}$ and $\alpha_{34}$.
The parameters $\alpha_{21}$ and $\alpha_{34}$ are related to $\alpha_{w1}$ in our model,
whereas $\alpha_{24}$ and $\alpha_{31}$ are related to $\alpha_{w2}$.

Figures~\ref{fig:IBM_NYSE_Hawkes_alpha11_44}~and~\ref{fig:IBM_NYSE_Hawkes_alpha12_43} present the estimates of $\alpha_{11}, \alpha_{44}, \alpha_{12},$ and $\alpha_{43}$.
The parameters $\alpha_{11}$ and $\alpha_{44}$ are typical self-exciting terms in the Hawkes model and their values are similar.
The parameters $\alpha_{12}$ and $\alpha_{43}$ are typical mutually exciting terms in the Hawkes model and, again, they have similar values.
Figure~\ref{fig:IBM_NYSE_Hawkes_alpha13_42} illustrates the other mutually exciting terms $\alpha_{13}$ and $\alpha_{42}$, showing that these values are also positive and similar.

Figure~\ref{fig:IBM_NYSE_Hawkes_alpha14_41} shows the estimates of $\alpha_{14}$ and $\alpha_{41}$.
These values are much lower than those of the other parameters.
Thus, if the ask price is up, then the tendency of the bid price to decline is relatively less significant.
We can apply the same argument to the case in which the bid price is down.
Because these values are relatively small, we assume that the corresponding parameters in our proposed model are zeros for model parsimony.
We discuss more details in the next model comparison part.

\begin{figure}[!hbt]
	\begin{subfigure}{.5\textwidth}
		\centering
		\includegraphics[width=0.94\textwidth]{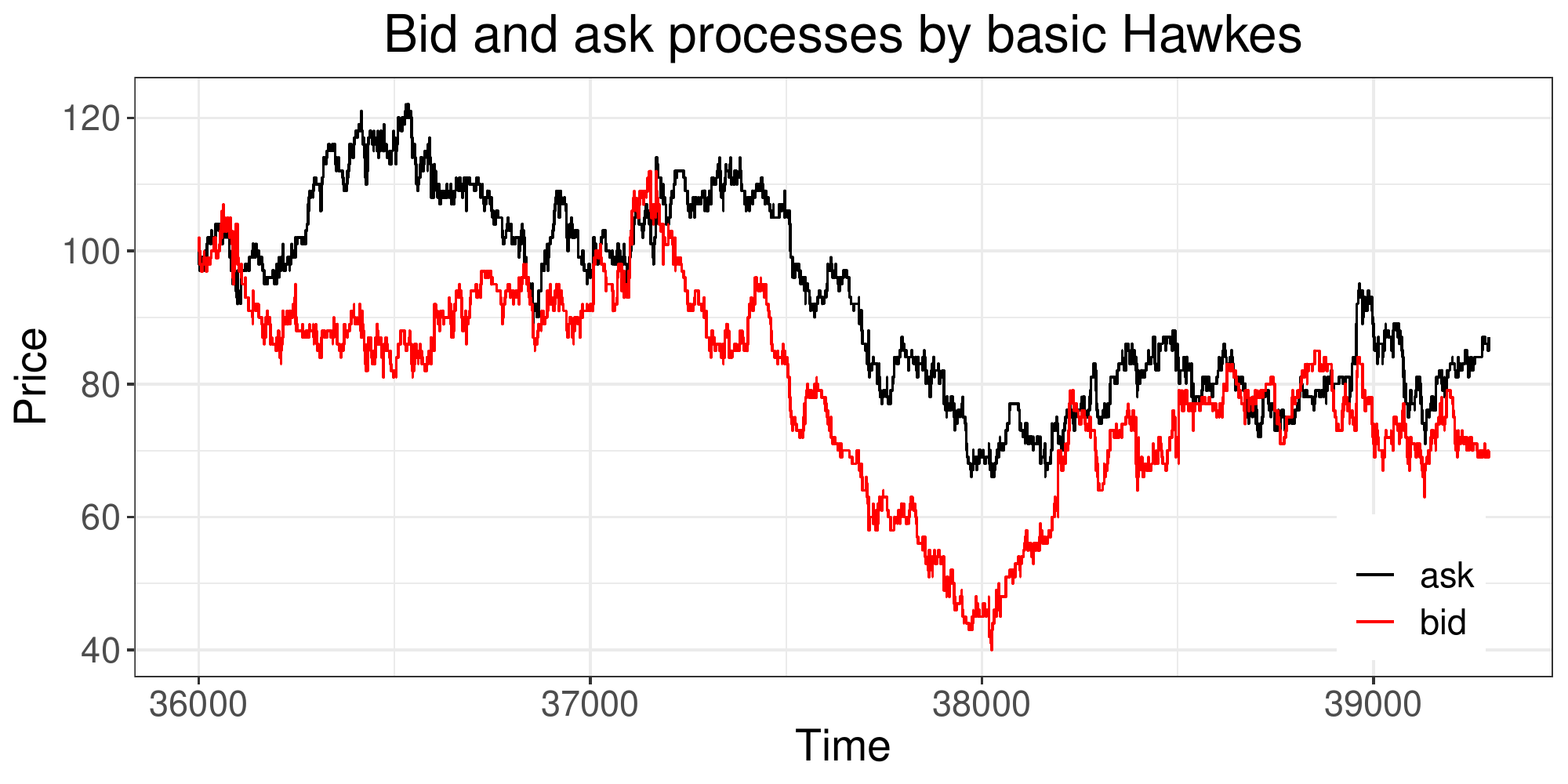}
		\caption{basic Hawkes}
		\label{fig:basic}
	\end{subfigure}
	\begin{subfigure}{.5\textwidth}
		\centering
		\includegraphics[width=0.94\textwidth]{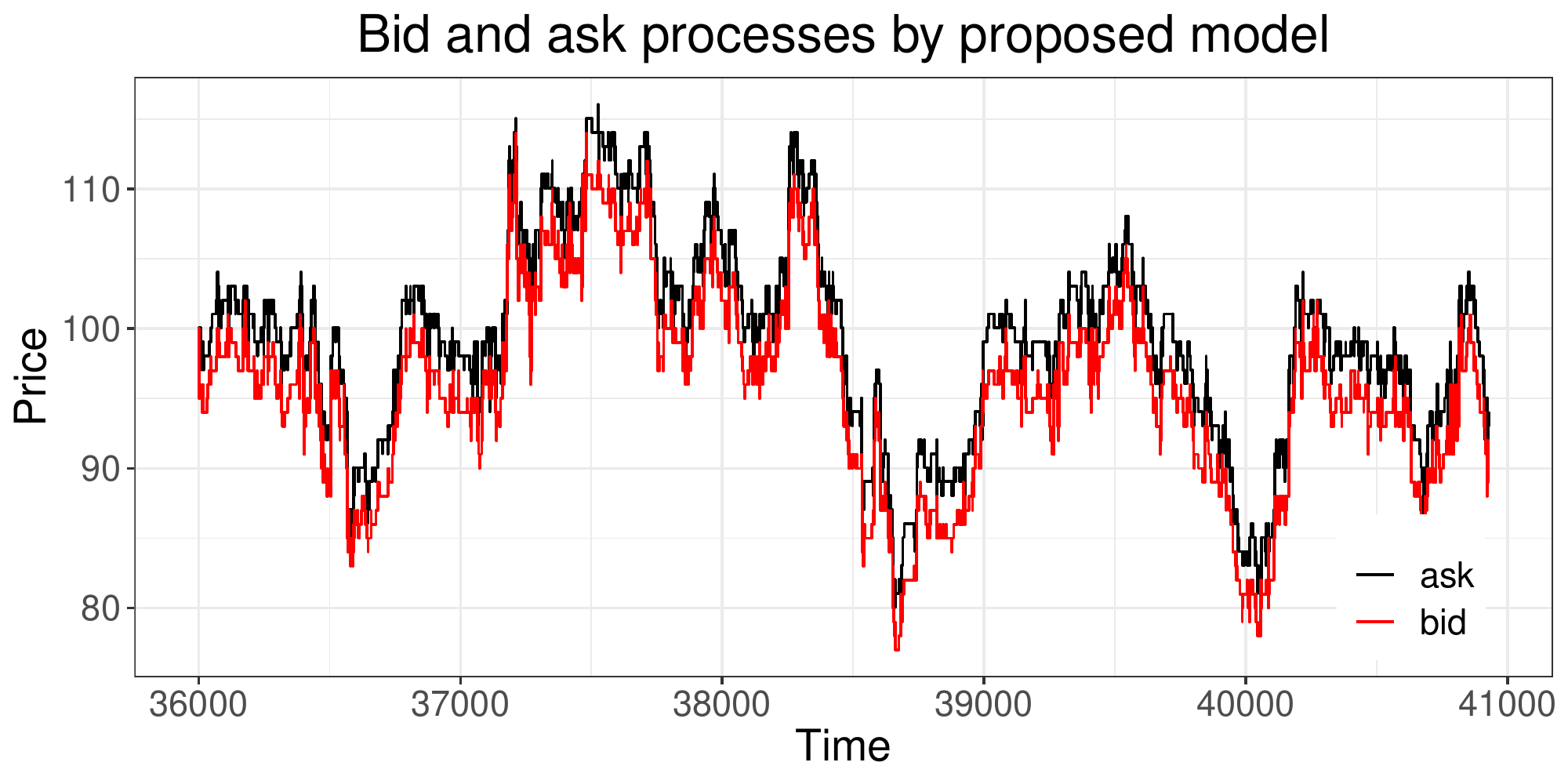}
		\caption{proposed model}
		\label{fig:our}
	\end{subfigure}
	\caption{Comparison of the bid and ask processes between the basic Hawkes model and proposed model}
	\label{Fig:simulation_path}
\end{figure}

To highlight the difference between the basic Hawkes model and proposed model, Figure~\ref{Fig:simulation_path} shows the sample paths for the bid and ask processes.
For the simulation, the parameter values are assumed to be one of the estimates obtained in the empirical analysis.
As shown in Figure~\ref{fig:basic}, the basic Hawkes model has negative or unrealistically wide spreads.
The proposed model, on the contrary, incorporates the basic nature of the bid-ask spread.

\begin{figure}[!hbt]
	\begin{subfigure}{.5\textwidth}
		\centering
		\includegraphics[width=0.94\textwidth]{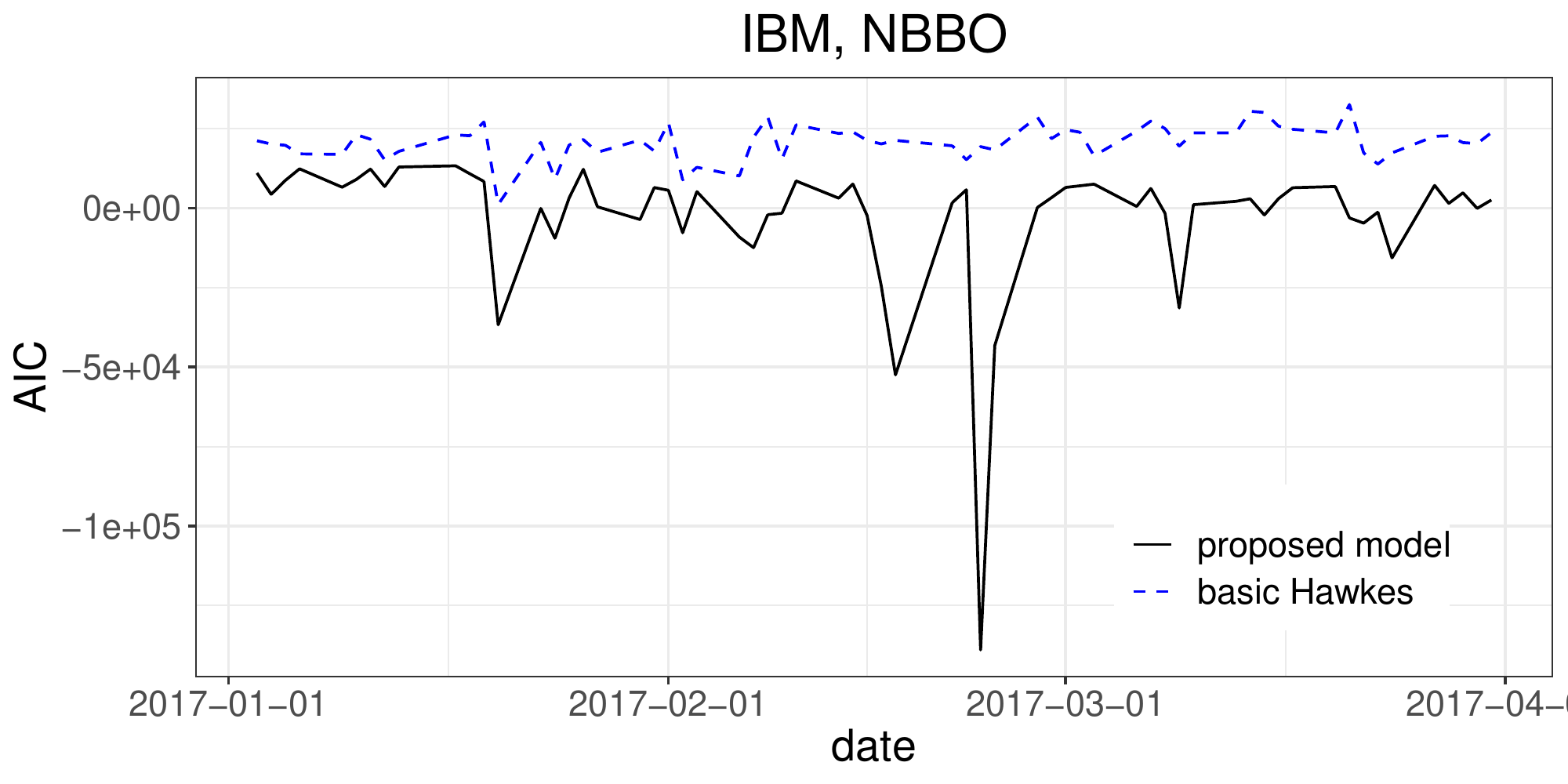}
		\caption{AIC}
		\label{fig:AIC}
	\end{subfigure}
	\begin{subfigure}{.5\textwidth}
		\centering
		\includegraphics[width=0.94\textwidth]{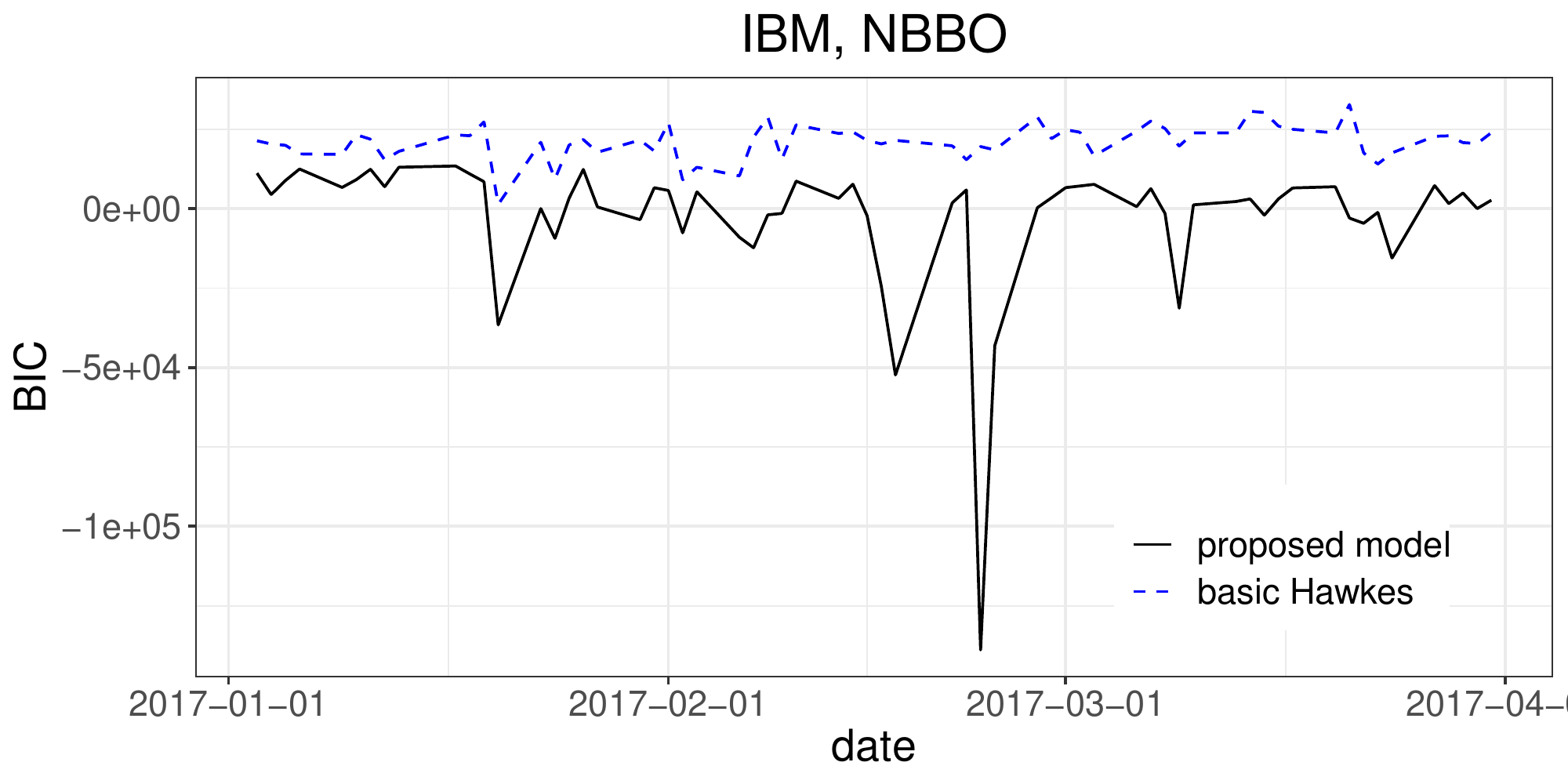}
		\caption{BIC}
		\label{fig:BIC}
	\end{subfigure}
	\caption{Comparison of the AIC and BIC between the proposed model and basic Hawkes model}
	\label{Fig:AIC_BIC}
\end{figure}

Figure~\ref{Fig:AIC_BIC} shows the Akaike information criterion (AIC) and Bayesian information criterion (BIC) using IBM's NBBO data for the first quarter of 2017.
The proposed model (solid black line) is preferred, as it has lower values than the basic Hawkes model (dashed blue line) in this period.

\begin{figure}[!hbt]
	\begin{subfigure}{.5\textwidth}
		\centering
		\includegraphics[width=0.94\textwidth]{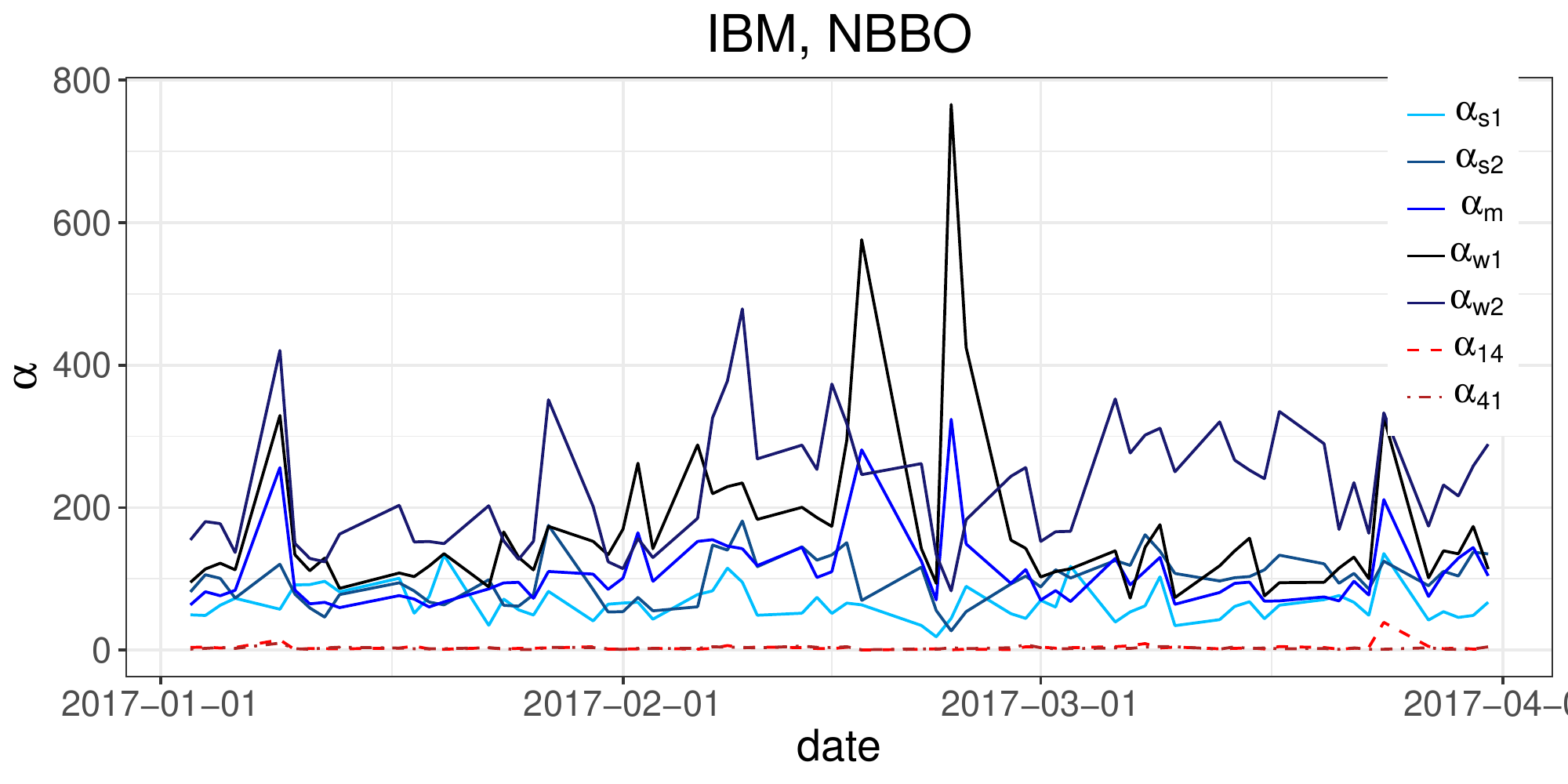}
		\caption{$\alpha$s}
		\label{fig:alphas}
	\end{subfigure}
	\begin{subfigure}{.5\textwidth}
		\centering
		\includegraphics[width=0.94\textwidth]{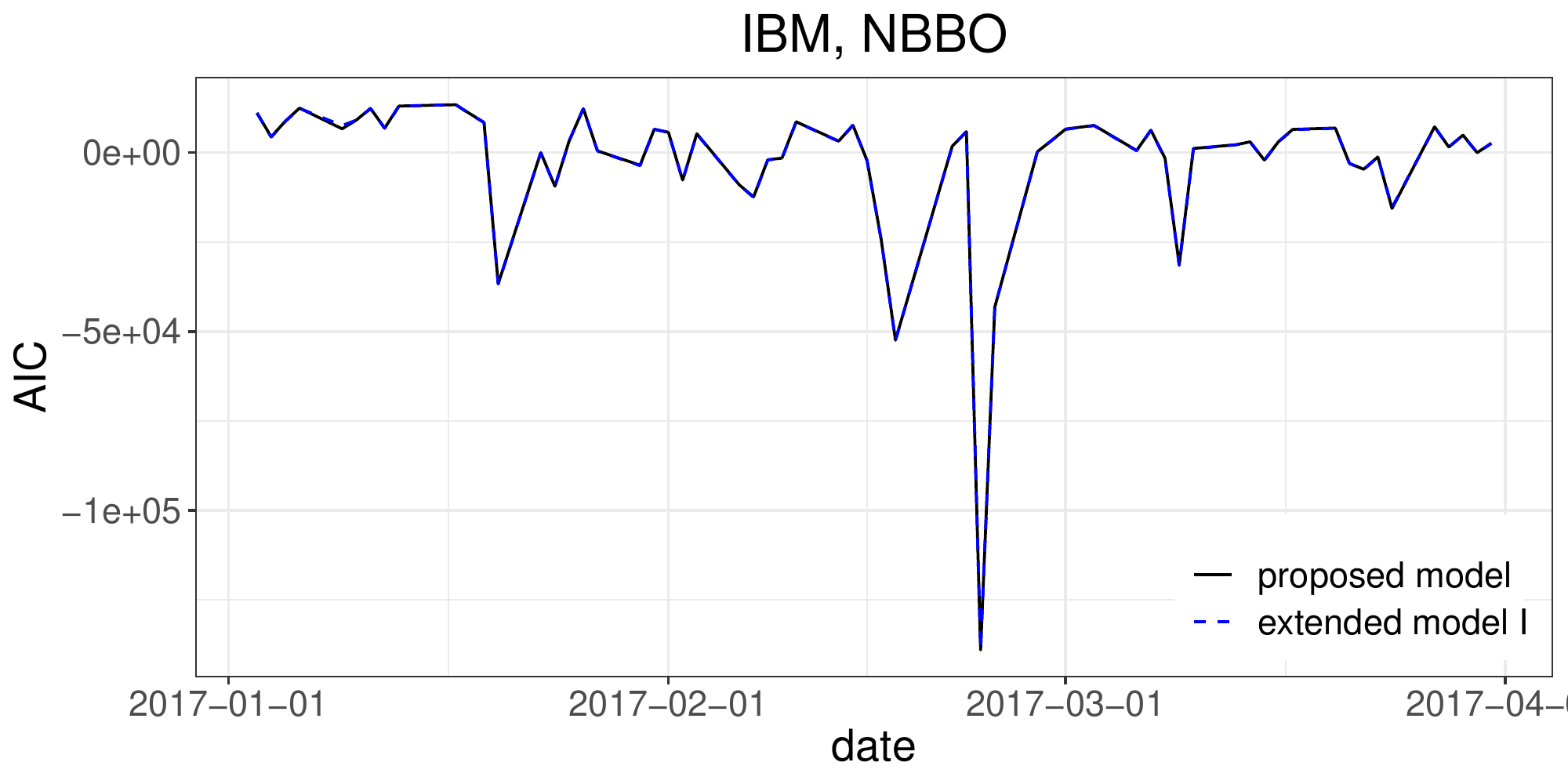}
		\caption{AIC}
		\label{fig:AIC_em1}
	\end{subfigure}
	\caption{Estimates of the $\alpha$s under extended model I (left) and comparison of the AIC between the proposed model and extended model I (right)}
	\label{Fig:AIC1}
\end{figure}

Second, we examine the model assuming that the coefficients of the top right and bottom left of $\mathbf h$ are not zeros.
$$
\textrm{Extended model I: }\quad
\bm{h}(t, u) =
\e^{-\beta(t-u)}
\begin{bmatrix}
\alpha_{s1} & \alpha_{m} & \alpha_{s2} & \alpha_{14} \\
\alpha_{w_1} & \alpha_{n1}(u) & \alpha_{n1}(u) & \alpha_{w_2} \\
\alpha_{w_2} & \alpha_{n2}(u) & \alpha_{n2}(u) & \alpha_{w_1} \\
\alpha_{41} & \alpha_{s2} & \alpha_{m} & \alpha_{s1} \\
\end{bmatrix}
$$
and the other parts are as in Definition~\ref{Def:model}.
The assumption in our model that the corner coefficients of $\mathbf h$ are zeros does not imply
that these coefficients are always zeros.
Based on the empirical results analyzed daily, 
on some of the days, the null hypothesis that the corner coefficients are zeros is not rejected
while on other days this hypothesis is rejected.
However, even on the day that this hypothesis is rejected, the coefficients have smaller values than the other coefficients in the matrix $\bm h$, as shown in Figure~\ref{fig:alphas}.
The estimates of $\alpha_{14}$ and $\alpha{41}$ are shown as the red lines in the figure.
These parameters may only be significant on certain days when unexpected news causes a liquidity shock with sudden spread widening \citep{jiang2011information,christensen2014fact}
and can reflect the bilateral spread-widening event.
However, the inclusion of these two parameters does not significantly improve the model from the AIC point of view,
as shown in Figure~\ref{fig:AIC_em1} on typical days.

\begin{figure}[!hbt]
	\begin{subfigure}{.5\textwidth}
		\centering
		\includegraphics[width=0.94\textwidth]{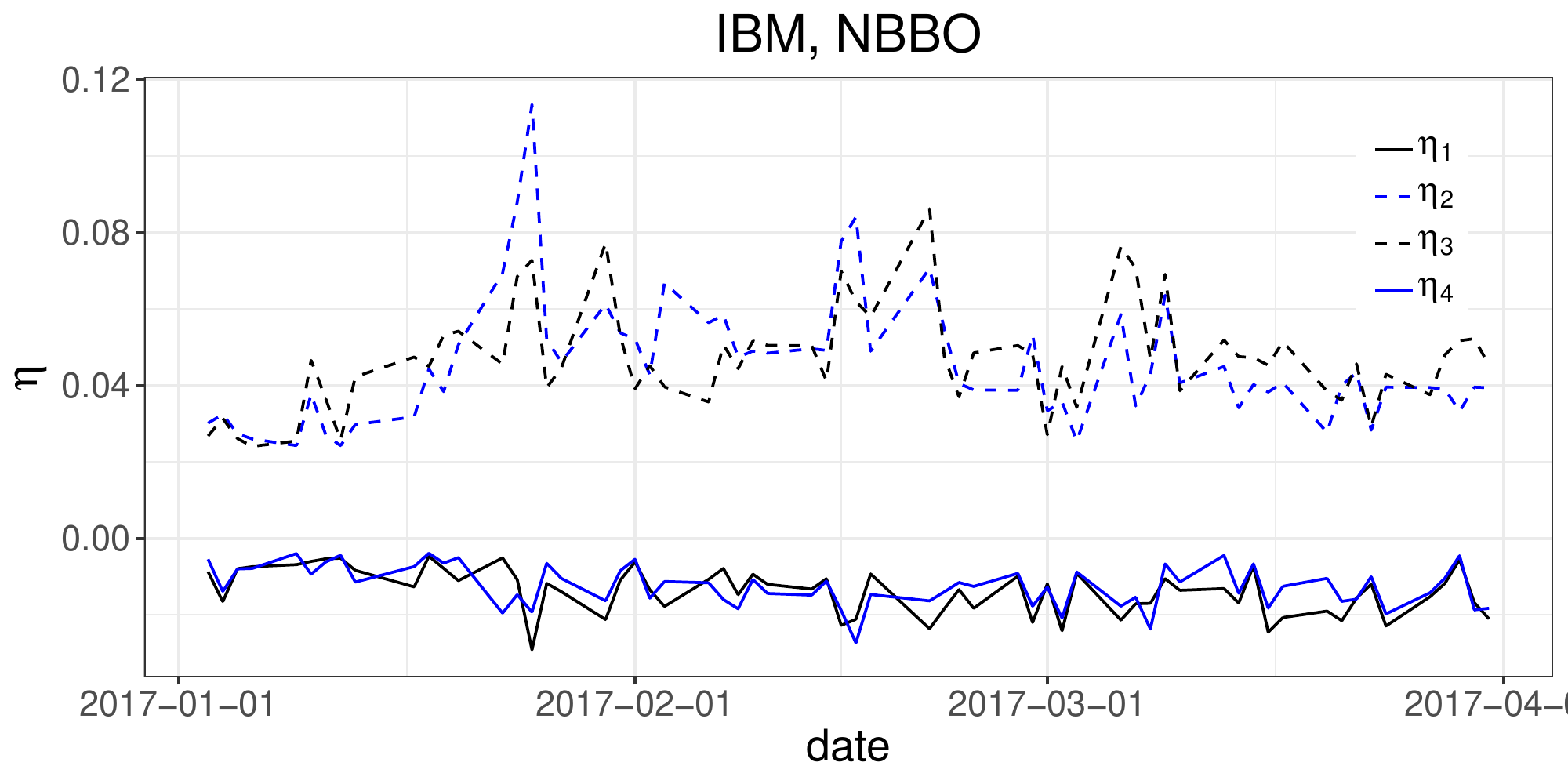}
		\caption{$\alpha$s}
		\label{fig:etas}
	\end{subfigure}
	\begin{subfigure}{.5\textwidth}
		\centering
		\includegraphics[width=0.94\textwidth]{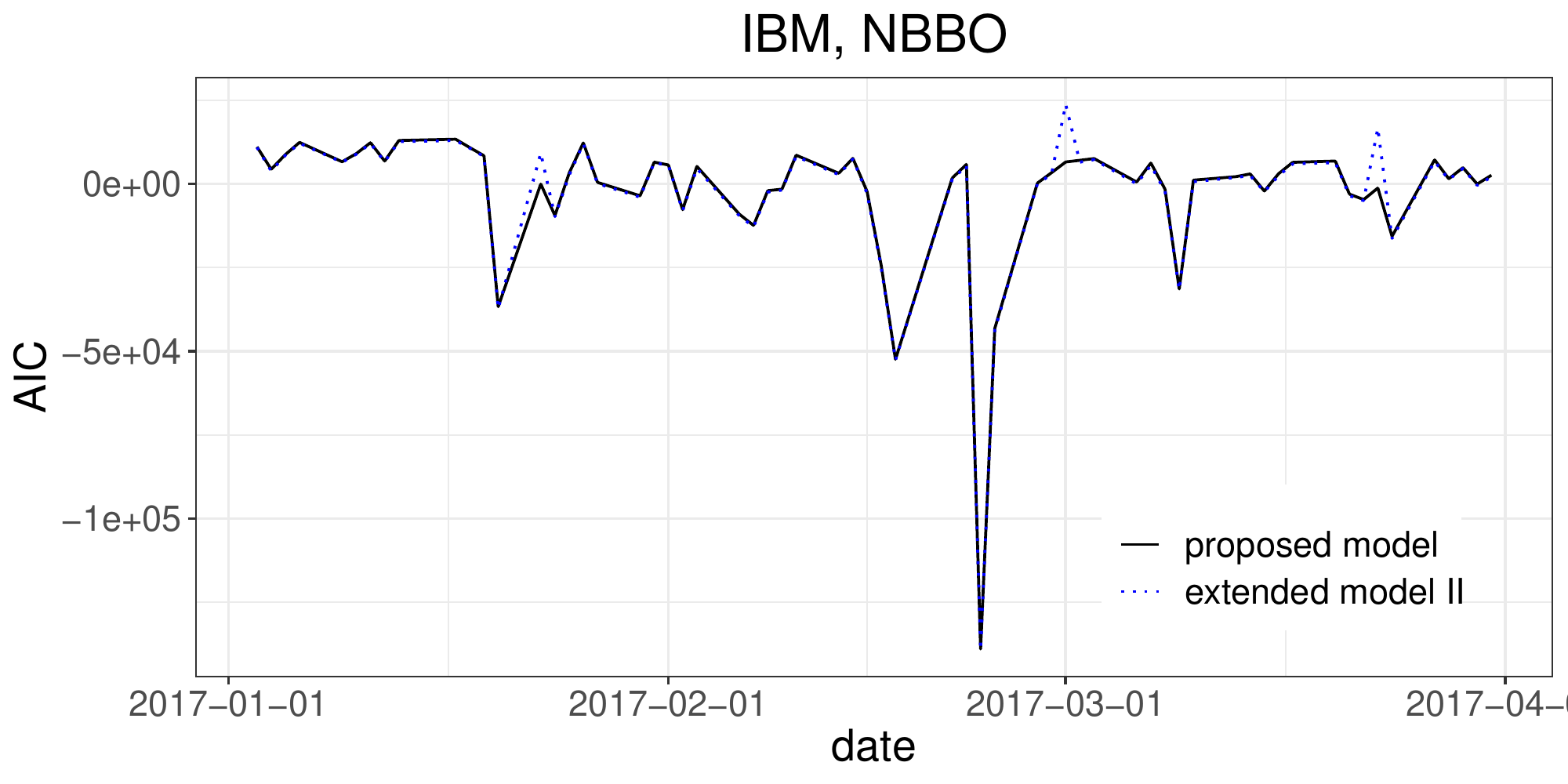}
		\caption{AIC}
		\label{fig:AIC_em2}
	\end{subfigure}
	\caption{Estimates of the $\alpha$s under extended model II (left) and comparison of the AIC between the proposed model and extended model II (right)}
	\label{Fig:AIC2}
\end{figure}

Third, we examine the model with different $\mu$s and $\eta$s in $\mathbf \mu$:
$$
\textrm{Extended model II: }\quad
\bm{\mu}_t = 
\begin{bmatrix} 
\mu_1 + \eta_1 \ell(t-) \\
\eta_2 \ell(t-)\\
\eta_3 \ell(t-) \\
\mu_4 + \eta_4 \ell(t-)
\end{bmatrix}
\quad
$$
and the other parts are as in Definition\ref{Def:model}.
One may think that when the spreads increase 
and market makers provide liquidity more actively, 
the intensities of bid-ask narrowing $A_d$ and $B_u$ increase
and the intensities of bid-ask widening $A_u$ and $B_d$ decrease.
Thus, one can assume that the base intensities of $A_d$ and $B_u$ are also functions of the spread.
Indeed, this negative relationship between the spread and intensities of the spread-widening processes is observed to some extent, as shown in Figure~\ref{fig:etas}, but is close to zero compared with the other coefficients.
It also does not significantly improve the model from the perspective of the AIC, as shown in Figure~\ref{fig:AIC_em2}.
We do not show a graph of $\mu_1$ and $\mu_4$ because of a lack of space, 
but their estimates are found to be similar.

\begin{figure}[!hbt]
	\begin{subfigure}{.5\textwidth}
		\centering
		\includegraphics[width=0.94\textwidth]{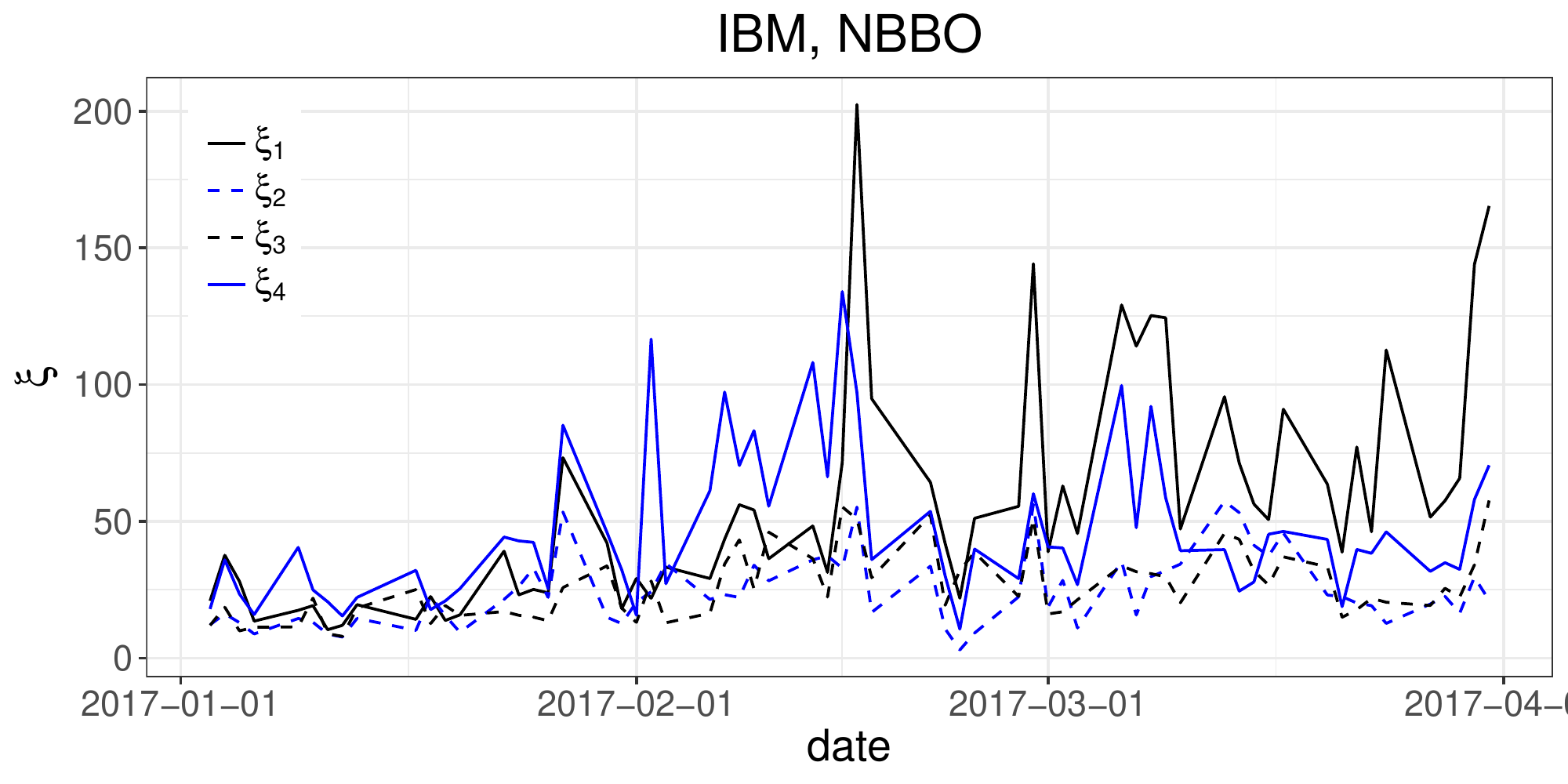}
		\caption{$\xi$s}
		\label{fig:xis}
	\end{subfigure}
	\begin{subfigure}{.5\textwidth}
		\centering
		\includegraphics[width=0.94\textwidth]{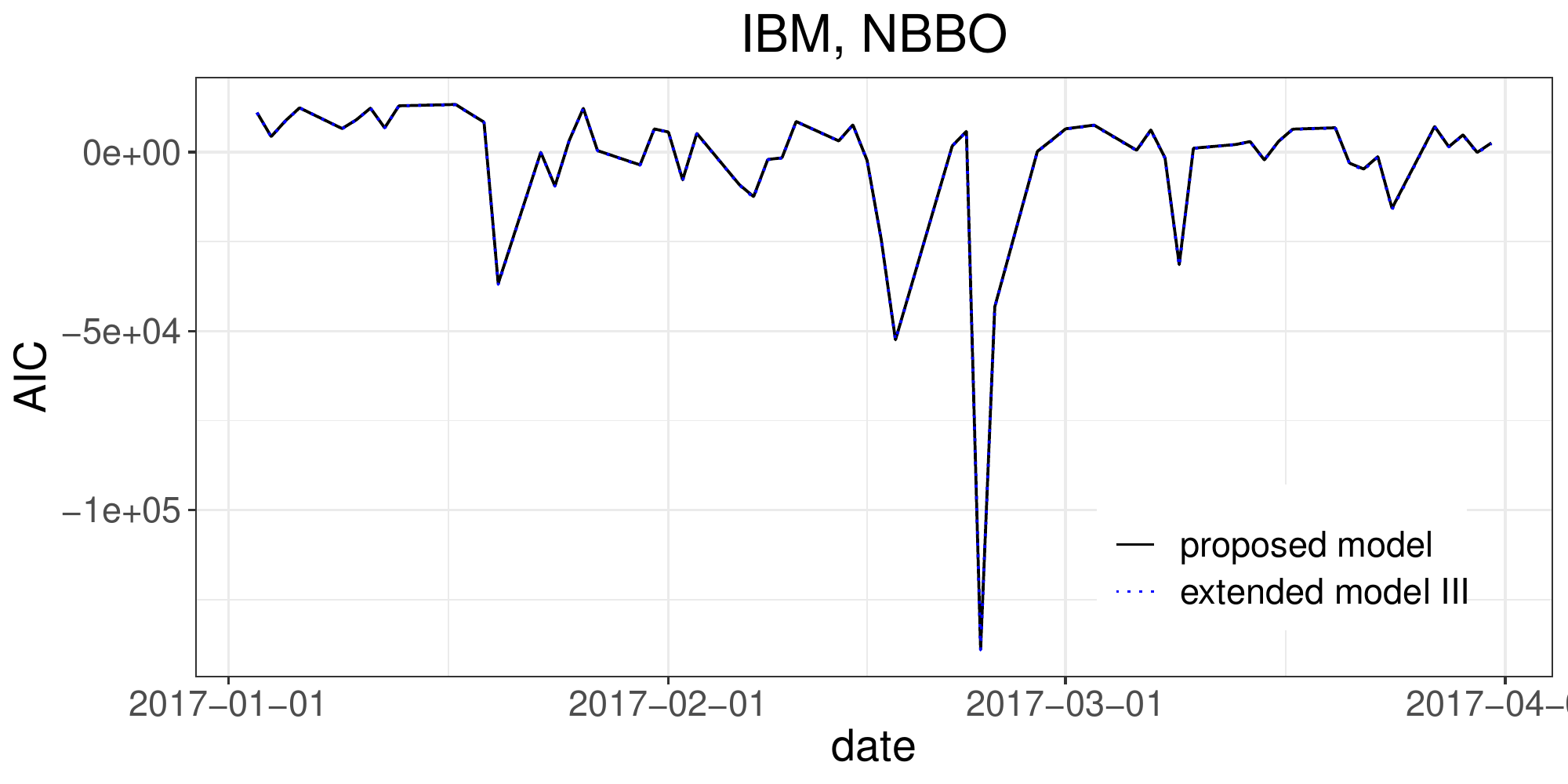}
		\caption{AIC}
		\label{fig:AIC_em3}
	\end{subfigure}
	\caption{Estimates of the $\xi$s under extended model III (left) and comparison of the AIC between the proposed model and extended model III (right)}
	\label{Fig:AIC3}
\end{figure}

Fourth, we examine the model with different $\xi$s.
To see if it can be justified to assume that the $\alpha_n$s in the middle of the matrix $\bm h$ are equal, 
we extend the model so that the $\alpha_n$s have different $\xi$s:
$$
\textrm{Extended model III: }\quad
\bm{h}(t, u) =
\e^{-\beta(t-u)}
\begin{bmatrix}
\alpha_{s1} & \alpha_{m} & \alpha_{s2} & 0 \\
\alpha_{w_1} & \alpha_{n1}(u) & \alpha_{n2}(u) & \alpha_{w_2} \\
\alpha_{w_2} & \alpha_{n3}(u) & \alpha_{n4}(u) & \alpha_{w_1} \\
0 & \alpha_{s2} & \alpha_{m} & \alpha_{s1} \\
\end{bmatrix}
$$
where
$$
\alpha_{ni}(t) = - \sum_{j=1}^4 \lambda_{kj}(t) + \xi_i \ell(t)
$$
and the other parts are as in Definition~\ref{subsec:def}.
As in Figure~\ref{fig:xis}, $\xi_1 \approx \xi_4$ and $\xi_2 \approx \xi_3$
and hence the model extension can serve as an alternative.
However, no significant benefit can be obtained from the perspective of the AIC, as shown in Figure~\ref{fig:AIC_em3}.

\begin{figure}[!hbt]
	\begin{subfigure}{.5\textwidth}
		\centering
		\includegraphics[width=0.94\textwidth]{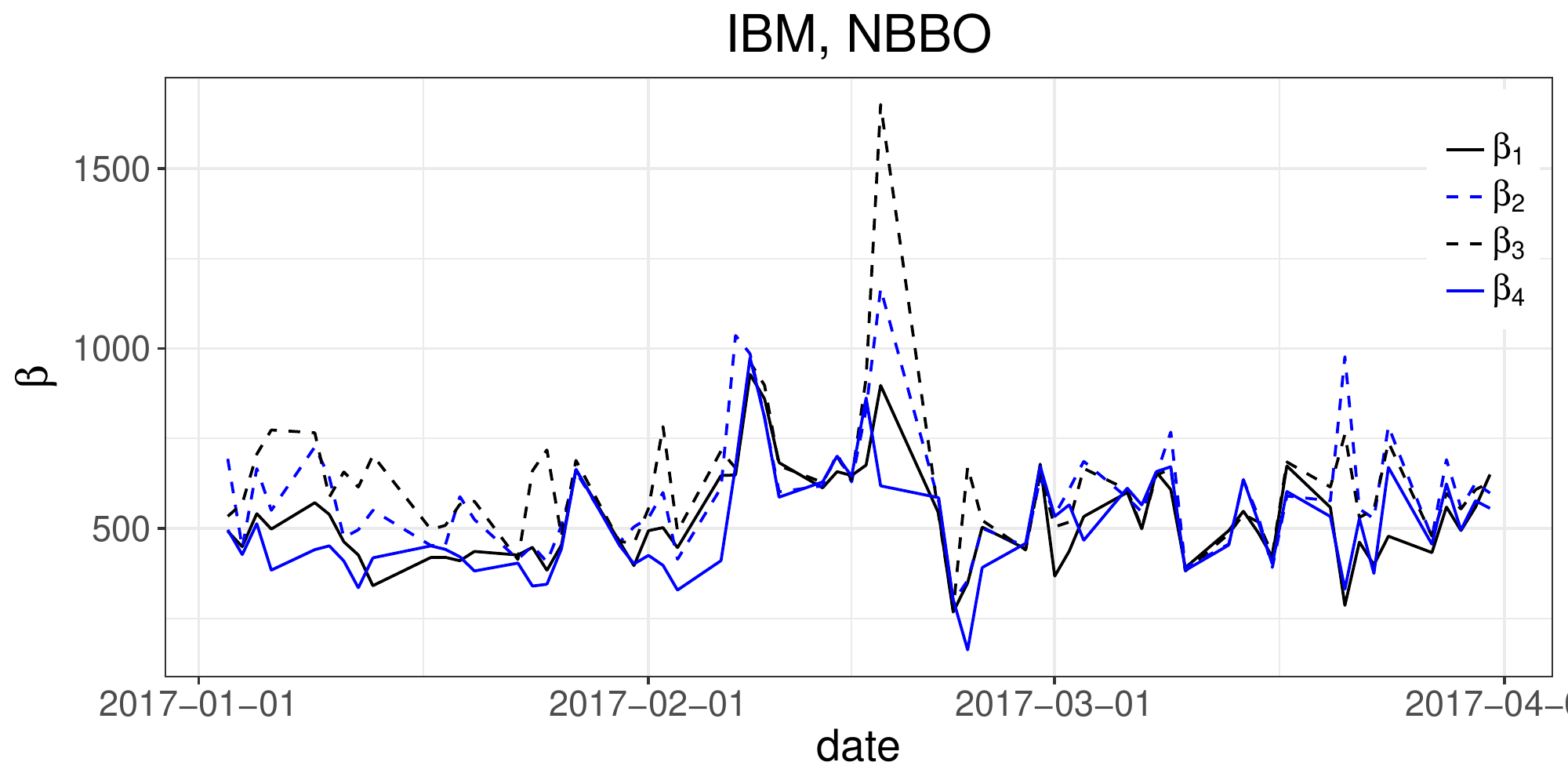}
		\caption{$\beta$s}
		\label{fig:betas_em4}
	\end{subfigure}
	\begin{subfigure}{.5\textwidth}
		\centering
		\includegraphics[width=0.94\textwidth]{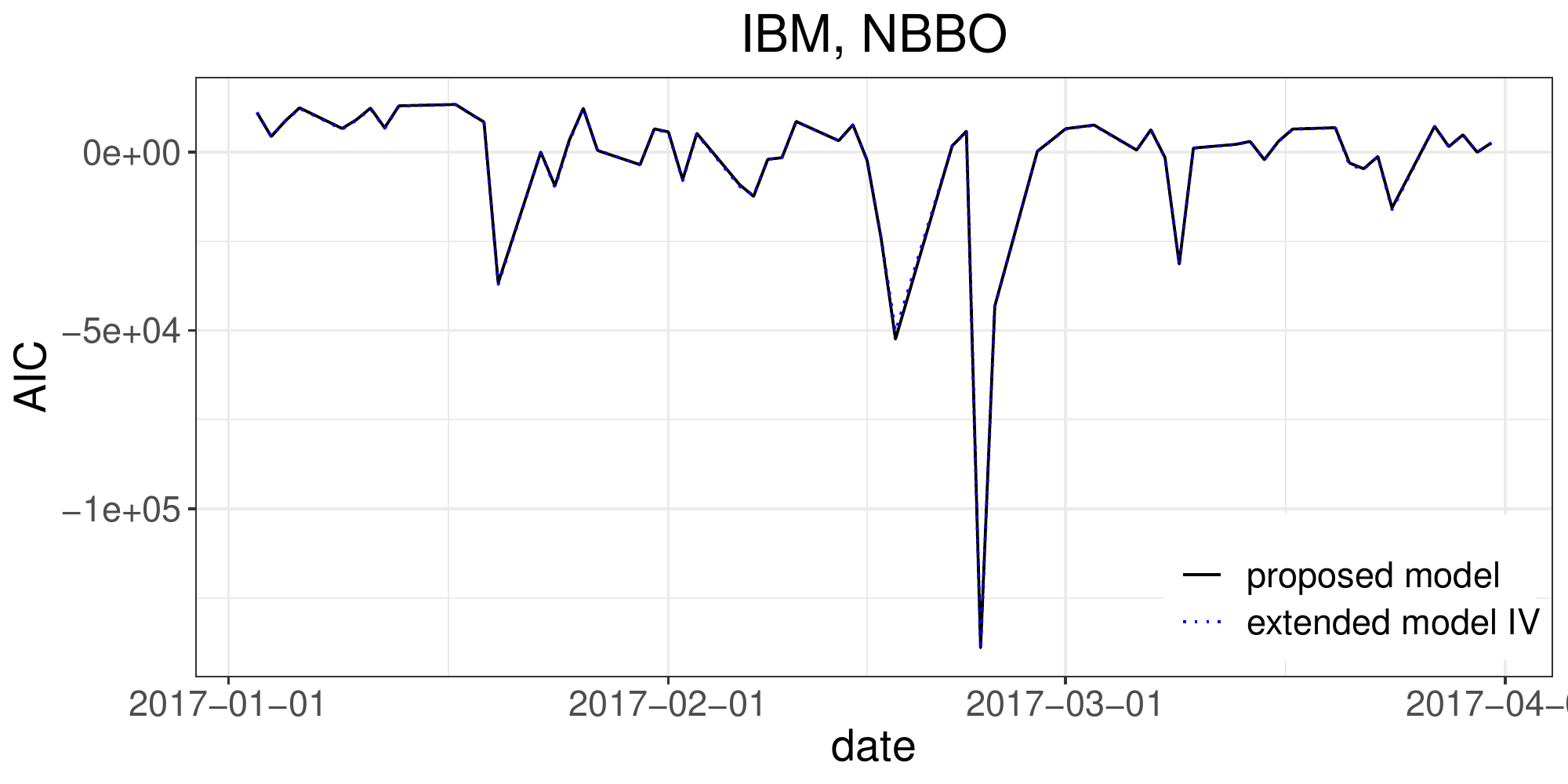}
		\caption{AIC}
		\label{fig:AIC_em4}
	\end{subfigure}
	\caption{Estimates of the $\beta$s under extended model IV (left) and comparison of the AIC between the proposed model and extended model IV}
	\label{Fig:AIC4}
\end{figure}

Fifth, one may think that the decaying parameter $\beta$ does not have to be the same
and we compare the proposed model with an extended model with a column-wise different $\beta$:
$$
\textrm{Extended model IV: }\quad
\bm{h}(t,u) = 
\mathbf B(t-u)
\circ
\begin{bmatrix}
\alpha_{s1} & \alpha_{m} & \alpha_{s2} & 0 \\
\alpha_{w_1} & \alpha_{n1}(u) & \alpha_{n2}(u) & \alpha_{w_2} \\
\alpha_{w_2} & \alpha_{n3}(u) & \alpha_{n4}(u) & \alpha_{w_1} \\
0 & \alpha_{s2} & \alpha_{m} & \alpha_{s1} \\
\end{bmatrix}
$$
where $\circ$ denotes the element-wise multiplication (the Hadamard product),
$$
\mathbf B(t-u)
= \begin{bmatrix}
\e^{-\beta_1(t-u)} & \e^{-\beta_2(t-u)} & \e^{-\beta_3(t-u)} & \e^{-\beta_4(t-u)} \\
\e^{-\beta_1(t-u)} & \e^{-\beta_2(t-u)} & \e^{-\beta_3(t-u)} & \e^{-\beta_4(t-u)} \\
\e^{-\beta_1(t-u)} & \e^{-\beta_2(t-u)} & \e^{-\beta_3(t-u)} & \e^{-\beta_4(t-u)} \\
\e^{-\beta_1(t-u)} & \e^{-\beta_2(t-u)} & \e^{-\beta_3(t-u)} & \e^{-\beta_4(t-u)} \\
\end{bmatrix} $$
and the other parts are as in Definition~\ref{subsec:def}.
As shown in Figure~\ref{Fig:AIC4}, the column-wise $\beta$s seem to have the same value 
and there is no improvement in terms of the AIC.

Sixth, we compare the proposed model with an extended model with a row-wise different $\beta$:
$$
\textrm{Extended model V: }\quad
\bm{h}(t, u) = 
\mathbf B(t-u)
\circ
\begin{bmatrix}
\alpha_{s1} & \alpha_{m} & \alpha_{s2} & 0 \\
\alpha_{w_1} & \alpha_{n1}(u) & \alpha_{n2}(u) & \alpha_{w_2} \\
\alpha_{w_2} & \alpha_{n3}(u) & \alpha_{n4}(u) & \alpha_{w_1} \\
0 & \alpha_{s2} & \alpha_{m} & \alpha_{s1} \\
\end{bmatrix}
$$
where
$$
\mathbf B(t-u)
= \begin{bmatrix}
\e^{-\beta_1(t-u)} & \e^{-\beta_1(t-u)} & \e^{-\beta_1(t-u)} & \e^{-\beta_1(t-u)} \\
\e^{-\beta_2(t-u)} & \e^{-\beta_2(t-u)} & \e^{-\beta_2(t-u)} & \e^{-\beta_2(t-u)} \\
\e^{-\beta_3(t-u)} & \e^{-\beta_3(t-u)} & \e^{-\beta_3(t-u)} & \e^{-\beta_3(t-u)} \\
\e^{-\beta_4(t-u)} & \e^{-\beta_4(t-u)} & \e^{-\beta_4(t-u)} & \e^{-\beta_4(t-u)} \\
\end{bmatrix} $$
and the other parts are as in Definition~\ref{subsec:def}.
Extended model V also shows slightly improved performance in terms of the AIC,
even though $\beta_1 \approx \beta_4$ and $\beta_2 \approx \beta_3$,
which is consistent with the results reported by \cite{lee2017modeling}.
This slight improvement is insufficient to use the extended model at the expense of increased model complexity.
Therefore, the empirical studies presented hereafter are conducted based on the proposed model in Definition~\ref{Def:model}.

\begin{figure}[!hbt]
	\begin{subfigure}{.5\textwidth}
		\centering
		\includegraphics[width=0.94\textwidth]{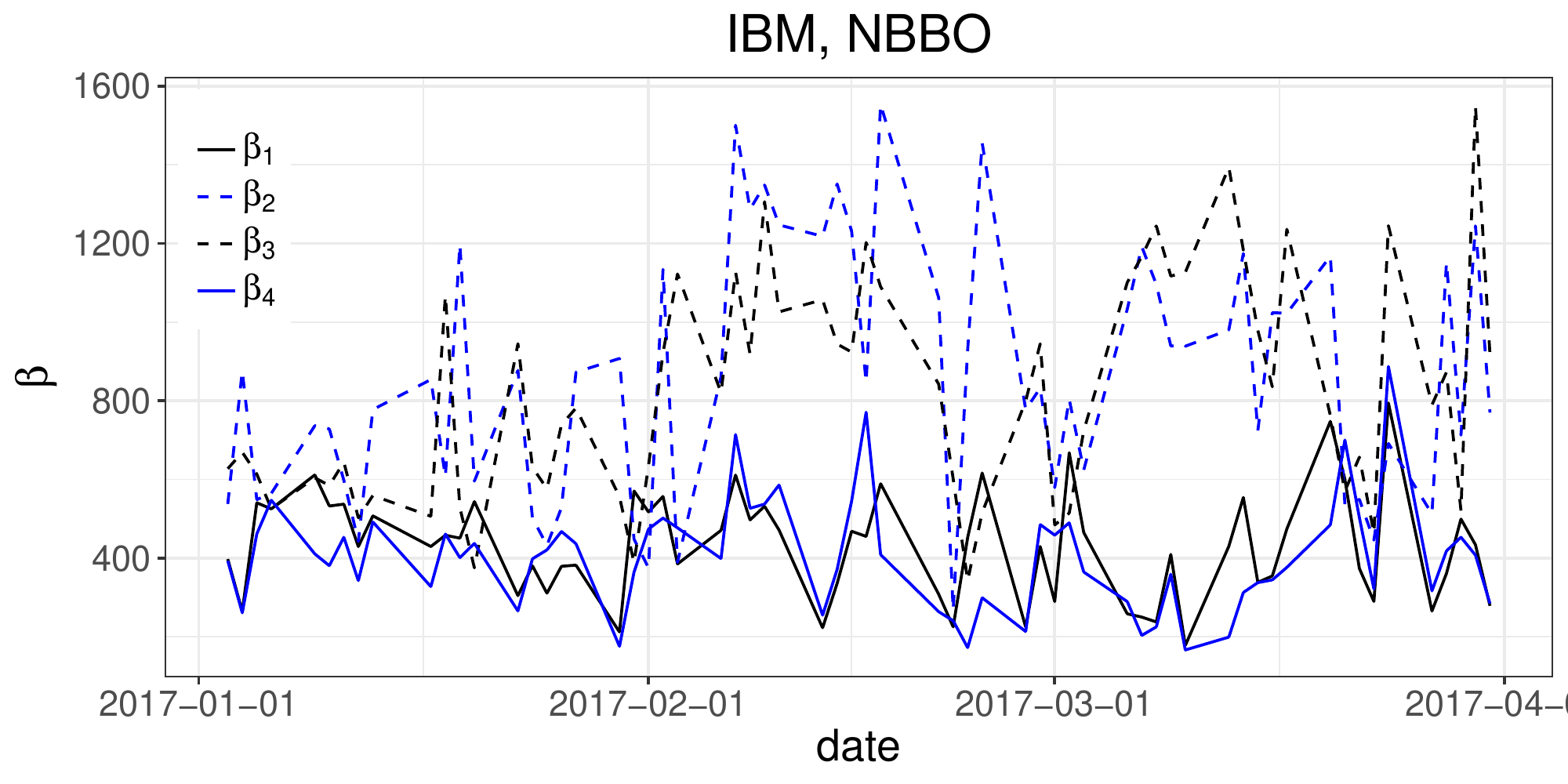}
		\caption{$\beta$s}
		\label{fig:betas}
	\end{subfigure}
	\begin{subfigure}{.5\textwidth}
		\centering
		\includegraphics[width=0.94\textwidth]{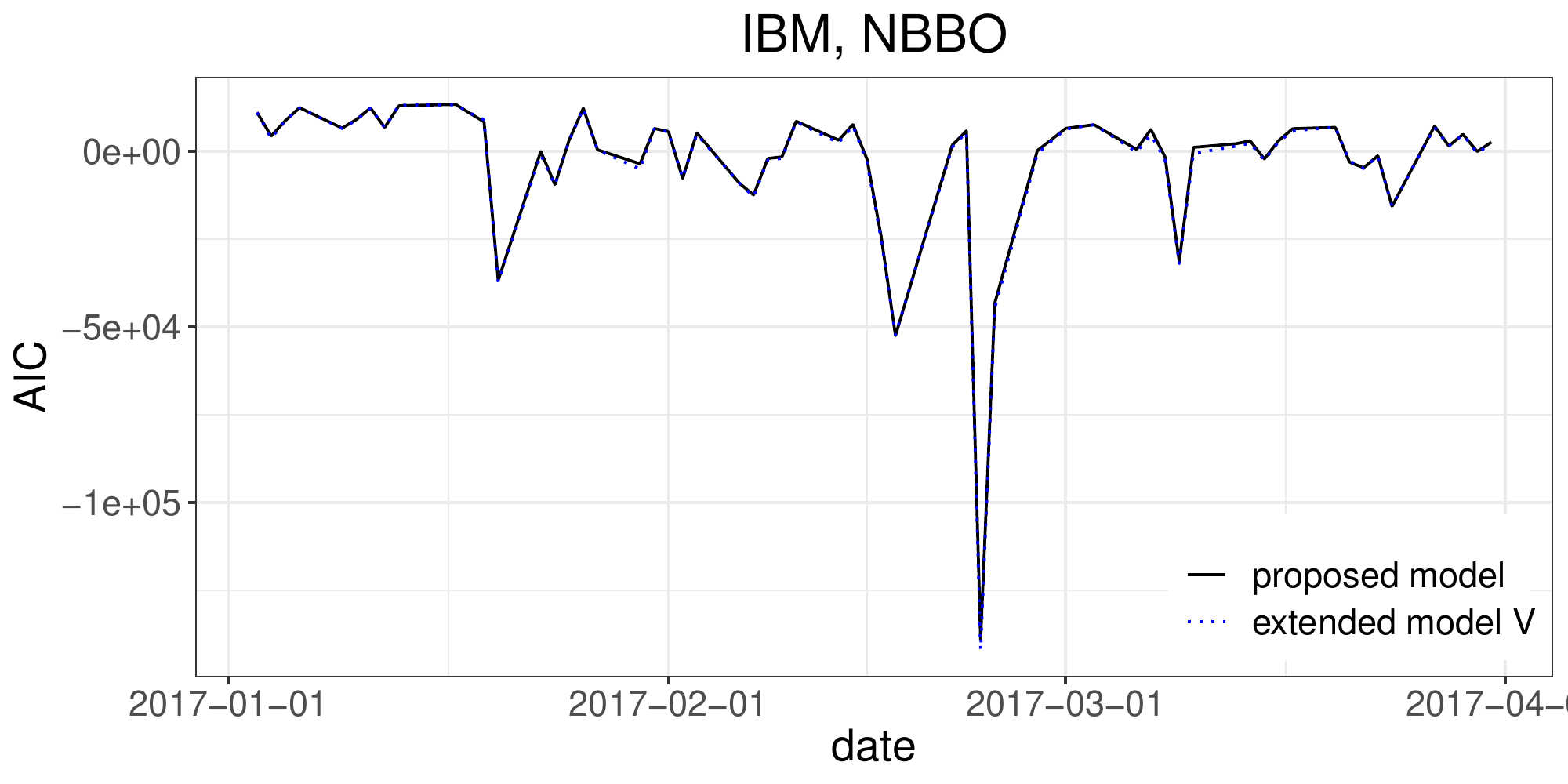}
		\caption{AIC}
		\label{fig:AIC_em5}
	\end{subfigure}
	\caption{Estimates of the $\beta$s under extended model V (left) and comparison of the AIC between the proposed model and extended model V}
	\label{Fig:AIC5}
\end{figure}

\begin{figure}[!hbt]
	\begin{subfigure}{.5\textwidth}
		\centering
		\includegraphics[width=0.85\textwidth]{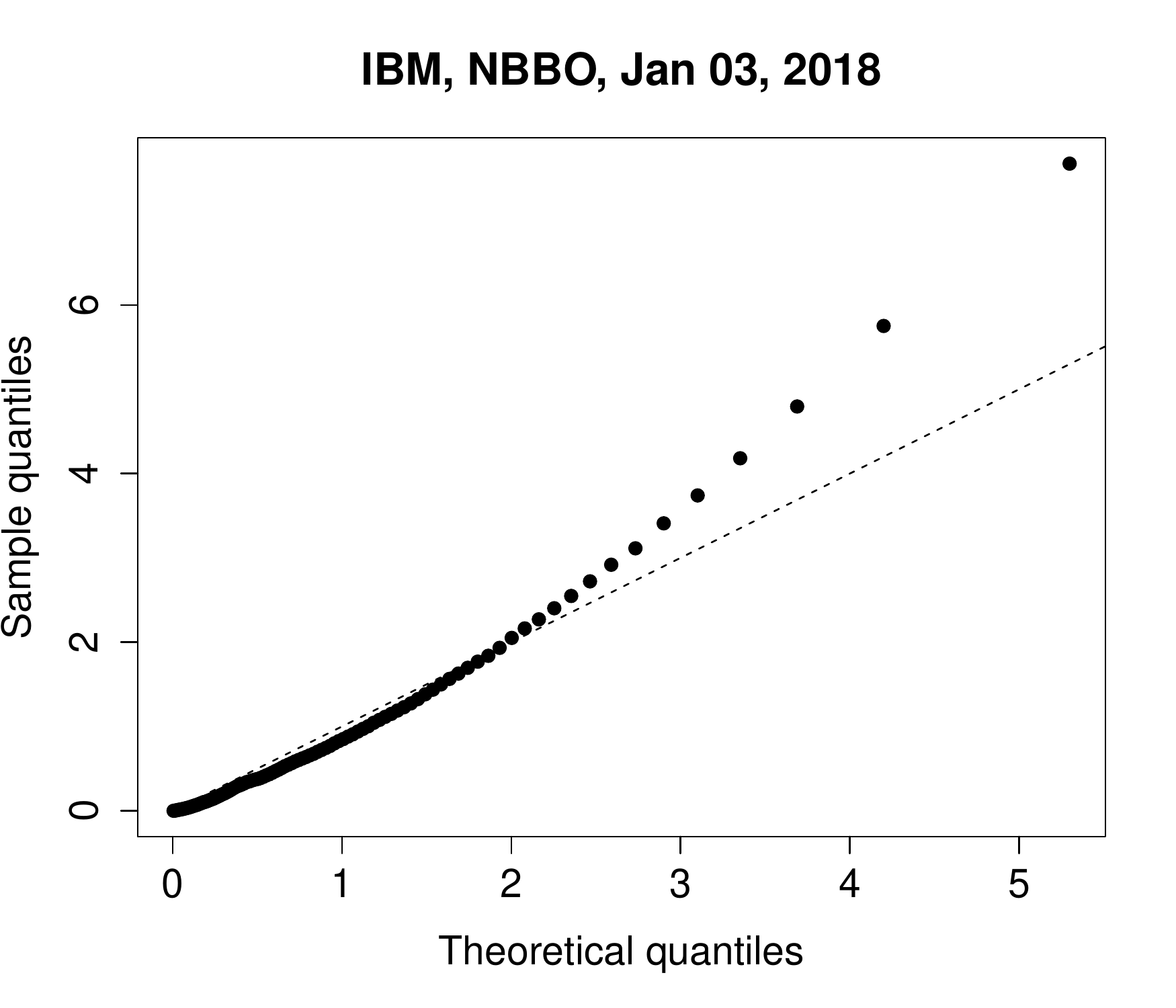}
		\caption{IBM}
	\end{subfigure}
	\begin{subfigure}{.5\textwidth}
		\centering
		\includegraphics[width=0.85\textwidth]{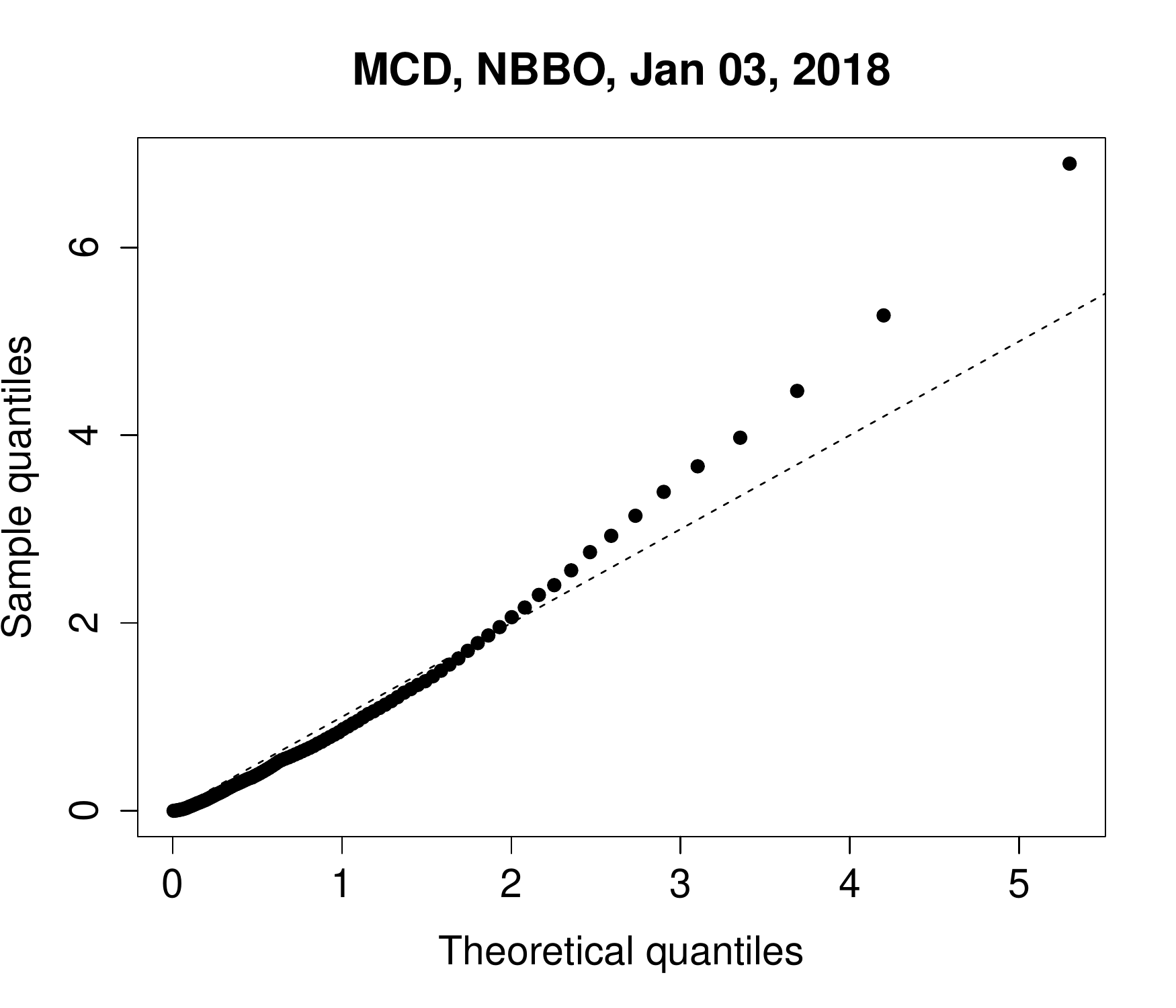}
		\caption{MCD}
	\end{subfigure}
	\caption{Q-Q plot of residual process of our model}
	\label{Fig:qqplot}
\end{figure}

In Figure~{\ref{Fig:qqplot}}, 
quantile-quantile plots of the residuals versus the unit exponential distribution are presented.
The residuals are defined by
$$\bigcup_{1\leq i \leq 4} \left\{ \int_{t_{i, j}}^{t_{i, j+1}} \hat \lambda_i (u) \D u \right\}$$
where $\hat \lambda_i$ are fitted intensities of our model and $t_{i,j}$ are event times.
The plot shows that the residuals follow a slightly fatter tail distribution.
In order to obtain a better fitting result, it is necessary to introduce a multi-kernel with different $\beta$s, which leads more complex model.
This is beyond our scope.
Instead we focus on describing the activities at the ultra-high-frequency level, so the extension to multiple kernels is left as a future work.

\subsubsection{Basic results}~\label{subsec:basic}

\begin{figure}[hbt!]
	\begin{subfigure}{.5\textwidth}
		\centering
		\includegraphics[width=0.94\textwidth]{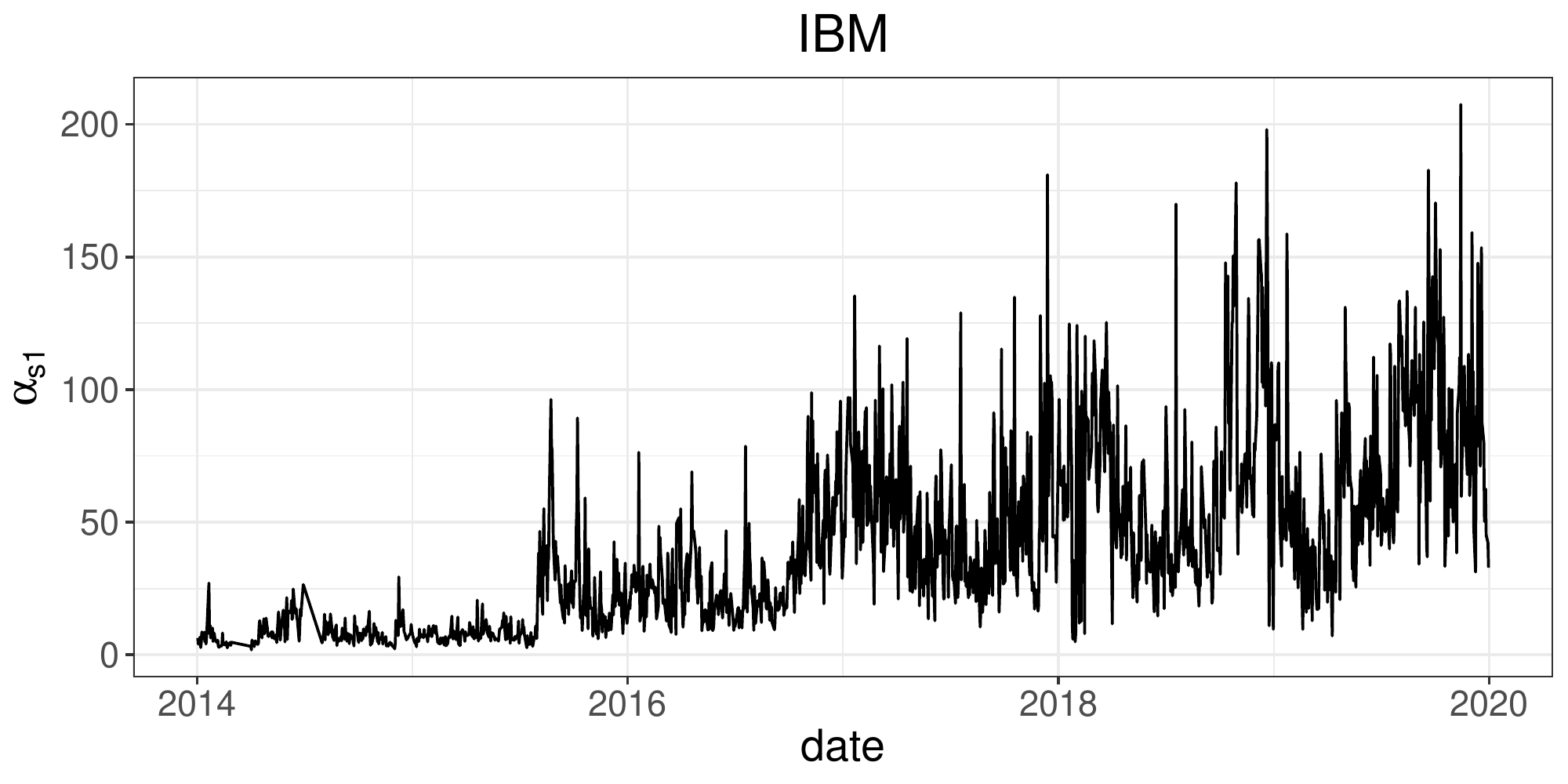}
		\caption{$\alpha_{s1}$}
		\label{fig:IBM_NBBO_one_kernel_alpha_s1}
	\end{subfigure}
	\begin{subfigure}{.5\textwidth}
		\centering
		\includegraphics[width=0.94\textwidth]{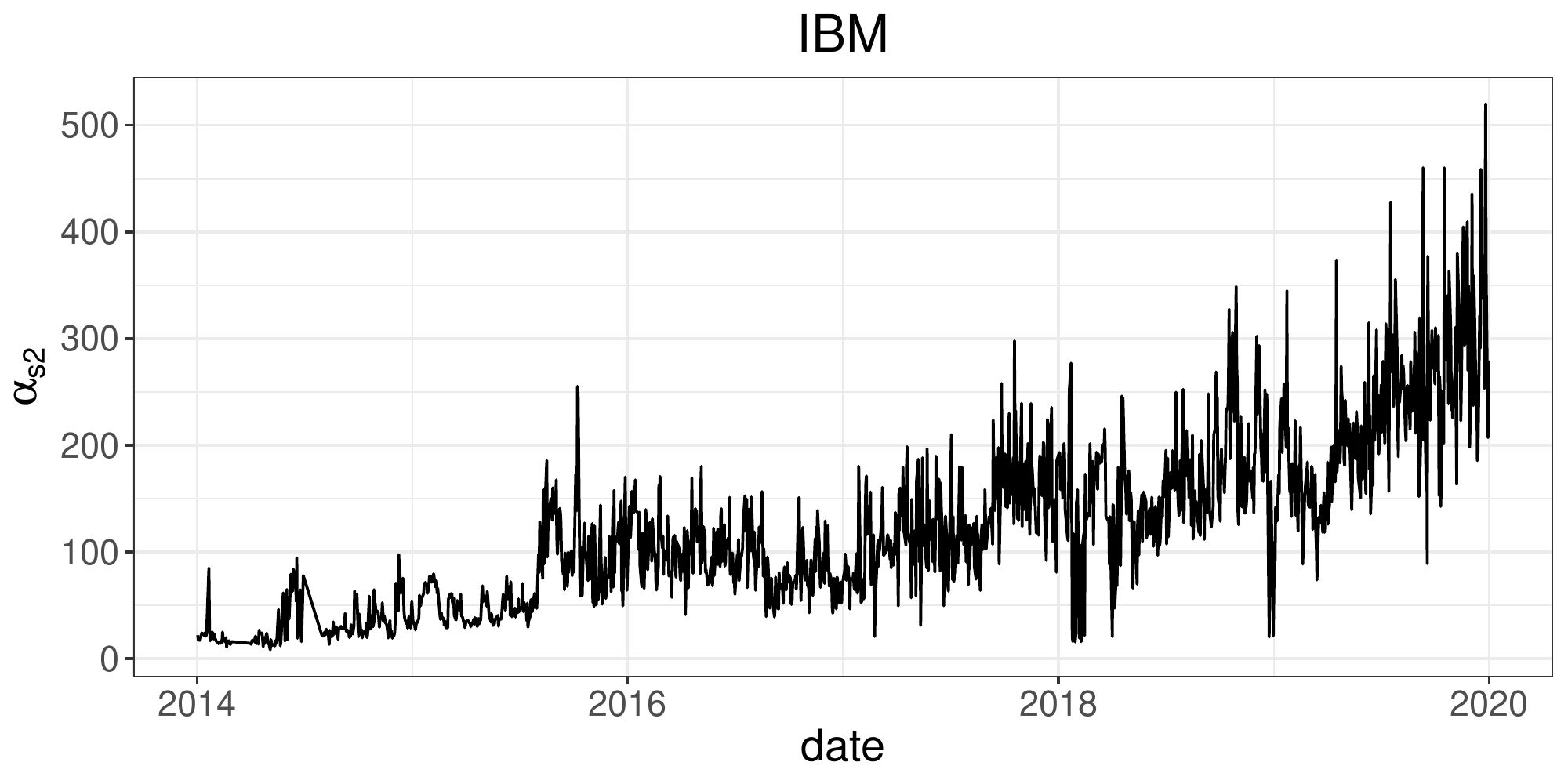}
		\caption{$\alpha_{s2}$}
		\label{fig:IBM_NBBO_one_kernel_alpha_s2}
	\end{subfigure}
	
	\begin{subfigure}{.5\textwidth}
		\centering
		\includegraphics[width=0.94\textwidth]{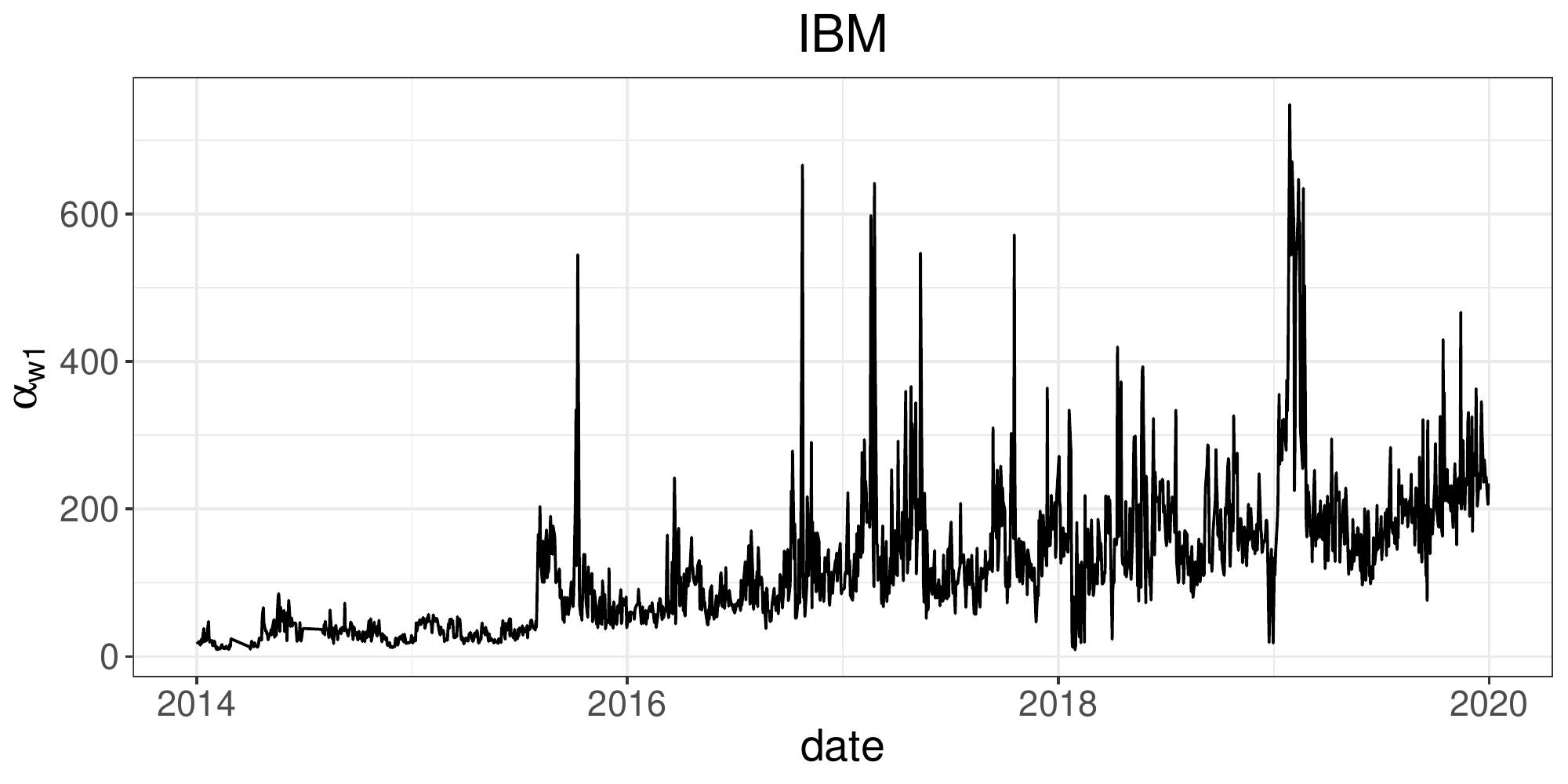}
		\caption{$\alpha_{w1}$}
		\label{fig:IBM_NBBO_one_kernel_alpha_w1}
	\end{subfigure}
	\begin{subfigure}{.5\textwidth}
		\centering
		\includegraphics[width=0.94\textwidth]{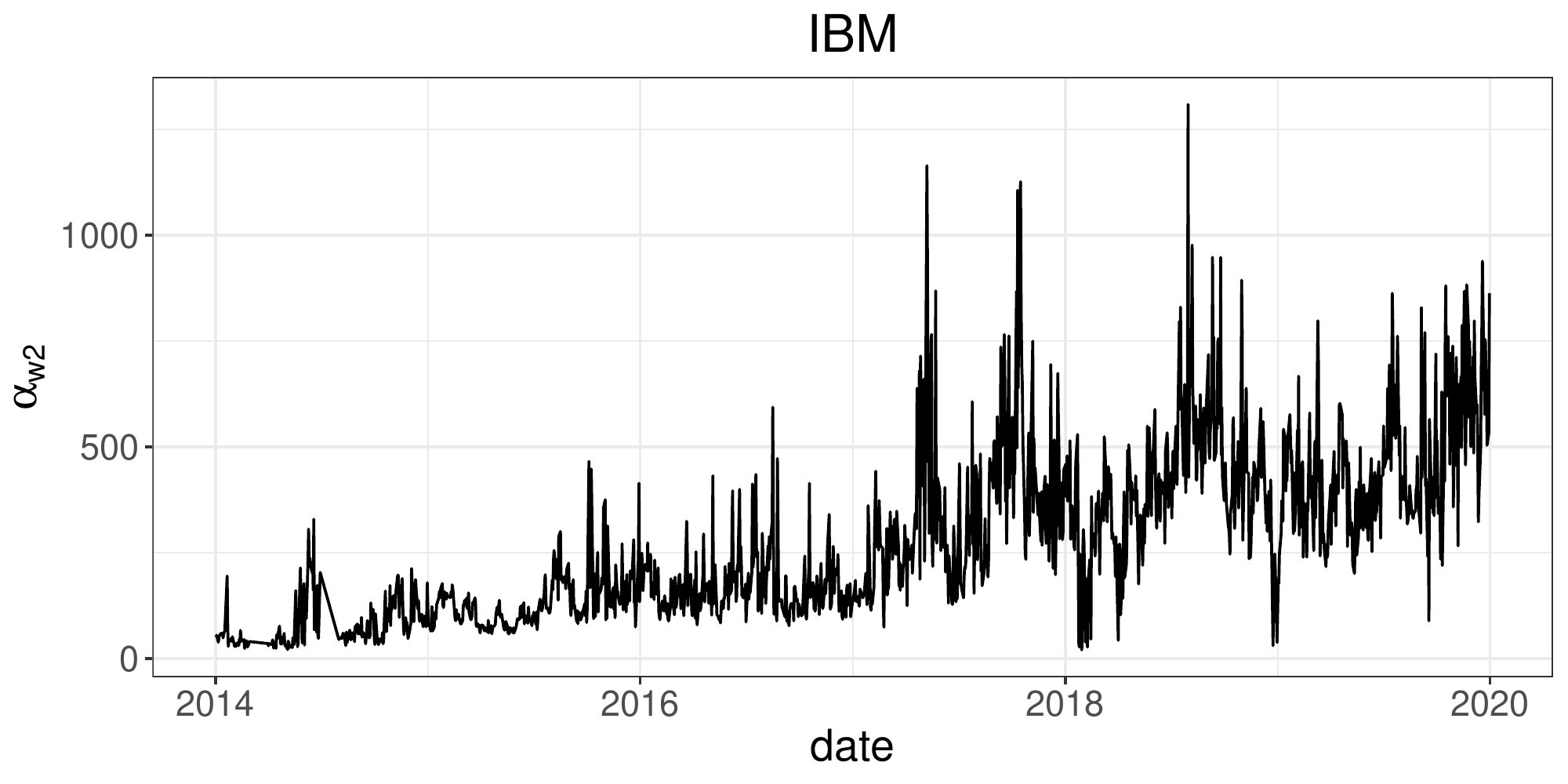}
		\caption{$\alpha_{w2}$}
		\label{fig:IBM_NBBO_one_kernel_alpha_w2}
	\end{subfigure}
	
	\begin{subfigure}{.5\textwidth}
		\centering
		\includegraphics[width=0.94\textwidth]{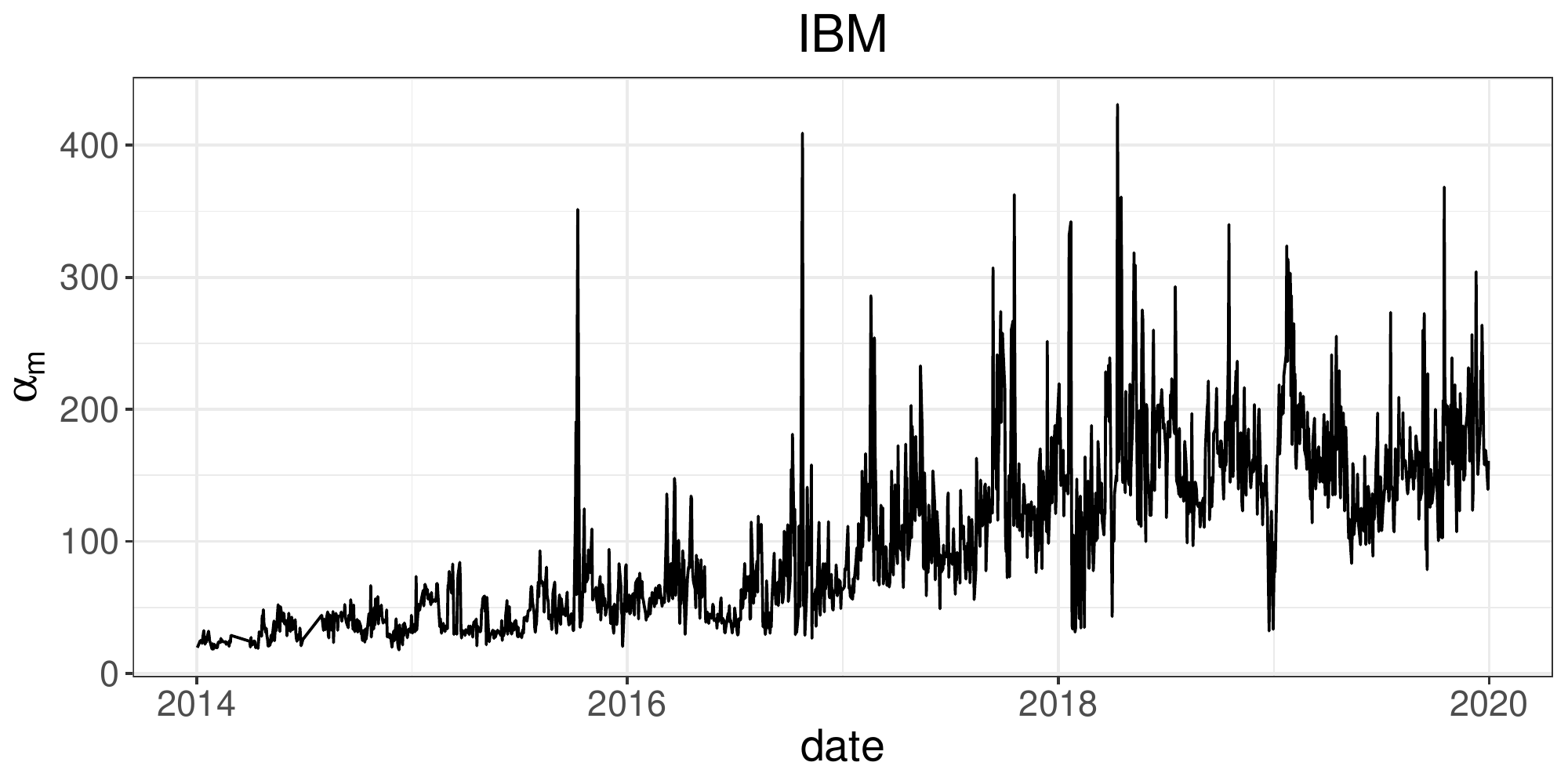}
		\caption{$\alpha_m$}
		\label{fig:IBM_NBBO_one_kernel_alpha_m}
	\end{subfigure}
	\begin{subfigure}{.5\textwidth}
		\centering
		\includegraphics[width=0.94\textwidth]{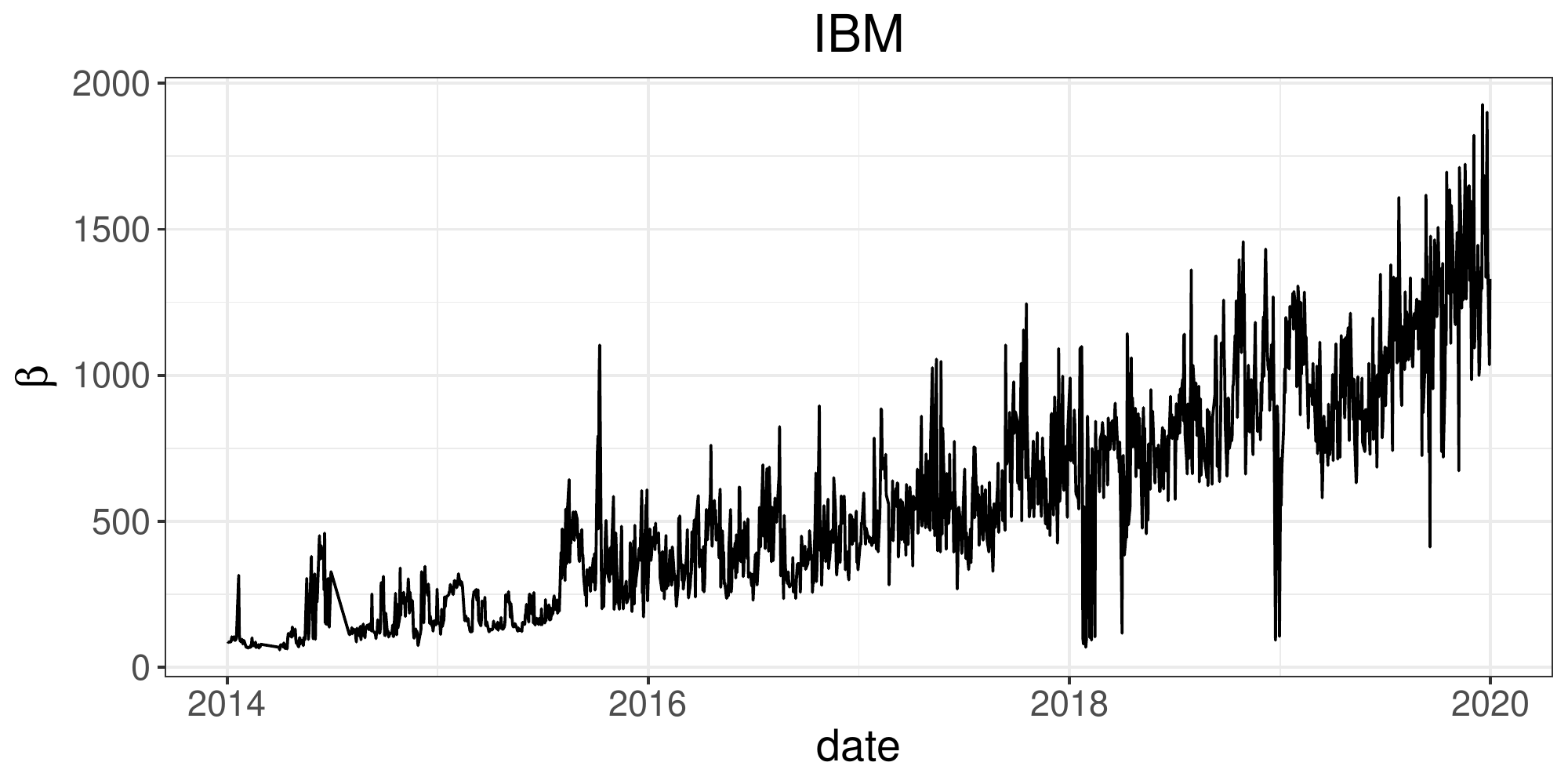}
		\caption{$\beta$}
		\label{fig:IBM_NBBO_one_kernel_beta}
	\end{subfigure}
	
	\begin{subfigure}{.5\textwidth}
		\centering
		\includegraphics[width=0.94\textwidth]{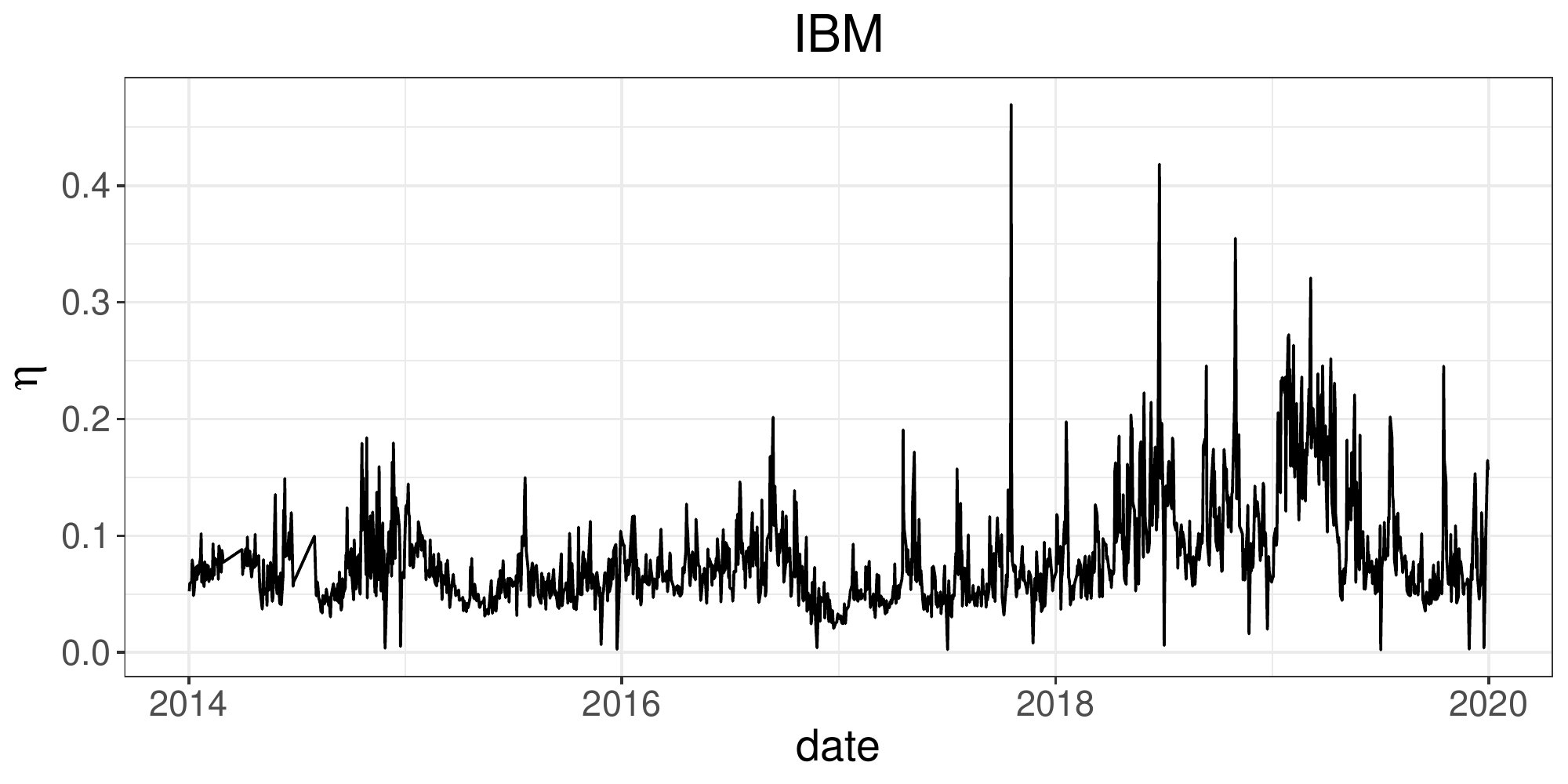}
		\caption{$\eta$}
		\label{fig:IBM_NBBO_one_kernel_alpha_eta}
	\end{subfigure}
	\begin{subfigure}{.5\textwidth}
		\centering
		\includegraphics[width=0.94\textwidth]{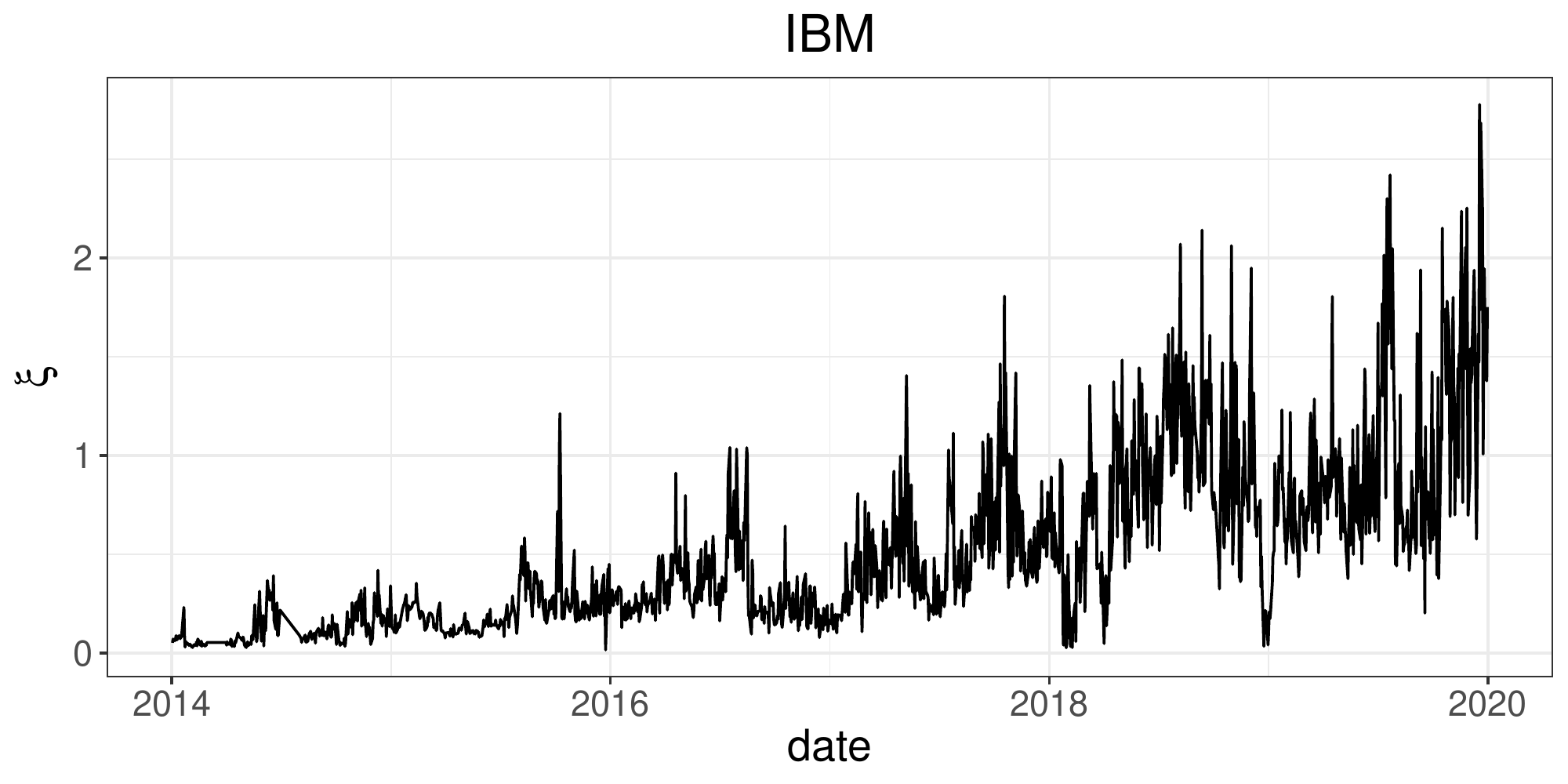}
		\caption{$\xi$}
		\label{fig:IBM_NBBO_one_kernel_alpha_xi}
	\end{subfigure}
	
	\caption{Dynamics of the daily estimated parameters for the NBBO of IBM from 2014 to 2019}
	\label{Fig:IBM_NBBO_one_kernel}
\end{figure}

This section reports the basic results obtained by the NBBO. 
Some of these results may seem obvious, but they are worth checking, while other results are new.
Figure~\ref{Fig:IBM_NBBO_one_kernel} illustrates the dynamics of the estimates from the NBBO of the IBM stock 
under the model defined in Definition~\ref{subsec:def} from 2014 to 2019.
We compute all the estimates on a daily basis, and hence have around 250 estimates for each parameter in each year.
The time unit for the intensities is one second.
For the detailed results of the estimations with standard errors, see Appendix~\ref{Appendix:estimation}.

First, the results show that the narrowing tendency of the bid-ask spread depends on the spread.
The estimates of liquidity parameter $\eta$ are significantly positive throughout the period, as Figure~\ref{fig:IBM_NBBO_one_kernel_alpha_eta} shows.
Recall that we use $\eta \ell(t-)$ in the base intensity of the spread-narrowing processes (i.e., downward movements of ask prices and upward movements of bid prices).
A positive $\eta$ implies that with a larger bid-ask spread, it is more likely that the spread will narrow
as the base intensities of the narrowing movements increase.
This finding is in line with our intuition because the larger the spread is, the more likely it is that market makers will earn profits and will submit aggressive limit orders.
Note that $\eta$ exists only in the base intensity, but not in the kernel; hence, is not related to the previous events, but only to the current spread.
In addition, this parameter changes little over time and has a stable value in general.

Not every day shows a strong spread-narrowing tendency.
If liquidity providers are concerned about adverse selection (i.e., they expect the realized spread to be negative), then they will not try to reduce the spread.
For example, on January 5, 2017, IBM's liquidity parameter $\eta$ is 0.025, which is less than half of the average $\eta$s estimated in 2017,
and we can assume that market makers were particularly careful about adverse selection on this day.
In addition, when market disruption occurs, liquidity parameter $\eta$ can come close to zero during the day, which we discuss in Subsection~\ref{subsec:Crash}.

Second, the figure shows a large excitement parameter $\alpha$, which implies an abundance of automated high-frequency quotes and trading.
In our model, some limit order arrivals depend on the spread described by $\eta$, 
whereas the rest relate to the excitations caused by past events,
as the parameter $\alpha$s also affect the arrivals or removals of limit orders, which changes the best bid or ask prices.
The estimates of the $\alpha$s range in magnitude from tens to hundreds in general, 
which means that in the event of a limit order arrival or removal that changes the best bid or ask price, 
the intensities increase by tens to hundreds, and hence the probability of subsequent events dramatically increases.
Because the decaying parameter $\beta$ is also large, the increased probability also decreases sharply over time.
For example, if an event increases intensity to 500, then approximately five successive events can also occur in the next 0.01 seconds.
Intensity also tends to converge to the background rate $\mu$ owing to the large decaying parameter $\beta$,
provided that no additional events occur.

This instantaneous responsiveness is high; as human behavior is unlikely to be the cause, it may be closely related to high-frequency automated algorithmic trading.
Many trading venues ensure that transactions are fast, especially in quote revisions.
\citet{chan2017machine} demonstrates that better financial services and technologies such as colocation systems in which the trading servers are in the same facilities as the execution venues can reduce the latency to microseconds.

Third, the excitement parameters $\alpha$s show an increasing trend (similar to the decaying parameter) in general.
One reason for this increasing trend is the more accurate time resolution.
As mentioned before, from August 2015 to around September 2018, the data are recorded in microseconds compared with the previous frequency of milliseconds for the stocks reported by the CTS.
This change is noticeable on the graph, as we observe peaks around August 2015 for the excitement parameters.
However, even when excluding the effect of the improved resolution, the overall estimates increase,
which means lower latencies for order submissions and the data feed in high-frequency trading over time.
In addition, when the time resolution improves to nanoseconds (around September 2018), we find no notable abrupt changes. 
This result implies that the resolution of microseconds is sufficient for this analysis.

\begin{figure}[hbt!]
	\centering
	\includegraphics[width=0.7\textwidth]{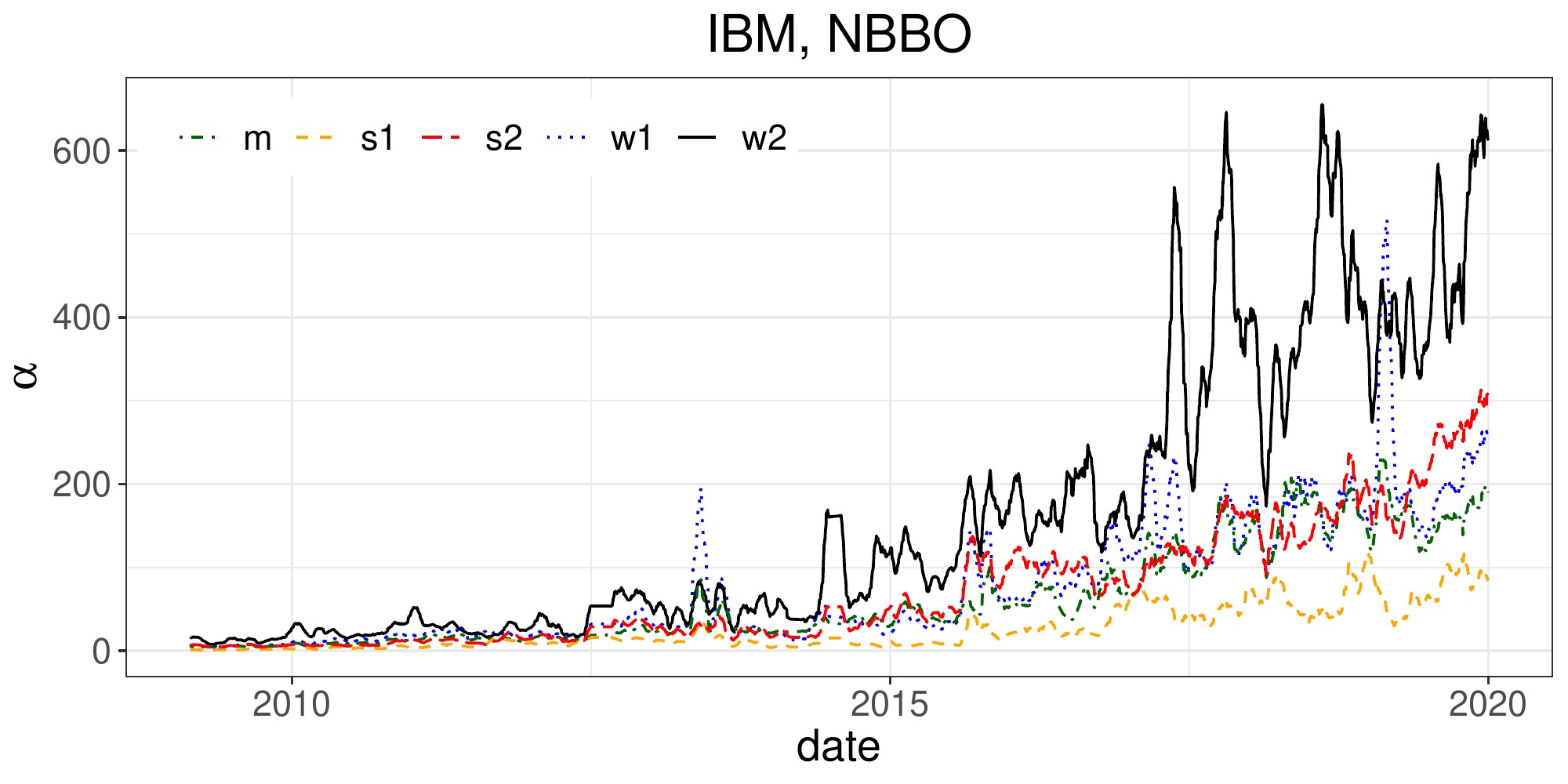}
	\caption{Estimates of the $\alpha$s for IBM's NBBO, smoothed by the moving average of 20 successive days}
	\label{Fig:IBM_alphas_smooth}
\end{figure}

Because the daily estimates of the excitement parameter $\alpha$s are too variable over time, 
Figure~\ref{Fig:IBM_alphas_smooth} presents each $\alpha$ estimated from IBM's NBBO as a moving average.
We compute the moving average using 20 successive days from 2009 to 2019. The increasing trends for all the parameters are clear.
Overall, $\alpha_{w2}$ is the largest, $\alpha_{s1}$ is the smallest, and the rest are similar.
The large $\alpha_{w}$s mean that liquidity-providing actions occur with fast reactions in response to spread-widening events, indicating that market making is active.

\begin{figure}[!hbt]
	\centering
	\includegraphics[width=0.7\textwidth]{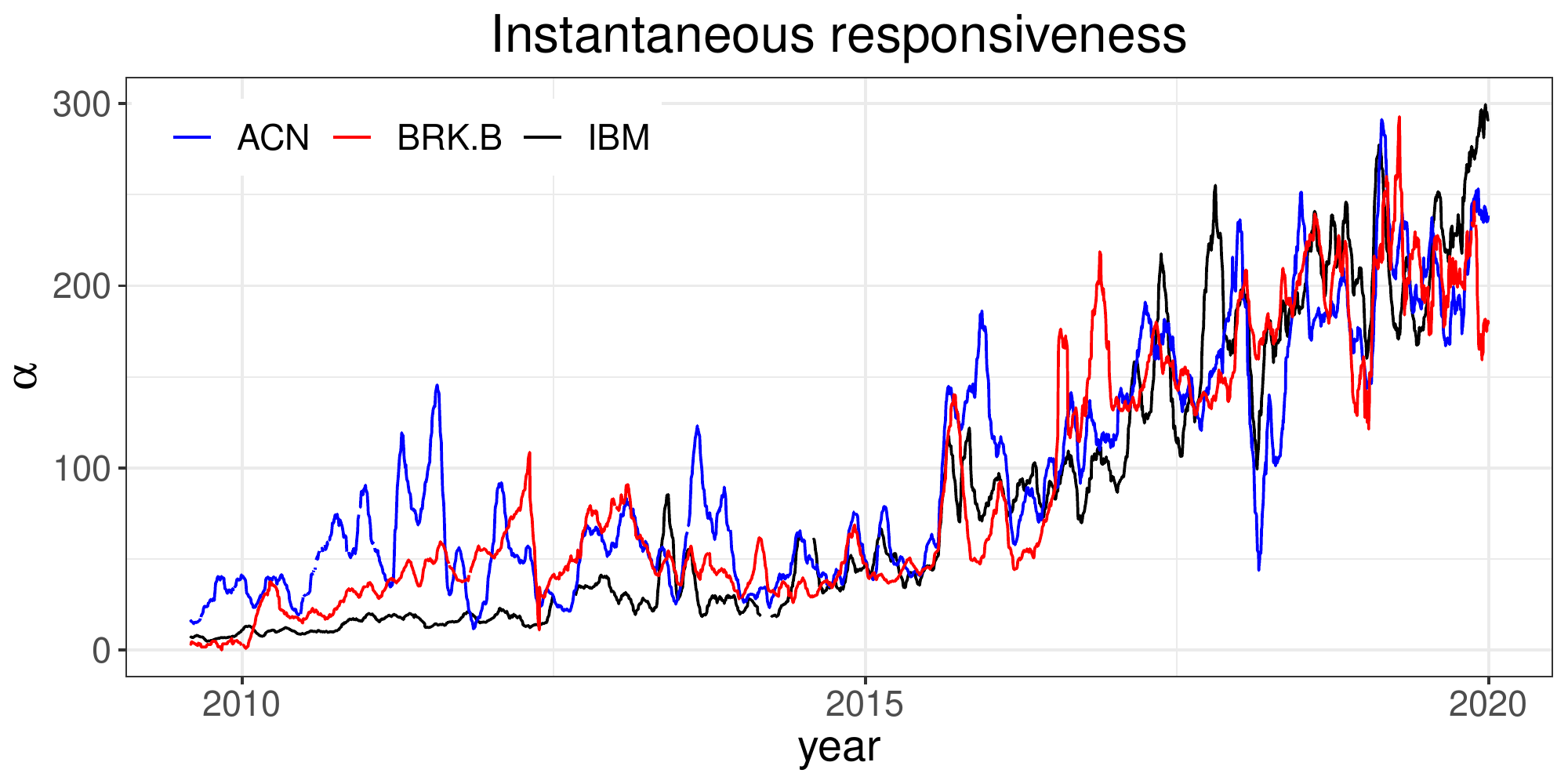}
	\caption{Moving average of $\bar \alpha$ for the NBBOs of Accenture, Berkshire Hathaway Inc. Class B, and IBM}
	\label{Fig:NBBO_ma_alpha}
\end{figure}

We define 
$$\bar \alpha =  \frac{\alpha_{s1} + \alpha_{s2} + \alpha_{m} + \alpha_{w1} + \alpha_{w2}}{5}$$
and consider this value to be a measure of the overall instantaneous responsiveness of each stock.
Of course, we cannot assert that successive events occur solely based on some decision making or algorithm depending on preceding events, 
but this instantaneous responsiveness can measure how fast a series of successive potentially related events occurs.
Figure~\ref{Fig:NBBO_ma_alpha} plots the moving average of $\bar \alpha$ from 2009 to 2019 for the NBBO of Accenture, Berkshire Hathaway Inc. Class B, and IBM.
Instantaneous responsiveness generally increases over time owing to technological innovations such as enhanced and cheaper hardware, faster communication speeds, new software, and better financial services.
Including these stocks, most stocks show a typical increasing pattern, 
but not all equities show monotone increases over time. We discuss other examples in the next subsection.

\subsubsection{Liquidity provision versus depletion (removal)}\label{subsec:compare}

\begin{figure}[hbt!]
	\centering
	\begin{subfigure}{.49\textwidth}
		\centering
		\includegraphics[width=0.99\textwidth]{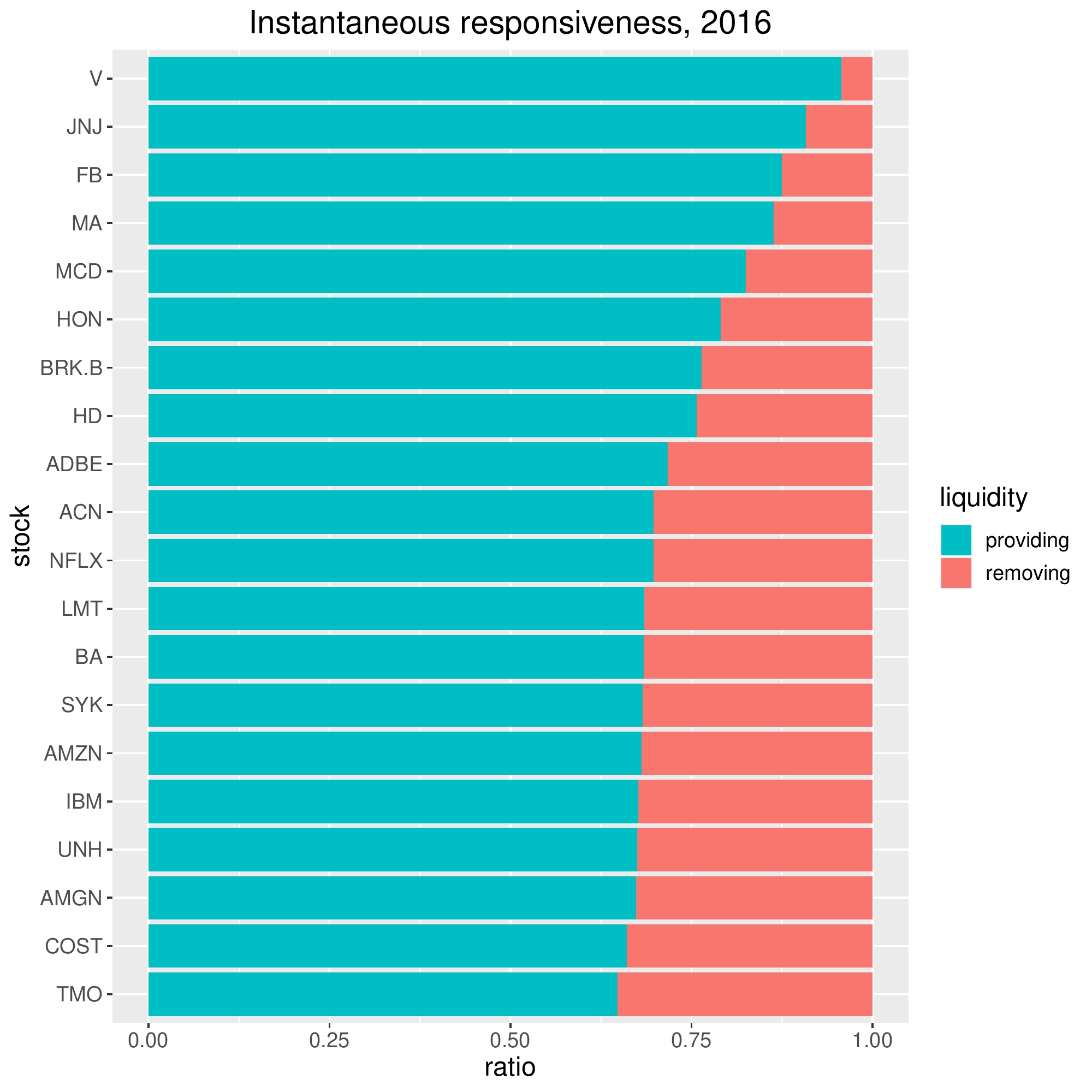}
		\caption{2016}
		\label{fig:liquidity_rem_vs_prov_2016}
	\end{subfigure}
	\begin{subfigure}{.49\textwidth}
		\centering
		\includegraphics[width=0.99\textwidth]{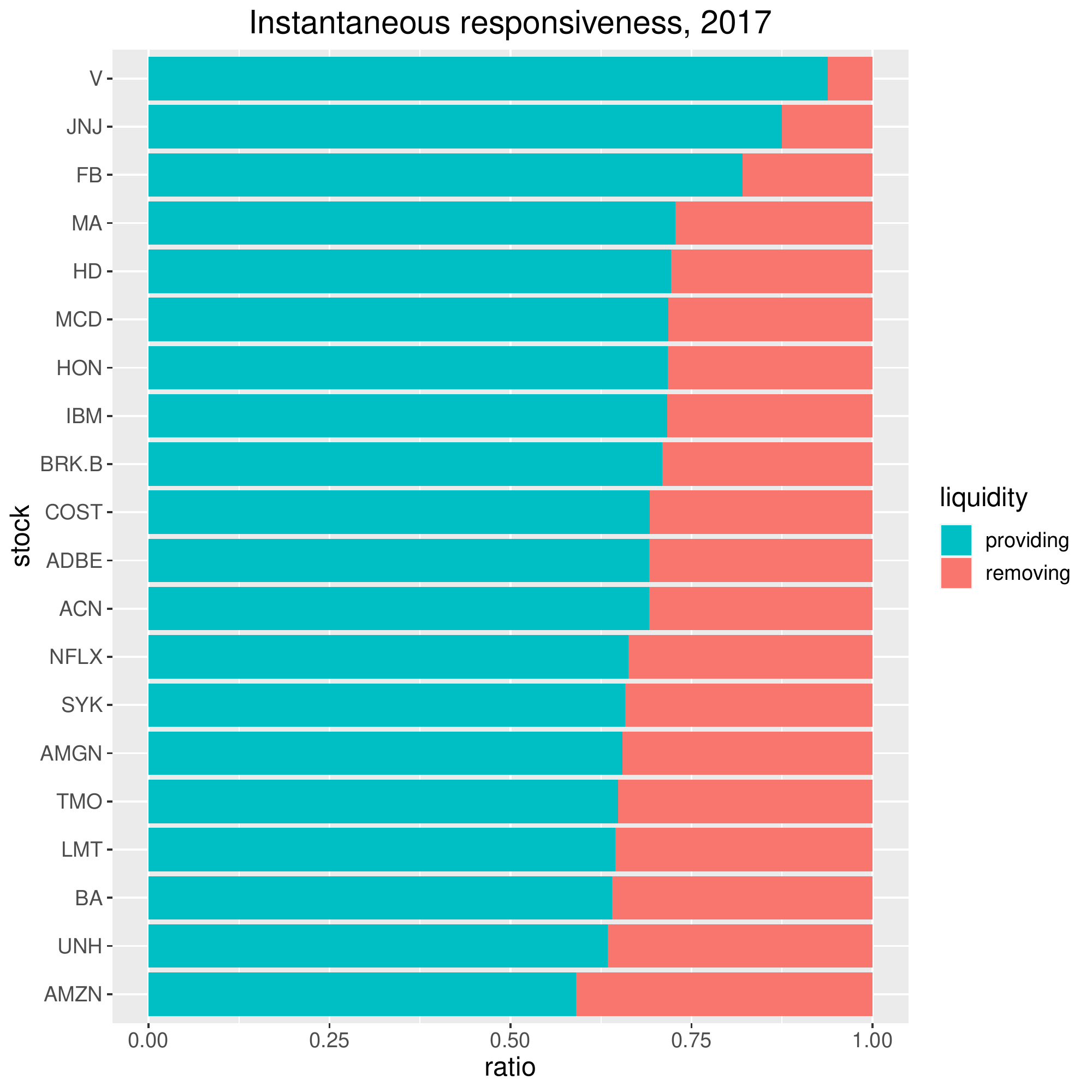}
		\caption{2017}
		\label{fig:liquidity_rem_vs_prov_2017}
	\end{subfigure}
	\caption{Responsiveness ratio measured using the liquidity provision and depletion parameters for various stocks, 2016 and 2017}
	\label{Fig:liquidity}
\end{figure}

Figure~\ref{Fig:liquidity} shows the estimation results of the ratio between the parameters related to liquidity provision and those related to liquidity depletion for various stocks. The parameter for the former is the mean of $\alpha_{w1}$ and $\alpha_{w2}$, whereas that for the latter is the mean of $\alpha_{s1}$, $\alpha_{s2}$, and $\alpha_{m}$.
The figure shows the average values of the parameter estimates in 2016 and 2017.
In all the examined stocks, the average estimates of the liquidity provision parameters are larger than the average liquidity depletion parameters.
This result implies that actions to provide liquidity generally occur faster than do actions that remove liquidity in response to previous events.

Among those, Amazon in 2017 has the smallest ratio of liquidity provision, at slightly greater than 0.5, 
and Visa has a value close to 1 in 2016.
Market markers' speed competition in providing liquidity can be well understood, 
since the fastest provider of limit orders preempts the advantageous position in the limit order queue.
For brevity, we illustrate only the estimates for 2016 and 2017, but the ratio changes over time.
We can also clearly see that the ratios of liquidity provision are generally larger in 2016.

\begin{figure}[hbt!]
	\centering
	\begin{subfigure}{.49\textwidth}
		\centering
		\includegraphics[width=0.99\textwidth]{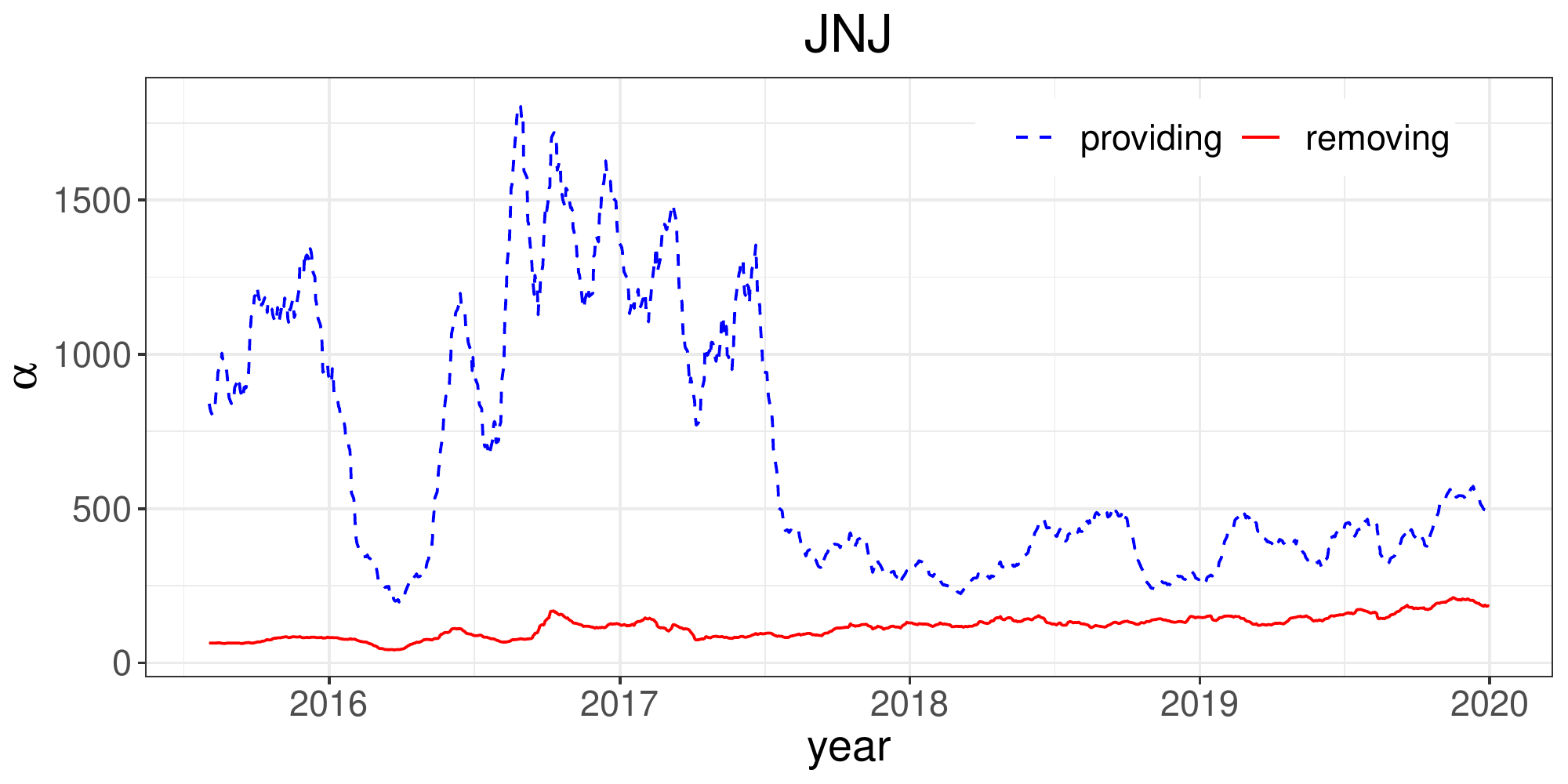}
		\caption{Johnson \& Johnson}
		\label{fig:jnj_liquidity}
	\end{subfigure}
	\begin{subfigure}{.49\textwidth}
		\centering
		\includegraphics[width=0.99\textwidth]{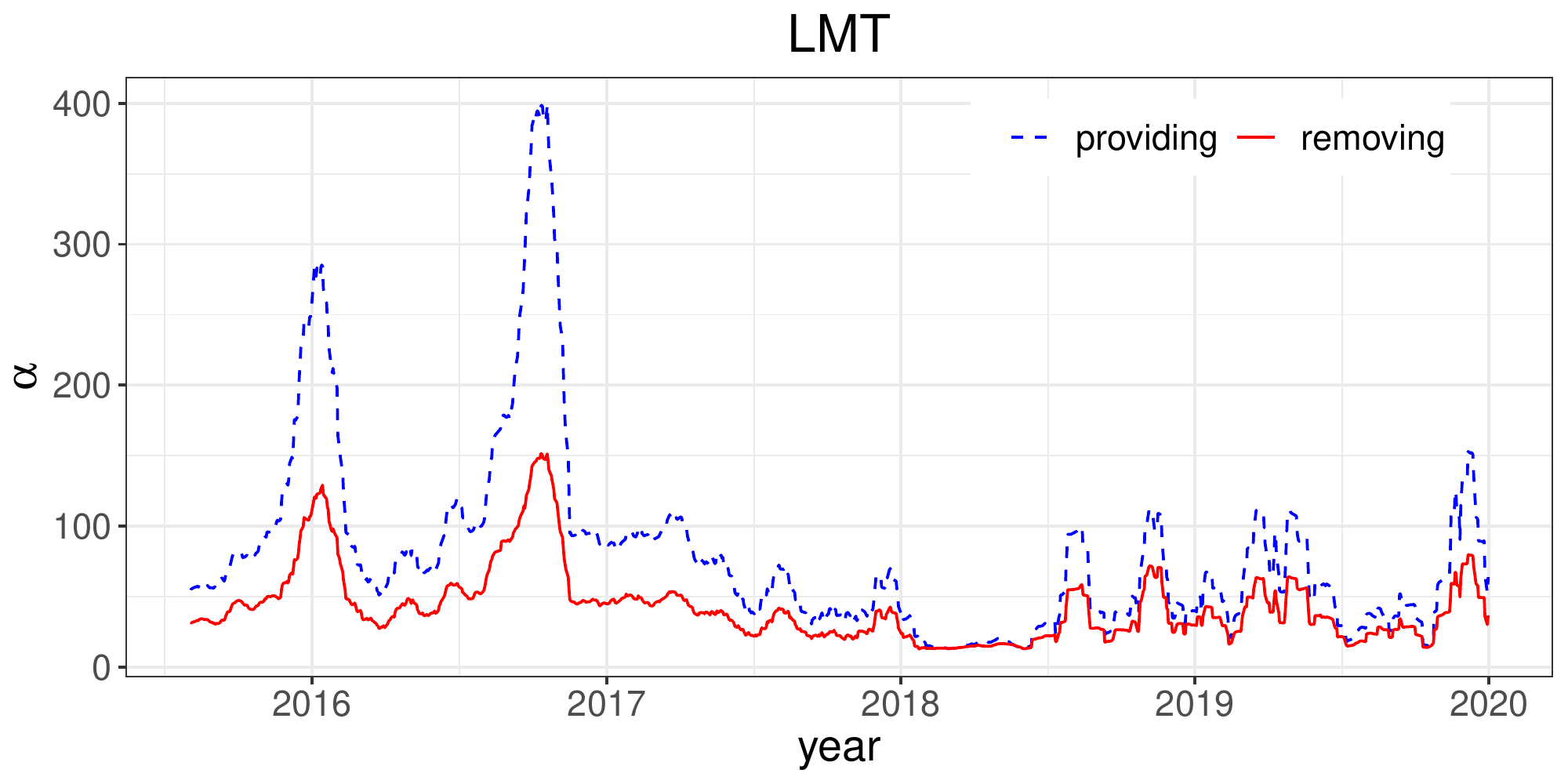}
		\caption{Lockheed Martin}
		\label{fig:fb_liquidity}
	\end{subfigure}
	
	\begin{subfigure}{.49\textwidth}
		\centering
		\includegraphics[width=0.99\textwidth]{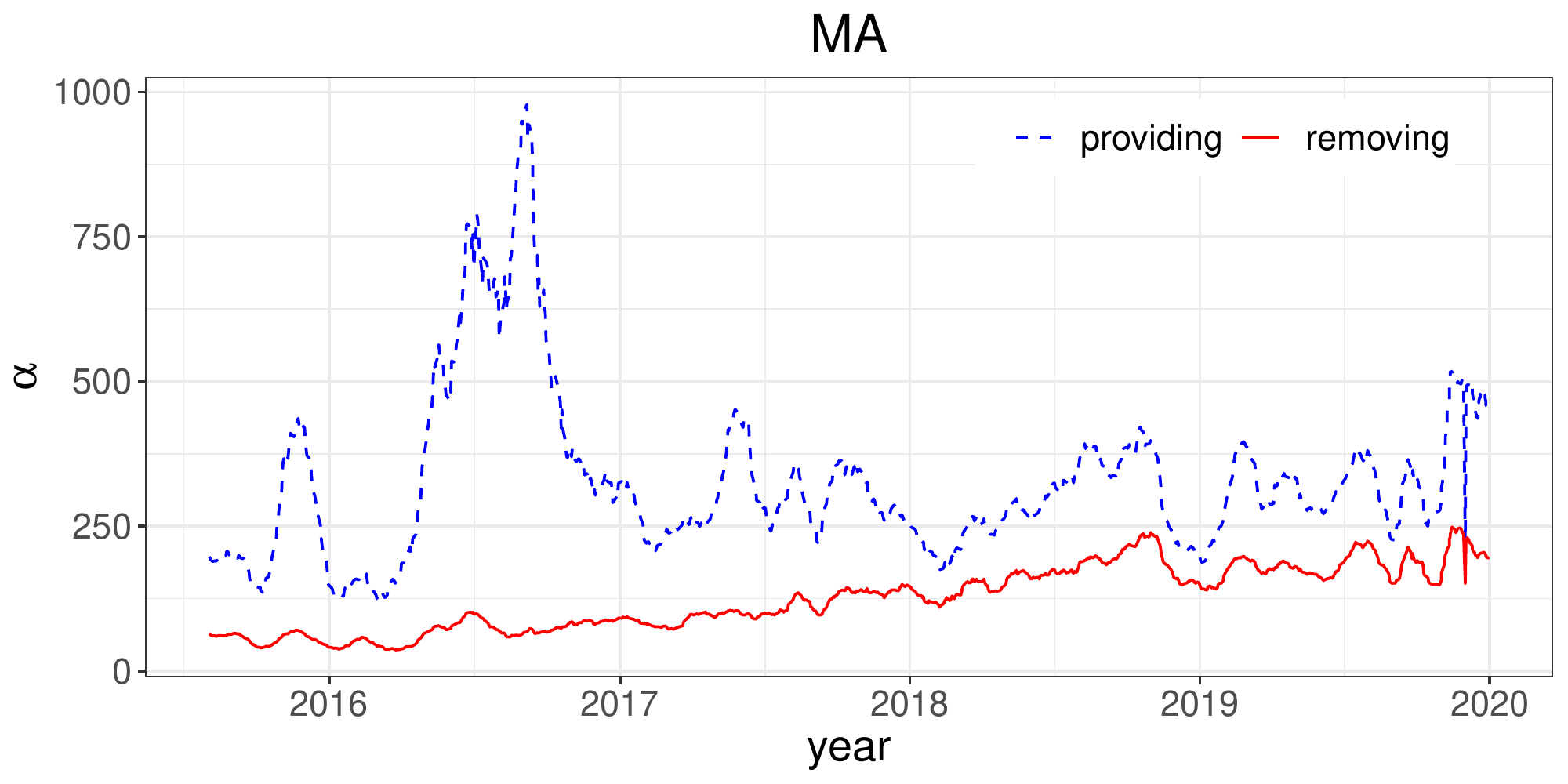}
		\caption{Mastercard}
		\label{fig:ma_liquidity}
	\end{subfigure}
	\begin{subfigure}{.49\textwidth}
		\centering
		\includegraphics[width=0.99\textwidth]{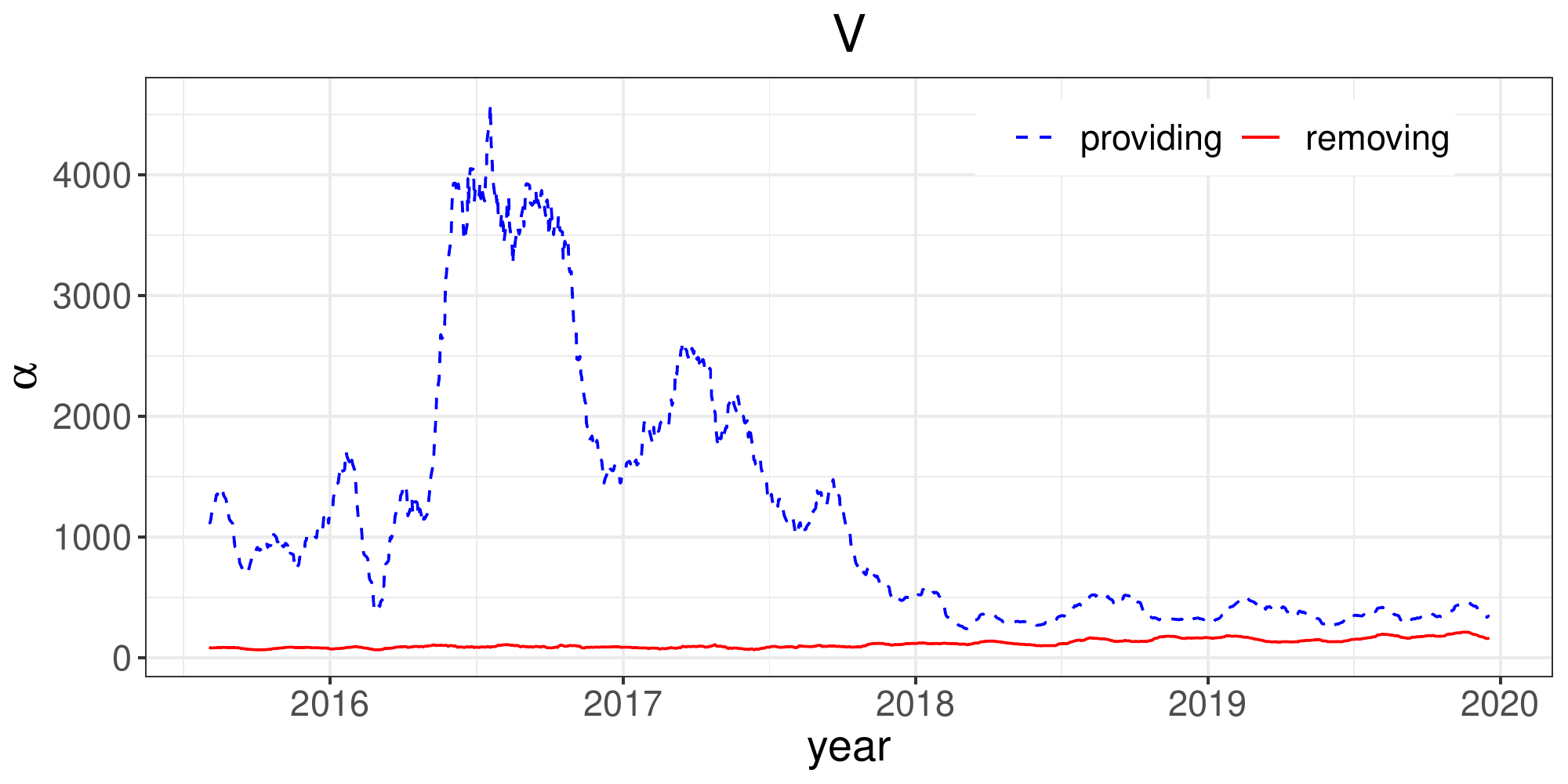}
		\caption{Visa}
		\label{fig:v_liquidity}
	\end{subfigure}
	
	\caption{The 2016 peak and diminishing pattern in the liquidity provision parameter}
	\label{Fig:stock_liquidity}
\end{figure}

Some stocks show an interesting pattern that peaks around 2016 and diminishes in the liquidity provision parameter in Figure~\ref{Fig:stock_liquidity}.
This figure shows the smoothed dynamics of the liquidity provision (blue dashed line) and depletion $\alpha$ parameters (red solid line).
For example, the largest liquidity provision parameter for Visa is over 4,000, 
which is large compared with the $\alpha$ values in Figure~\ref{Fig:NBBO_ma_alpha},
suggesting that ultra-high-frequency market makers paid much attention to these stocks during this period.
However, market makers might conclude that such ultra-fast reaction times do not necessarily improve profits. Indeed, to reduce costs,
the liquidity provision parameters converge to much lower values after 2018,
which are similar to the levels of $\alpha$ in Figure~\ref{Fig:NBBO_ma_alpha}.
We discuss the reaction speed for liquidity provision, which is particularly high in 2016, in 
Section~\ref{Subsubsect:BYX} in more detail.

As we show, for most stocks, $\bar \alpha$, 
or the liquidity-providing $\alpha$, which represents the average reaction time, reaches levels of around tens to hundreds by 2019.
In some cases, it increases steadily to that level in Figure~\ref{Fig:NBBO_ma_alpha}; on the contrary, in other cases, it reaches this level after several years of volatile liquidity provision speeds, as in Figure~\ref{Fig:stock_liquidity}.
The behavior of high-frequency traders on most stocks appears to eventually reach similar levels over time in terms of speed and risk attitude.
However, not all the stocks show the same pattern.
In 2018 and 2019, a few stocks have fierce market-making competition in ultra-high-frequency trading with strong confidence in adverse selection.

\subsubsection{Outliers}\label{subsec:outlier}

\begin{figure}[hbt!]
	\centering
	\begin{subfigure}{.49\textwidth}
		\centering
		\includegraphics[width=0.99\textwidth]{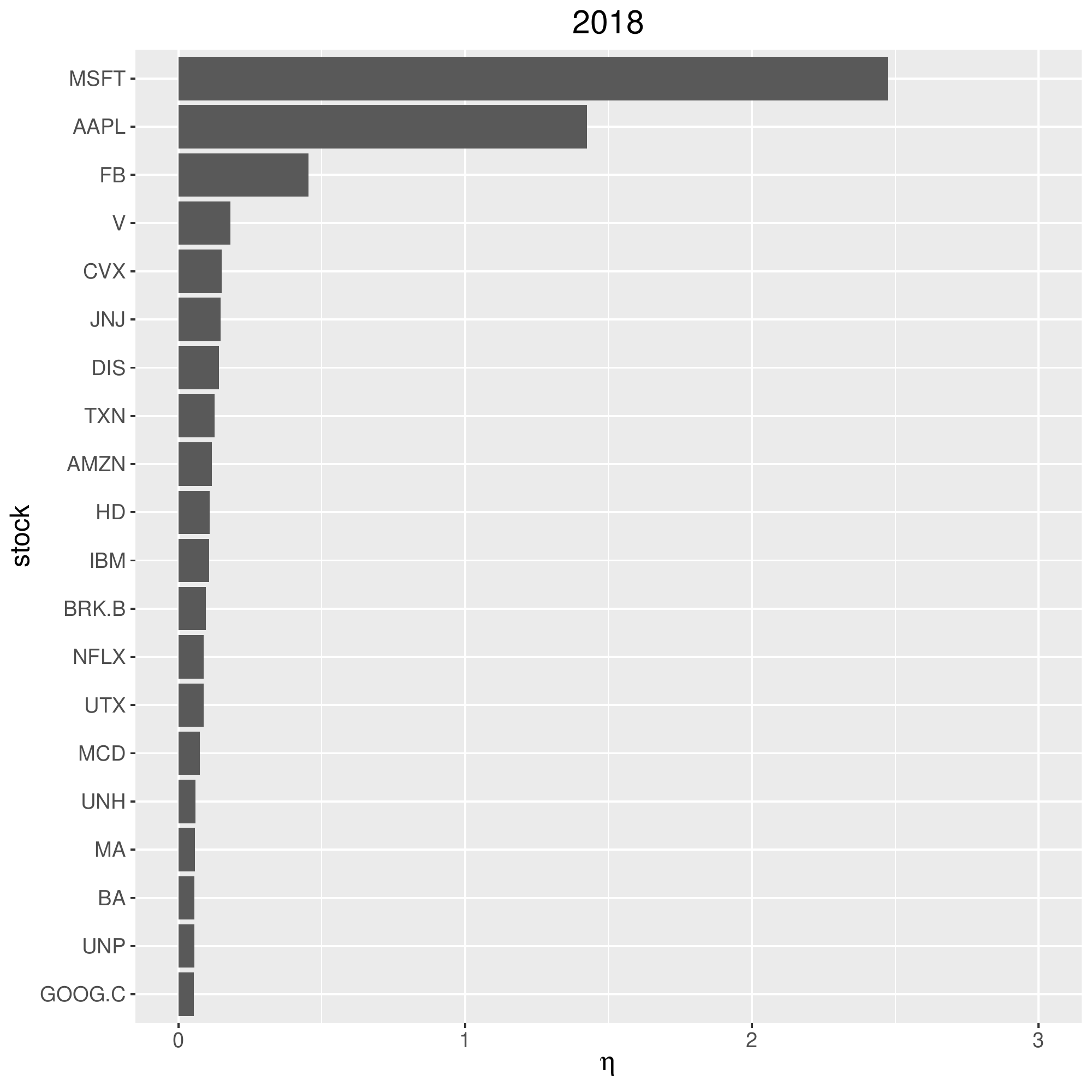}
		\caption{2018}
		\label{fig:eta_2018}
	\end{subfigure}
	\begin{subfigure}{.49\textwidth}
		\centering
		\includegraphics[width=0.99\textwidth]{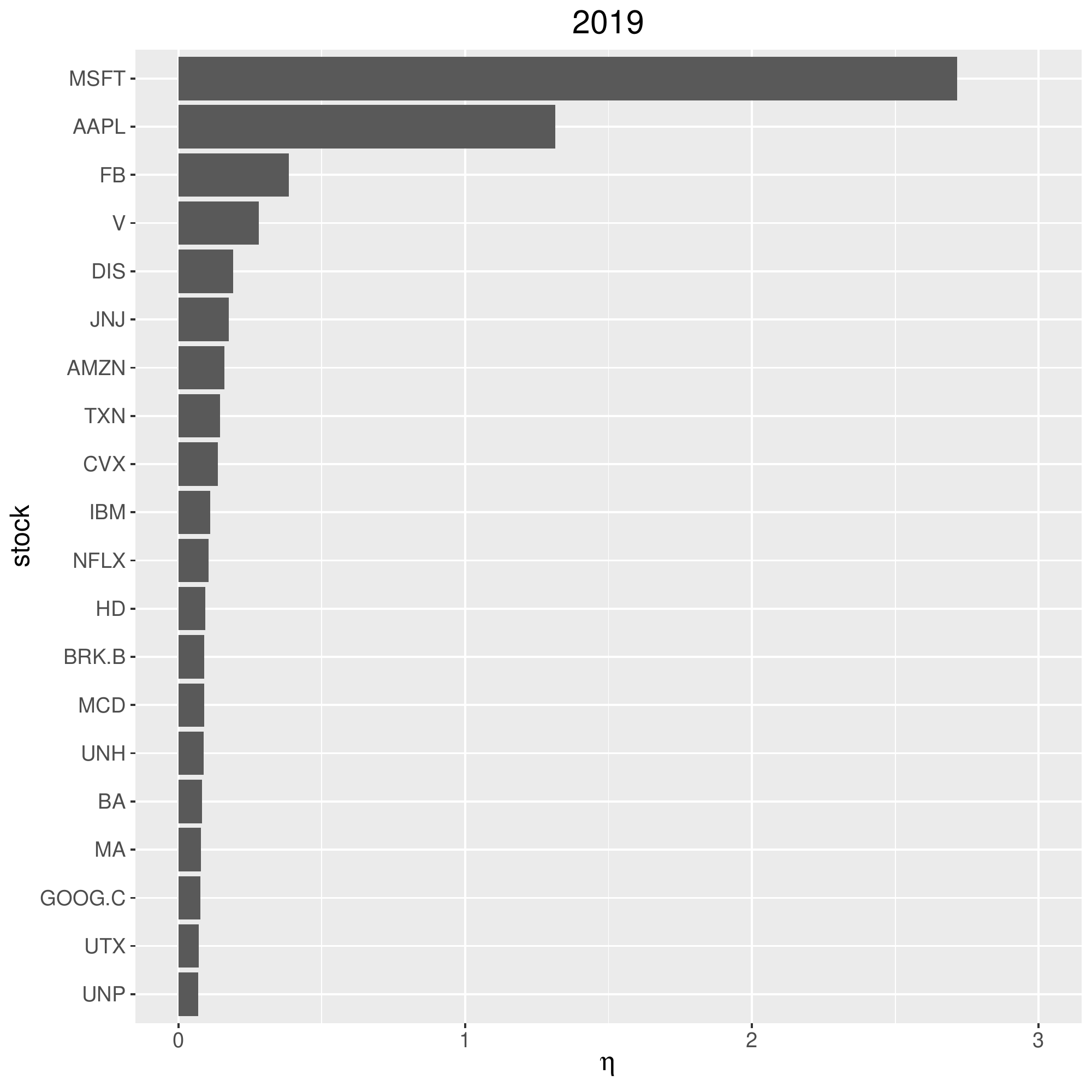}
		\caption{2019}
		\label{fig:eta_2019}
	\end{subfigure}
	\caption{The average $\eta$ for various stocks, 2018 and 2019}
	\label{Fig:eta}
\end{figure}

Some of the stocks we analyze seem to have received particular attention from ultra-high-frequency liquidity providers in terms of the liquidity parameter $\eta$. 
These are IT giants such as Microsoft, Apple, and Facebook.
The larger the $\eta$ is, 
the stronger market makers tend to narrow the spread through aggressive limit orders 
when spreads widen.
This happens when market makers are confident and less concerned about adverse selection.
Figure~\ref{Fig:eta} shows bar graphs of the $\eta$s estimated in 2018 and 2019 on average for 20 stocks.
The $\eta$ of Microsoft is large, followed by those of Apple and Facebook.
For the other stocks, the $\eta$s are at a similar level, implying a similar assessment of adverse selection risk on average.

\begin{figure}[hbt!]
	\centering
	\begin{subfigure}{.49\textwidth}
		\centering
		\includegraphics[width=0.99\textwidth]{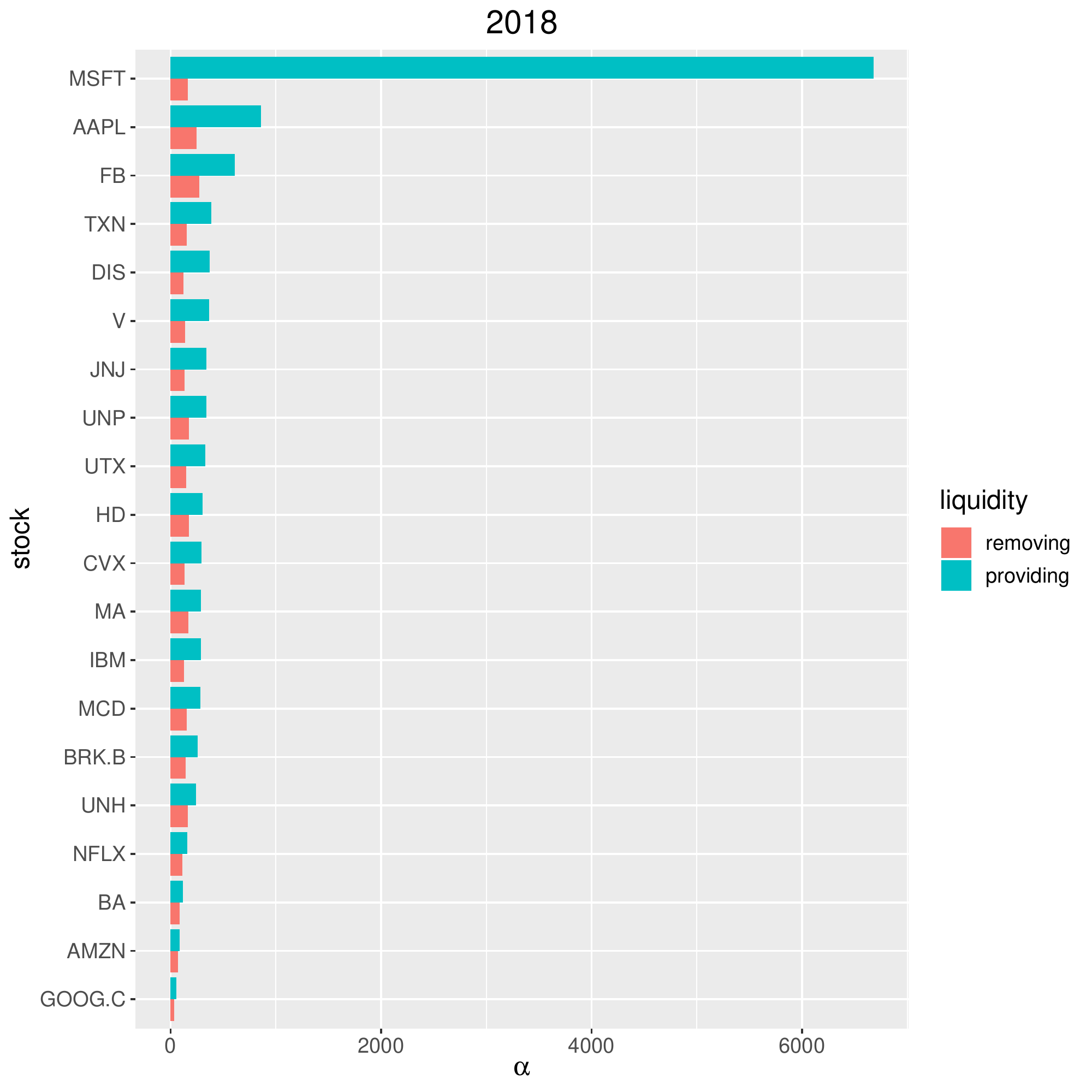}
		\caption{2018}
		\label{fig:liq_2018}
	\end{subfigure}
	\begin{subfigure}{.49\textwidth}
		\centering
		\includegraphics[width=0.99\textwidth]{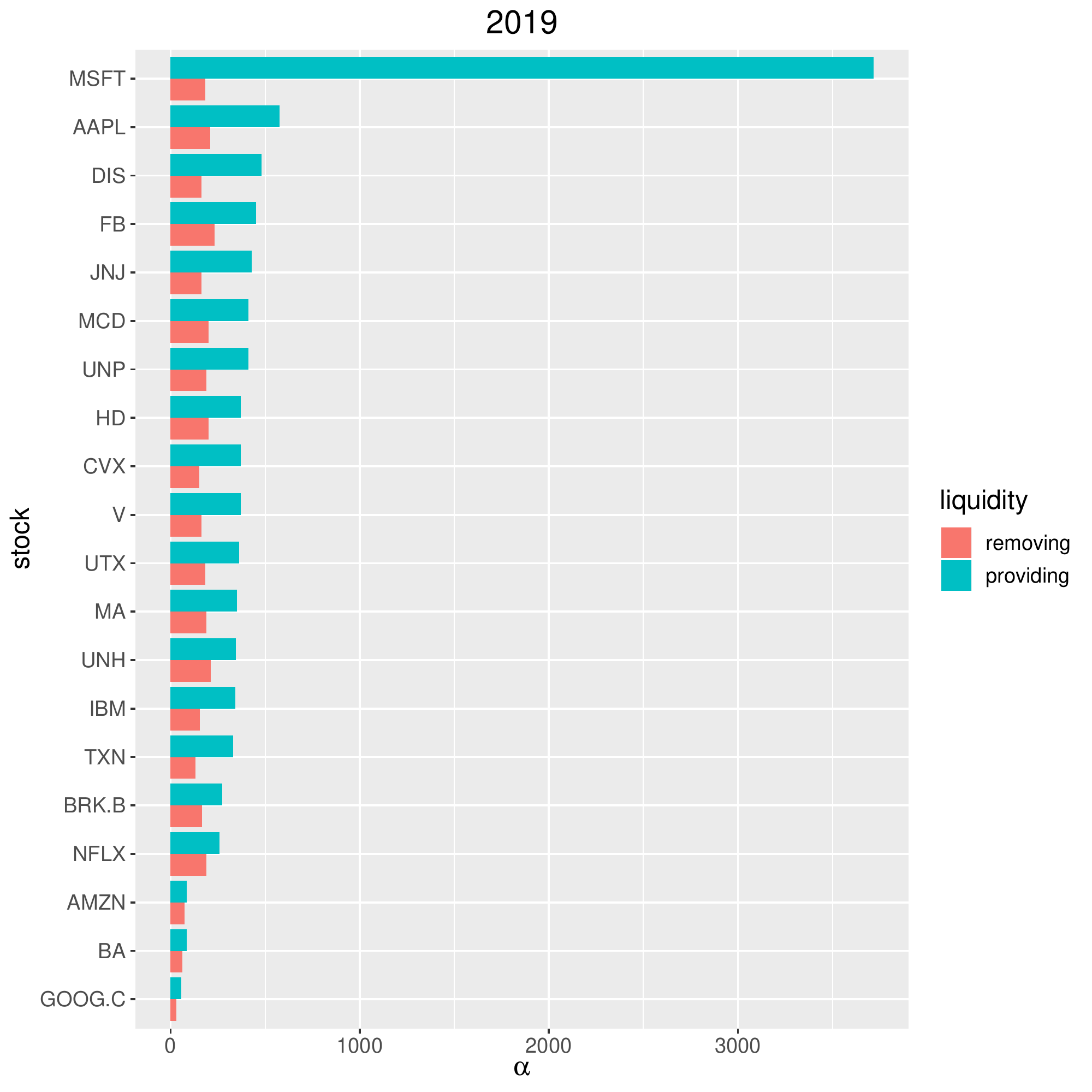}
		\caption{2019}
		\label{fig:liq_2019}
	\end{subfigure}
	\caption{The average liquidity-providing and -depleting $\alpha$s for various stocks, 2018 and 2019}
	\label{Fig:liq}
\end{figure}

The competition for these stocks, particularly for Microsoft, to occupy the top priority of the limit order books is intense, as Figure~\ref{Fig:liq} shows.
This figure presents the yearly averages of the liquidity-providing $\alpha$s, namely, $\alpha_{w1}$ and $\alpha_{w2}$, with green bars
and the yearly averages of the liquidity-depleting $\alpha$s with red bars.
Microsoft, Apple, and Facebook has the top liquidity provision speed in 2018, but responsiveness decreases from 2018 to 2019.
Especially for Apple and Facebook, by 2019, the relative gap between the liquidity provision speed, measured by the average $\alpha_w$s, and the other stocks narrows.
However, Microsoft is still the only outlier in 2019, at least in our sample, and seems to attract more attention from high-frequency liquidity providers than the other stocks do.
Although we do not know the reason for Microsoft's popularity among high-frequency traders,
this result means that not all the stocks receive the same interest from high-frequency traders.

\subsection{Estimation results by exchange}\label{subsec:exchange}

\subsubsection{NYSE versus BZX}

Using our model, we can capture the varying characteristics of the different exchanges by analyzing the quote data for each trading venue.

\begin{figure}[hbt!]
	\centering
	\includegraphics[width=0.48\textwidth]{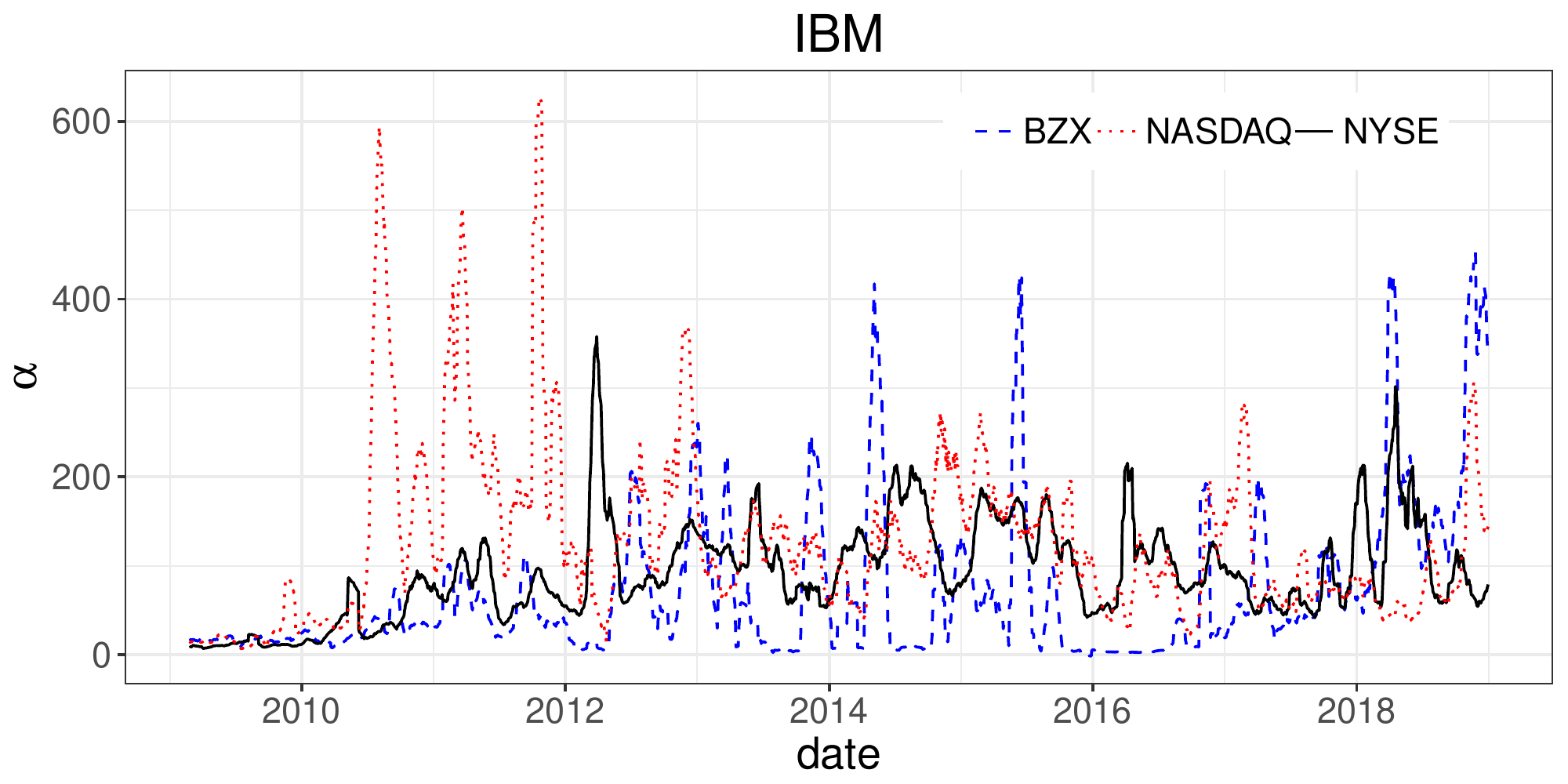}\quad
	\includegraphics[width=0.48\textwidth]{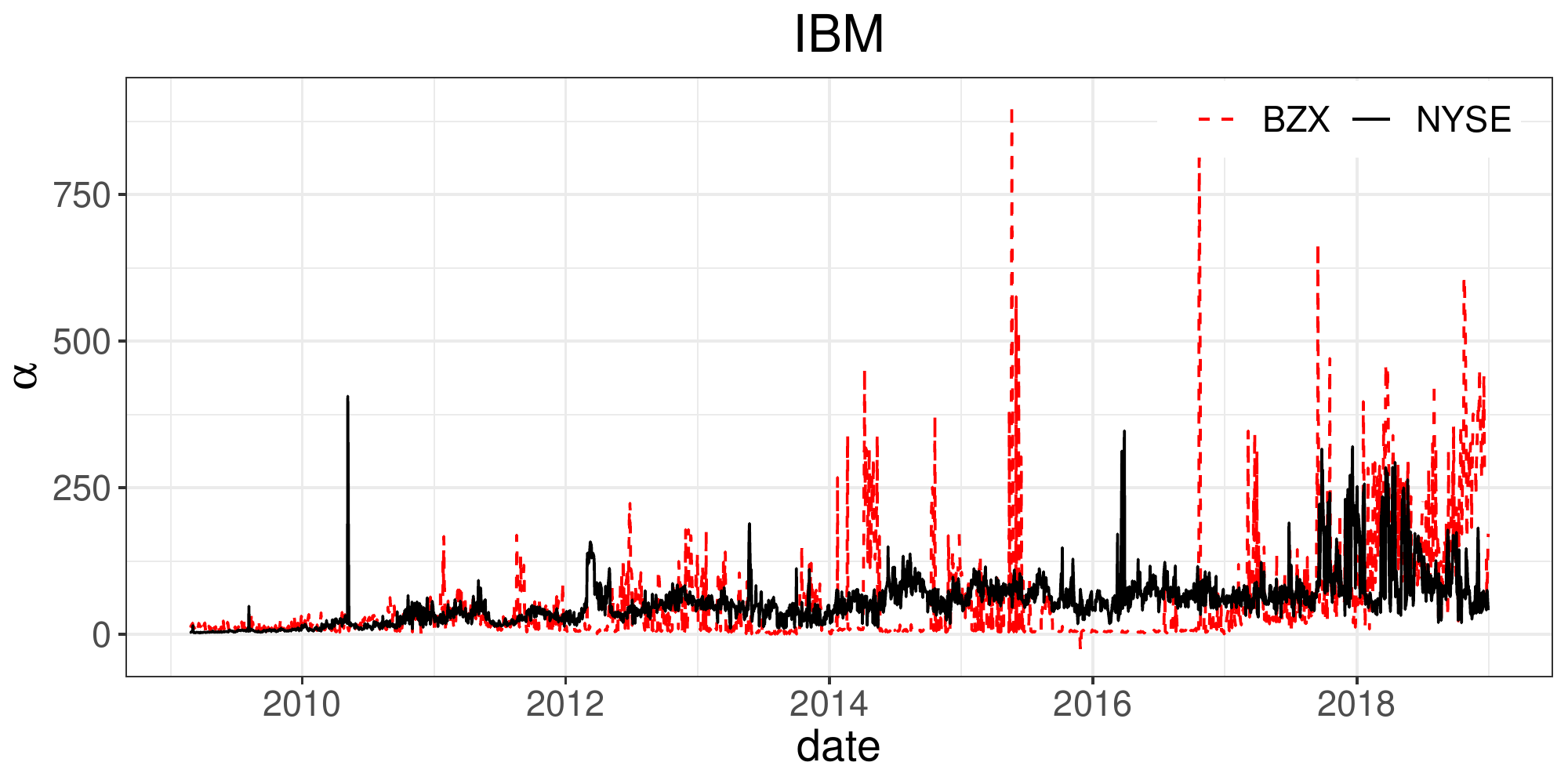}
	\caption{Smoothed $\bar \alpha$ for IBM from the NYSE, NSDQ, and BZX (left) and nonsmoothed $\bar \alpha$ for IBM from the NYSE and BZX (right)}\label{Fig:IBM_per_exchange}
\end{figure}

The left-hand side of Figure~\ref{Fig:IBM_per_exchange} plots the dynamics of the moving average $\bar \alpha$ for the BBO of IBM in several of the stock exchanges.
In 2010 and 2011, the responsiveness of the Nasdaq, shown by the red dotted line, is the fastest, whereas the responsiveness of the Nasdaq, NYSE (black solid line), and BZX (blue dashed line) is similar.
Although omitted, the Arca has similar $\bar \alpha$ values to the NYSE, which is not surprising as it operates as a subsidiary of the NYSE Group.
Indeed, the Arca, formerly the Archipelago, merged with the NYSE in 2006.
The Archipelago had electronic stock trading technology, which the NYSE did not at the time.
However, from 2009, the starting point of our data analysis, we see 
no substantial performance difference in the speeds of the NYSE and Arca.

The Nasdaq’s electronic stock trading technology originated from one of the first electronic communication networks, Island, which had a good reputation for fast electronic trading in the early 2000s~\citep{patterson2012dark}. 
By adopting this technology, the Nasdaq shows relatively large $\bar \alpha$ values in the early 2010s.
We also include a graph of the Cboe BZX Exchange, which claims to be designed for cheap and fast transactions 
after consecutive mergers between traditional stock exchanges and electronic trading systems.
The Cboe BZX Exchange had a 6\% market share in 2017 and is considered to be among top-tier exchanges such as the NYSE, NYSE Arca, and Nasdaq \citep{Elaine}.

The right-hand side of Figure~\ref{Fig:IBM_per_exchange} shows the time-varying shape of IBM's overall responsiveness $\bar \alpha$ for the NYSE and Cboe BZX Exchange.
We do not smooth the estimates in this figure.
The black solid line represents the $\bar \alpha$ estimated daily by the best bid and ask prices quoted in the NYSE 
and the red dashed line represents the $\bar \alpha$ estimated by the prices in the BZX.
When comparing the two exchanges, the NYSE has relatively little variation in $\bar \alpha$, 
but $\bar \alpha$ changes more drastically over time for the BZX.
The maximum $\bar \alpha$ of the BZX is almost twice that of the NYSE,
implying that the BZX can operate under very low latency compared with the NYSE
and can attract high-frequency traders in this respect.
This result is in line with the BZX's original claim of high-speed technology.

\begin{figure}
	\centering
	\includegraphics[width=0.48\textwidth]{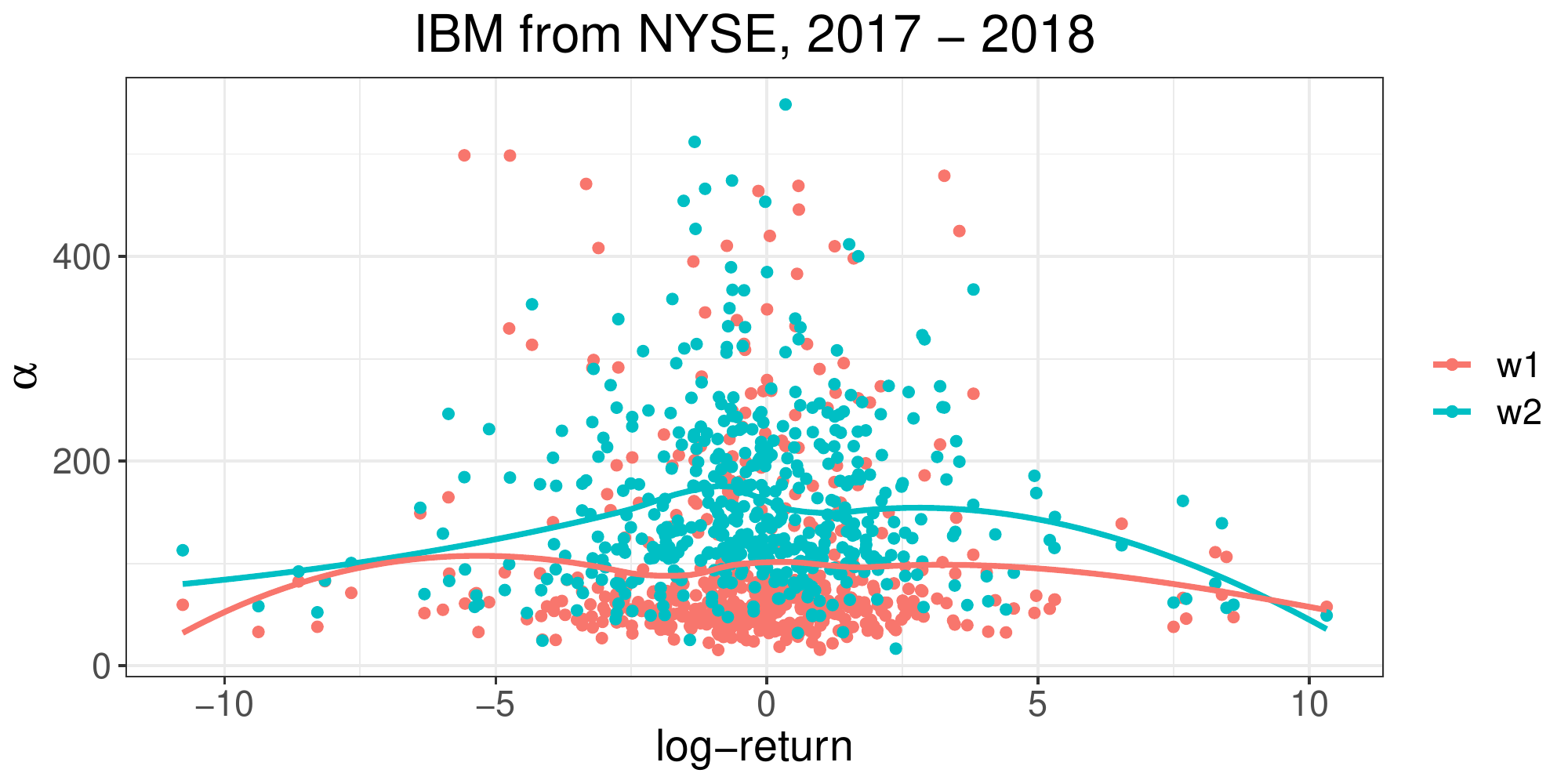}\quad
	\includegraphics[width=0.48\textwidth]{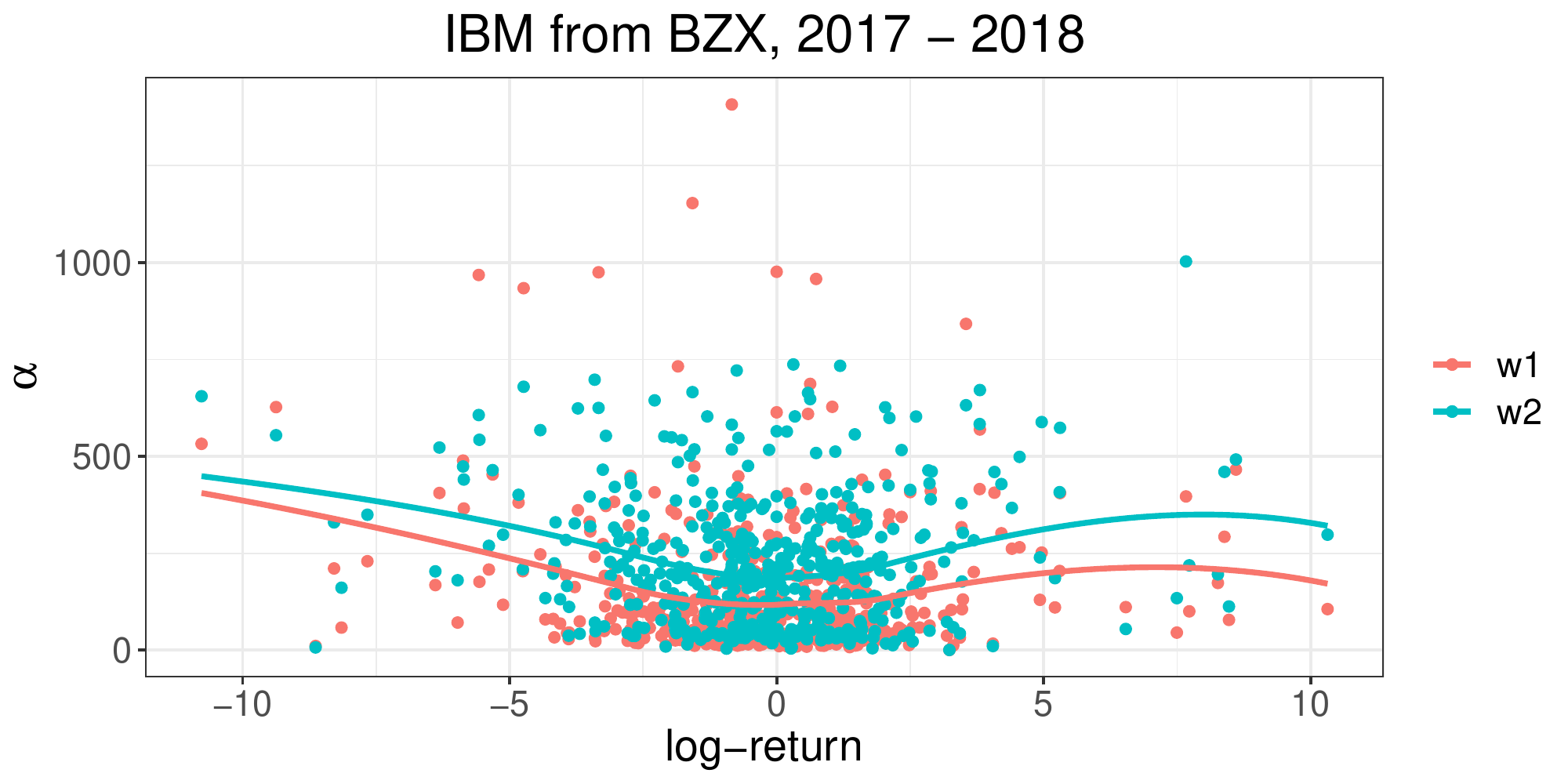}
	\caption{Scatterplot of the $\alpha$s versus the daily log-return, IBM, 2017--2018, from the NYSE (left) and BZX (right)}
	\label{Fig:IBM_scatter_alpha_return}
\end{figure}

We can also see the different behaviors between NYSE and BZX traders in volatile markets in Figure~\ref{Fig:IBM_scatter_alpha_return} based on data for IBM from 2017 to 2018.
The figure presents the scatterplots of $\alpha_{w1}$ and $\alpha_{w2}$, the liquidity provision parameters, versus daily log-returns.
The other excitement parameters have similar patterns.
In the NYSE (left), when the absolute value of the log-return is large (i.e., on days when the market is volatile),
liquidity provision parameters are relatively small, and the absolute return and excitement parameters have a negative correlation.
Thus, high-frequency traders in the NYSE, especially liquidity providers, are less active in volatile conditions, perhaps concerned about adverse selection.
Conversely, we can observe the opposite situation in the BZX.
High-frequency quotes, indicated by the high liquidity provision parameter $\alpha_w$ values, remain active and become even more active in a volatile market in the BZX.
The absolute return and liquidity provision parameters are positively correlated. 
Thus, high-frequency traders in the BZX tend to take aggressive positions, even in volatile markets.

\subsubsection{The BZX versus BYX and the EDGA versus EDGX}~\label{Subsubsect:BYX}

\begin{figure}[hbt!]
	\centering
	\includegraphics[width=0.48\textwidth]{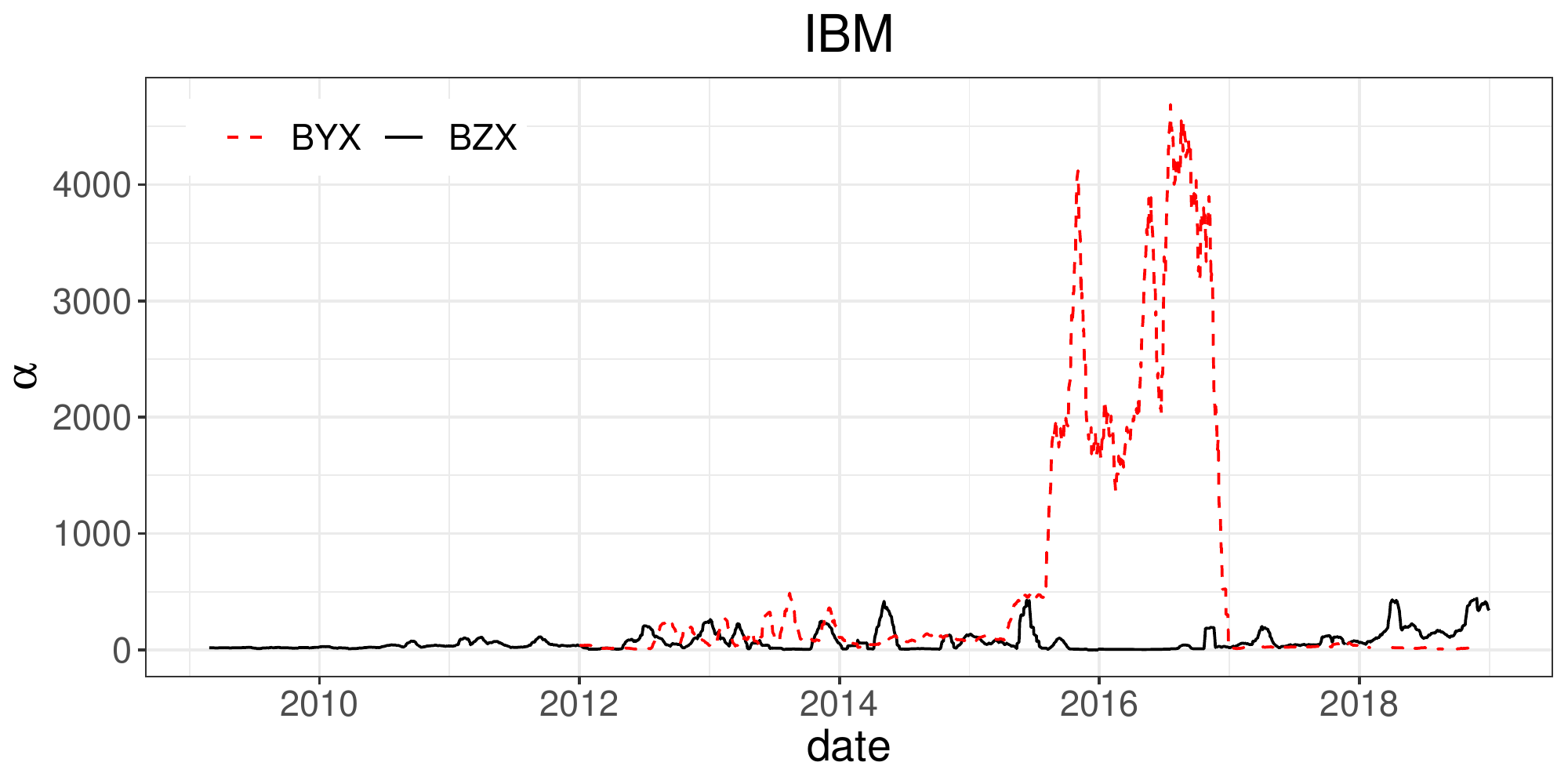}\quad
	\includegraphics[width=0.48\textwidth]{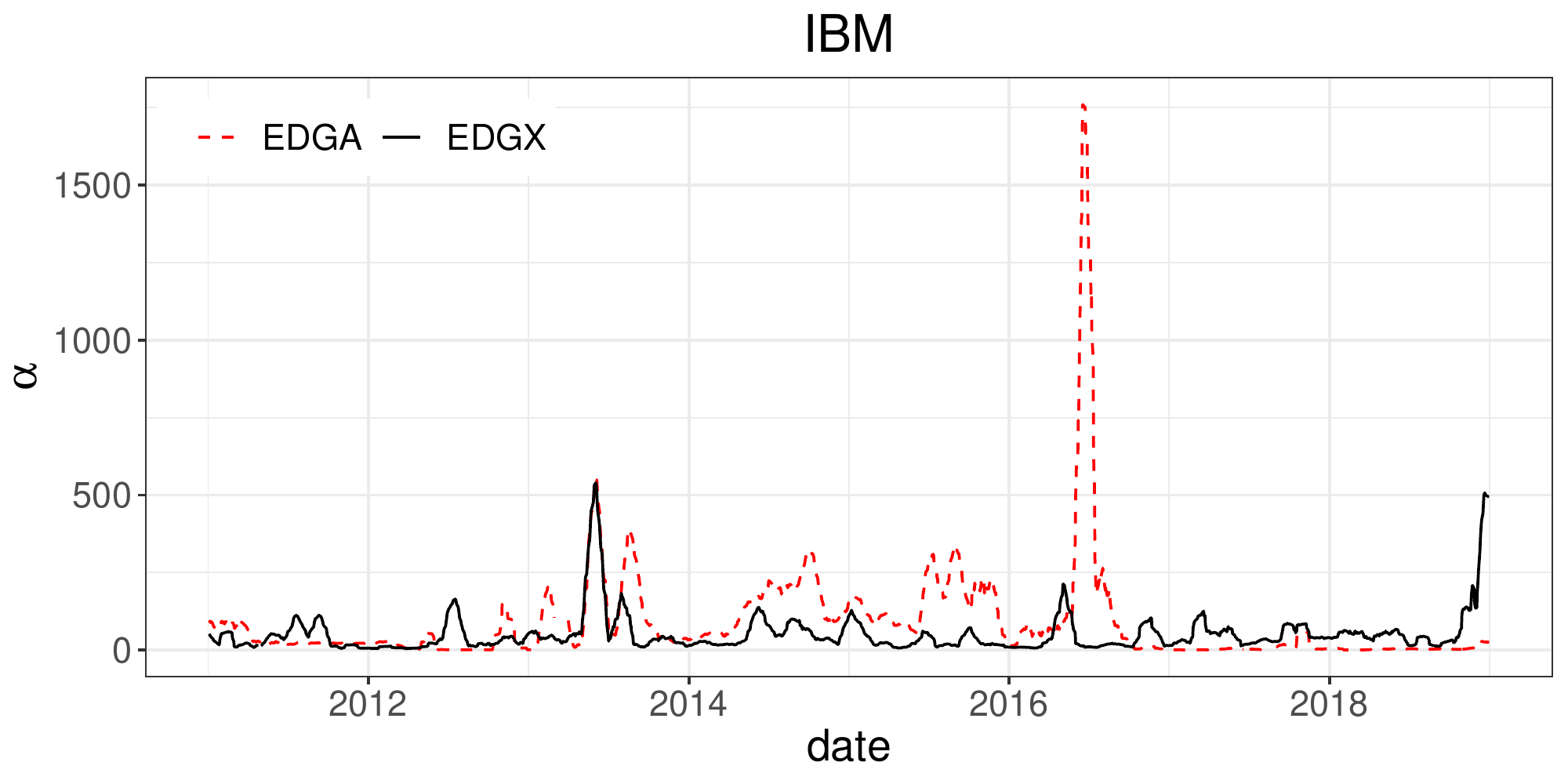}
	\caption{Smoothed dynamics of the $\bar \alpha$s of the BZX and BYX (left) and the EDGA and EDGX (right)}
	\label{Fig:IBM_BATS_BATSY}
\end{figure}

In this subsection, we compare instantaneous responsiveness depending on the fee structure.
The right-hand side of Figure~\ref{Fig:IBM_per_exchange} shows that the averaged excitement parameters, $\bar \alpha$, of the BZX approaching 2016 is close to zero; 
thus, the left-hand side of Figure~\ref{Fig:IBM_BATS_BATSY} plots the $\bar \alpha$s of the BZX and BYX.
The reason for the low averaged excitement parameter values in this period is unknown,
but we suppose that the high-frequency traders, especially liquidity providers, active in the BZX before 2016 moved to the BYX at this time.
The averaged excitement parameters of the BYX rises sharply during the period when the ones of the BZX is small.
Conversely, after 2017, the averaged excitement parameters of the BZX rises steadily, whereas the ones of the BYX remains close to zero.
The BZX provides some rebates when traders provide liquidity through a limit order
and takes fees when transacting market orders.
Conversely, the BYX uses an inverted fee schedule, providing rebates to traders removing liquidity through market orders
and taking fees for quoting limit orders. 

We can see a similar relationship for the EDGA and EDGX.
The EDGX's averaged excitement parameters increases from the second half of 2018, 
whereas the EDGA's parameter is close to zero in later years, as shown on the right-hand side of Figure~\ref{Fig:IBM_BATS_BATSY}.
In addition, the EDGA's averaged excitement parameters rises sharply in 2016.
The EDGA provides rebates for removing liquidity, whereas the EDGX provides rebates for adding liquidity.

\begin{figure}[hbt!]
	\centering
	\includegraphics[width=0.48\textwidth]{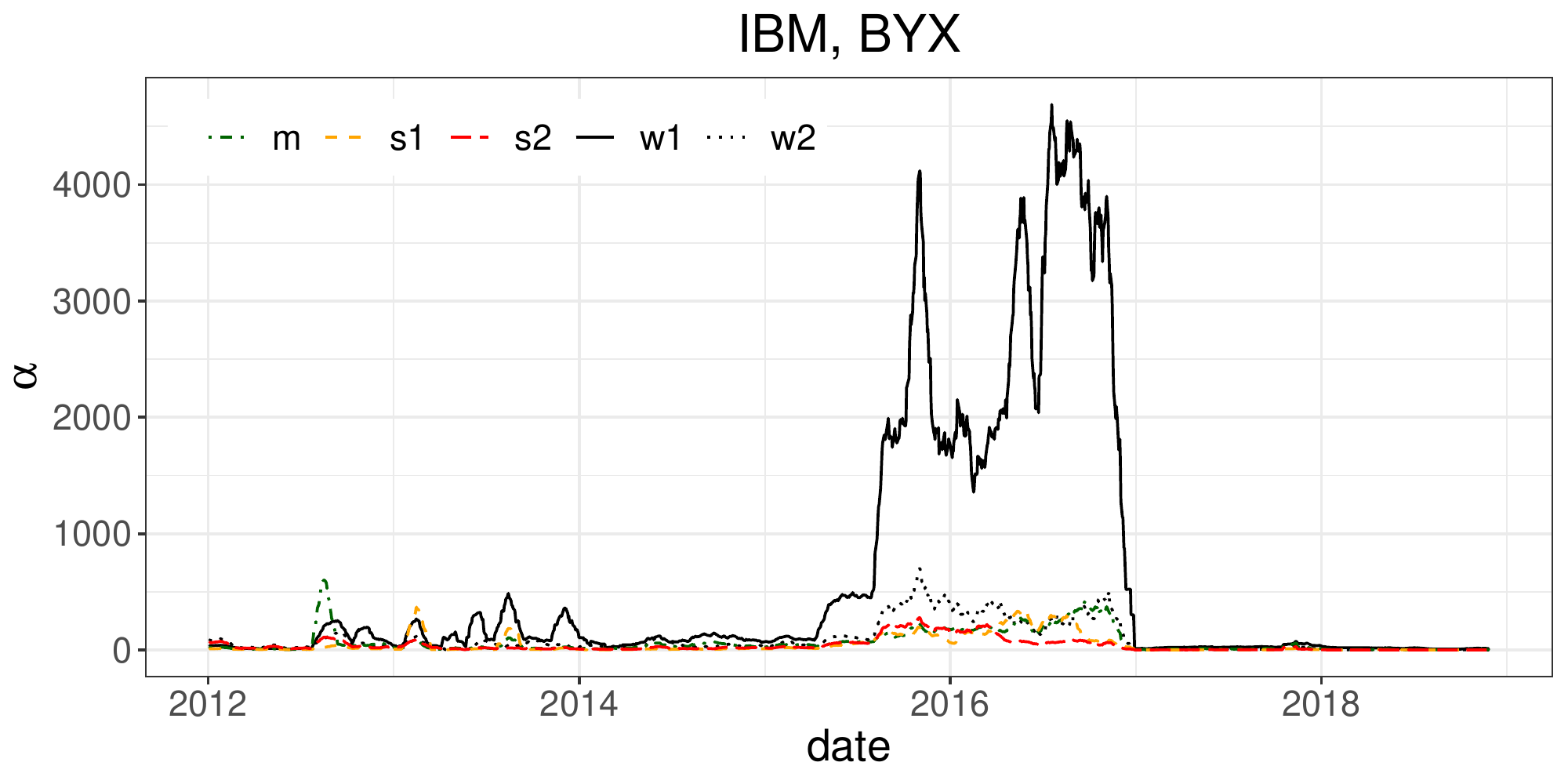}\quad
	\includegraphics[width=0.48\textwidth]{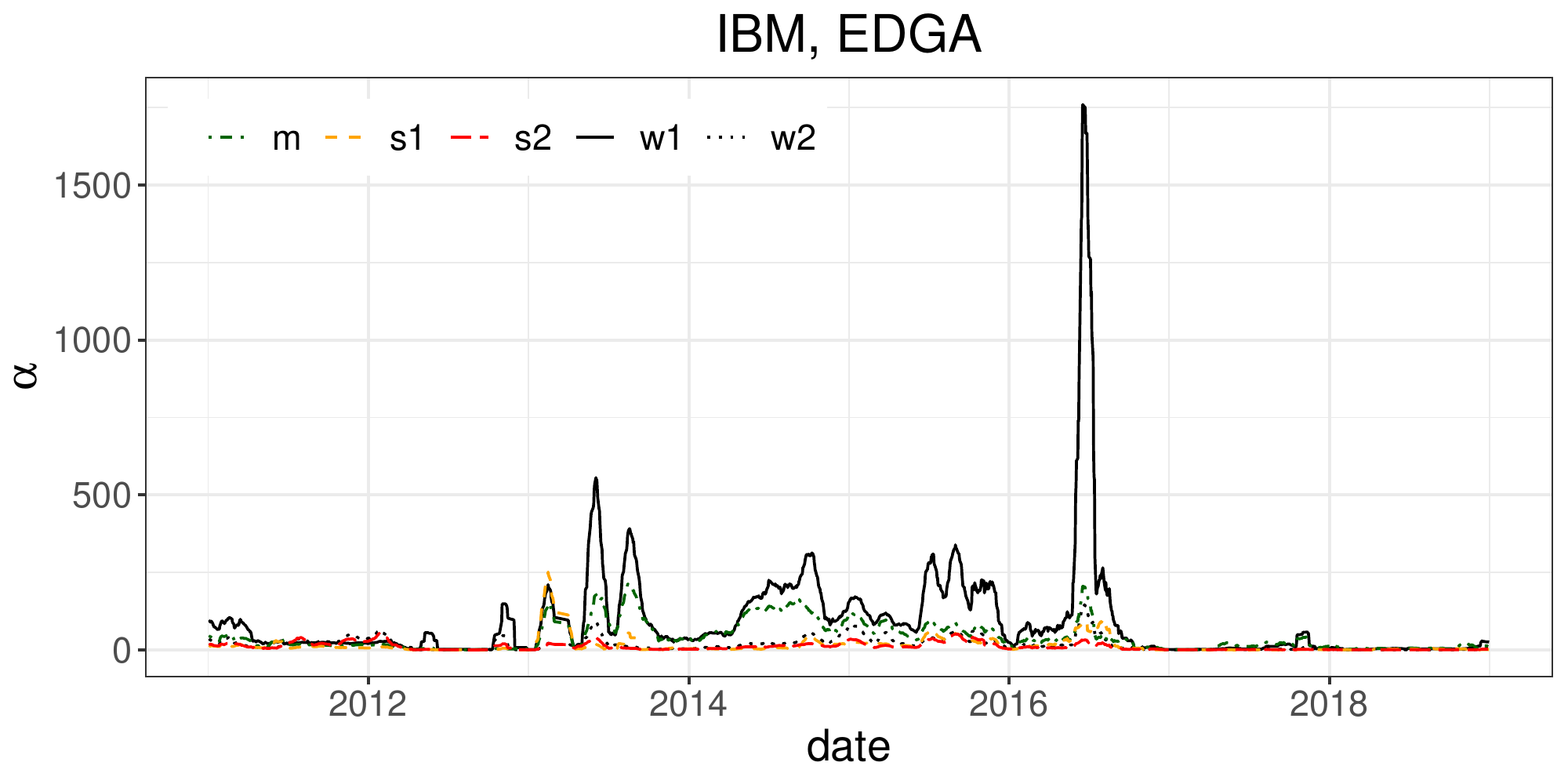}
	\caption{Smoothed dynamics of the $\alpha$s of IBM from the BYX (left) and EDGA (right)}
	\label{Fig:IBM_alphas_BYX}
\end{figure}

The BYX adopted an inverted fee schedule, and the EDGA also had an inverted fee structure until June 1, 2017.
These exchanges have very high excitement parameter values for about one year or a few months around 2016.
Thus, we discuss whether we can assume that limit order quotes and trading were more active in both exchanges during this period.
Indeed, not all the excitement parameters of the BYX and EDGA are large.
The estimated $\alpha_{w1}$ dominates all the other values on the left-hand side of Figure~\ref{Fig:IBM_alphas_BYX} for the BYX; the case is similar for the EDGA on the right-hand side of the figure.
Recall that $\alpha_{w1}$ is related to the replenishment rate when the best limit orders are removed,
which we can consider to be the basic job of market makers.
We assume that market making that provides liquidity was active, with low latency in the BYX (and EDGA) around 2016.
Thus, we can assume that market participants actively tried to profit through market making during this period.
While it may seem strange to see markets on exchanges with inverted fee schedules
that charge a fee to provide liquidity,
inverted structures attract market orders, and hence the limit orders in these exchanges are likely to be concluded first.
However, these activities were discouraged after 2017, or at least market making slowed.
Although we do not show the results in detail, the $\alpha_{w1}$ for IBM in the BYX is still profound in 2018, with an average around 15; however, it is much smaller than the value in 2016 is.
Perhaps the BYX's market makers determined that market making at a higher latency than before would be sufficiently competitive given that low latency is costly.
Conversely, the $\alpha_{w1}$ of the EDGA is small in 2018 (average around 0.7), meaning that high-frequency market making should not have occurred in that year.

\begin{figure}[hbt!]
	\centering
	\includegraphics[width=0.48\textwidth]{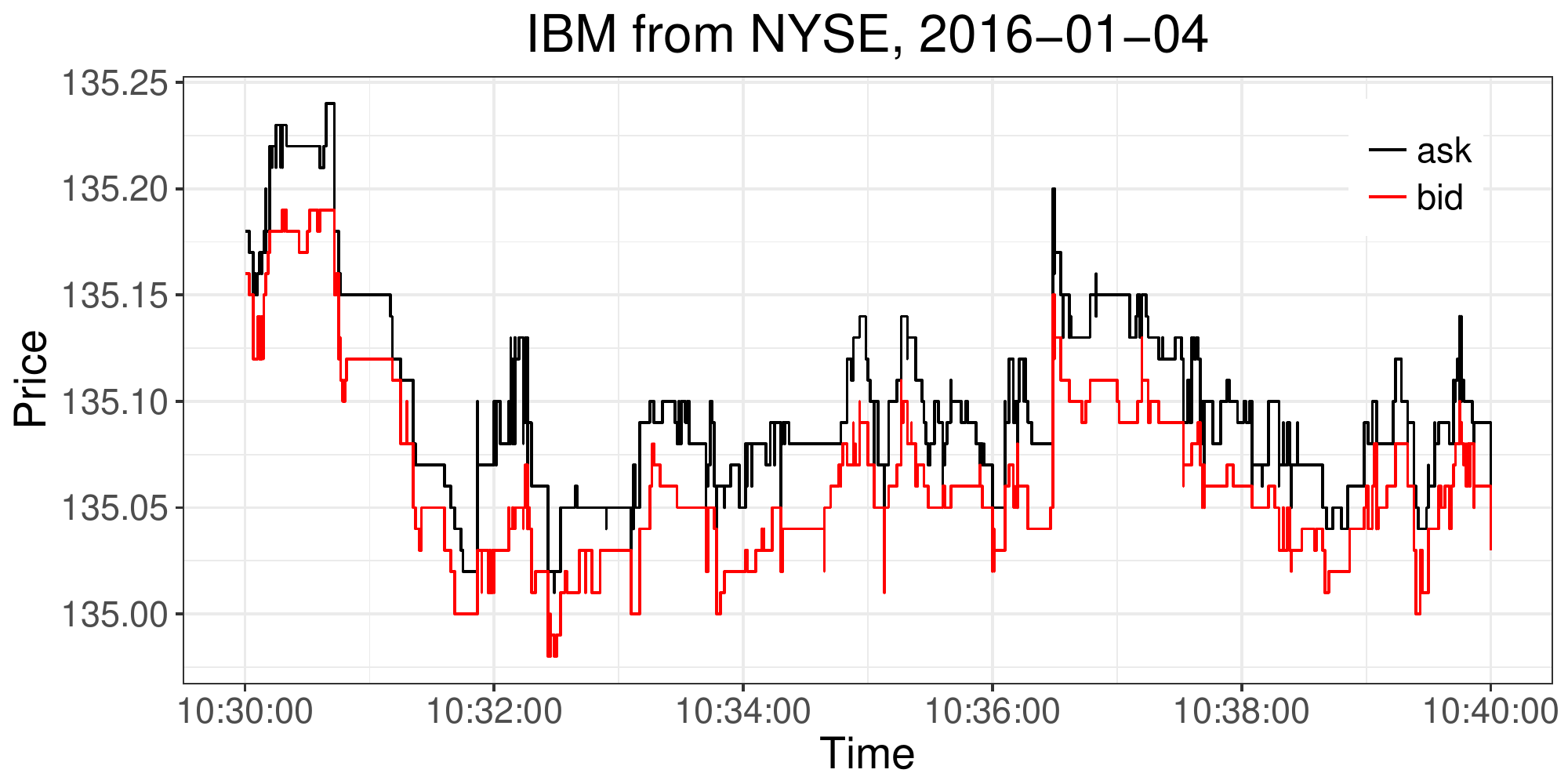}\quad
	\includegraphics[width=0.48\textwidth]{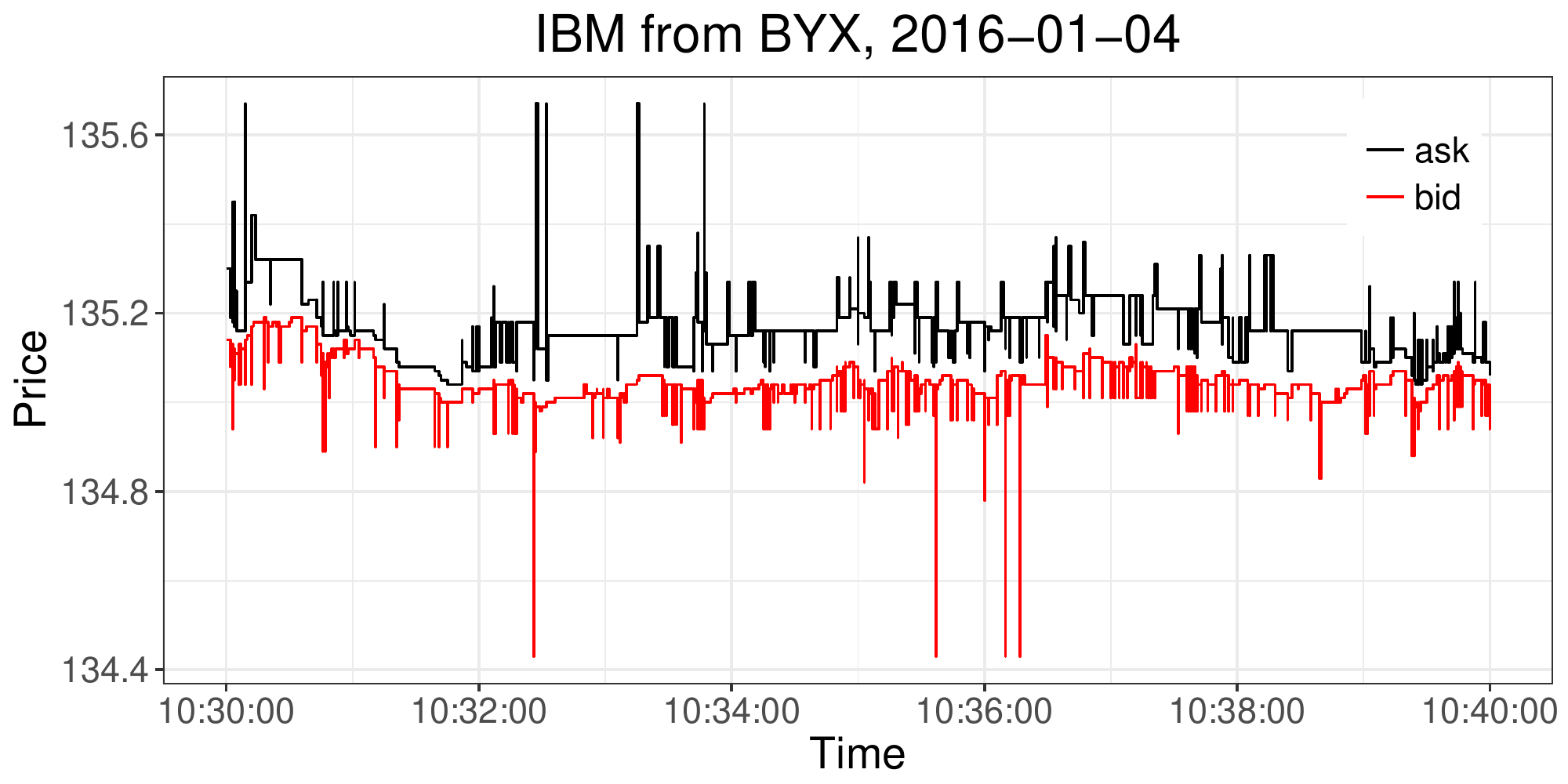}
	\caption{Comparison of IBM's tick dynamics in the NYSE (left) and BYX (right)}
	\label{Fig:IBM_tick}
\end{figure}

Around 2016, the best bid and ask price processes of the BYX (and EDGA) differ from the typical dynamics of liquid exchanges.
The left-hand side of Figure\ref{Fig:IBM_tick} shows the dynamics of the best bid and ask prices in the NYSE and the right-hand side shows the best bid and ask prices in the BYX.
Although we do not show the NBBO's movements, the NYSE's movements closely resemble them.
Conversely, the bid-ask spread in the BYX changes rapidly in jump size 
and the overall movement is different from the NBBO dynamics.
We observe similar movements in the EDGA.
On the right-hand side of the figure, 
when a transaction occurs and the existing best quote is removed,
we momentarily observe a rather large spread.
However, responsible market makers seem to supplement quotes quickly
and the spread narrows.
Except for these quotes, it appears that limit orders lack sufficient depth.
The best bid and ask prices are far from the mid-price
when the market maker does not quote the limit orders sufficiently near the mid-price.
This shallow depth is consistent with the fact that the BYX has a maker-taker fee schedule.

\begin{figure}[hbt!]
	\centering
	\begin{subfigure}{.49\textwidth}
		\centering
		\includegraphics[width=0.99\textwidth]{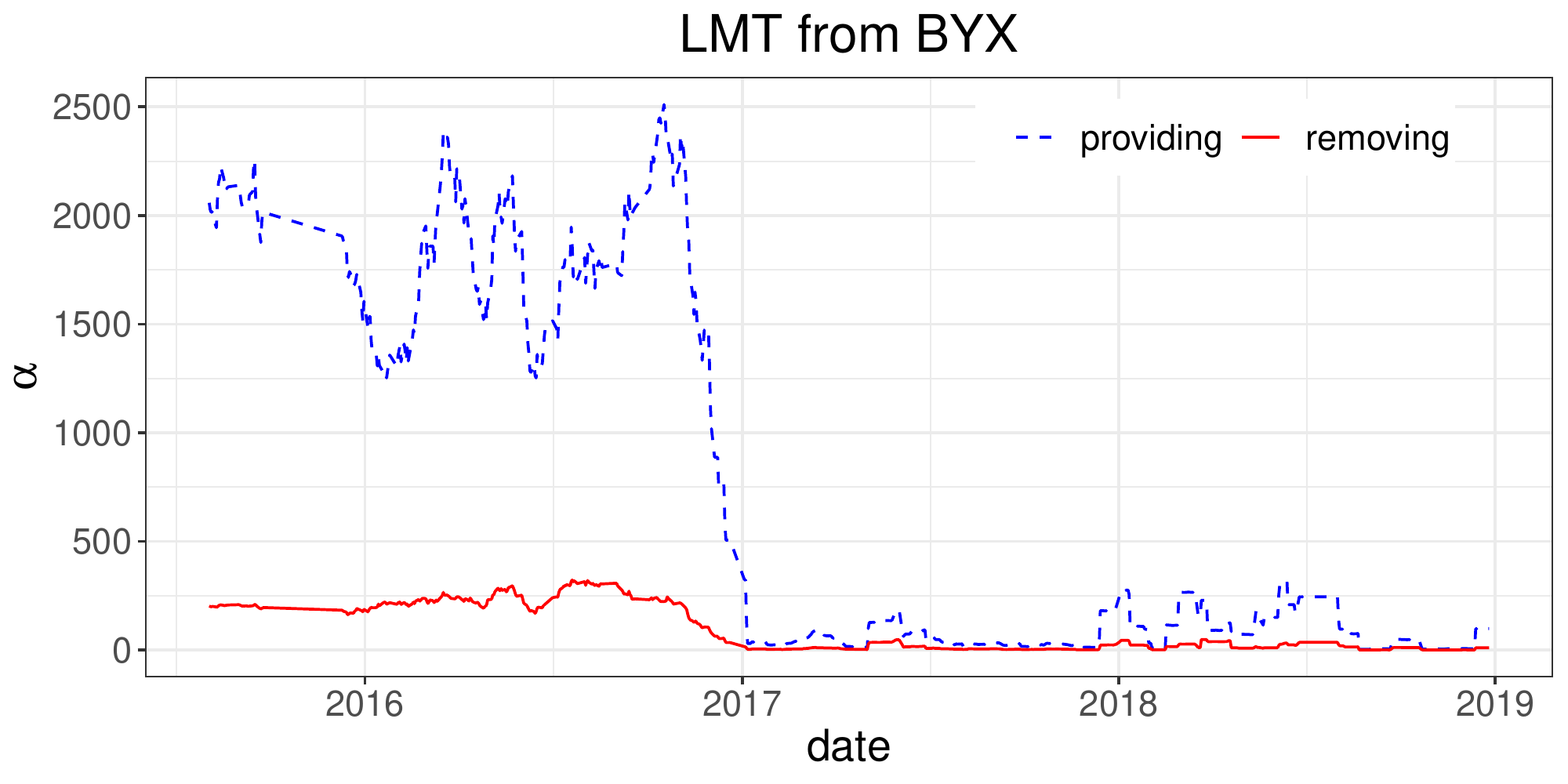}
		\caption{Lockheed Martin from the BYX}
		\label{fig:LMT_BYX}
	\end{subfigure}
	\begin{subfigure}{.49\textwidth}
		\centering
		\includegraphics[width=0.99\textwidth]{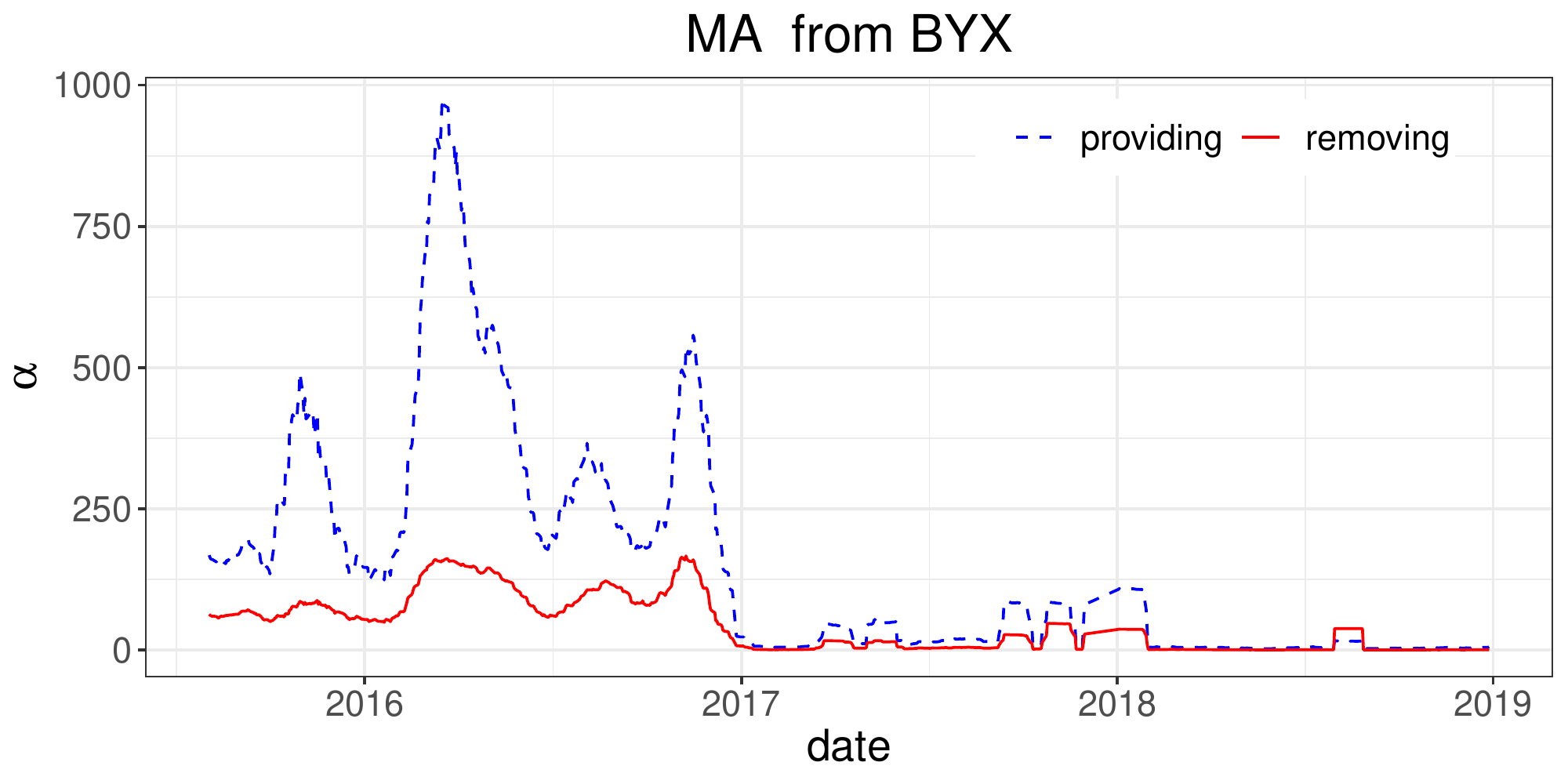}
		\caption{Mastercard from the BYX}
		\label{fig:MA_BYX}
	\end{subfigure}
	
	\begin{subfigure}{.49\textwidth}
		\centering
		\includegraphics[width=0.99\textwidth]{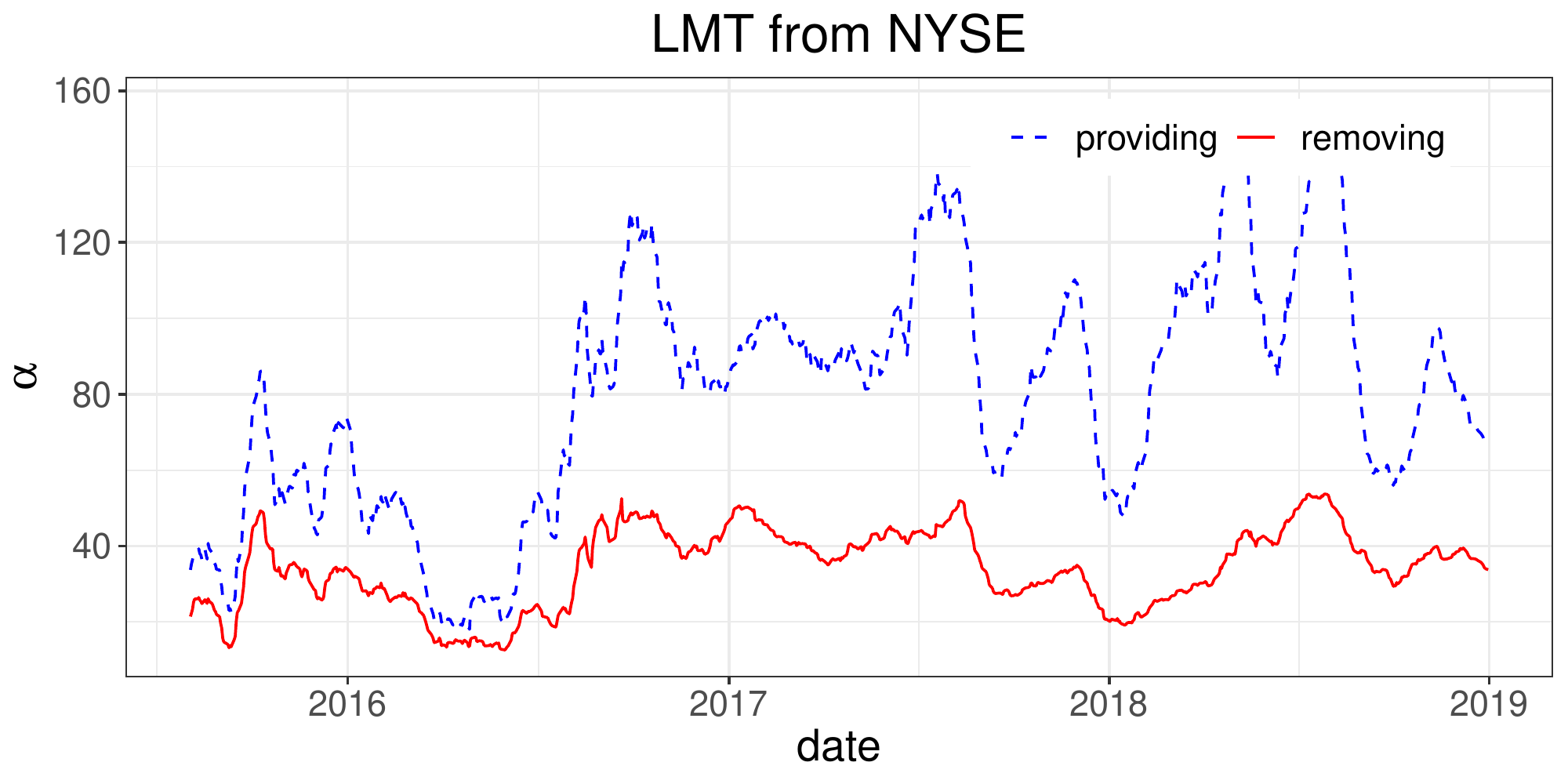}
		\caption{Lockheed Martin from the NYSE}
		\label{fig:LMT_NYSE}
	\end{subfigure}
	\begin{subfigure}{.49\textwidth}
		\centering
		\includegraphics[width=0.99\textwidth]{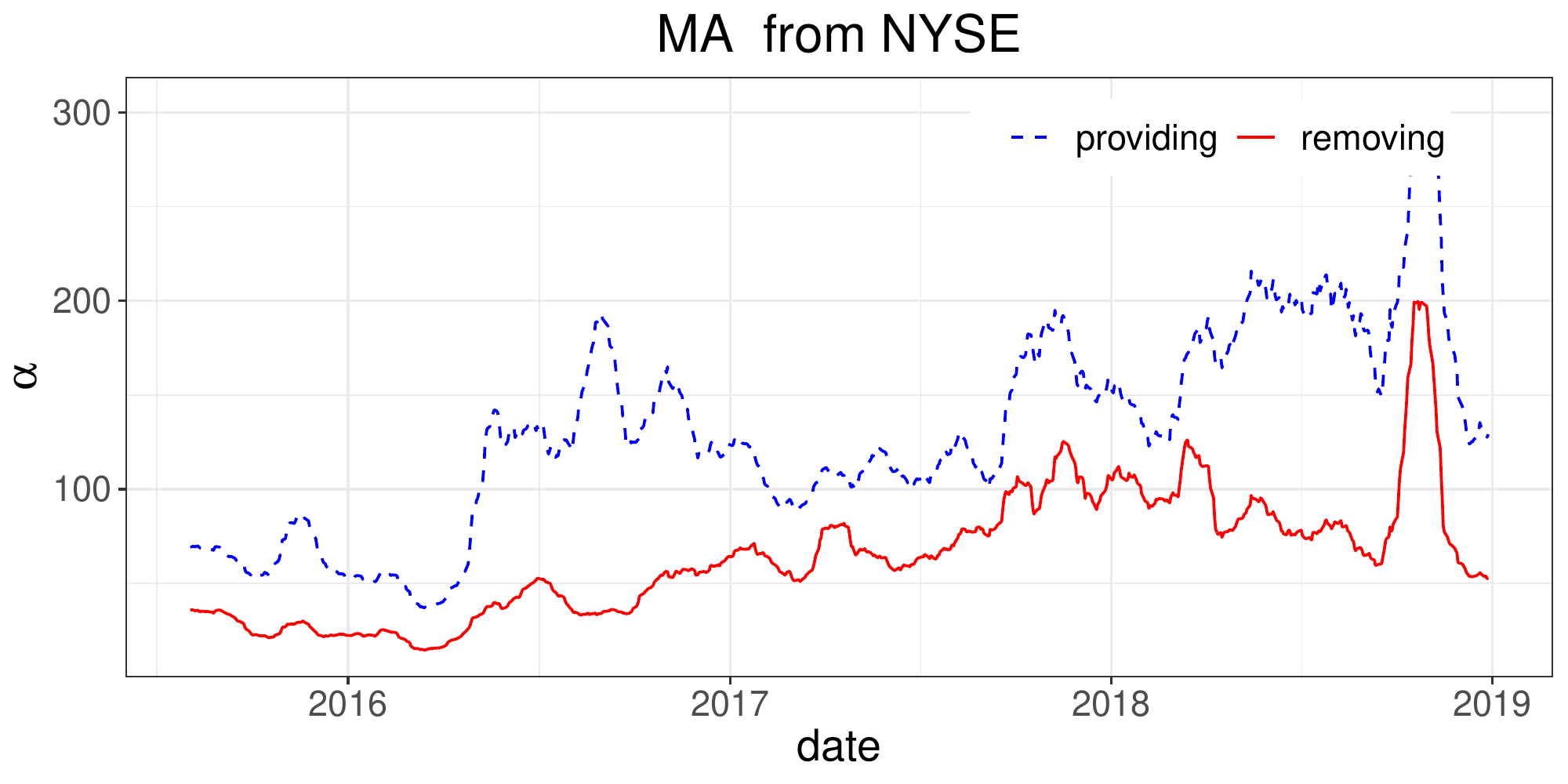}
		\caption{Mastercard from the NYSE}
		\label{fig:MA_NYSE}
	\end{subfigure}
	
	\caption{Smoothed dynamics of the liquidity-providing and -depleting $\alpha$s from 2015 to 2018}
	\label{Fig:BYX_NYSE}
\end{figure}

Figure~\ref{Fig:stock_liquidity} shows the large liquidity-providing $\alpha$s (the mean of $\alpha_{w1}$ and $\alpha_{w2}$) around 2016.
We strongly suspect that the main source of this phenomenon is an exchange such as the BYX.
For the Lockheed Martin and Mastercard stocks, Figures~\ref{fig:LMT_BYX}~and~\ref{fig:MA_BYX} plot the time-varying liquidity-providing and -depleting excitement parameter, respectively,
using the bid and ask prices published by the BYX.
The liquidity-providing $\alpha$s in both stocks from the BYX remain high until the second half of 2016, but suddenly fall close to zero after 2017.
On the contrary, Figures~\ref{fig:LMT_NYSE} and \ref{fig:MA_NYSE} show the results estimated using the data from the NYSE (the main exchange for these stocks), indicating that the $\alpha$s remain relatively stable.

These two results suggest that the peak of the liquidity-providing excitement parameters around 2016 observed from the NBBO in Figure~\ref{Fig:stock_liquidity} did not exist in all the exchanges, 
but particularly occurred in exchanges such as the BYX.
This result is consistent with the previous discussion. 
High-frequency traders thought that they would profit from ultra-high-frequency trading on specific stocks such as Lockheed Martin and Mastercard on specific exchanges such as the BYX.
The exact reason is unknown; however, after 2017, it seems that high-frequency traders stopped making these types of trades.

\subsubsection{The IEX}

Recently, the IEX introduced the speed bump technique, which intentionally slows order processing to 350 microseconds to protect low-frequency traders \citep{bishop2017evolution}.
Exchanges can use this technique to slow ultra-fast market orders to prevent orders from snipping stale quotes that do not yet reflect the state of the national bid and offer prices.
The IEX can also use it to help protect market orders executed with the rest of the orders to complete on other exchanges from phantom liquidity by delaying the execution and transmission of the results.
In addition, with the crumbling quote detection algorithm and discretionary pegged orders,
in which orders aim to access liquidity more aggressively in a stable market and become passive when quotes crumble,
traders in the IEX are less concerned about speed.

\begin{figure}[hbt!]
	\centering
	\includegraphics[width=0.45\textwidth]{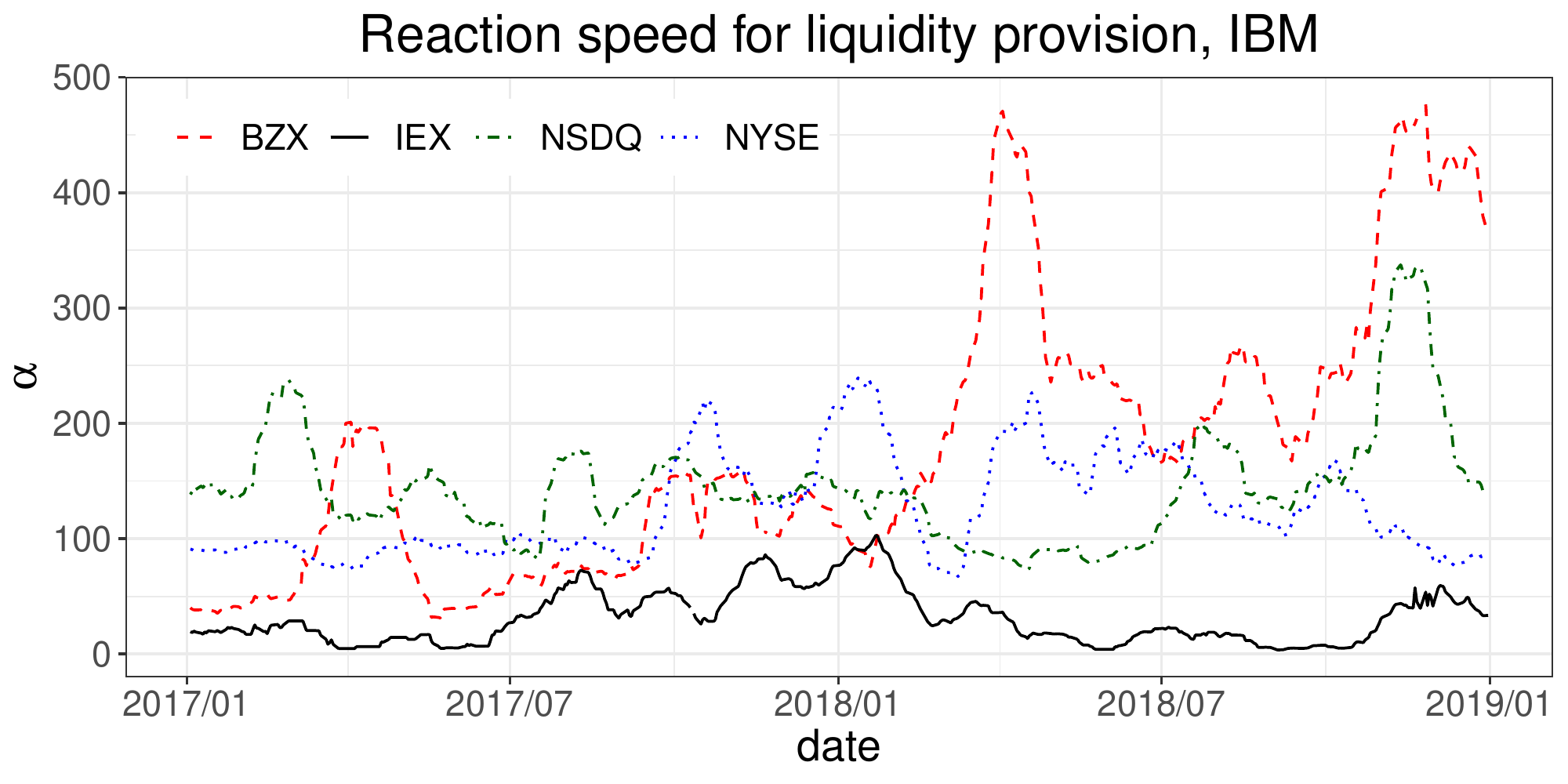}\quad
	\includegraphics[width=0.45\textwidth]{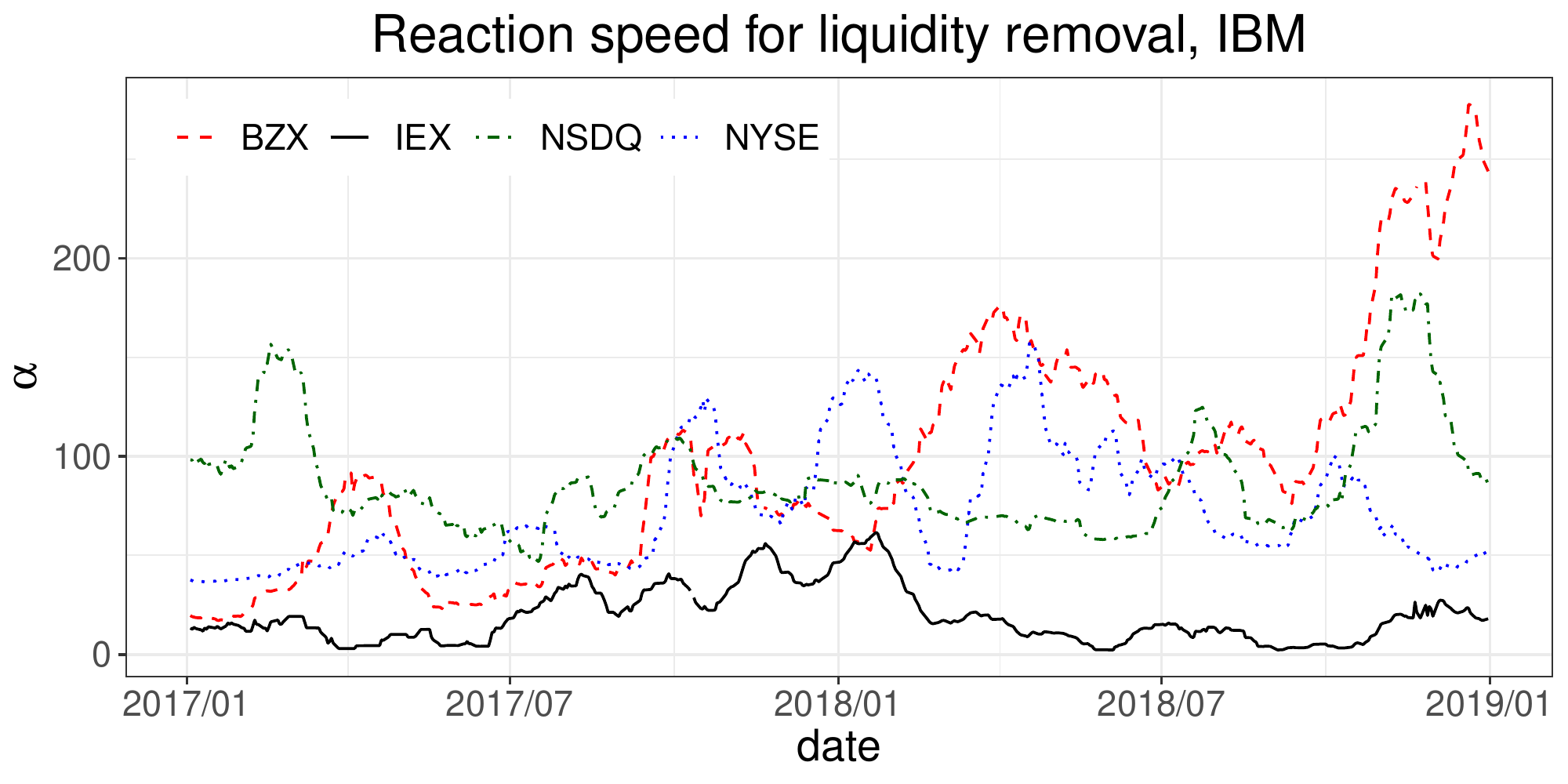}
	\caption{Estimates of IBM's $\alpha$s from the IEX, smoothed by a moving average of 20 successive days}
	\label{Fig:IBM_IEX_smoothed}
\end{figure}

Unfortunately, we cannot obtain information on the invisible pegged orders of the IEX. 
Instead, we perform the estimation using visible limit orders, which represents about 20\% of all the orders of the IEX \citep{aldrich2019order}.
The responsiveness we measure using the excitement parameter $\alpha$s in this exchange is lower than that in the other exchanges, 
as Figure~ \ref{Fig:IBM_IEX_smoothed} shows, for both liquidity depletion and liquidity provision.
For a better visualization, we use the moving averages of the excitement parameter $\alpha$s' estimates.
We can interpret these results in two ways.
The IEX's liquidity providers are relatively unconcerned about improving responsiveness because they expect the speed bump technique to protect their stale quotes.
Alternatively, the IEX's meticulous traders already use invisible pegged orders, meaning that we cannot observe their exact behaviors and can only present the behaviors of those traders relatively insensitive to increasing the speed of quote processing.

\subsection{The Flash Crash}\label{subsec:Crash}

\begin{figure}[hbt!]
	\centering
	\includegraphics[width=0.45\textwidth]{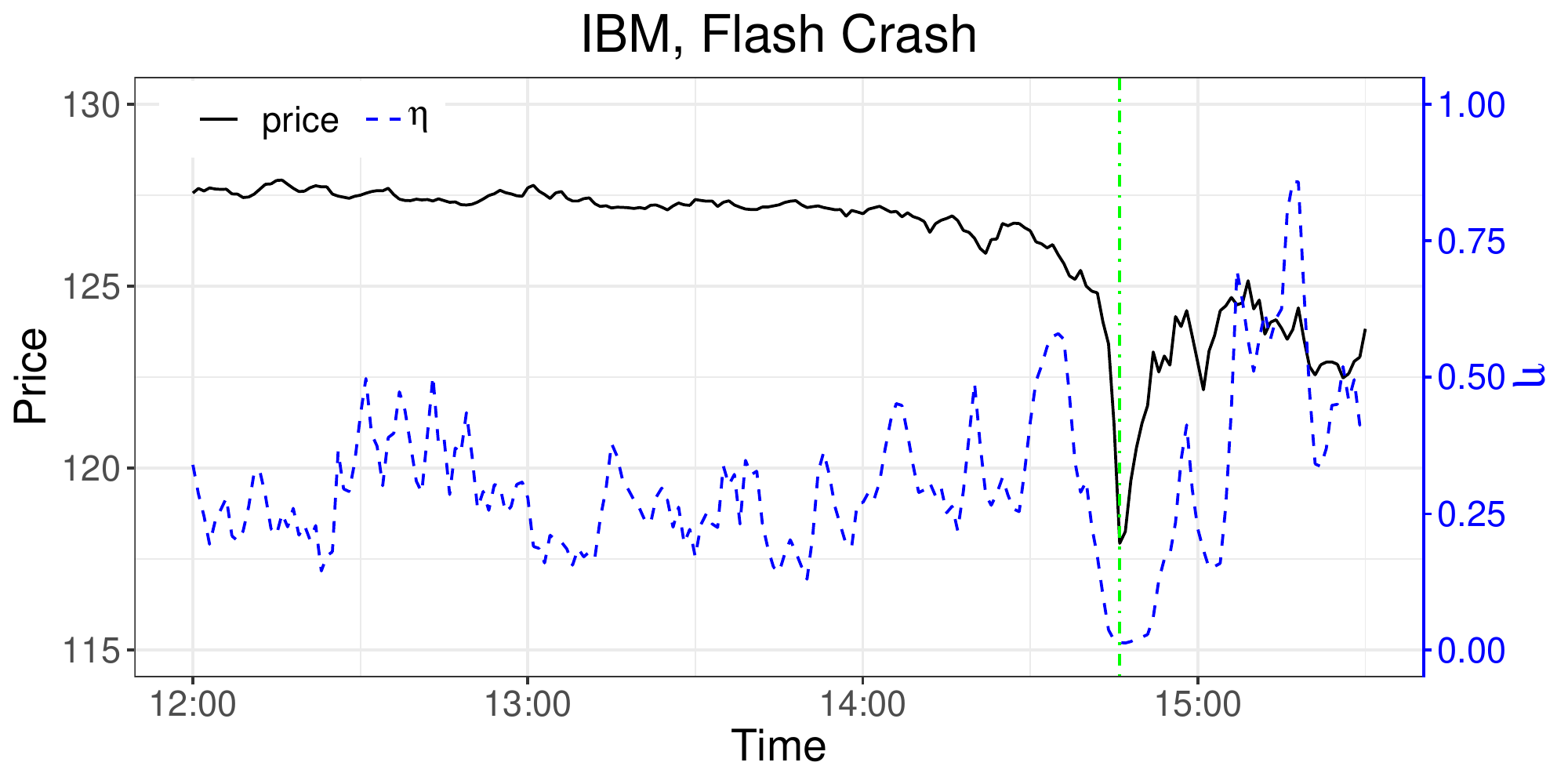}\quad
	\includegraphics[width=0.45\textwidth]{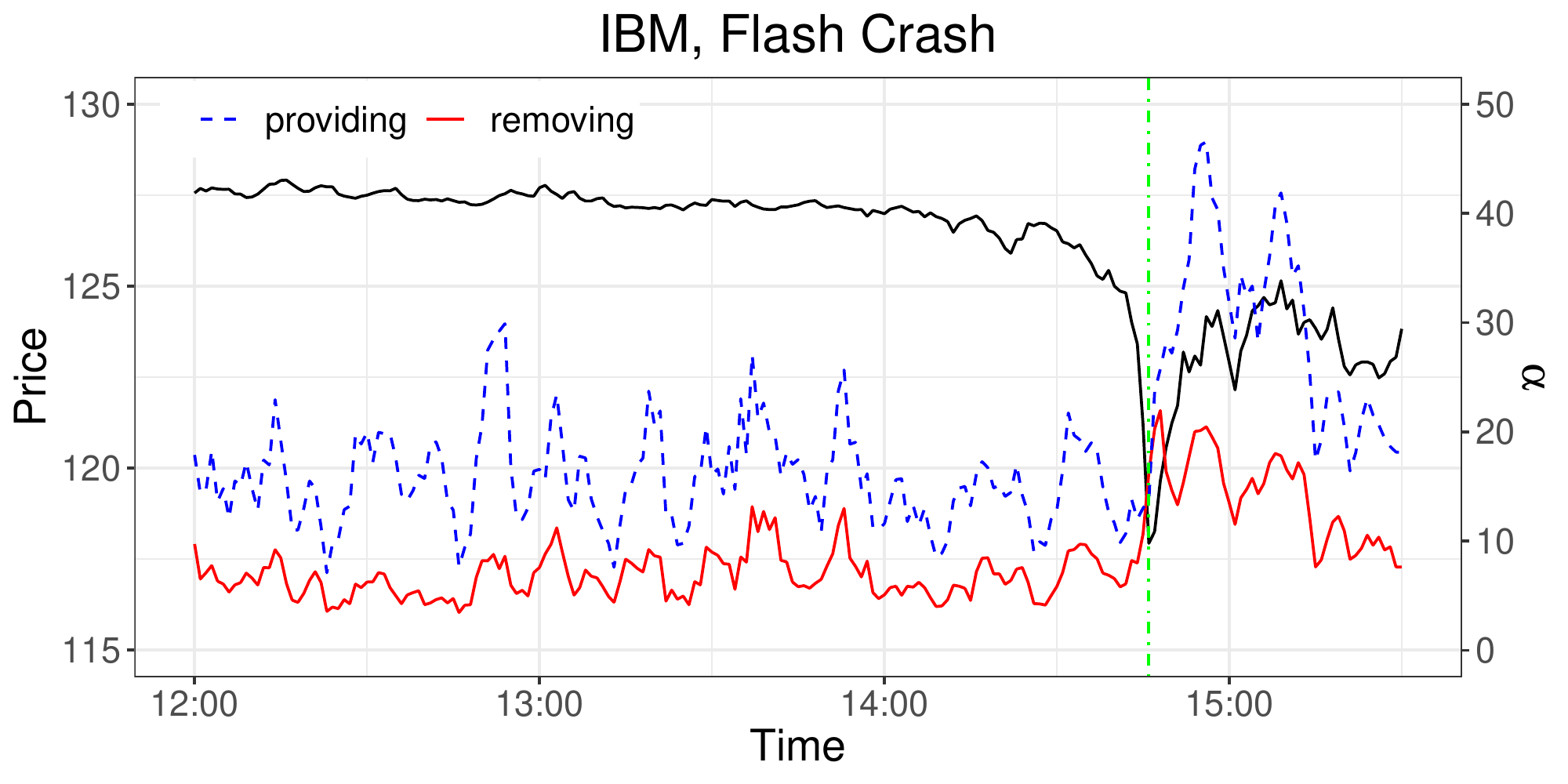}
	\caption{Stock price dynamics during the Flash Crash and intraday changes in $\eta$ and $\alpha$}
	\label{Fig:Flash_Crash}
\end{figure}

On May 6, 2010, the Flash Crash occurred during which stock prices fell sharply due to large sell orders and then recovered quickly.
Our model can investigate the details of the behavior of market participants on this day.
The left-hand side of Figure~\ref{Fig:Flash_Crash} plots the IBM stock price (black solid line) and dynamics of the liquidity parameter $\eta$ (blue dashed line).
Recall that $\eta$ represents the tendency to narrow when the bid-ask spread widens.

Unlike the previous cases, to examine the intraday changes, we perform the estimations every minute and base them on three-minute overlapping intervals using IBM’s NBBO data on bid and ask prices on May 6, 2010.
The Flash Crash occurred around 14:32. 
As the figure shows, the high $\eta$ value indicates that liquidity providers continued to provide liquidity confidently for a few minutes immediately after the start of the price collapse.
However, $\eta$ then drops to almost zero, implying that liquidity providers became seriously concerned about adverse selection due to the unexpectedly large price drop (i.e., they did not try to narrow the spread).
The value of $\eta$ begins to rise almost at the same time as the price recovery
and then returns to its original level with synchronized price fluctuations.

The right-hand side of the figure depicts the liquidity-providing $\alpha$ (blue dashed line) and -depleting $\alpha$ (red solid line) with price changes (black solid line).
Interestingly, both the excitement parameters start to rise at the point the price reaches its lowest value.
In addition, the two parameters are almost the same near to and immediately after the time of the lowest price.
In a stable market, the liquidity-providing parameter $\alpha$ is greater than the liquidity-depleting $\alpha$, as shown in the previous figures.
Immediately after the Flash Crash, these parameter values rise; that is,
market participants are more likely to respond faster to market changes in both liquidity provision and liquidity depletion,
which continued and strengthened until the next day.

\begin{figure}[hbt!]
	\centering
	\includegraphics[width=0.45\textwidth]{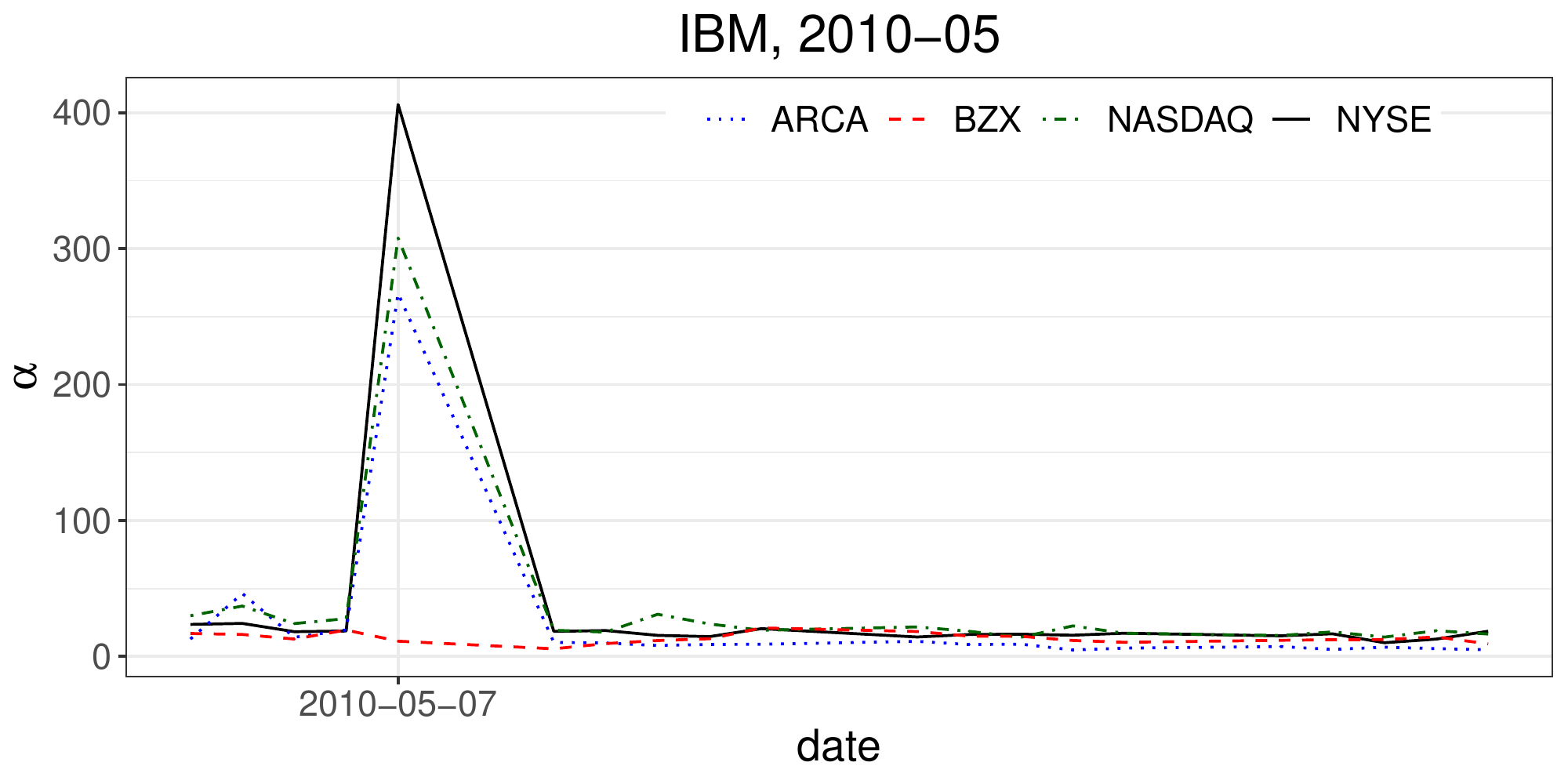}\quad
	\includegraphics[width=0.45\textwidth]{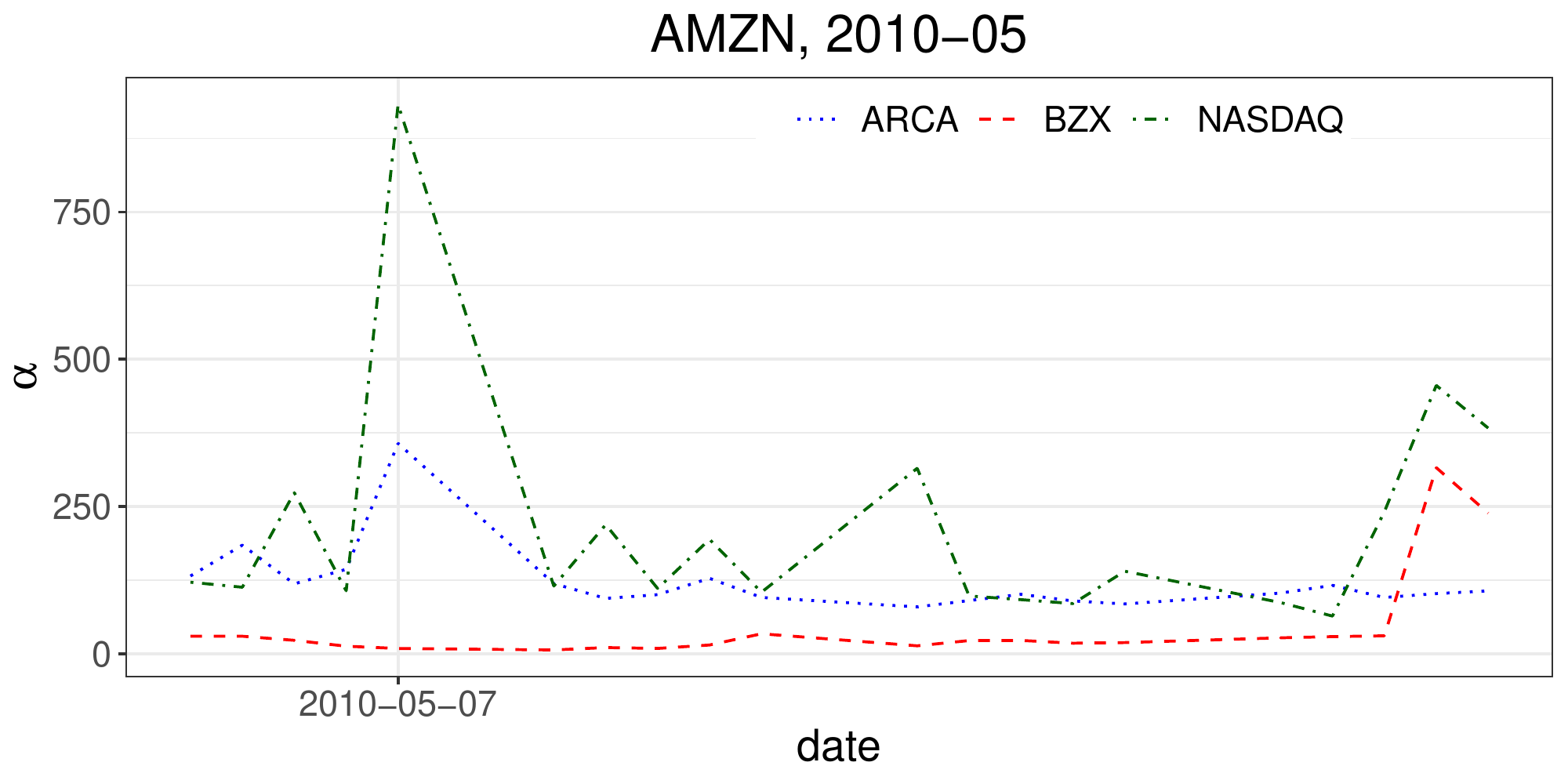}
	\caption{Dynamics of $\bar \alpha$ around the Flash Crash for IBM (left) and AMZN (right). The peak occurs the day after the Flash Crash.}
	\label{Fig:After_Flash_Crash}
\end{figure}

Returning to Figure~\ref{Fig:IBM_per_exchange}, we can see that the average of excitement parameters, $\bar \alpha$, of the NYSE reaches its maximum value in May 2010.
Figure~\ref{Fig:After_Flash_Crash} provides an enlarged picture of this part of the graph together with the $\bar \alpha$ of AMZN.
Interestingly, the peaked day with the largest excitement parameters (i.e., May 7, 2010) is the day after the Flash Crash.
On this day, the excitement parameter values estimated for the NYSE and Arca are much higher than those on the other days; that is, 
it seems to be the busiest day of 2010 for the high-frequency traders on these exchanges.
Additionally, when the market closes on that day without any particular event, 
their responsiveness returns to their original state after May 7, 2010.
However, the BZX data indicate no such busy activities on May 7, 2010.

\begin{figure}[hbt!]
	\centering
	\includegraphics[width=0.45\textwidth]{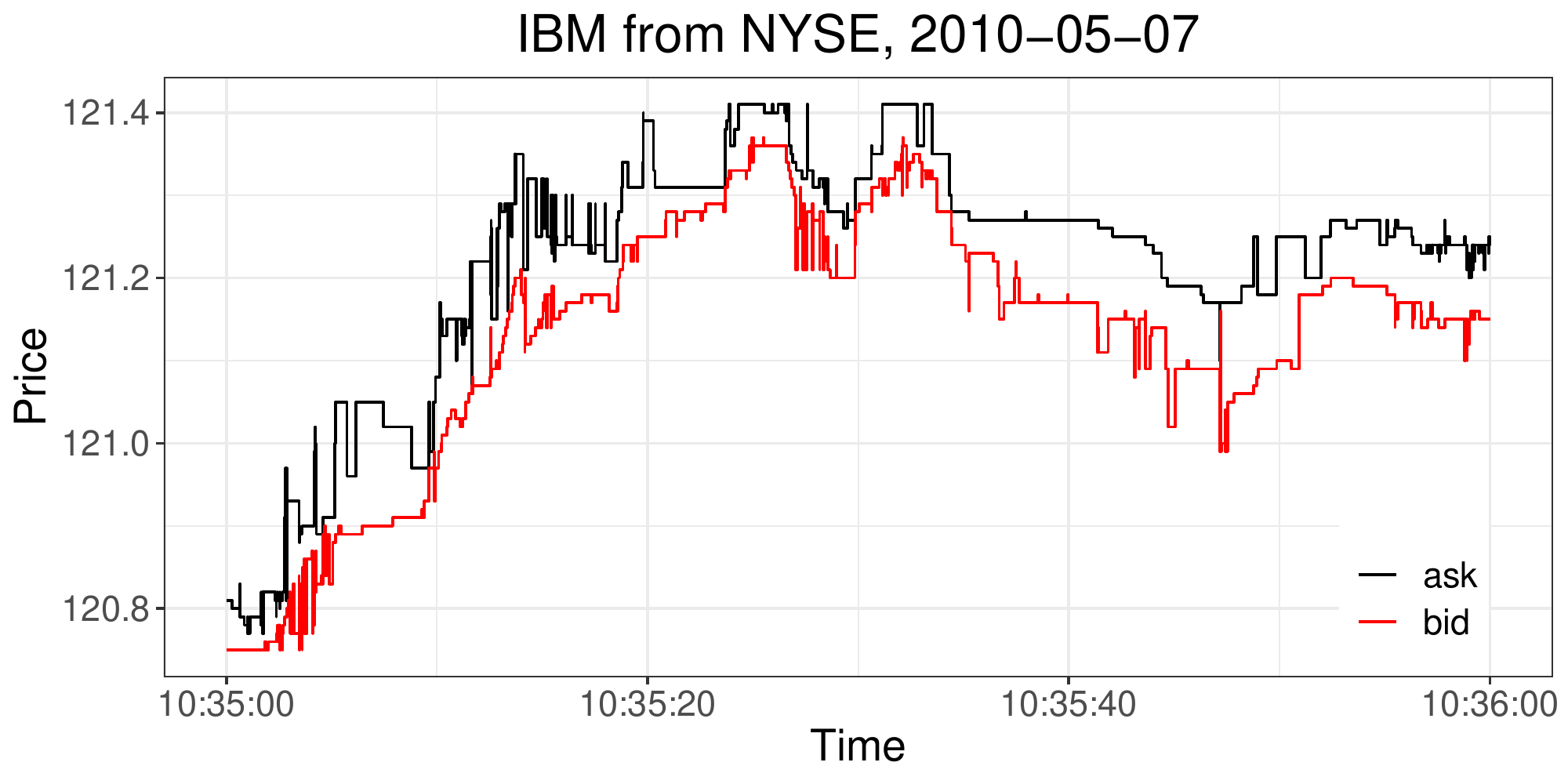}\quad
	\includegraphics[width=0.45\textwidth]{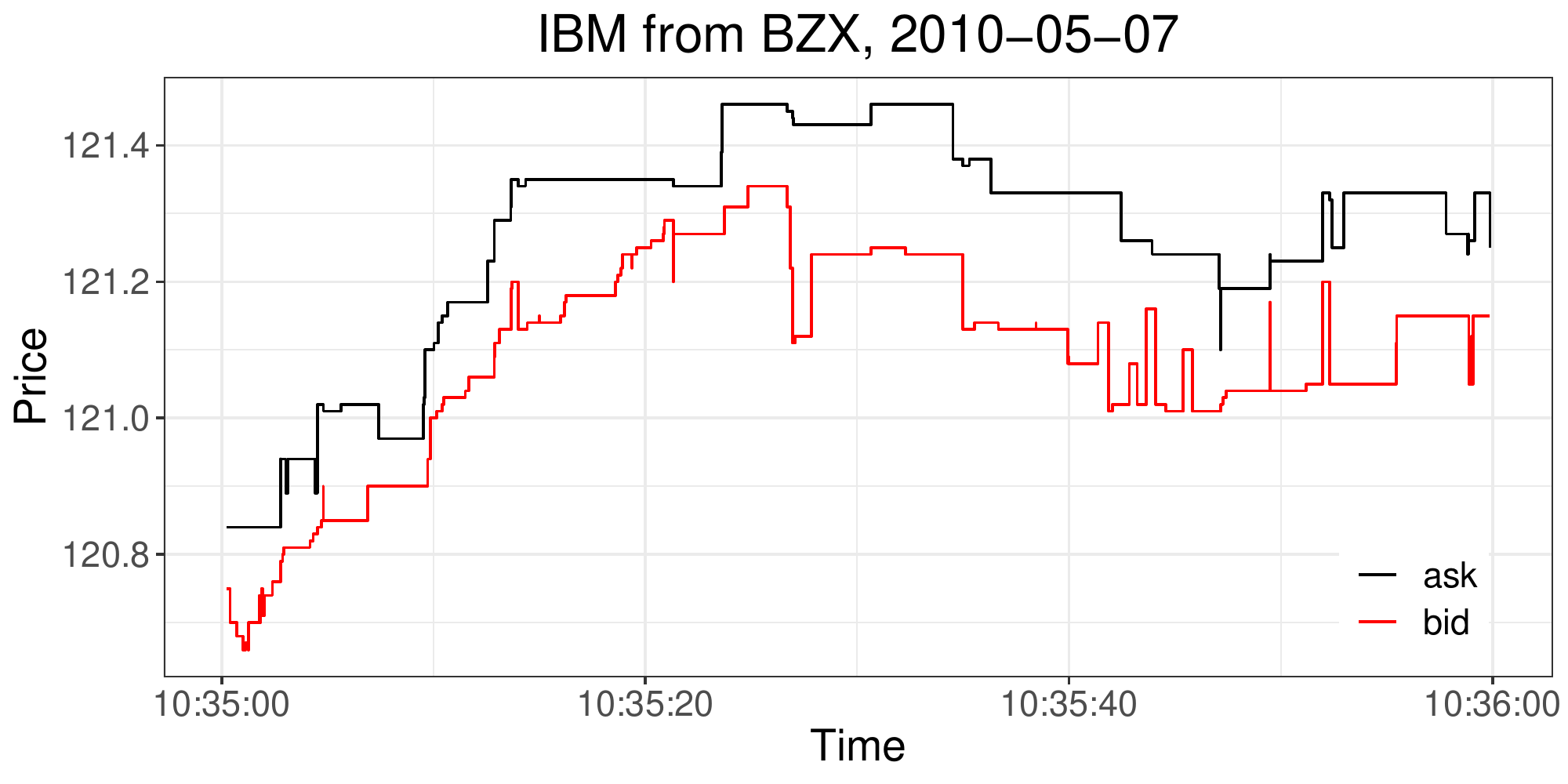}
	\caption{IBM stock tick structure the day after the Flash Crash on the NYSE (left) and BZX (right)}
	\label{Fig:Flash_Crash_tick}
\end{figure}

The reason behind the above estimation results becomes clearer when we observe the best bid and ask price processes on this day.
The left-hand side of Figure~\ref{Fig:Flash_Crash_tick} shows the best bid and ask price processes reported by the NYSE the day after the Flash Crash, May 7, 2010.
The price process moves much more irregularly in the NYSE than in the BZX, as the right-hand side of the figure shows.
The NYSE best bid and ask processes on this day are more irregular than on the other days, although we do not show them to save space.
The liquidity providers on the NYSE were extra careful to post limit orders on this day based on the previous Flash Crash experience,
and they tried to avoid adverse selection by repeating the process of submitting and then canceling limit orders.
The two largest estimates of excitement parameters on this day are $\alpha_{w1} = 883$ and $\alpha_m = 554$,
representing the fast repetitions of liquidity supply and removal.
Note that $\alpha_{w1}$ is related to supplementing quotes and $\alpha_{m}$ is related to submitting and immediately canceling (or executing) limit orders.

\section{Conclusion}\label{Sec:concl}

We propose an extended Hawkes model to describe the best bid and ask price processes.
The model aims to rigorously reflect the nature of the price processes and
capture the various aspects of high-frequency stock price data observed in micro- or milliseconds.

The empirical analysis shows that larger spreads tend to reduce spreads in general,
but that this tendency decreases greatly when adverse selection is a concern. In addition, the movements of the bid and ask price processes not only depend on the spread, 
but also are excited by recent events that cause price changes.
With the abundance of automated high-frequency trading, the estimated responsiveness and corresponding decay rate have become high in recent years.
For most stocks, we observe increasing responsiveness over the past decade and find that some stocks show unique behavior. 
Specifically, responsiveness to liquidity provision increased dramatically around 2016 and then declined.
Moreover, some stocks, namely Microsoft, Apple, and Facebook, have become of particular interest to high-frequency traders in recent years.
These stocks, particularly Microsoft, show high responsiveness in terms of liquidity provision and a high tendency to replenish 
when spreads widen.

Exchanges have different responsiveness patterns.
For example, the responsiveness of the NYSE is more stable than that of the Cboe BZX, which changes more drastically.
The proposed model also allows us to observe the different characteristics of high-frequency traders across exchanges.
We find that the speed of liquidity provision peaked around 2016, stemming from the speed of stock exchanges such as the BYX.
However, the new IEX has low responsiveness relative to other exchanges.

Focusing on intraday changes in the parameters during the Flash Crash in 2010,
we observe a change in confidence related to adverse selection among market makers along with a sharp decline and recovery in a stock price.
We also find that responsiveness peaked on the day after the Flash Crash in response to the previous market anomaly, as traders repeatedly submitted and canceled limit orders.

The application of the proposed model is not limited to those presented here.
Owing to its nature, which can incorporate a variety of properties, our proposed model, or variations of it, is able to reveal some of the information hidden in high-frequency financial data.
Hence, the presented findings should provide a better understanding of the market and aid decision making and policymaking.

\appendix

\section{Estimation result: Example}~\label{Appendix:estimation}

Table~\ref{Table:se} presents the estimates and standard errors in parentheses from our model for IBM for January 2018. We calculate the standard error numerically.
To save space, we report only some of the results. However, the other cases show similar standard errors.

\begin{table}[hbt!]
	\centering
	\footnotesize
	\caption{Estimates and standard errors (in parentheses) from our model for IBM, January 2018}
	\label{Table:se}
	\begin{tabular}{cccccccccc}
		\hline
		Date  &  $\mu$  & $\eta$ &  $\alpha_{s1}$  & $\alpha_{s2}$ & $\alpha_{m}$ & $\alpha_{w1}$  & $\alpha_{w2}$ & $\beta$ & $\xi$ \\
		\hline
		2018-01-02 & 0.067    & 0.0693   & 96.32     & 193.2     & 219.5    & 271.6     & 459.6     & 991.1    & 61.44    \\
		& (0.0013) & (0.0014) & (0.3570)  & (0.3306)  & (0.3929) & (0.2521)  & (0.2169)  & (0.0668) & (0.1012) \\
		2018-01-03 & 0.1004   & 0.1184   & 69.34     & 192.6     & 173.5    & 210.1     & 381.2     & 707.0    & 71.82    \\
		& (0.0016) & (0.0018) & (0.0642)  & (0.0605)  & (0.0818) & (0.1111)  & (0.0990)  & (0.0444) & (0.1146) \\
		2018-01-04 & 0.0787   & 0.1165   & 66.79     & 158.3     & 193.5    & 201.4     & 479.0     & 718.3    & 89.27    \\
		& (0.0014) & (0.0022) & (0.1439)  & (0.1829)  & (0.2113) & (0.1049)  & (0.1261)  & (0.0410) & (0.1658) \\
		2018-01-05 & 0.0736   & 0.0758   & 63.81     & 185.0     & 142.2    & 126.2     & 371.6     & 739.4    & 48.41    \\
		& (0.0014) & (0.0014) & (0.3090)  & (0.7946)  & (0.5339) & (0.3335)  & (0.0165)  & (0.0719) & (0.1475) \\
		2018-01-08 & 0.0757   & 0.0886   & 69.77     & 150.3     & 158.9    & 157.6     & 395.6     & 699.0    & 47.31    \\
		& (0.0014) & (0.0016) & (0.1229)  & (0.1692)  & (0.3506) & (0.4053)  & (0.1474)  & (0.0543) & (0.2686) \\
		2018-01-09 & 0.0768   & 0.0911   & 50.70     & 178.1     & 169.1    & 174.7     & 514.5     & 881.1    & 71.06    \\
		& (0.0014) & (0.0019) & (0.1341)  & (0.1719)  & (0.2671) & (1.1020)  & (1.1630)  & (0.0869) & (0.1191) \\
		2018-01-10 & 0.0447   & 0.0369   & 71.22     & 201.6     & 119.3    & 135.6     & 281.6     & 831.4    & 43.54    \\
		& (0.0011) & (0.0008) & (0.4306)  & (0.4451)  & (0.6818) & (0.9216)  & (0.4001)  & (0.1445) & (0.0683) \\
		2018-01-11 & 0.0513   & 0.0635   & 50.83     & 190.2     & 150.2    & 160.2     & 344.6     & 683.5    & 60.97    \\
		& (0.0012) & (0.0013) & (0.2936)  & (0.2478)  & (0.1626) & (0.3738)  & (0.0587)  & (0.1526) & (0.4125) \\
		2018-01-12 & 0.0625   & 0.0669   & 64.45     & 139.4     & 137.8    & 166.9     & 362.0     & 595.0    & 54.68    \\
		& (0.0013) & (0.0014) & (0.2930)  & (0.1787)  & (0.1425) & (0.3041)  & (0.1897)  & (0.0838) & (0.1723) \\
		2018-01-16 & 0.1099   & 0.1124   & 51.86     & 113.7     & 150.6    & 142.8     & 265.1     & 576.0    & 42.94    \\
		& (0.0017) & (0.0017) & (0.3374)  & (0.2102)  & (0.2304) & (0.2500)  & (0.6593)  & (0.1097) & (0.4373) \\
		2018-01-17 & 0.1051   & 0.0931   & 55.83     & 110.5     & 135.9    & 141.3     & 256.4     & 541.7    & 40.56    \\
		& (0.0017) & (0.0014) & (0.0950)  & (0.0968)  & (0.1000) & (0.1193)  & (0.1199)  & (0.0337) & (0.2273) \\
		2018-01-18 & 0.1211   & 0.135    & 97.30     & 166.6     & 159.9    & 186.7     & 386.7     & 750.5    & 55.04    \\
		& (0.0018) & (0.0019) & (0.0918)  & (0.1867)  & (0.0113) & (0.0626)  & (0.0604)  & (0.1009) & (0.0402) \\
		2018-01-19 & 0.1791   & 0.1977   & 124.8     & 251.1     & 331.5    & 334.2     & 466.9     & 1091     & 97.98    \\
		& (0.0021) & (0.0022) & (0.1484)  & (0.1087)  & (0.0670) & (0.1072)  & (0.0495)  & (0.0112) & (0.0487) \\
		2018-01-22 & 0.0835   & 0.1042   & 80.68     & 276.9     & 342.1    & 280.2     & 529.5     & 1098     & 94.90     \\
		& (0.0015) & (0.0016) & (0.0517)  & (0.0723)  & (0.0333) & (0.0955)  & (0.0500)  & (0.0508) & (0.0647) \\
		2018-01-23 & 0.0781   & 0.0645   & 72.98     & 173.7     & 139.6    & 141.8     & 264.7     & 760.0    & 37.03    \\
		& (0.0014) & (0.0011) & (0.0704)  & (0.0602)  & (0.1765) & (0.1796)  & (0.2232)  & (0.0419) & (0.1726) \\
		2018-01-24 & 0.0736   & 0.0621   & 12.04     & 20.31     & 36.15    & 19.56     & 37.39     & 99.01    & 4.355    \\
		& (0.0014) & (0.0008) & (0.2121)  & (0.0651)  & (0.0849) & (0.1409)  & (0.1630)  & (0.2076) & (0.0836) \\
		2018-01-25 & 0.0544   & 0.0565   & 5.861     & 15.96     & 33.48    & 12.35     & 27.12     & 79.13    & 3.944    \\
		& (0.0012) & (0.0009) & (0.2745)  & (1.1480)  & (1.4180) & (0.4565)  & (0.5676)  & (0.1170) & (0.6511) \\
		2018-01-26 & 0.0534   & 0.0407   & 35.69     & 117.1     & 104.9    & 81.64     & 247.1     & 550.5    & 39.55    \\
		& (0.0012) & (0.0009) & (0.4864)  & (0.3846)  & (0.4970) & (1.1400)  & (0.7538)  & (0.1820) & (0.2034) \\
		2018-01-29 & 0.046    & 0.0519   & 4.861     & 15.46     & 31.30    & 8.930     & 20.29     & 68.35    & 2.682    \\
		& (0.0011) & (0.0008) & (0.4999)  & (1.772)  & (1.883) & (0.2002)  & (0.5845)  & (0.1414) & (0.2195) \\
		2018-01-30 & 0.0567   & 0.0575   & 8.437     & 22.42     & 34.29    & 14.94     & 35.35     & 90.93    & 5.069    \\
		& (0.0012) & (0.0009) & (0.2398)  & (0.4787)  & (0.6678) & (0.4286)  & (0.1209)  & (0.0723) & (0.1531) \\
		2018-01-31 & 0.0839   & 0.0453   & 70.94     & 146.2     & 98.07    & 94.21     & 253.4     & 649.5    & 32.98    \\
		& (0.0015) & (0.0008) & (0.2563)  & (0.2819)  & (0.3680) & (0.1556)  & (0.2291)  & (0.0819) & (0.2457) \\
		\hline
	\end{tabular}
\end{table}


\begin{thebibliography}{72}
\providecommand{\natexlab}[1]{#1}
\providecommand{\url}[1]{\texttt{#1}}
\expandafter\ifx\csname urlstyle\endcsname\relax
  \providecommand{\doi}[1]{doi: #1}\else
  \providecommand{\doi}{doi: \begingroup \urlstyle{rm}\Url}\fi

\bibitem[Aldrich and Friedman(2019)]{aldrich2019order}
E.~M. Aldrich and D.~Friedman.
\newblock Order protection through delayed messaging.
\newblock \emph{Available at SSRN 2999059}, 2019.

\bibitem[Andersen et~al.(2003)Andersen, Bollerslev, Diebold, and Labys]{ABDL}
T.~G. Andersen, T.~Bollerslev, F.~X. Diebold, and P.~Labys.
\newblock Modeling and forecasting realized volatility.
\newblock \emph{Econometrica}, 71:\penalty0 579--625, 2003.

\bibitem[Avellaneda and Stoikov(2008)]{Avellaneda2008}
M.~Avellaneda and S.~Stoikov.
\newblock {High-frequency trading in a limit order book}.
\newblock \emph{Quantitative Finance}, 8\penalty0 (3):\penalty0 217--224, 2008.

\bibitem[Bacry and Muzy(2014)]{Bacry2014}
E.~Bacry and J.-F. Muzy.
\newblock Hawkes model for price and trades high-frequency dynamics.
\newblock \emph{Quantitative Finance}, 14:\penalty0 1147--1166, 2014.

\bibitem[Bacry et~al.(2013)Bacry, Delattre, Hoffmann, and Muzy]{Bacryetal2013}
E.~Bacry, S.~Delattre, M.~Hoffmann, and J.-F. Muzy.
\newblock Modelling microstructure noise with mutually exciting point
  processes.
\newblock \emph{Quantitative Finance}, 13:\penalty0 65--77, 2013.

\bibitem[Bacry et~al.(2015)Bacry, Mastromatteo, and Muzy]{Barcry2015}
E.~Bacry, I.~Mastromatteo, and J.-F. Muzy.
\newblock Hawkes processes in finance.
\newblock \emph{Market Microstructure and Liquidity}, 1\penalty0 (1):\penalty0
  1550005, 2015.

\bibitem[Bacry et~al.(2016)Bacry, Jaisson, and Muzy]{Bacry2016}
E.~Bacry, T.~Jaisson, and J.~Muzy.
\newblock Estimation of slowly decreasing {Hawkes} kernels: application to
  high-frequency order book dynamics.
\newblock \emph{Quantitative Finance}, 16:\penalty0 1179--1201, 2016.

\bibitem[Barndorff-Nielsen and Shephard(2002)]{Barndorff2002a}
O.~E. Barndorff-Nielsen and N.~Shephard.
\newblock Econometric analysis of realized volatility and its use in estimating
  stochastic volatility models.
\newblock \emph{Journal of the Royal Statistical Society: Series B (Statistical
  Methodology)}, 64:\penalty0 253--280, 2002.

\bibitem[Bershova and Rakhlin(2013)]{Bershova2013}
N.~Bershova and D.~Rakhlin.
\newblock High-frequency trading and long-term investors: a view from the
  buy-side.
\newblock \emph{Journal of Investment Strategies}, 2:\penalty0 25--69, 2013.

\bibitem[Biais et~al.(2015)Biais, Foucault, and Moinas]{Biais2015}
B.~Biais, T.~Foucault, and S.~Moinas.
\newblock Equilibrium fast trading.
\newblock \emph{Journal of Financial Economics}, 116:\penalty0 292 -- 313,
  2015.
\newblock ISSN 0304-405X.

\bibitem[Bishop(2017)]{bishop2017evolution}
A.~Bishop.
\newblock The evolution of the crumbling quote signal.
\newblock \emph{Available at SSRN 2956535}, 2017.

\bibitem[Bollen and Whaley(2015)]{Bollen2015}
N.~P. Bollen and R.~E. Whaley.
\newblock Futures market volatility: What has changed?
\newblock \emph{Journal of Futures Markets}, 35:\penalty0 426--454, 2015.

\bibitem[Bouchaud et~al.(2002)Bouchaud, M{\'e}zard, Potters,
  et~al.]{bouchaud2002statistical}
J.-P. Bouchaud, M.~M{\'e}zard, M.~Potters, et~al.
\newblock Statistical properties of stock order books: empirical results and
  models.
\newblock \emph{Quantitative finance}, 2\penalty0 (4):\penalty0 251--256, 2002.

\bibitem[Bowsher(2007)]{Bowsher2007}
C.-G. Bowsher.
\newblock Modelling security market events in continuous time: Intensity based,
  multivariate point process models.
\newblock \emph{Journal of Econometrics}, 141:\penalty0 876--912, 2007.

\bibitem[Brogaard et~al.(2014)Brogaard, Hendershott, and Riordan]{Brogaard2014}
J.~Brogaard, T.~Hendershott, and R.~Riordan.
\newblock {High-Frequency Trading and Price Discovery}.
\newblock \emph{The Review of Financial Studies}, 27\penalty0 (8):\penalty0
  2267--2306, 2014.

\bibitem[Brogaard et~al.(2015)Brogaard, Hagstr{\"o}mer, Nord{\'e}n, and
  Riordan]{brogaard2015trading}
J.~Brogaard, B.~Hagstr{\"o}mer, L.~Nord{\'e}n, and R.~Riordan.
\newblock Trading fast and slow: Colocation and liquidity.
\newblock \emph{The Review of Financial Studies}, 28\penalty0 (12):\penalty0
  3407--3443, 2015.

\bibitem[Brémaud and Massoulié(1996)]{Bremaud1996}
P.~Brémaud and L.~Massoulié.
\newblock {Stability of nonlinear {Hawkes} processes}.
\newblock \emph{The Annals of Probability}, 24:\penalty0 1563 -- 1588, 1996.

\bibitem[Cartea and Jaimungal(2013)]{Cartea2013}
A.~Cartea and Jaimungal.
\newblock Modelling asset prices for algorithmic and high-frequency trading.
\newblock \emph{Applied Mathematical Finance}, 20:\penalty0 512--547, 2013.

\bibitem[Cartea et~al.(2018)Cartea, Jaimungal, and Ricci]{Cartea2018}
A.~Cartea, S.~Jaimungal, and J.~Ricci.
\newblock Algorithmic trading, stochastic control, and mutually exciting
  processes.
\newblock \emph{SIAM Review}, 60:\penalty0 673--703, 2018.

\bibitem[Chaboud et~al.(2014)Chaboud, Chiquoine, Hjalmarsson, and
  Vega]{CHABOUD2014}
A.~P. Chaboud, B.~Chiquoine, E.~Hjalmarsson, and C.~Vega.
\newblock Rise of the machines: Algorithmic trading in the foreign exchange
  market.
\newblock \emph{The Journal of Finance}, 69:\penalty0 2045--2084, 2014.

\bibitem[Chan(2017)]{chan2017machine}
E.~P. Chan.
\newblock \emph{Machine trading: Deploying computer algorithms to conquer the
  markets}.
\newblock John Wiley \& Sons, Hoboken, New Jersey, 2017.

\bibitem[Chavez-Demoulin and McGill(2012)]{Chavez2012}
V.~Chavez-Demoulin and J.~McGill.
\newblock High-frequency financial data modeling using {Hawkes} processes.
\newblock \emph{Journal of Banking \& Finance}, 36:\penalty0 3415 -- 3426,
  2012.

\bibitem[Choi et~al.(2021)Choi, Jang, Lee, and Zheng]{choi2021}
S.~E. Choi, H.~J. Jang, K.~Lee, and H.~Zheng.
\newblock Optimal market-making strategies under synchronised order arrivals
  with deep neural networks.
\newblock \emph{Journal of Economic Dynamics and Control}, 125:\penalty0
  104098, 2021.

\bibitem[Christensen et~al.(2014)Christensen, Oomen, and
  Podolskij]{christensen2014fact}
K.~Christensen, R.~C. Oomen, and M.~Podolskij.
\newblock Fact or friction: Jumps at ultra high frequency.
\newblock \emph{Journal of Financial Economics}, 114\penalty0 (3):\penalty0
  576--599, 2014.

\bibitem[Chung and Lee(2016)]{Chung2016}
K.~H. Chung and A.~J. Lee.
\newblock High-frequency trading: Review of the literature and regulatory
  initiatives around the world.
\newblock \emph{Asia-Pacific Journal of Financial Studies}, 45\penalty0
  (1):\penalty0 7--33, 2016.

\bibitem[Cont and de~Larrard(2013)]{Cont2013}
R.~Cont and A.~de~Larrard.
\newblock Price dynamics in a markovian limit order market.
\newblock \emph{SIAM Journal on Financial Mathematics}, 4:\penalty0 1--25,
  2013.

\bibitem[Cvitanic and Kirilenko(2010)]{Cvitanic2010}
J.~Cvitanic and A.~A. Kirilenko.
\newblock High frequency traders and asset prices.
\newblock Available at SSRN: https://ssrn.com/abstract=1569067, 2010.

\bibitem[Da~Fonseca and Zaatour(2014)]{Fonseca2014}
J.~Da~Fonseca and R.~Zaatour.
\newblock Clustering and mean reversion in a {Hawkes} microstructure model.
\newblock \emph{Journal of Futures Markets}, 35:\penalty0 813--838, 2014.
\newblock ISSN 1096-9934.

\bibitem[Dassios et~al.(2013)Dassios, Zhao, et~al.]{dassios2013exact}
A.~Dassios, H.~Zhao, et~al.
\newblock Exact simulation of {Hawkes} process with exponentially decaying
  intensity.
\newblock \emph{Electronic Communications in Probability}, 18, 2013.

\bibitem[Ding et~al.(2014)Ding, Hanna, and Hendershott]{ding2014slow}
S.~Ding, J.~Hanna, and T.~Hendershott.
\newblock How slow is the {NBBO}? a comparison with direct exchange feeds.
\newblock \emph{Financial Review}, 49\penalty0 (2):\penalty0 313--332, 2014.

\bibitem[Foucault and Moinas(2018)]{Foucault2018}
T.~Foucault and S.~Moinas.
\newblock Is trading fast dangerous?
\newblock In W.~Mattli, editor, \emph{Global Algorithmic Capital Markets: High
  Frequency Trading, Dark Pools, and Regulatory Challenges}, chapter~2, pages
  9--27. Oxford University Press, 2018.

\bibitem[Fox et~al.(2019)Fox, Glosten, and Rauterberg]{fox2019new}
M.~B. Fox, L.~Glosten, and G.~Rauterberg.
\newblock \emph{The New Stock Market: Law, Economics, and Policy}.
\newblock Columbia University Press, 2019.

\bibitem[Guilbaud and Pham(2013)]{Guilbaud2013}
F.~Guilbaud and H.~Pham.
\newblock Optimal high-frequency trading with limit and market orders.
\newblock \emph{Quantitative Finance}, 13\penalty0 (1):\penalty0 79--94, 2013.

\bibitem[Hagstr{\"o}mer and Nord{\'e}n(2013)]{hagstromer2013}
B.~Hagstr{\"o}mer and L.~Nord{\'e}n.
\newblock The diversity of high-frequency traders.
\newblock \emph{Journal of Financial Markets}, 16\penalty0 (4):\penalty0
  741--770, 2013.

\bibitem[Hainaut and Goutte(2019)]{Hainaut2019}
D.~Hainaut and S.~Goutte.
\newblock A switching microstructure model for stock prices.
\newblock \emph{Mathematics and Financial Economics}, 13\penalty0 (3):\penalty0
  459--490, 2019.

\bibitem[Hansen et~al.(2015)Hansen, Reynaud-Bouret, and Rivoirard]{Hansen2015}
N.~R. Hansen, P.~Reynaud-Bouret, and V.~Rivoirard.
\newblock {Lasso and probabilistic inequalities for multivariate point
  processes}.
\newblock \emph{Bernoulli}, 21:\penalty0 83 -- 143, 2015.

\bibitem[Harris(1999)]{Harris}
L.~Harris.
\newblock Trading in pennies: a survey of the issues.
\newblock Unpublished working paper, University of Southern California, 1999.

\bibitem[Hasbrouck(2018)]{hasbrouck2018}
J.~Hasbrouck.
\newblock High-frequency quoting: Short-term volatility in bids and offers.
\newblock \emph{Journal of Financial and Quantitative Analysis}, 53:\penalty0
  613–641, 2018.

\bibitem[Hasbrouck and Saar(2009)]{hasbrouck2009technology}
J.~Hasbrouck and G.~Saar.
\newblock Technology and liquidity provision: The blurring of traditional
  definitions.
\newblock \emph{Journal of financial Markets}, 12\penalty0 (2):\penalty0
  143--172, 2009.

\bibitem[Hawkes(1971{\natexlab{a}})]{Hawkes1}
A.~G. Hawkes.
\newblock Point spectra of some mutually exciting point processes.
\newblock \emph{Journal of the Royal Statistical Society. Series B
  (Methodological)}, 33:\penalty0 438--443, 1971{\natexlab{a}}.

\bibitem[Hawkes(1971{\natexlab{b}})]{Hawkes2}
A.~G. Hawkes.
\newblock Spectra of some self-exciting and mutually exciting point processes.
\newblock \emph{Biometrika}, 58:\penalty0 83--90, 1971{\natexlab{b}}.

\bibitem[Hawkes(2018)]{Hawkes2018}
A.~G. Hawkes.
\newblock Hawkes processes and their applications to finance: a review.
\newblock \emph{Quantitative Finance}, 18\penalty0 (2):\penalty0 193--198,
  2018.

\bibitem[Hawkes and Oakes(1974)]{Hawkes&Oakes}
A.~G. Hawkes and D.~Oakes.
\newblock A cluster process representation of a self-exciting process.
\newblock \emph{Journal of Applied Probability}, 11\penalty0 (3):\penalty0
  493--503, 1974.

\bibitem[Hewlett(2006)]{Hewlett2006}
P.~Hewlett.
\newblock Clustering of order arrivals, price impact and trade path
  optimisation.
\newblock In \emph{Workshop on Financial Modeling with Jump processes, Ecole
  Polytechnique}, 2006.

\bibitem[Hollifield et~al.(2004)Hollifield, Miller, and
  Sand{\aa}s]{hollifield2004empirical}
B.~Hollifield, R.~A. Miller, and P.~Sand{\aa}s.
\newblock Empirical analysis of limit order markets.
\newblock \emph{The Review of Economic Studies}, 71\penalty0 (4):\penalty0
  1027--1063, 2004.

\bibitem[Huang et~al.(2015)Huang, Lehalle, and Rosenbaum]{huang2015simulating}
W.~Huang, C.-A. Lehalle, and M.~Rosenbaum.
\newblock Simulating and analyzing order book data: The queue-reactive model.
\newblock \emph{Journal of the American Statistical Association}, 110:\penalty0
  107--122, 2015.

\bibitem[Jang et~al.(2020)Jang, Lee, and Lee]{Jang2020}
H.~J. Jang, K.~Lee, and K.~Lee.
\newblock Systemic risk in market microstructure of crude oil and gasoline
  futures prices: A {Hawkes} flocking model approach.
\newblock \emph{Journal of Futures Markets}, 40:\penalty0 247--275, 2020.

\bibitem[Jarrow and Protter(2012)]{Jarrow2012}
R.~A. Jarrow and P.~Protter.
\newblock A dysfunctional role of high frequency trading in electronic markets.
\newblock \emph{International Journal of Theoretical and Applied Finance},
  15:\penalty0 1250022, 2012.

\bibitem[Jiang et~al.(2011)Jiang, Lo, and Verdelhan]{jiang2011information}
G.~J. Jiang, I.~Lo, and A.~Verdelhan.
\newblock Information shocks, liquidity shocks, jumps, and price discovery:
  Evidence from the us treasury market.
\newblock \emph{Journal of Financial and Quantitative Analysis}, 46\penalty0
  (2):\penalty0 527--551, 2011.

\bibitem[Kelejian and Mukerji(2016)]{Kelejian2016}
H.~H. Kelejian and P.~Mukerji.
\newblock Does high frequency algorithmic trading matter for non-at investors?
\newblock \emph{Research in International Business and Finance}, 37:\penalty0
  78 -- 92, 2016.

\bibitem[Kirilenko et~al.(2017)Kirilenko, Kyle, Samadi, and
  Tuzin]{Kirilenko2017}
A.~Kirilenko, A.~S. Kyle, M.~Samadi, and T.~Tuzin.
\newblock The {Flash} crash: High-frequency trading in an electronic market.
\newblock \emph{The Journal of Finance}, 72\penalty0 (3):\penalty0 967--998,
  2017.

\bibitem[Large(2007)]{Large2007}
J.~Large.
\newblock Measuring the resiliency of an electronic limit order book.
\newblock \emph{Journal of Financial Markets}, 10:\penalty0 1--25, 2007.

\bibitem[Law and Viens(2015)]{Law2015}
B.~Law and F.~Viens.
\newblock Hawkes processes and their applications to high-frequency data
  modeling.
\newblock In I.~Florescu, M.~C. Mariani, H.~E. Stanley, and F.~G. Viens,
  editors, \emph{Handbook of High-Frequency Trading and Modeling in Finance},
  chapter~6, pages 183--219. John Wiley \& Sons, 2015.

\bibitem[Lee and Seo(2017{\natexlab{a}})]{lee2017marked}
K.~Lee and B.~K. Seo.
\newblock Marked {Hawkes} process modeling of price dynamics and volatility
  estimation.
\newblock \emph{Journal of Empirical Finance}, 40:\penalty0 174--200,
  2017{\natexlab{a}}.

\bibitem[Lee and Seo(2017{\natexlab{b}})]{lee2017modeling}
K.~Lee and B.~K. Seo.
\newblock Modeling microstructure price dynamics with symmetric {Hawkes} and
  diffusion model using ultra-high-frequency stock data.
\newblock \emph{Journal of Economic Dynamics and Control}, 79:\penalty0
  154--183, 2017{\natexlab{b}}.

\bibitem[Manahov and Hudson(2014)]{Manahov2014}
V.~Manahov and R.~Hudson.
\newblock The implications of high-frequency trading on market efficiency and
  price discovery.
\newblock \emph{Applied Economics Letters}, 21\penalty0 (16):\penalty0
  1148--1151, 2014.

\bibitem[Menkveld(2013)]{Menkveld2013}
A.~J. Menkveld.
\newblock High frequency trading and the new market makers.
\newblock \emph{Journal of Financial Markets}, 16\penalty0 (4):\penalty0 712 --
  740, 2013.

\bibitem[Menkveld(2016)]{Menkveld2016}
A.~J. Menkveld.
\newblock The economics of high-frequency trading: Taking stock.
\newblock \emph{Annual Review of Financial Economics}, 8:\penalty0 1--24, 2016.

\bibitem[Morariu-Patrichi and Pakkanen(2018)]{morariu2018state}
M.~Morariu-Patrichi and M.~S. Pakkanen.
\newblock State-dependent {Hawkes} processes and their application to limit
  order book modelling.
\newblock \emph{arXiv preprint arXiv:1809.08060}, 2018.

\bibitem[Nash(2014)]{nash2014nonlinear}
J.~C. Nash.
\newblock \emph{Nonlinear parameter optimization using R tools}.
\newblock John Wiley \& Sons, 2014.

\bibitem[Ogata(1978)]{ogata}
Y.~Ogata.
\newblock The asymptotic behaviour of maximum likelihood estimators for
  stationary point processes.
\newblock \emph{Annals of the Institute of Statistical Mathematics},
  30:\penalty0 243--261, 1978.

\bibitem[Ogata(1981)]{ogata1981lewis}
Y.~Ogata.
\newblock On {Lewis}' simulation method for point processes.
\newblock \emph{IEEE Transactions on Information Theory}, 27:\penalty0 23--31,
  1981.

\bibitem[Ozaki(1979)]{ozaki1979maximum}
T.~Ozaki.
\newblock Maximum likelihood estimation of {Hawkes}' self-exciting point
  processes.
\newblock \emph{Annals of the Institute of Statistical Mathematics},
  31:\penalty0 145--155, 1979.

\bibitem[Patterson(2012)]{patterson2012dark}
S.~Patterson.
\newblock \emph{Dark pools: The rise of AI trading machines and the looming
  threat to Wall Street}.
\newblock Random House, London, 2012.

\bibitem[Scholtus and van Dijk(2015)]{scholtus2014}
M.~Scholtus and D.~van Dijk.
\newblock High-frequency activity on {NASDAQ}.
\newblock In G.~N. Gregoriou, editor, \emph{Handbook of high frequency
  trading}, chapter~1, pages 3--23. Academic Press, 2015.

\bibitem[Stoll(2014)]{stoll2014}
H.~R. Stoll.
\newblock High speed equities trading: 1993--2012.
\newblock \emph{Asia-Pacific Journal of Financial Studies}, 43:\penalty0
  767--797, 2014.

\bibitem[Toke(2015)]{Toke2015}
I.~M. Toke.
\newblock The order book as a queueing system: average depth and influence of
  the size of limit orders.
\newblock \emph{Quantitative Finance}, 15:\penalty0 795--808, 2015.

\bibitem[van Kervel(2015)]{Kervel2015}
V.~van Kervel.
\newblock {Competition for Order Flow with Fast and Slow Traders}.
\newblock \emph{The Review of Financial Studies}, 28:\penalty0 2094--2127,
  2015.

\bibitem[Virgilio(2019)]{Virgilio2019}
G.~P.~M. Virgilio.
\newblock High-frequency trading: a literature review.
\newblock \emph{Financial Markets and Portfolio Management}, 33\penalty0
  (2):\penalty0 183--208, 2019.

\bibitem[Wah et~al.(2018)Wah, Feldman, Chung, Bishop, and Aisen]{Elaine}
E.~Wah, S.~Feldman, F.~Chung, A.~Bishop, and D.~Aisen.
\newblock A comparison of execution quality across {US} stock exchanges.
\newblock In W.~Mattli, editor, \emph{Global Algorithmic Capital Markets: High
  Frequency Trading, Dark Pools, and Regulatory Challenges}, chapter~5, pages
  91--146. Oxford University Press, 2018.

\bibitem[Zheng et~al.(2014)Zheng, Roueff, and Abergel]{Zheng2014}
B.~Zheng, F.~Roueff, and F.~Abergel.
\newblock Modelling bid and ask prices using constrained {Hawkes} processes:
  Ergodicity and scaling limit.
\newblock \emph{SIAM Journal on Financial Mathematics}, 5:\penalty0 99--136,
  2014.

\bibitem[Zhou(1996)]{Zhou1996}
B.~Zhou.
\newblock High-frequency data and volatility in foreign-exchange rates.
\newblock \emph{Journal of Business \& Economic Statistics}, 14:\penalty0
  45--52, 1996.

\end{thebibliography}

\end{document}